\documentclass[twocolumn,tighten]{aastex631}
\usepackage[figure,figure*]{hypcap}
\usepackage{gensymb}
\usepackage{longtable}
\usepackage{float}
\usepackage{comment}
\usepackage{wrapfig}
\usepackage{appendix}
\usepackage{rotating}    % for sidewaystable
\usepackage{silence}
\usepackage{textcomp}

\usepackage{graphicx}
\usepackage{stackengine}
\usepackage{xcolor}
\usepackage[percent]{overpic}
\usepackage{xcolor}
\renewcommand{\d}{\ensuremath{\rm d}}

\newcommand{\overlay}[2]{%
  \stackinset{l}{2pt}{v}{4pt}{\color{white}\scalebox{1}{#1}}{\includegraphics[width=0.4\linewidth]{#2}}%
}

\usepackage{tabularx}
\usepackage{array}

\graphicspath{{./}{figures/}}

\begin{document}
% \nolinenumbers

% Author list: Camilo, Shobita, Gaby, Nathan, Ryan, Thomas Bohn, Kristina, Archana, Michael, Jeffrey, Matt Malkan, Sara (alphabetical)

\title{A High Resolution Search for Dual AGN Candidates in Mergers: A Pre-Selection Strategy using Keck AO}

\author[0000-0002-2121-8426]{Camilo Vazquez}
\altaffiliation{Presidential Scholar}
\affiliation{Department of Physics and Astronomy, George Mason University, MS3F3, 4400 University Drive, Fairfax, VA 22030, USA}

\author[0000-0003-2277-2354]{S. Satyapal}
\affiliation{Department of Physics and Astronomy, George Mason University, MS3F3, 4400 University Drive, Fairfax, VA 22030, USA}

\author[0000-0003-4693-6157]{G. Canalizo}
\affiliation{Department of Physics and Astronomy, University of California, Riverside, 900 University Avenue, Riverside, CA 92521, USA}

\author[0000-0002-4902-8077]{N. J. Secrest}
\affiliation{U.S. Naval Observatory, 3450 Massachusetts Avenue NW, Washington, DC 20392-5420, USA}

\author[0000-0001-8640-8522]{R. W. Pfeifle}
\altaffiliation{NASA Postdoctoral Program Fellow}
\affiliation{X-ray Astrophysics Laboratory, NASA Goddard Space Flight Center, Code 662, Greenbelt, MD 20771, USA}
% \affiliation{Oak Ridge Associated Universities, NASA NPP Program, Oak Ridge, TN 37831, USA}

\author[0000-0002-4375-254X]{T. Bohn}
\affil{Research Center for Space and Cosmic Evolution, Ehime University, Bunkyo-cho 2-5, Matsuyama, Ehime 790-8577, Japan}

\author[0000-0003-1991-370X]{K. Nyland}
\affiliation{U.S. Naval Research Laboratory, 4555 Overlook Ave. SW, Washington, DC 20375, USA}

\author[0000-0001-7578-2412]{A. Aravindan}
\affiliation{Department of Physics and Astronomy, University of California, Riverside, 900 University Avenue, Riverside, CA 92521, USA}

%Alphabetical

\author[0000-0002-2183-1087]{L. Blecha}
\affiliation{Department of Physics, University of Florida, 2001 Museum Rd., Gainesville, FL 32611, USA}

\author[0000-0003-1051-6564]{Jenna M.\ Cann}
\affiliation{X-ray Astrophysics Laboratory, NASA Goddard Space Flight Center, Code 662, Greenbelt, MD 20771, USA}
\affiliation{Center for Space Science and Technology, University of Maryland, Baltimore County, 1000 Hilltop Circle, Baltimore, MD 21250, USA}
\affiliation{Center for Research and Exploration in Space Science and Technology, NASA/GSFC, Greenbelt, MD 20771}
% \email{jenna.cann@nasa.gov}

\author[0000-0003-3152-4328]{S. Doan}
\affiliation{Department of Physics and Astronomy, George Mason University, MS3F3, 4400 University Drive, Fairfax, VA 22030, USA}

\author[0000-0002-4457-5733]{E. K. S. Hicks}
\affiliation{Department of Physics and Astronomy, University of Alaska Anchorage, Anchorage , AK 99508-4664, USA}

\author[0000-0002-8816-5146]{P. Kurczynski}
\affiliation{Astrophysics Science Division Code 660, NASA Goddard Space Flight Center, Greenbelt MD 207781, USA}

\author[0000-0002-0000-2394]{S. Juneau}
\affiliation{NSF’s National Optical-Infrared Astronomy Research Laboratory, 950 N. Cherry Avenue, Tucson, AZ 85719, USA}

\author[0000-0001-6919-1237]{M. Malkan}
\affiliation{Department of Physics \& Astronomy, 430 Portola Plaza, University of California, Los Angeles, CA 90095, USA}

\author{M. McDonald}
\affiliation{Department of Physics and Astronomy, University of California, Riverside, 900 University Avenue, Riverside, CA 92521, USA}

\author[0000-0002-0913-3729]{J. McKaig}
\altaffiliation{NASA Postdoctoral Program Fellow}
\affiliation{X-ray Astrophysics Laboratory, NASA Goddard Space Flight Center, Code 662, Greenbelt, MD 20771, USA}
\affiliation{Oak Ridge Associated Universities, NASA NPP Program, Oak Ridge, TN 37831, USA}

\author[0000-0001-7069-4026]{P. Nair}
\affiliation{Department of Physics and Astronomy, The University of Alabama, Tuscaloosa, AL 35487, USA}

\author[0000-0003-2283-2185]{B. Rothberg}
\affiliation{U.S. Naval Observatory, 3450 Massachusetts Avenue NW, Washington, DC 20392-5420, USA}
\affiliation{Department of Physics and Astronomy, George Mason University, MS3F3, 4400 University Drive, Fairfax, VA 22030, USA}

\author[0000-0002-2713-0628]{F. Muller-Sanchez}
\affiliation{Department of Physics and Materials Science, The University of Memphis, 3720 Alumni Avenue, Memphis, TN 38152, USA}

\author[0000-0002-6454-861X]{E. Schwartzman}
\affiliation{U.S. Naval Research Laboratory, 4555 Overlook Ave. SW, Washington, DC 20375, USA}

\author[0000-0003-3432-2094]{R. Sexton}
\affiliation{Department of Physics and Astronomy, George Mason University, MS3F3, 4400 University Drive, Fairfax, VA 22030, USA}
\affiliation{U.S. Naval Observatory, 3450 Massachusetts Avenue NW, Washington, DC 20392-5420, USA}

\author[0000-0002-1912-0024]{V. U}
\affiliation{IPAC, Caltech, 1200 E. California Blvd., Pasadena, CA 91125, USA}
\affiliation{Department of Physics and Astronomy, 4129 Frederick Reines Hall, University of California, Irvine, CA 92697, USA}
% \email{vivianu@ipac.caltech.edu}

\correspondingauthor{Camilo Vazquez}
\email{cvazquez4@gmu.edu}

\begin{abstract}
% \nolinenumbers
Accreting supermassive black holes (SMBHs) in galaxy mergers with separations $<$ 1 kpc are crucial to our understanding of SMBH growth, galaxy evolution, and SMBH binary evolution. Despite their importance, few are known, and most have been discovered serendipitously. In this work, we develop and test a method to systematically pre-select candidate advanced mergers likely to contain unresolved sub-kpc nuclear substructure constituting high-priority dual-AGN candidates for follow-up spectroscopy.. By exploiting the survey area and astrometric precision of the Wide-field Infrared Survey Explorer (WISE) and the Sloan Digital Sky Survey (SDSS), we identified 46 nearby advanced mergers that have red WISE colors ($W_1-W_2>0.5$) indicative of accretion activity and significant sub-arcsecond offsets between their optical and infrared coordinates as measured by SDSS and WISE. We conducted high-resolution adaptive optics (AO) observations with the Keck NIRC2 camera in the $K_p$ band ($2.124 \mu m$, $\Delta\lambda = 0.351 \mu m$) to search for unresolved substructure suggested by the optical-to-infrared offsets. We find that 20/46 (43\%) of the sample shows substructure tracing the SDSS/WISE offset and unresolved by SDSS , representing a higher yield than previous pre-selection techniques such as double-peaked [O III] or hard X-ray selection. These results demonstrate that the SDSS/WISE offset method provides an efficient pathway for identifying late-stage mergers and dual-AGN candidates for spectroscopic confirmation. Archival optical Hubble Space Telescope (HST) imaging reveals that substructure identified with Keck is often missed in the optical or erroneously identified due to partial obscuration, underscoring the importance of infrared studies of late-stage mergers.

\end{abstract}
% \setkeys{Gin}{draft}
%
\section{Introduction}\label{sec:intro}

Galaxy mergers play a fundamental role in shaping the growth of supermassive black holes (SMBHs) and the evolution of their host galaxies. During the late stages of a merger, when the two galactic nuclei reach sub-kiloparsec (sub-kpc) separations, gravitational torques efficiently drive gas inflows toward the nuclear regions, triggering intense star formation and accretion onto the central SMBHs \citep{Van_Wassenhove+2012,capelo+2015,capelo+2017,blecha+2018,chen+2023a}. At these separations, both SMBHs are theoretically expected to be actively accreting, forming a dual active galactic nucleus (AGN) \citep{blecha+2018}. Dual AGNs trace a critical phase of SMBH growth and represent the immediate progenitors of bound SMBH binaries, which ultimately merge and generate low-frequency gravitational waves \citep{merritandmilosavljevic+2005}. Constraining the occurrence and properties of mergers at sub-kpc separations is therefore essential for understanding SMBH growth, galaxy evolution, and SMBH binary formation.

Despite their importance, dual accreting SMBHs remain extremely rare observationally, with most systems discovered serendipitously and the vast majority of known mergers exhibiting nuclear separations greater than 1~kpc \citep{satyapal+2017}. Only a small number of dual AGNs are known at projected separations below 10~kpc \citep[e.g.][]{bianchi+2008,fu+2011b,koss+2012,liu+2013,comerford+2015,mullersanchez+2015,satyapal+2017,hou+2019,pfeifle+2019b}, and only a handful have been identified at sub-kpc separations (e.g.\ \citealp{komossa+2003,U+2013,iwasawa+2018,mullersanchez+2015,voggel+2022,koss+2023}). With such a small and heterogeneously selected sample, the frequency and properties of dual AGNs at separations $<1$~kpc remain poorly constrained \citep{pfeifle+2024}. These constraints are critical for improving predictions of SMBH merger rates detectable by pulsar timing array experiments such as NANOGrav \citep{agazie+2023}. In particular, the efficiency of dynamical friction at separations between $\sim$1~kpc and $\sim$0.1~kpc plays a key role in setting SMBH merger timescales \citep{kelley+2017,li+2021,li+2022}.

Detecting late-stage mergers at sub-kpc separations is, however, observationally challenging. The spatial resolution required to directly resolve such systems is beyond the reach of large-area optical surveys such as the Sloan Digital Sky Survey \citep[SDSS;][]{abazajian+2009}, which are seeing-limited. Even very long baseline interferometry (VLBI) is incapable of directly identifying SMBH binaries at parsec and sub-parsec separations in large samples. As a result, the dual AGN population at separations below $\sim$1~kpc remains statistically unconstrained.

To overcome these spatial-resolution limitations, a variety of indirect techniques have been developed to identify closely separated late-stage mergers and potential dual AGNs. One widely used approach relies on the detection of double-peaked [O~III] $\lambda5007$ emission in SDSS spectra, which has been proposed as a signature of the orbital motion of kpc-scale dual SMBHs \citep[e.g.][]{wang+2009,liu+2010a,smith+2010,lyu_liu+2016,comerford+2018,kim+2020,maschmann+2020}. However, high spatial resolution follow-up studies have shown that only a small fraction ($\sim$2\%) of such systems host bona fide dual AGNs, with the majority of double-peaked profiles arising from outflows or rotating narrow-line regions associated with single AGNs \citep[e.g.][]{liu+2010b,comerford+2011,fu+2011a,fu+2011b,comerford+2013,liu+2013,comerford+2015,mullersanchez+2015,rosario+2010,fu+2012,shen+2013,mcgurk2015}.

A complementary strategy exploits Gaia's high-precision astrometry to search for unresolved dual AGNs through ``varstrometry,'' in which variability-induced shifts in the photocenter signal the presence of multiple accreting sources \citep[e.g.][]{shen+2019,hwang+2020,chen+2022,chen+2023b,wang+2023,schwartzman+2024,uppal+2024}. While promising, this method is limited by its sensitivity to optical obscuration and depends strongly on parameters such as flux ratios, variability amplitudes, and projected separations. Enhanced obscuration is both theoretically predicted and observationally common in late-stage mergers \citep{hopkins+2006,snyder+2013,blecha+2018,satyapal+2014,ricci+2017,treister+2018}, further reducing the completeness of optical astrometric techniques.

An alternative and complementary approach is to exploit the all-sky coverage of the Wide-field Infrared Survey Explorer \citep[WISE;][]{WISE+2010} to identify obscured AGNs in merging systems. Importantly, simulations predict that red mid-infrared (MIR) AGN colors in mergers emerge preferentially in the late stages of the interaction, when nuclear separations have decreased to scales where dual AGNs are most likely to be triggered \citep{blecha+2018}. In earlier merger phases, enhanced star formation dominates the MIR emission and dilutes the AGN contribution, preventing such systems from satisfying MIR AGN color criteria. This picture is supported observationally by \emph{Chandra} follow-up studies, which show that MIR-selected AGNs in mergers are preferentially late-stage systems, including confirmed and candidate dual AGNs \citep{satyapal+2017,pfeifle+2019a,pfeifle+2019b}.

In this work, we combine MIR AGN color selection with sub-arcsecond astrometric offsets between optical and infrared centroids measured by SDSS and WISE to develop a systematic pre-selection strategy for identifying late-stage mergers likely to host unresolved sub-kpc nuclear substructure. Although neither SDSS nor WISE can directly resolve nuclei separated by less than $\sim1.5''$, their astrometric precision enables the detection of positional offsets that trace unresolved structure in merging systems. We refer to these positional discrepancies as SDSS/WISE offsets. Using this approach, we identify a sample of nearby ($z<0.244$) advanced mergers with MIR colors indicative of accretion activity and significant SDSS/WISE offsets.

We present high-resolution adaptive optics imaging of 46 such systems obtained with the Keck Near Infrared Camera~2 (NIRC2) in the $K_p$ band. The goal of this paper is to test the effectiveness of the SDSS/WISE offset method by quantifying how often it successfully identifies unresolved sub-kpc substructure in late-stage mergers. While the ultimate scientific motivation of this work is to enable the discovery of closely separated dual AGNs, confirmation of AGN multiplicity requires spatially resolved spectroscopy and is beyond the scope of this paper. Such follow-up will be presented in future work.

This paper is organized as follows. In Section~\ref{sec:sampleselection} we describe the sample selection and SDSS/WISE offset methodology. Section~\ref{sec:obsredux} presents the Keck/NIRC2 observations and data reduction. In Section~\ref{sec:results} we analyze the NIRC2 images and evaluate the performance of the offset selection technique. We summarize our conclusions in Section~\ref{conclusions}. Throughout this paper, we assume a $\Lambda$CDM cosmology with $H_0=70~\mathrm{km~s^{-1}~Mpc^{-1}}$, $\Omega_\mathrm{M}=0.3$, and $\Omega_\Lambda=0.7$.

\section{Sample selection}\label{sec:sampleselection}

The goal of this work is to test whether SDSS/WISE astrometric offsets in disturbed galaxy mergers reliably trace unresolved near-infrared substructure associated with late-stage mergers and potential dual-AGN hosts. To do so, we construct a hierarchical set of samples drawn from a common parent population and observed with different datasets, each designed to robustly test the reliability of the offset selection method in identifying sub-arcsecond substructure unresolved by both SDSS and WISE. In this section, we describe (i) the construction of the parent merger sample, (ii) the primary Keck AO merger sample used for high-resolution validation, (iii) an expanded-offset sample incorporating UKIDSS imaging to test offset behavior and establish its reliability over a wider range of angular separations, and (iv) comparison samples used to assess the role of merger status and astrometric offsets in identifying sub-arcsecond substructure identified in Keck AO data. A schematic overview of the selection steps is shown in Figure~\ref{flowchart}. The general properties of our Primary Keck AO sample are listed in Table~\ref{table:sampleproperties} and a summary of all samples and their relationships is provided in Table~\ref{table:sampletable}.

\subsection{Parent Merger Sample and Primary Keck AO Merger Sample}\label{sec:sampleselection:keckmergersample}
\begin{figure}[t!]
    \includegraphics[width = \linewidth]{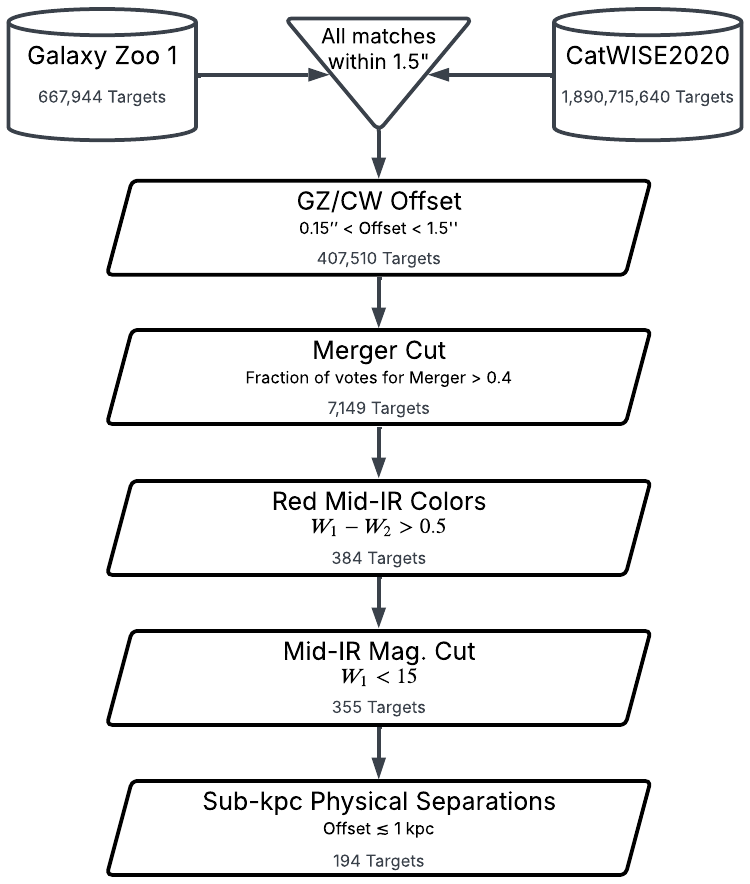}
    \caption{Selection steps for sample of 194 galaxy merger systems with red WISE colors and significant SDSS/WISE offsets ($0.15'' <$ offset $ < 1.5''$) corresponding to physical separations $\lesssim 1\,$kpc.}
    \label{flowchart}
\end{figure}

We begin by constructing a parent sample of likely galaxy mergers using the Galaxy Zoo DR1 catalog \citep{lintott+2008}, which provides visual classifications of morphologically disturbed systems. The full sequence of selection steps is summarized schematically in Figure~\ref{flowchart}. We cross-match all Galaxy Zoo targets with the CatWISE2020 catalog \citep{marocco+2021} using a 1.5$''$ matching radius, where Galaxy Zoo positions are derived from SDSS DR7. To ensure that the SDSS/WISE positional offsets probe the late-stage merger regime that remains largely unexplored observationally, we restrict the redshift range of the sample such that the adopted offset range corresponds to projected physical separations $\lesssim1$~kpc.

From this matched catalog, we select systems satisfying the following criteria:
\begin{enumerate}
   \item a high merger vote fraction ($f_m > 0.4$), indicating a strongly disturbed merger \citep{darg+2010};
   \item red mid-infrared colors ($W1-W2 > 0.5$), indicative of accretion activity in mergers \citep{blecha+2018};
   \item robust CatWISE2020 detections with $W1 < 15$; and
   \item significant SDSS/WISE positional offsets in the range $0.15'' <$ offset $ < 1.5''$.
\end{enumerate}

The lower limit of the offset range corresponds to the WISE astrometric precision, while the upper limit is set by the approximate spatial resolution of SDSS. Over the redshift range of the sample, these offsets correspond to projected physical separations $\lesssim1$~kpc. The $W1<15$ magnitude cut ensures high-reliability CatWISE2020 detections, for which the completeness and positional accuracy exceed $\sim98\%$ \citep{marocco+2021}, and minimizes spurious offsets arising from low signal-to-noise centroid measurements. The requirement of red WISE colors further enhances the likelihood that the systems host dust-heated nuclear emission, which may be associated with obscured AGN activity. While MIR color selection alone is not sufficient to uniquely identify AGNs—particularly at the relatively inclusive threshold adopted here ($W_1 - W_2 > 0.5$), where contamination from star-forming galaxies is expected \citep{stern+2012,assef+2013,assef+2018}—the color cut serves as an efficient pre-selection for dusty, late-stage mergers. When combined with the SDSS/WISE positional offset criterion, this approach preferentially selects systems with unresolved and potentially obscured nuclear structure that are strong candidates for hosting accreting SMBHs.

Applying these criteria yields 355 highly disturbed mergers with red MIR colors and statistically significant optical–infrared centroid offsets. From this set, we define the SDSS/WISE merger parent population by selecting systems in which the observed angular offsets correspond to projected separations $\lesssim1$~kpc, resulting in 194 merger systems. These offsets are measurable because both SDSS and CatWISE2020 provide high astrometric precision even when the nuclei themselves remain unresolved. In such cases, positional discrepancies between the optical and infrared centroids can arise from unresolved dual nuclei, compact nuclear star clusters, or obscured accreting SMBHs, making this parent population an ideal starting point for testing whether SDSS/WISE offsets trace previously unresolved substructure.

From this parent population, we selected targets accessible to Keck adaptive optics observations based on the availability of nearby guide stars suitable for AO correction. This resulted in 36 systems observed with Keck/NIRC2 as part of this program. We additionally identified 10 mergers in the Keck archive that satisfy the same parent-sample selection criteria, yielding a total Primary Keck AO Merger Sample of 46 systems. These systems form the core sample used to directly test whether SDSS/WISE positional offsets trace previously unresolved near-infrared substructure at sub-kpc scales.

We note that the inclusion of archival Keck targets may introduce modest selection biases, as archival observations can preferentially target systems with distinctive morphologies or enhanced nuclear activity. However, all archival targets satisfy the same parent-sample selection criteria and therefore provide a consistent extension of the primary sample for evaluating the offset-selection methodology.

Galaxies in our sample are classified as AGNs or star-forming (SF) galaxies based on their optical narrow emission line ratios, using the Baldwin-Philips-Terlevich diagram \citep[BPT;][]{Baldwin+1981}. Classifications follow the criteria of \citet[$K_{03}$]{kauffmann+2003a} and \citet[$K_{01}$]{kewley+2001} applied to emission line fluxes from the MPA-JHU DR8 Catalog \citep{kauffmann+2003b,brinchmann+2004,tremonti+2004}. Only spectra with signal-to-noise ratios greater than 5 in the relevant narrow lines are classified with the BPT diagram. We also use the three-band AGN color cut from \cite{jarrett+2011} to identify MIR AGNs in the sample. The general properties of our Primary Keck AO sample are listed in Table~\ref{table:sampleproperties}.

% --- Primary Keck AO Merger Sample table (multipage, single table number) ---
\startlongtable
\begin{deluxetable*}{lccccccccc}
\tabletypesize{\scriptsize}
\tablecaption{General properties of the Primary Keck AO Merger Sample, consisting of 46 advanced mergers selected from the SDSS/WISE offset parent sample and observed with Keck adaptive optics imaging. These targets form the core dataset used to evaluate whether SDSS/WISE astrometric offsets reliably trace unresolved near-infrared substructure at sub-kpc scales.\label{table:sampleproperties}}
\tablewidth{0pt}

\tablehead{
\colhead{SDSS Name} &
\colhead{Name} &
\colhead{R.A.} &
\colhead{Dec} &
\colhead{$W_1-W_2$} &
\colhead{Offset} &
\colhead{$z$} &
\colhead{$E(B-V)$} &
\colhead{Optical Class} &
\colhead{Morph.}\\
\colhead{} &
\colhead{} &
\colhead{(deg)} &
\colhead{(deg)} &
\colhead{(mag)} &
\colhead{(arcsec)} &
\colhead{} &
\colhead{(mag)} &
\colhead{}&
\colhead{}
}

\startdata
J001707.95+011506.3 & \ldots & 4.283   & 1.252  & 1.014 & 0.511 & 0.106 & 1.021 & {AGN} & 1 \\
J010942.21-004417.4 & \ldots & 17.426  & -0.738 & 0.713 & 0.443 & 0.089 & 0.486 & {Comp.} & 1\\
J013720.02+005722.2 & \ldots & 24.333  & 0.956  & 0.514 & 0.280 & 0.198 & 0.832 & {Comp.} & 3 \\
J015028.40+130858.4 & \ldots & 27.618  & 13.150 & 1.258 & 0.275 & 0.147 & 0.299 & {SF} & 2\\
J033037.40-003304.0 & \ldots & 52.656  & -0.551 & 0.552 & 0.169 & 0.147 & 0.470 & N/A & 0\\
J082312.61+275139.8 & \ldots & 125.803 & 27.861 & 0.972 & 0.294 & 0.168 & 0.659 & {Comp.} & 2 \\
J083824.00+254516.3 & NGC 2623 & 129.600 & 25.755 & 0.718 & 0.912 & 0.018 & 1.190 & {AGN} & 0\\
J084344.98+354942.0 & \ldots & 130.937 & 35.828 & 0.709 & 0.387 & 0.054 & 0.438 & {AGN} & 0\\
J090025.37+390353.7 & IRAS 08572+3915 & 135.106 & 39.065 & 2.293 & 0.510 & 0.058 & 0.858 & {Comp.} & 0\\
J091529.62+090601.5 & \ldots & 138.873 & 9.100  & 1.439 & 0.431 & 0.143 & 0.662 & {AGN} & 3\\
J093314.80+630945.0 & \ldots & 143.312 & 63.163 & 0.822 & 0.167 & 0.118 & 0.564 & {AGN} & 1\\
J093551.60+612111.3 & UGC 5101 & 143.965 & 61.353 & 1.695 & 0.318 & 0.040 & 1.418 & {AGN} & 0\\
J101653.82+002857.0 & \ldots & 154.224 & 0.483  & 1.082 & 0.275 & 0.116 & 0.185 & {AGN} & 2\\
J101833.64+361326.6 & \ldots & 154.640 & 36.224 & 0.663 & 0.189 & 0.054 & 0.496 & {AGN} & 0\\
J105705.40+580437.4 & VII Zw 353 & 164.273 & 58.077 & 0.606 & 0.308 & 0.140 & 0.064 & {Comp.} & 0\\
J110213.01+645924.8 & \ldots & 165.554 & 64.990 & 1.193 & 0.276 & 0.078 & 0.742 & {AGN} & 1\\
J111519.98+542316.7$^a$ & MCG +09-19-015 NED02 & 168.833 & 54.388 & 0.790 & 0.294 & 0.070 & 0.671 & {AGN} & 0\\
J120149.75-015327.5 & \ldots & 180.457 & -1.891 & 0.574 & 0.316 & 0.091 & 0.499 & {AGN} & 0\\
J120408.82+132117.0 & UGC 7043 & 181.037 & 13.355 & 0.546 & 0.486 & 0.076 & 0.744 & {AGN} & 1\\
J121522.77+414620.9 & \ldots & 183.845 & 41.773 & 0.901 & 0.321 & 0.196 & 0.271 & {AGN} & 0\\
J123915.40+531414.6$^b$ & \ldots & 189.814 & 53.237 & 0.599 & 0.306 & 0.202 & 0.396 & {AGN} & 1\\
J124046.40+273353.6$^a$ & KUG 1238+278A & 190.193 & 27.565 & 1.103 & 0.315 & 0.057 & 0.889 & {Comp.} & 0\\
J124117.64+273252.3 & \ldots & 190.324 & 27.548 & 0.598 & 0.411 & 0.140 & 0.685 & {Comp.} & 2\\
J124136.22+614043.4 & \ldots & 190.401 & 61.679 & 0.746 & 0.259 & 0.135 & 0.538 & {AGN} & 2\\
J124406.61+652925.2 & \ldots & 191.028 & 65.490 & 0.794 & 0.537 & 0.107 & 0.612 & {AGN} & 0\\
J131517.26+442425.5$^c$ & MRK 248 & 198.822 & 44.407 & 1.142 & 0.689 & 0.036 & 0.580 & {Comp.} & 1\\
J131535.06+620728.6$^d$ & UGC 8335B & 198.896 & 62.125 & 0.762 & 0.585 & 0.031 & 0.659 & {Comp.} & 3\\
J131701.41+004509.7 & \ldots & 199.256 & 0.753  & 0.682 & 0.187 & 0.138 & 0.479 & {AGN} & 0\\
J132035.40+340821.5$^d$ & UGC 8387 & 200.148 & 34.139 & 0.687 & 1.209 & 0.023 & 1.265 & {Comp.} & 1\\
J133032.00-003613.5 & VIII Zw 318 & 202.634 & -0.604 & 0.578 & 0.300 & 0.054 & 0.585 & {AGN} & 3\\
J134442.16+555313.5$^e$ & MRK 273 & 206.176 & 55.887 & 1.165 & 0.799 & 0.037 & 1.031 & {AGN} & 1\\
J134651.82+371231.0 & \ldots & 206.716 & 37.209 & 0.692 & 0.230 & 0.215 & 0.440 & {AGN} & 3\\
J135255.67+252859.6 & KUG 1350+257 & 208.232 & 25.483 & 0.686 & 0.198 & 0.064 & 0.276 & {SF} & 0 \\
J135429.05+132757.2$^a$ & \ldots & 208.621 & 13.466 & 1.016 & 0.184 & 0.063 & 0.117 & {AGN} & 0\\
J135646.10+102609.0$^b$ & \ldots & 209.192 & 10.436 & 1.347 & 0.258 & 0.123 & 0.213 & {AGN} & 1\\
J142742.68+050318.4 & \ldots & 216.928 & 5.055  & 0.764 & 0.400 & 0.159 & 0.571 & {AGN} & 0\\
J150517.88+080912.7 & \ldots & 226.325 & 8.154  & 0.923 & 0.194 & 0.039 & 0.779 & {Comp.} & 2\\
J151747.47+334740.4 & \ldots & 229.448 & 33.795 & 0.704 & 0.494 & 0.119 & 0.493 & {AGN} & 0\\
J151806.13+424445.0 & VV 705 NED01 & 229.526 & 42.746 & 0.629 & 0.500 & 0.040 & 0.891 & {Comp.} & 2 \\
J151907.33+520606.0 & SBS 1517+522 & 229.781 & 52.102 & 1.066 & 0.332 & 0.138 & 0.705 & {AGN} & 0\\
J152659.44+355837.0 & IRAS 15250+3609 & 231.748 & 35.977 & 1.437 & 0.376 & 0.055 & 0.452 & {Comp.} & 2\\
J153142.10+051801.6 & \ldots & 232.925 & 5.300  & 0.617 & 0.232 & 0.123 & N/A   & {Comp.} & 0\\
J153954.78+272758.7 & \ldots & 234.978 & 27.466 & 0.847 & 0.267 & 0.119 & 0.723 & {AGN} & 0\\
J161708.95+222627.9 & \ldots & 244.287 & 22.441 & 0.872 & 0.672 & 0.066 & 0.484 & {AGN} & 0\\
J163115.52+235257.5$^a$ & \ldots & 247.815 & 23.883 & 0.736 & 0.168 & 0.059 & 0.521 & {AGN} & 0\\
J165306.55+262638.0 & \ldots & 253.277 & 26.444 & 1.087 & 0.261 & 0.119 & 0.340 & {AGN}& 0 \\
\enddata

\tablecomments{Optical classifications are based on the BPT criteria of \citet{kewley+2001} and \citet{kauffmann+2003a}. The Morph. column indicates near-infrared morphology as defined in Section~\ref{sec:results:subsID}: (0) single nucleus; (1) dual nuclei; (2) compact secondary source consistent with a star-forming region; (3) prominent spiral arms or tidal features. Superscripts indicate archival Keck observations from previous programs: (a) Koss (H230N2L, H208N2L), (b) Djorgovski (C304N2L), (c) Urry (Y076), (d) Ryder (Z229), (e) Max (U063N2L).}
\end{deluxetable*}
% --- end table ---

\begin{figure}[t]
    \includegraphics[width = \linewidth]{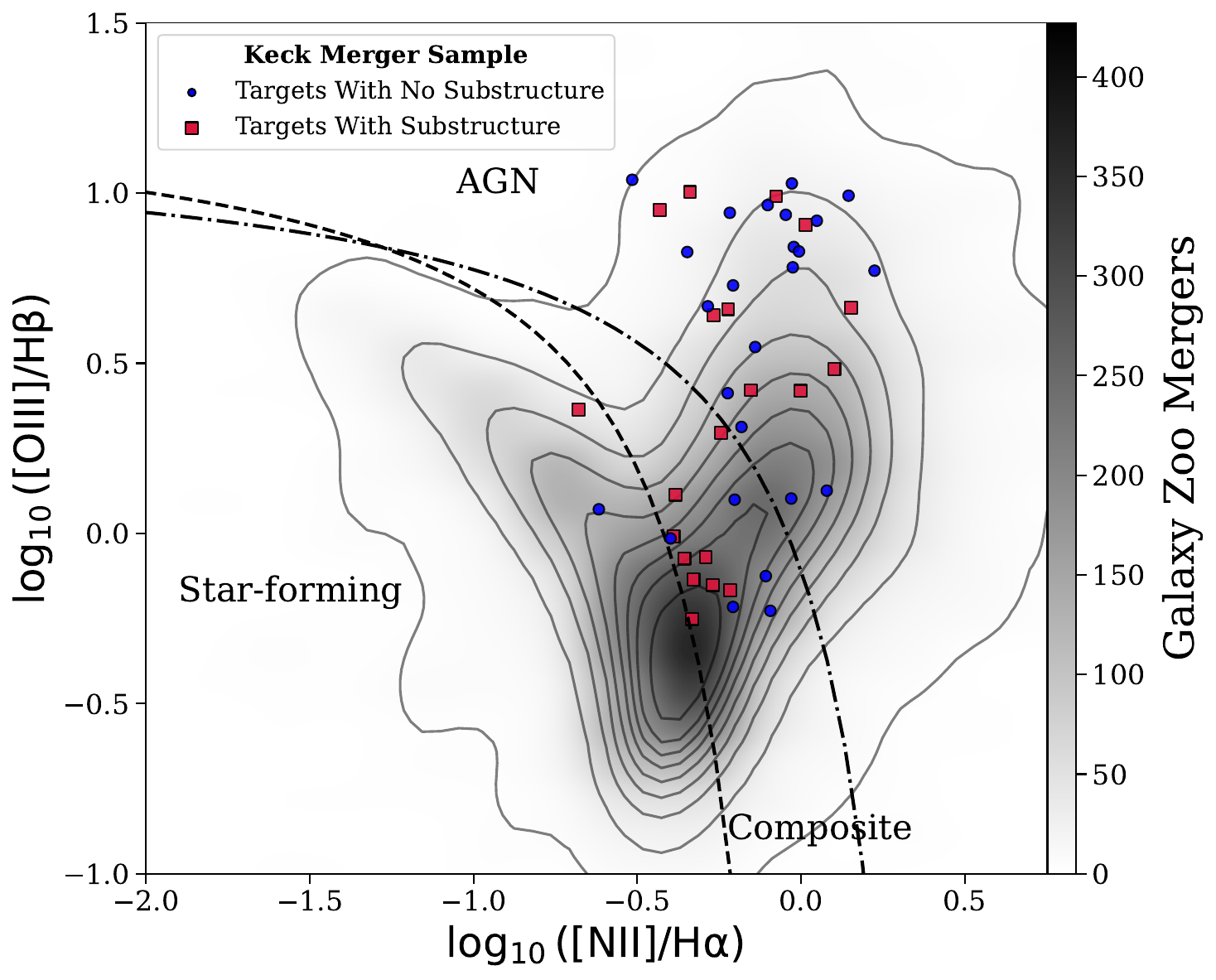}
    \caption{BPT diagram for the Primary Keck AO Merger Sample (46 systems with SDSS/WISE offsets $<1.5''$), shown relative to the general population of highly disturbed Galaxy Zoo mergers ($f_m>0.4$; shaded contours). Narrow emission-line ratios are drawn from the MPA–JHU DR8 catalog \citep{kauffmann+2003b,brinchmann+2004,tremonti+2004}. Crimson squares denote systems in which Keck AO imaging reveals previously unresolved near-infrared substructure, while blue points indicate systems with no detected substructure. The empirical star-forming demarcation from \citet{kauffmann+2003a} (dashed line) and the theoretical maximum starburst boundary from \citet{kewley+2001} (dash-dotted line) are shown for reference; galaxies between these curves are classified as composite systems. This figure demonstrates that the SDSS/WISE offset-selected sample is preferentially composed of AGN and composite classifications compared to the general merger population, as expected given the mid-infrared AGN selection. The substructure classifications are shown here to provide diagnostic context for the sample and to illustrate where systems with and without detected substructure reside in optical emission-line parameter space. The definition of substructure and its relationship to the SDSS/WISE offsets are presented in Section~\ref{sec:results:subsID}. One target (J033037) is excluded due to insufficient signal-to-noise ($<5$) in the required emission lines.}
    \label{bptmergers}
\end{figure}

\begin{figure}[htbp]
    \includegraphics[width = \linewidth]{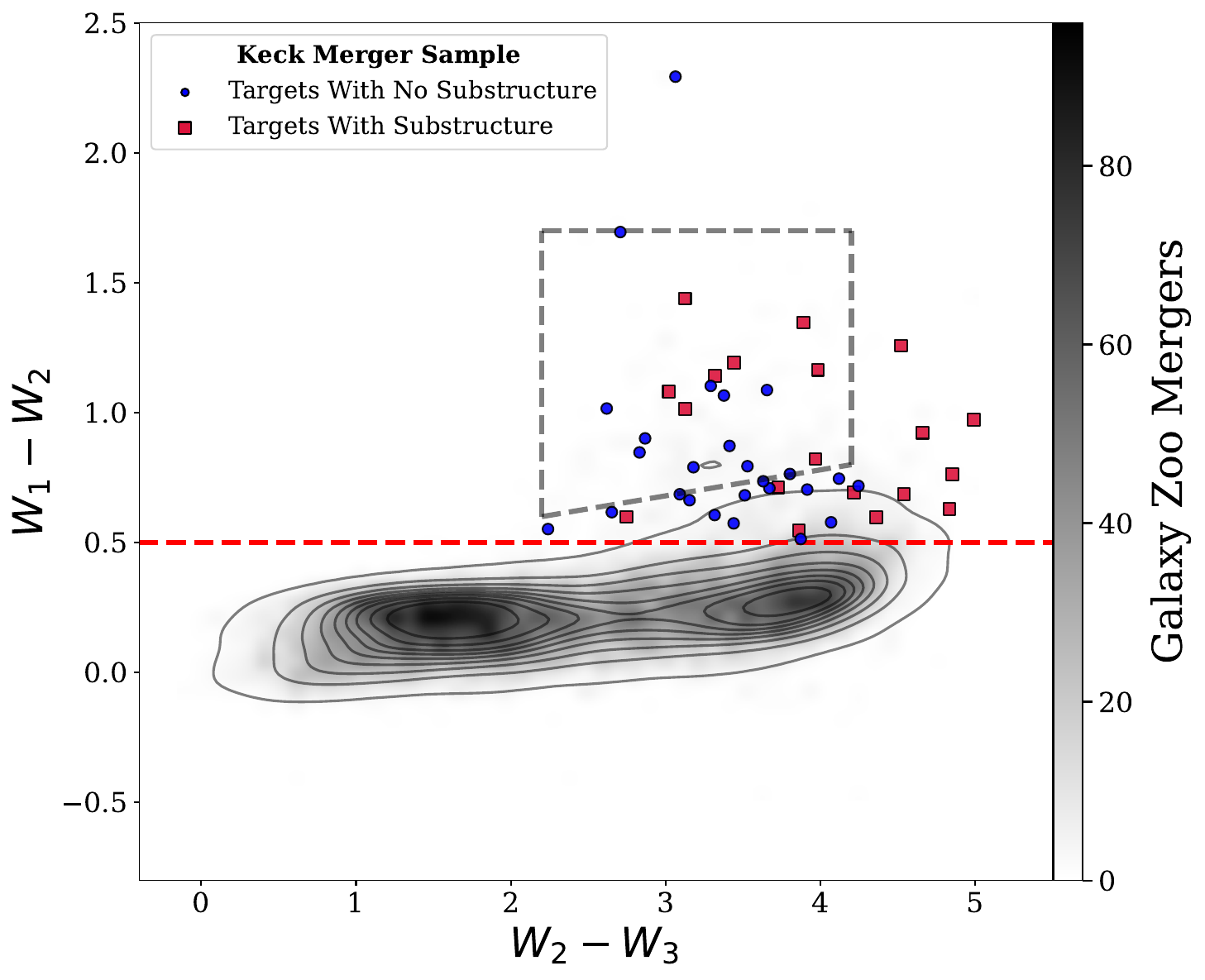}
    \caption{WISE mid-infrared color–color diagram ($W_1 - W_2$ versus $W_2 - W_3$) for the 46 targets in the Primary Keck AO Merger Sample, shown relative to the parent population of highly disturbed Galaxy Zoo mergers ($f_m>0.4$; shaded contours). Blue circles denote targets without detected near-infrared substructure, while crimson squares indicate targets in which substructure is identified in Keck AO imaging (see Section~\ref{sec:results:subsID}). The red dotted horizontal line shows the MIR color selection criterion ($W_1 - W_2 > 0.5$) used to construct the sample, which preferentially selects systems likely to host accreting SMBHs \citep{blecha+2018}. The dashed grey box outlines the stringent MIR AGN selection wedge defined by \citet{jarrett+2011}. This figure serves two purposes. First, it demonstrates that the SDSS/WISE offset-selected sample occupies a distinct region of MIR diagnostic space compared to the general merger population, consistent with the MIR AGN pre-selection criteria. Second, by identifying which targets contain resolved near-infrared substructure, this figure provides the MIR classification context needed to interpret the substructure detection rates presented in Section~\ref{sec:results:subsID}, enabling assessment of how substructure incidence depends on MIR AGN diagnostics. One target (J152659) is not included due to the absence of a reliable $W_3$ measurement.}
    \label{colorcolormergers}
\end{figure}

The Primary Keck AO Merger Sample consists of local galaxies, with a median redshift of 0.098 (corresponding to $1'' = 1.813$~kpc). In Figure~\ref{bptmergers}, we show the optical narrow-line ratios of the sample on the Baldwin--Phillips--Terlevich (BPT) diagram. Of the 46 targets, 29 lie in the AGN region, 14 in the composite region, and 2 in the star-forming region; one target is excluded due to a signal-to-noise ratio $<5$ in at least one required emission line. In Figure~\ref{colorcolormergers}, we show the mid-infrared colors of the sample ($W_1-W_2$ versus $W_2-W_3$), with 19 targets falling within the stringent AGN wedge defined by \citet{jarrett+2011}.

In both Figures~\ref{bptmergers} and~\ref{colorcolormergers}, we include the optical line ratios and MIR colors of the general population of highly disturbed Galaxy Zoo mergers ($f_m>0.4$; \citealt{darg+2010}) as background contours. These figures serve two purposes. First, they demonstrate that the SDSS/WISE offset-selected sample occupies a distinct region of optical and MIR diagnostic space compared to the general merger population, as expected given the MIR AGN pre-selection. Specifically, 93.5\% of the target sample lies in the AGN or composite regions of the BPT diagram, compared to 58.8\% of the general merger population, and 41\% of the targets fall within the MIR AGN wedge compared to only 1.92\% of all Galaxy Zoo mergers. Second, these figures establish the diagnostic context for interpreting the substructure results presented in Section~\ref{sec:results}, where we examine how the incidence and properties of resolved near-infrared substructure depend on optical and MIR classification.

\begin{figure}[htbp]
    \includegraphics[width = \linewidth]{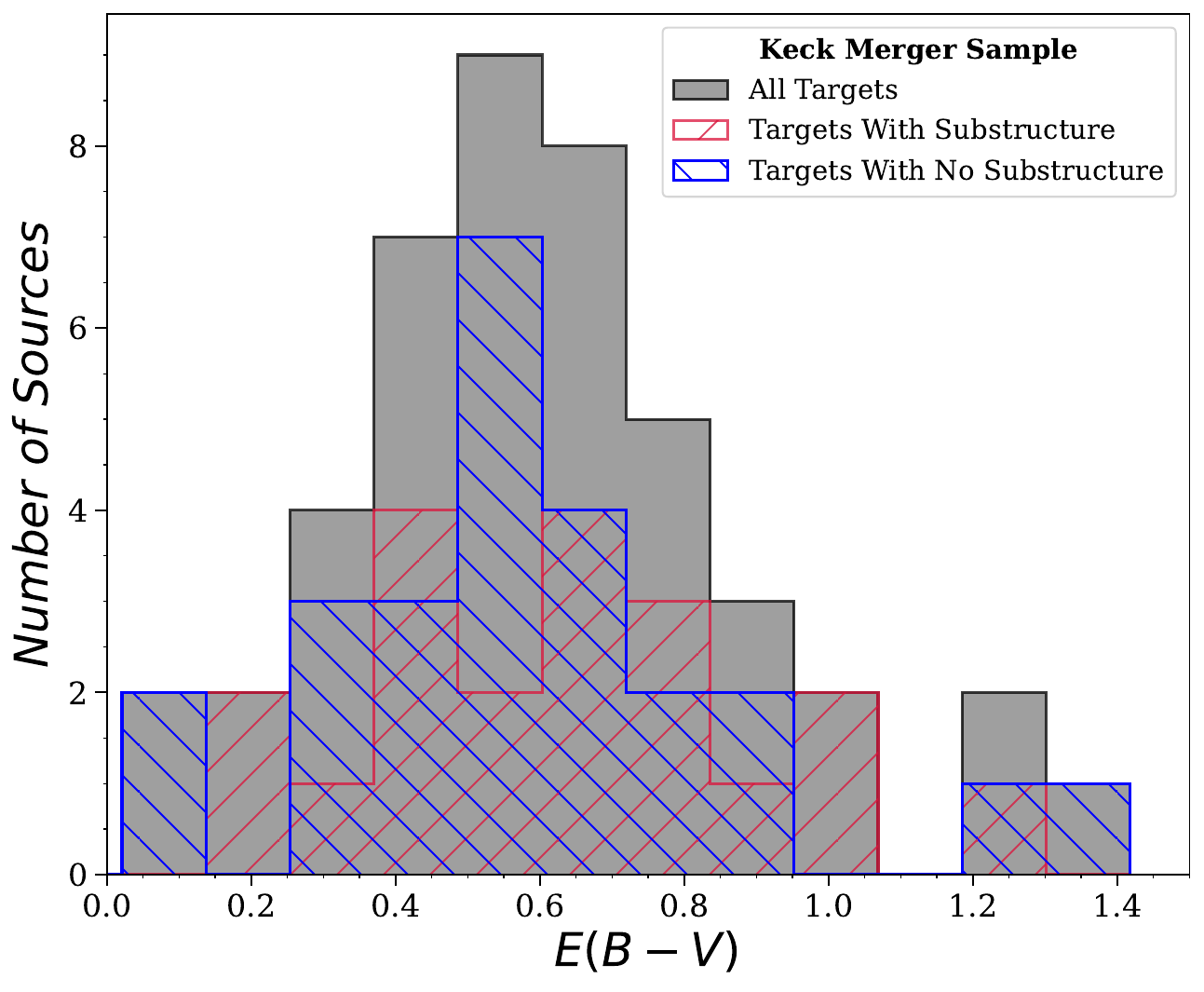}
\caption{
Distribution of optical extinction, parameterized by the color excess $E(B-V)$, for the Primary Keck AO Merger Sample. Extinction values are drawn from the MPA–JHU DR8 Catalog \citep{kauffmann+2003b,brinchmann+2004,tremonti+2004}. The grey histogram shows the full sample, while blue hatched bins and crimson hatched bins indicate systems without and with detected near-infrared substructure in Keck AO imaging, respectively (see Section~\ref{sec:results:subsID}). This figure characterizes the extinction properties of the offset-selected merger sample and shows where systems with and without detected substructure reside in extinction space. The implications of these distributions for substructure detection and their relationship to SDSS/WISE positional offsets are discussed in Section~\ref{sec:results:subsID}.
}

    \label{ebvmergers}
\end{figure}

We additionally show the distribution of color excess, $E(B-V)$, for the Primary Keck AO Merger Sample in Figure~\ref{ebvmergers}, separating systems with and without detected substructure. This figure is included to assess whether the success of the SDSS/WISE offset method is driven primarily by extinction effects, and to demonstrate that substructure is identified across a broad range of obscuration rather than only in the most highly reddened systems.

A cross-match with The Big Multi-AGN Catalog \citep[The Big MAC;][]{pfeifle+2024} reveals that the 46 targets in the Primary Keck AO Merger Sample (SDSS/WISE offsets $<1.5''$) include 13 previously identified candidate dual AGNs and 2 confirmed dual AGNs. However, only one of the confirmed systems has a nuclear separation small enough to plausibly correspond to the SDSS/WISE offset measured here; the second confirmed system has a much larger separation and is unlikely to drive the observed offset. Candidate or confirmed dual AGNs with very large separations are typically resolved galaxy pairs in which the SDSS/WISE offset traces substructure within one of the merging galaxies rather than the separation between the two galaxies themselves. Such systems may potentially host more than two accreting nuclei.

Throughout this work, we distinguish between systems in which Keck AO imaging reveals previously unresolved near-infrared substructure and those in which no such substructure is detected. These classifications are shown throughout the figures in this section to provide context for the physical and diagnostic properties of the sample. The identification and quantitative definition of substructure is presented in Section~\ref{sec:results:subsID}, where we assess whether the detected substructure spatially corresponds to the SDSS/WISE positional offsets and evaluate the effectiveness of the offset-selection method. Here, the substructure classifications are shown for reference only, to illustrate how the targets populate optical and MIR diagnostic space prior to the detailed structural analysis.

\par

\subsection{Expanded Offset Sample and UKIDSS Imaging}\label{sec:sampleselection:expandedkecksample}

The Primary Keck AO Merger Sample described in Section~\ref{sec:sampleselection:keckmergersample} is restricted to SDSS/WISE offsets $< 1.5''$ corresponding to projected separations $\lesssim1$~kpc. Although this regime is the primary focus of this work, it is important to validate this selection technique by exploring its reliability in tracing substructure across broader angular separations and identifying the limit of its robustness at smaller scales. In addition, the small field of view of the NIRC2 images and the absence of reference stars in most Keck frames prevent direct absolute astrometric registration of the Keck images to the SDSS and WISE coordinate systems. As a result, while Keck imaging reveals substructure, it is not possible to directly verify whether the detected components correspond spatially to the SDSS or WISE centroids.

To address this limitation and independently test the reliability of the SDSS/WISE offset selection method and explore its dependence on offset size, we constructed an expanded-offset sample extending to larger angular separations of $0.15'' < $ offset $ < 6''$, corresponding to the angular resolution limit of WISE. Applying the same Galaxy Zoo merger classification ($f_m>0.4$), MIR color selection ($W_1-W_2>0.5$), and magnitude cut ($W_1<15$) described in Section~\ref{sec:sampleselection:keckmergersample}, this expanded selection yields 540 merger systems.

\begin{figure}
    \centering
    \includegraphics[width=\linewidth]{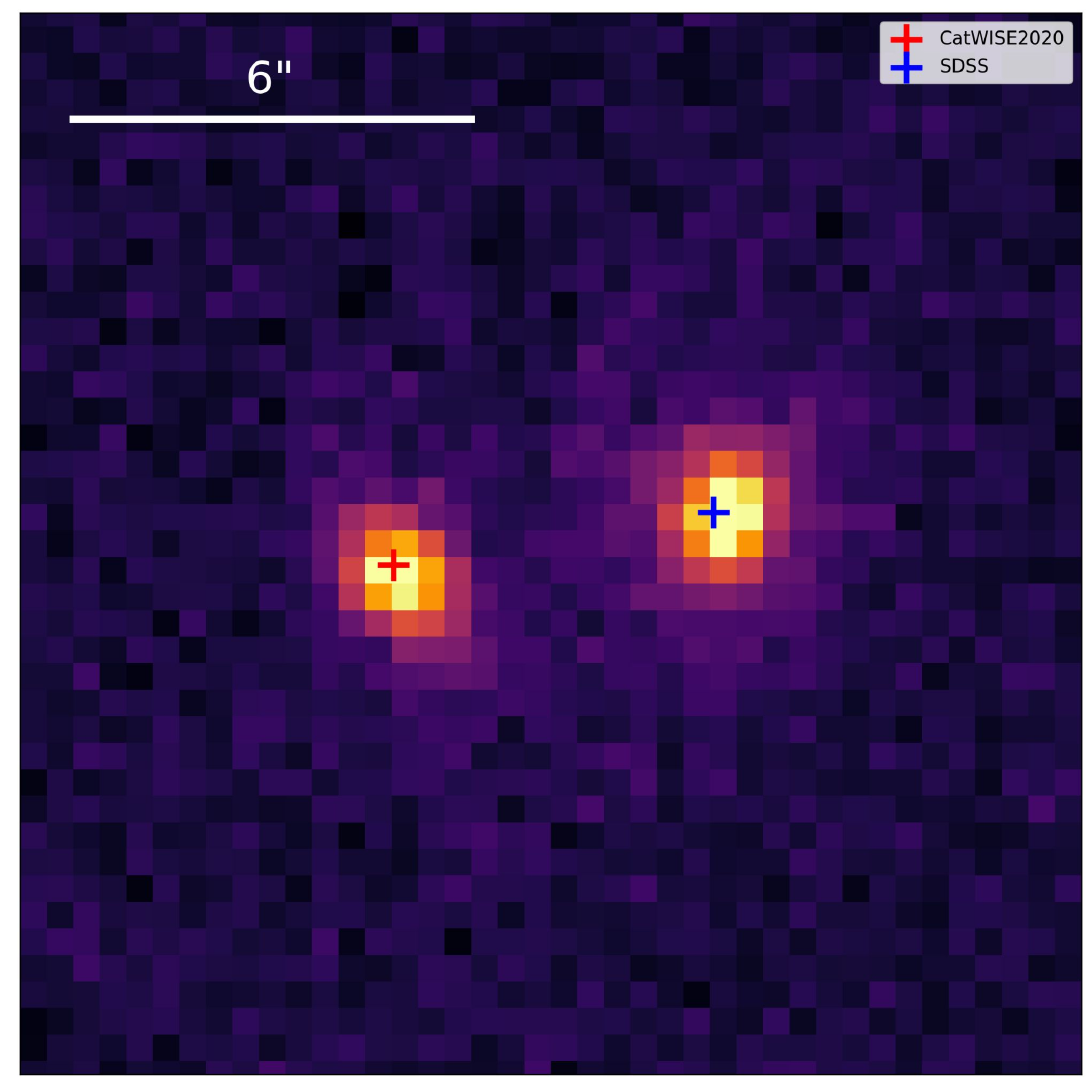}
   \caption{Example UKIDSS $K$-band image of J131012.68+001809.3 from the UKIDSS Merger Sample. UKIDSS imaging provides accurate absolute astrometry, enabling direct overlay of the SDSS and WISE centroid positions. The offset between these centroids corresponds to distinct near-infrared emission components resolved by UKIDSS but unresolved by WISE. This figure demonstrates that SDSS/WISE astrometric offsets trace physically distinct merger-driven substructure.}
    \label{ukidssim}
\end{figure}

We then identified 61 merger systems within this expanded-offset sample with publicly available near-infrared imaging from the UKIRT Infrared Deep Sky Survey \citep[UKIDSS;][]{lawrence+2007}. Unlike the Keck AO data, UKIDSS images provide reliable World Coordinate System (WCS) solutions, enabling direct comparison between the SDSS and WISE centroids and the near-infrared emission. Because the angular resolution of UKIDSS ($\sim0.8''$) is sufficient to resolve sources corresponding to larger SDSS/WISE offsets, these data provide an independent means of verifying whether the positional offsets correspond to physically distinct emission components. An example UKIDSS image with SDSS and WISE positions overlaid is shown in Figure~\ref{ukidssim}.

In addition, we identified 7 targets within the expanded-offset sample that have Keck AO imaging, which are included in the Expanded Keck AO Sample described in Table~\ref{table:sampletable}. Together, the UKIDSS and expanded Keck samples enable validation of the offset-selection method across a broader range of angular separations and provide an independent check on whether SDSS/WISE offsets trace physically distinct near-infrared emission components.

\subsection{Comparison Samples}\label{sec:sampleselection:comparisonsamples}

To evaluate whether the SDSS/WISE offset selection and merger classification independently contribute to the detection of substructure, we constructed two comparison samples drawn from the same Galaxy Zoo–CatWISE2020 parent population.

\subsubsection{Non-merger Comparison Sample}\label{sec:sampleselection:nonmergersample}

First, we constructed a sample of galaxies with significant SDSS/WISE offsets that are not classified as mergers. This sample was generated using the same selection criteria described in Section~\ref{sec:sampleselection:keckmergersample}, except that we required a low merger vote fraction ($f_m<0.4$). This selection yields 5,815 non-merger galaxies with red MIR colors, reliable CatWISE detections, and significant SDSS/WISE offsets corresponding to projected separations $\lesssim1$~kpc.

From this sample, we identified 40 systems with archival Keck AO imaging, forming the Keck Non-merger Comparison Sample. This sample allows us to assess whether SDSS/WISE offsets alone, independent of merger status, are sufficient to produce detectable substructure.

\begin{figure}[t]
\begin{flushleft}
    \includegraphics[width = \linewidth]{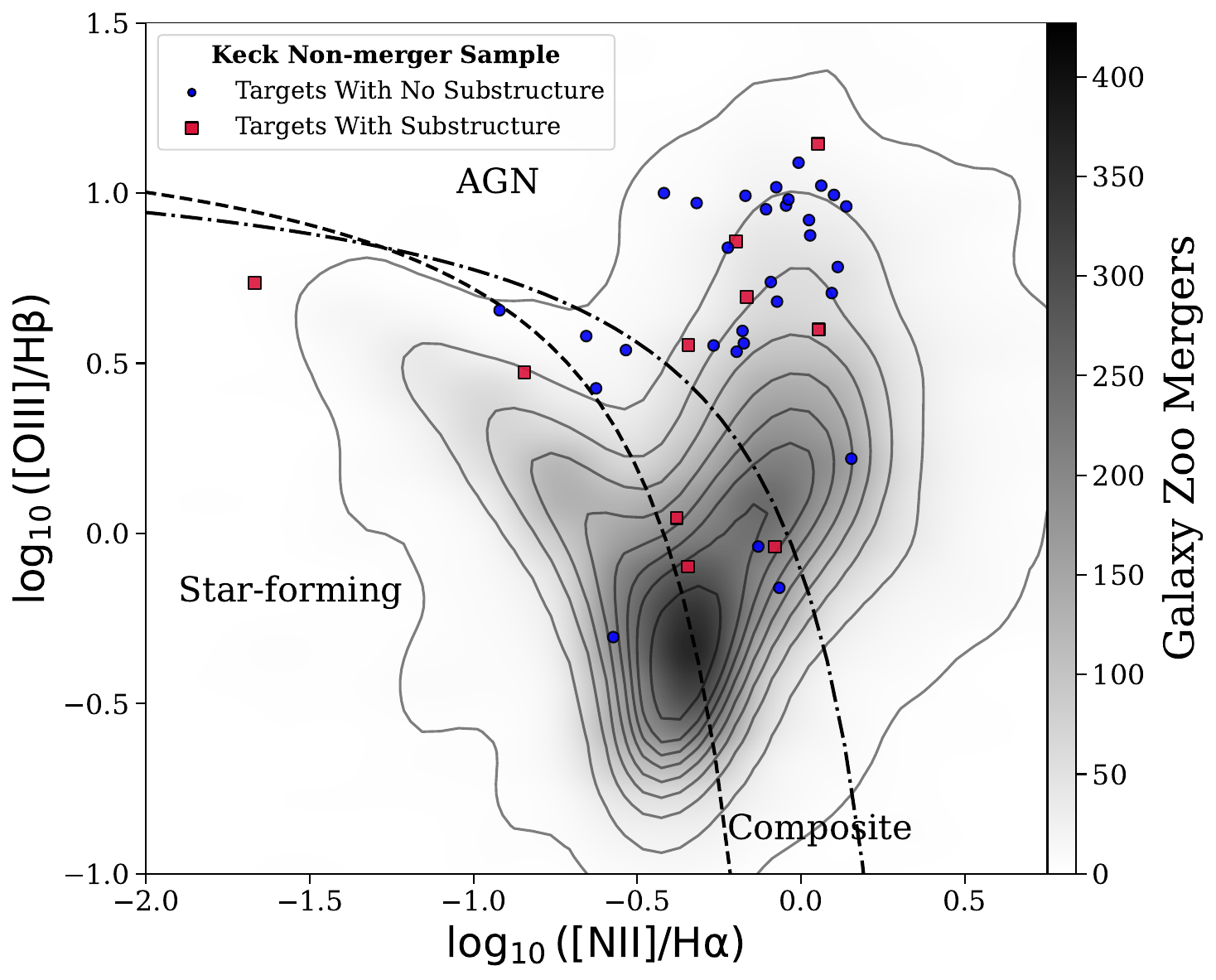}
\end{flushleft}
\caption{
BPT diagram for the Keck Non-merger Comparison Sample (40 galaxies with SDSS/WISE offsets but no merger classification, $f_m<0.4$), shown relative to the general population of Galaxy Zoo mergers ($f_m>0.4$; shaded contours). Crimson squares denote systems in which Keck AO imaging reveals near-infrared substructure, while blue points indicate systems with no detected substructure. The empirical star-forming demarcation from \citet{kauffmann+2003a} (dashed) and the theoretical maximum starburst boundary from \citet{kewley+2001} (dash-dotted) are shown for reference. This figure demonstrates that the Keck Non-merger Comparison Sample spans a similar range of optical spectroscopic classifications as the Primary Keck AO Merger Sample, allowing us to assess whether SDSS/WISE offsets alone, independent of merger status, are associated with unresolved near-infrared substructure.
}

\label{bptnonmergers}
\end{figure}

\begin{figure}[htbp]
\begin{flushleft}
    \includegraphics[width = \linewidth]{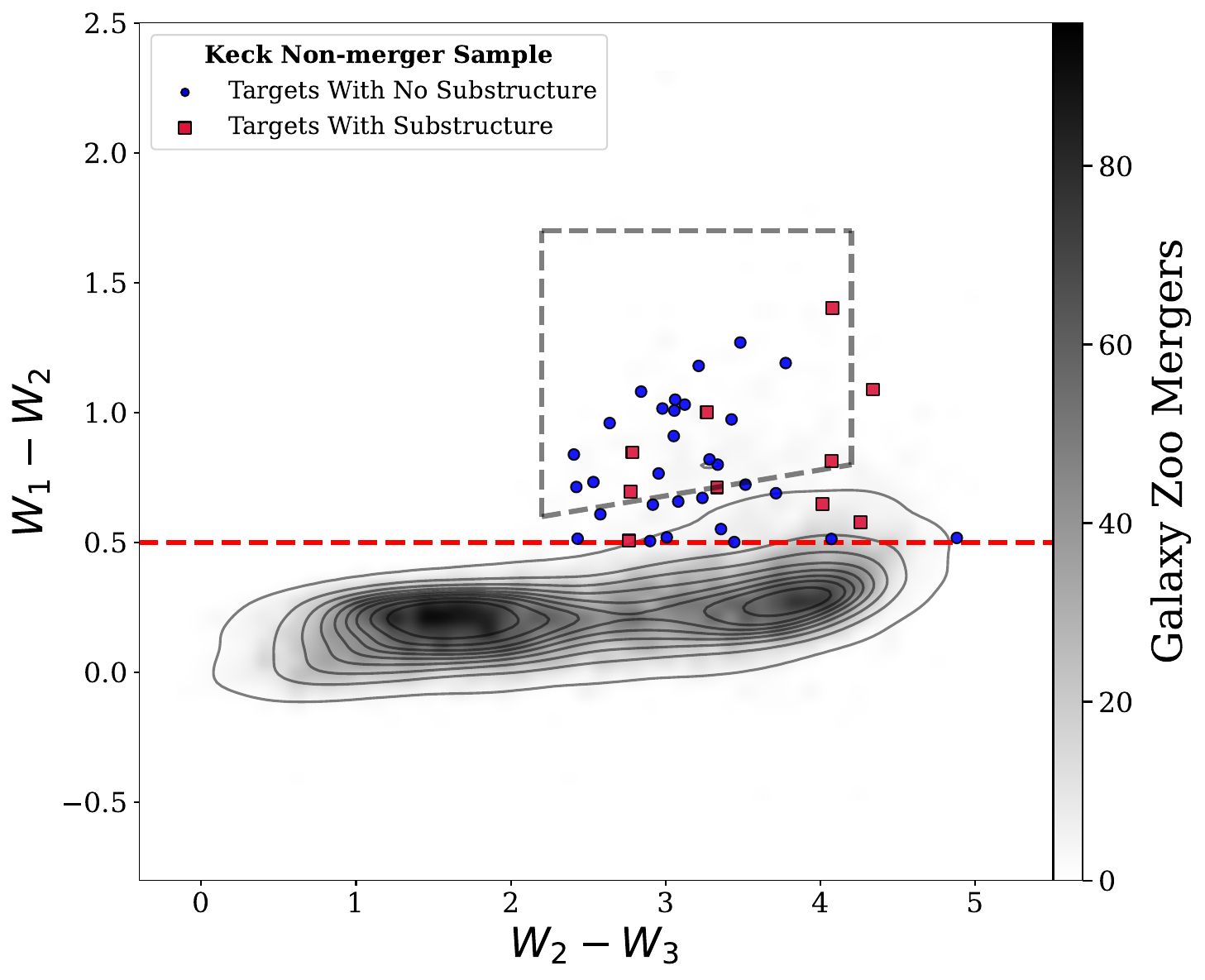}
\end{flushleft}
\caption{
WISE color–color diagram for the Keck Non-merger Comparison Sample, shown relative to the parent Galaxy Zoo merger population ($f_m>0.4$; shaded contours). Crimson squares denote systems with detected near-infrared substructure and blue points indicate systems without detected substructure. The horizontal dashed line marks the $W_1-W_2>0.5$ AGN color selection threshold \citep{blecha+2018}, and the dashed box indicates the AGN selection wedge from \citet{jarrett+2011}. This figure shows that the Keck Non-merger Comparison Sample occupies a similar region of MIR color space as the Primary Keck AO Merger Sample, enabling a direct test of whether SDSS/WISE positional offsets alone can identify systems with unresolved near-infrared substructure.
}

\label{colorcolornonmergers}
\end{figure}

\begin{figure}[t]
\begin{flushleft}
    \includegraphics[width = \linewidth]{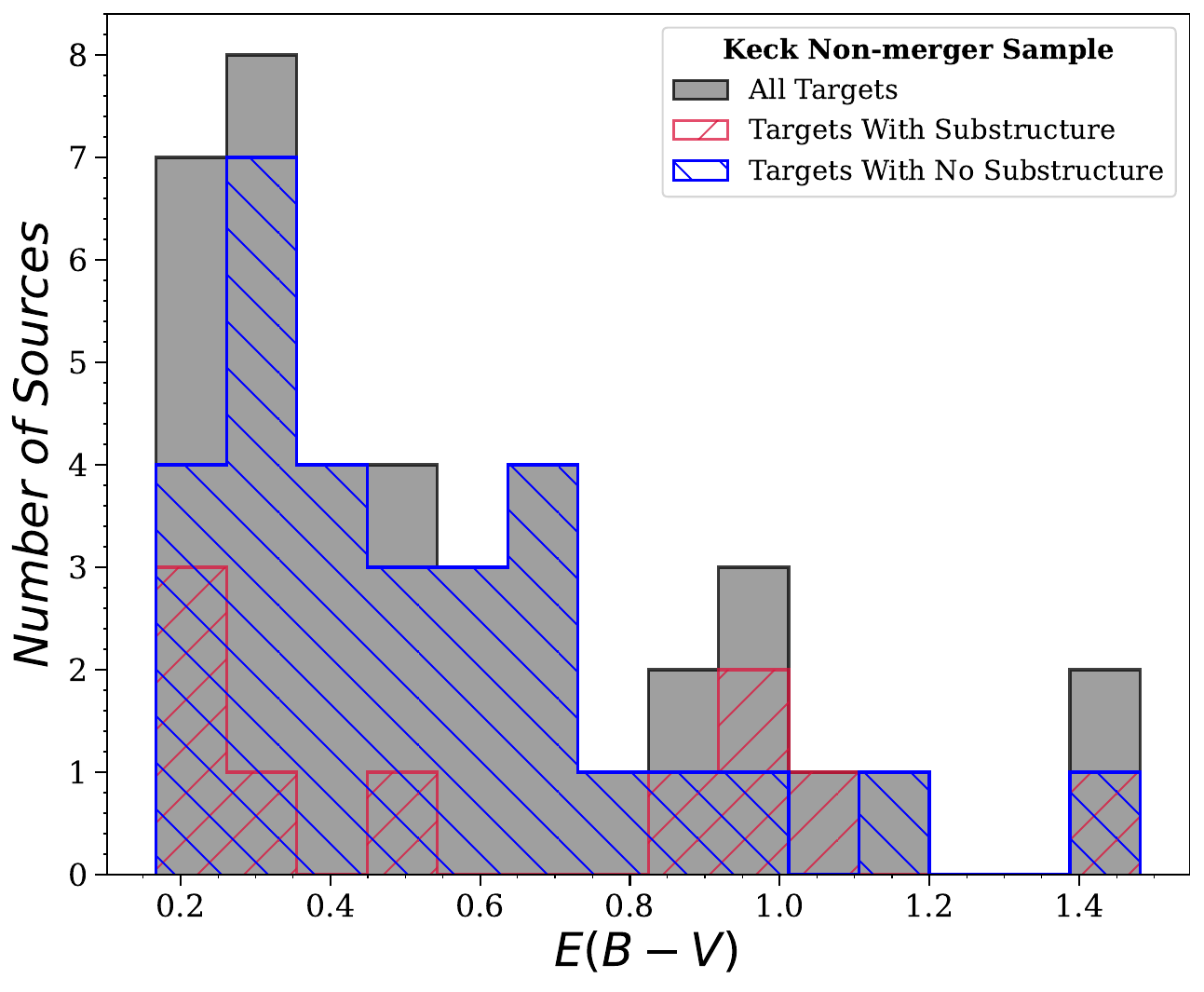}
\end{flushleft}
\caption{
Distribution of optical extinction, parameterized by the color excess $E(B-V)$, for the Keck Non-merger Comparison Sample. Extinction values are drawn from the MPA–JHU DR8 Catalog \citep{kauffmann+2003b,brinchmann+2004,tremonti+2004}. The grey histogram shows the full sample, while blue hatched bins and crimson hatched bins indicate systems without and with detected near-infrared substructure in archival Keck AO imaging, respectively (see Section~\ref{sec:results:subsID}). This figure provides a direct comparison to the Primary Keck AO Merger Sample by characterizing the extinction properties of non-mergers with SDSS/WISE positional offsets. The implications for substructure detection and the role of astrometric offsets are discussed in Section~\ref{sec:results:subsID}.
}

\label{ebvnonmergers}
\end{figure}

We compare the optical emission-line classifications mid-infrared colors, and optical extinctions of this sample with those of the Primary Keck AO Merger Sample in Figures~\ref{bptnonmergers}, \ref{colorcolornonmergers} and \ref{ebvnonmergers}, respectively. These figures, which are analogous to Figures~\ref{bptmergers}, \ref{colorcolormergers} and \ref{ebvmergers}, allow a direct comparison of the BPT classifications, WISE color and optical extinction distributions of offset-selected galaxies with and without merger signatures. This comparison shows that the Keck Non-merger Comparison Sample spans a similar range of AGN and composite classifications, MIR colors and optical extinctions as the Primary Keck AO Merger Sample, indicating that the two samples have comparable optical and MIR properties despite their different merger classifications.

We verified that the Keck Non-merger Comparison Sample is well matched to the Primary Keck AO Merger Sample in redshift and SDSS/WISE offset. The distributions of these quantities are shown in Appendix~\ref{appendix:samplecomparison}. A two-sample Kolmogorov–Smirnov test indicates no statistically significant differences between the samples, confirming that they are comparable in their observational properties. This ensures that any differences in substructure detection cannot be attributed to systematic differences in redshift or offset.

\subsubsection{Non-offset Merger Comparison Sample}\label{sec:sampleselection:nonoffsetsample}

Second, we constructed a comparison sample of mergers that satisfy all selection criteria except for the SDSS/WISE offset requirement. Applying the Galaxy Zoo merger classification, MIR color selection, and magnitude cut but requiring SDSS/WISE offsets $<0.15''$ yields 50 merger systems with no statistically significant SDSS/WISE offset.

Of these, 11 systems have archival Keck AO imaging and form the Keck Non-offset Comparison Sample. This sample enables us to test whether substructure detection is specifically associated with SDSS/WISE positional offsets, rather than being a generic property of late-stage mergers.

Together, these comparison samples provide independent tests of the relative roles of merger status and astrometric offsets in identifying unresolved near-infrared substructure. A summary of all samples and their hierarchical relationships is provided in Table~\ref{table:sampletable}.

\

\begin{table*}
\centering
\caption{Summary of samples analyzed in this work and their roles in testing the SDSS/WISE offset method.
All samples satisfy $W_1<15$ and $W_1-W_2>0.5$. The Primary Keck AO Merger Sample is the core dataset used to quantify unresolved near-infrared substructure at projected separations $\lesssim1$~kpc. The Expanded Keck AO Merger Sample, a superset of the Primary Keck AO Merger Sample, extends the offset range to $6''$ and is used to test whether the offset–substructure connection persists beyond the sub-kpc regime. The UKIDSS Merger Sample uses UKIDSS imaging (with reliable absolute astrometry) to verify whether SDSS/WISE offsets trace resolved near-infrared structure over a wider range of separations. Comparison samples modify either the merger classification or the offset requirement while keeping the MIR selection fixed. Note that the Expanded Keck AO Merger Sample includes the 46 targets in the Primary Keck AO Merger Sample}
\label{table:sampletable}

\vspace{0.15cm}
\footnotesize

\begin{tabular}{lccccccc}
\hline\hline
\tabletypesize{\scriptsize}
{Sample Name}
& {Imaging}
& {Merger}
& {Offset Range$^a$}
& {N}
& {HST Imaging}
& {Substructure$^b$}
& {Major$^c$} \\
\hline

\multicolumn{8}{l}{\textbf{Primary and Extended Merger Samples}}\\
\hline

Primary Keck AO Merger Sample$^d$
& Keck AO
& $f_m>0.4$
& $0.15'' < \Delta_{\rm off} < 1.5''$
& {46}
& {16 (35\%)}
& {20 (43\%)}
& {10 (22\%)} \\

Expanded Keck AO Merger Sample
& Keck AO
& $f_m>0.4$
& $0.15'' < \Delta_{\rm off} < 6''$
& {53}
& {18 (34\%)}
& {25 (47\%)}
& {\ldots} \\

UKIDSS Merger Sample
& UKIDSS
& $f_m>0.4$
& $0.8'' < \Delta_{\rm off} < 6''$
& {61}
& {6 (10\%)}
& {46 (75\%)}
& \nodata \\

\hline
\multicolumn{8}{l}{\textbf{Comparison Samples}}\\
\hline

Keck Non-merger Comparison Sample$^d$
& Keck AO
& $f_m<0.4$
& $0.15'' < \Delta_{\rm off} < 1.5''$
& {40}
& {13 (33\%)}
& {10 (25\%)}
& {7  (18\%)} \\

Keck Non-offset Comparison Sample
& Keck AO
& $f_m>0.4$
& $\Delta_{\rm off} < 0.15''$
& {11}
& {0 (0\%)}
& {1 (9\%)}
& {0 (0\%)} \\

\hline\hline
\end{tabular}

\vspace{0.12cm}
{\footnotesize
$^a$ $\Delta_{\rm off}$ denotes the SDSS/WISE astrometric offset.\\
$^b$ See Section~\ref{sec:results:subsID} for the definition of substructure. \\
$^c$ Fraction of systems with nuclear $K_p$-band flux ratios consistent with major mergers (Section~3.1); reported only for samples with Keck AO imaging.\\
$^d$ SDSS/WISE offsets correspond to projected physical separations $\lesssim1$~kpc. \\
}

\end{table*}

\section{Observations and Data Reduction}\label{sec:obsredux}

\subsection{Keck AO Observations}\label{sec:obsredux:intro}

We observed a total of 40 targets (36 targets from the Primary Keck AO Merger Sample and 4 additional targets from the Expanded Keck AO Sample) using the Keck II Telescope and the NIRC2 near-infrared camera \citep[PI: K.\ Matthews;][]{Wizinowich+2000,vanDam+2006,Wizinowich+2006}. Observations were conducted over three nights between UT 2021 September 19 and UT 2022 March 20 using Laser Guide Star and Natural Guide Star Adaptive Optics (LGSAO and NGSAO).

Images were obtained in the $K_p$ filter ($\lambda_c = 2.124\,\mu$m, $\Delta\lambda = 0.351\,\mu$m). Most targets were observed with the Narrow Camera (field of view: $10'' \times 10''$, pixel scale: $0.009942''$ pixel$^{-1}$), while eight targets were observed with the Wide Camera (field of view: $40'' \times 40''$, pixel scale: $0.039686''$ pixel$^{-1}$). Total exposure times ranged from $900$ s to $1620$ s per target under photometric conditions with natural seeing between $0.5''$ and $1.0''$.

A three-point dither pattern was used for all observations. Tip–tilt stars were observed contemporaneously with each science target to monitor AO performance and characterize the instantaneous PSF. The average PSF FWHM was $0.07''$ for the Narrow Camera and $0.13''$ for the Wide Camera. Flat fields, dark frames, and photometric standard stars were obtained for calibration.

\subsection{Data Reduction}\label{sec:obsredux:datredux}

Data reduction followed standard NIRC2 procedures, including dark subtraction, flat-field correction, bad-pixel masking, sky subtraction, and geometric distortion correction using the most recent distortion solution \citep{Service+2016}. Individual dithered frames were aligned using the centroid of the brightest nuclear component and median-combined to produce final science images. Flux calibration was performed using standard stars observed on the same nights.

\subsection{Identification of Compact Sources and Aperture Photometry}\label{sec:obsredux:phot}

Because the targets are advanced mergers with disturbed morphologies, strong tidal features, and diffuse stellar emission, conventional two-dimensional surface brightness profile fitting using S\'ersic or PSF-decomposition techniques was not feasible. The limited NIRC2 field of view further prevented robust modeling of the extended host galaxy light.

Instead, we identified compact emission sources visually in the AO images and measured their photometric properties using aperture photometry, following established techniques used in imaging studies of merging galaxies \citep[e.g.,][]{Troncoso+2025}. This approach is well suited for identifying compact, unresolved nuclear components in crowded or morphologically complex systems.

To minimize cross-contamination between closely separated sources, we adopted a small circular aperture with radius equal to 1.25 times the FWHM of the PSF, as measured from the tip–tilt star observed contemporaneously with each science target. Larger apertures were tested but resulted in significant blending between nearby sources in several systems (e.g., the two South West sources in J150517, see Figure~\ref{B:J1505}), preventing reliable isolation of individual components.

Because this aperture does not enclose the full PSF, aperture corrections were applied using a curve-of-growth analysis derived from the corresponding tip–tilt star. The tip–tilt star provides an empirical measurement of the instantaneous AO PSF, including both the diffraction-limited core and extended halo. For each observation, we measured the fraction of flux enclosed within the adopted aperture radius and applied the corresponding correction factor to estimate the total flux of each compact source.

This method provides a uniform methodology to obtain total flux estimates for compact, unresolved sources in the sample while avoiding aperture overlap between neighboring components. However, because of the complex morphologies of merging systems, overlapping stellar emission or diffuse light from foreground substructure may contribute to the measured flux in some cases. Therefore, the fluxes and flux ratios reported here should be interpreted as approximate indicators of the relative prominence of compact nuclear components rather than precise measurements of stellar mass.

For systems exhibiting two clearly resolved compact nuclei, we compute flux ratios using the aperture-corrected fluxes calculated as described above. We note that these flux ratios are consistent with those obtained by applying a human-guided aperture selection as was done in the study by \citet{Troncoso+2025}. Following standard practice in infrared studies of galaxy mergers, we use the $K$-band flux ratio as an approximate proxy for stellar mass ratio, adopting the conventional threshold of 4:1 (flux ratio $>0.25$) to distinguish major and minor mergers \citep[e.g.,][]{toomre+1972,hernandeztoledo+2005,conselice+2006,Troncoso+2025}. This classification is applied only to systems with clearly resolved compact nuclei and is intended to provide an empirical basis for comparison with previous studies but should not be interpreted as precise stellar mass measurements.

\subsection{Archival NIRC2 Imaging}\label{sec:obsredux:archivalimaging}

In addition to our observations, we identified 13 targets with publicly available NIRC2 imaging. These archival data were pre-calibrated using standard procedures including dark subtraction, flat-field correction, bad-pixel masking, and geometric distortion correction. We aligned and combined the archival frames following the same procedures described above.

Because flux calibration information was not available for all archival observations, we do not report absolute calibrated fluxes for these systems. However, these images provide valuable structural information and flux ratios are independent of absolute flux calibration and are included in the substructure analysis presented in Section~\ref{sec:results:subsID}.

\section{Results}\label{sec:results}

\subsection{Identification and Properties of Substructure}
\label{sec:results:subsID}

In Figures~\ref{keckstamps_duals}--\ref{keckstamps_singles}, we present the Keck/NIRC2 adaptive optics images of the Primary Keck AO Merger Sample, grouped according to their observed near-infrared morphology, including systems with apparent dual nuclei, compact secondary sources, spiral arms or tidal features, and systems dominated by a single compact nucleus (see Appendix~\ref{B} for descriptions of individual targets). These high-resolution images reveal intricate nuclear structure that is unresolved in SDSS imaging, with many systems containing two or more compact sources of emission separated by projected distances of $\lesssim 1$~kpc.

We define substructure as the presence of a secondary compact source of emission detected in the Keck AO imaging that is distinct from the primary nuclear component and unresolved in SDSS imaging. For each system with detected substructure, we identify a substructure pair consisting of the primary nucleus and the brightest secondary compact source. The selection of eligible secondary sources is guided by the SDSS/WISE positional offset: for systems with offsets $<1.5''$, the offset necessarily corresponds to structure unresolved in SDSS imaging, while systems with larger offsets may include sources that are marginally resolved in SDSS but remain unresolved by WISE.

Photometric measurements, separations, and flux ratios of the substructure pairs are listed in Table~\ref{table:substructure} (see Section~\ref{sec:obsredux:phot} for details of the photometric methodology). The detected substructure pairs span angular separations of $0.19''$--$1.25''$, corresponding to projected physical separations of 176~pc to 3.85~kpc. The distribution of substructure pair separations is shown in Figure~\ref{separations}. 
This figure illustrates the range of nuclear separations probed by the SDSS/WISE offset selection method and demonstrates that a substantial fraction of the detected substructure corresponds to compact nuclear pairs at sub-kpc scales.

Of the 46 mergers in the Primary Keck AO Merger Sample, 10 systems exhibit dual compact nuclei separated by $\lesssim 1\,$kpc with flux ratios greater than 0.25, consistent with the mass ratio criterion for a major merger (approximately 4:1), assuming that the $K$-band flux traces stellar mass (see section~\ref{sec:obsredux:phot} for details). While detailed surface brightness modeling is required to obtain precise stellar mass ratios, these flux ratios indicate that many of the detected compact sources are consistent with candidate merging galaxy nuclei.

In some systems, diffuse or asymmetric emission may also contribute to the observed SDSS/WISE positional offset. We emphasize that imaging alone cannot definitively determine whether the detected compact sources correspond to galaxy nuclei, star clusters, or AGN. Follow-up spectroscopy or multi-wavelength observations are required to establish their physical nature.

% ---------------- Figure 10a ----------------
\renewcommand{\thefigure}{9a}
\setcounter{figure}{9}
\begin{figure*}
\centering
\includegraphics[width=1.0\linewidth]{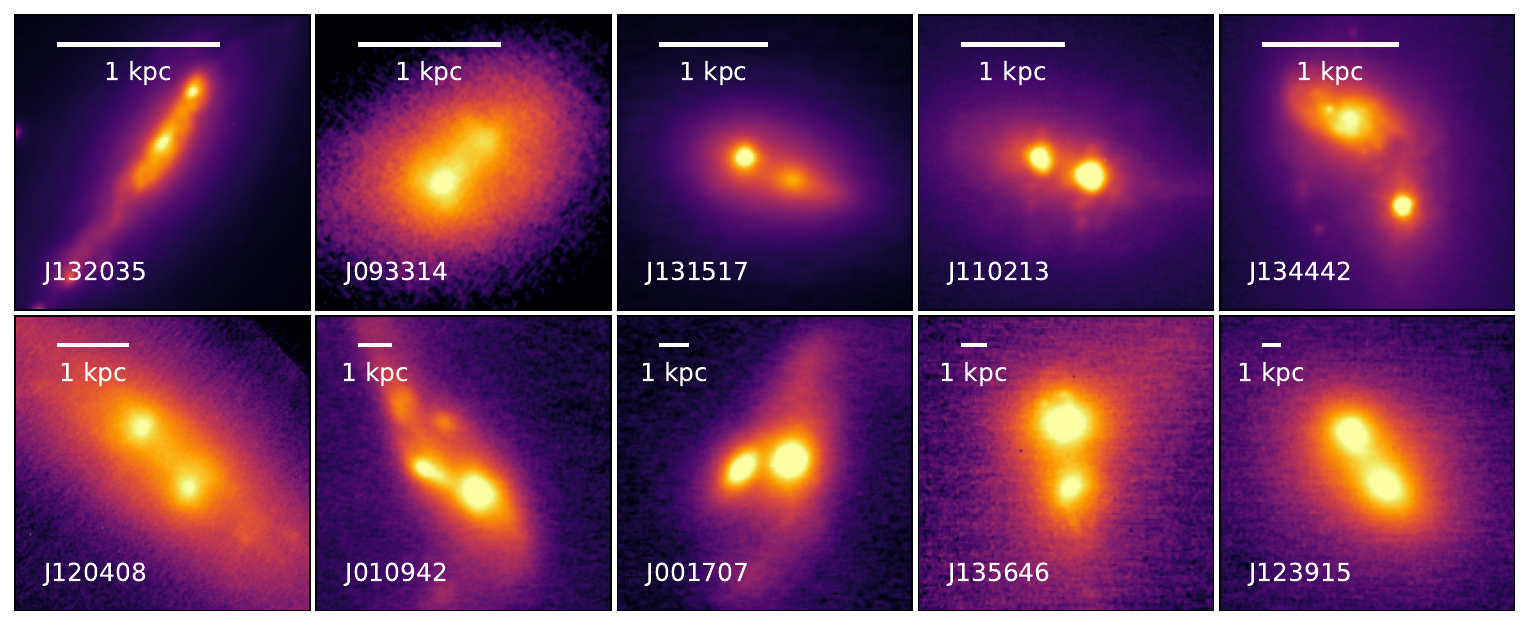}
\includegraphics[width=1.0\linewidth]{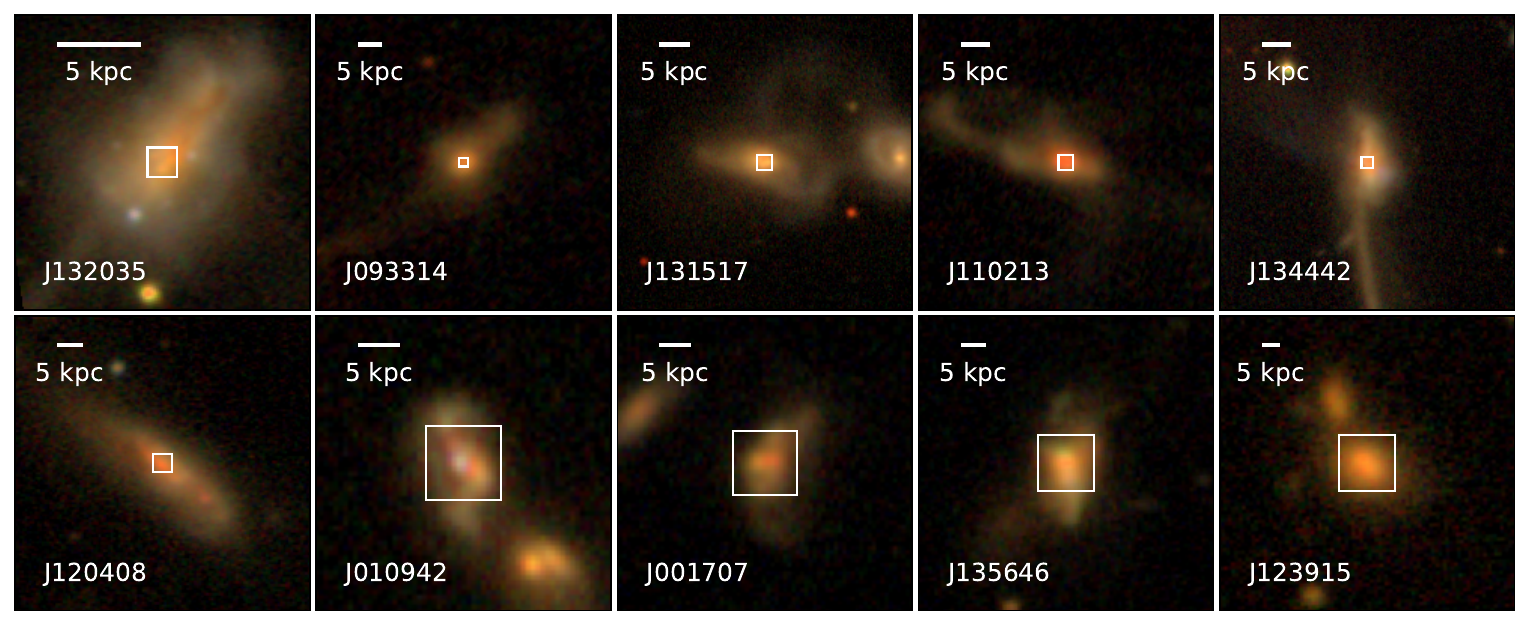}
\caption{
(Top) Keck/NIRC2 adaptive optics $K_p$-band cutout images of systems in the Primary Keck AO Merger Sample whose morphologies are consistent with two compact nuclei (i.e., apparent dual-nucleus systems). (Bottom) Corresponding SDSS \emph{gri} color cutout images of the same targets. White squares indicate the field of view of the Keck/NIRC2 images. Images are oriented north up and east to the left, and the systems are arranged in order of increasing projected nuclear separation. More detailed descriptions of individual targets are provided in Appendix~\ref{B}.
}
\label{keckstamps_duals}
\end{figure*}
\renewcommand{\thefigure}{\arabic{figure}} % restore default numbering

% ---------------- Figure 10b ----------------
\renewcommand{\thefigure}{9b}
\addtocounter{figure}{-1}  % Don't increment the counter
\begin{figure*}
% \captionsetup{labelformat=empty}
\centering
\includegraphics[width=0.805\linewidth]{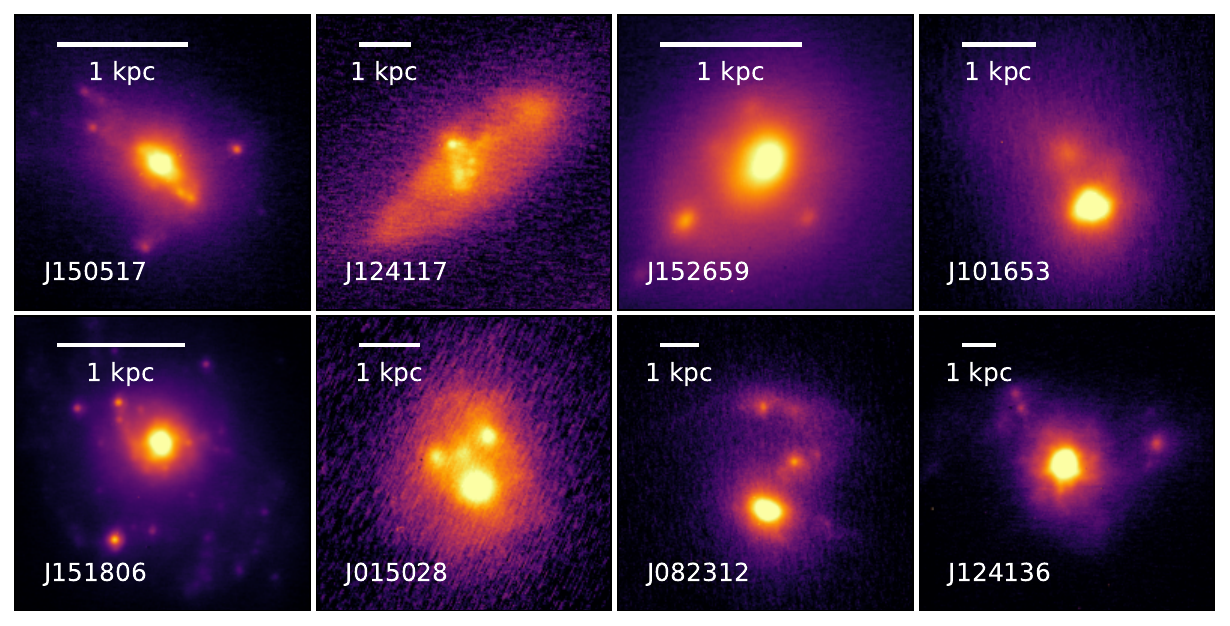}
\includegraphics[width=0.805\linewidth]{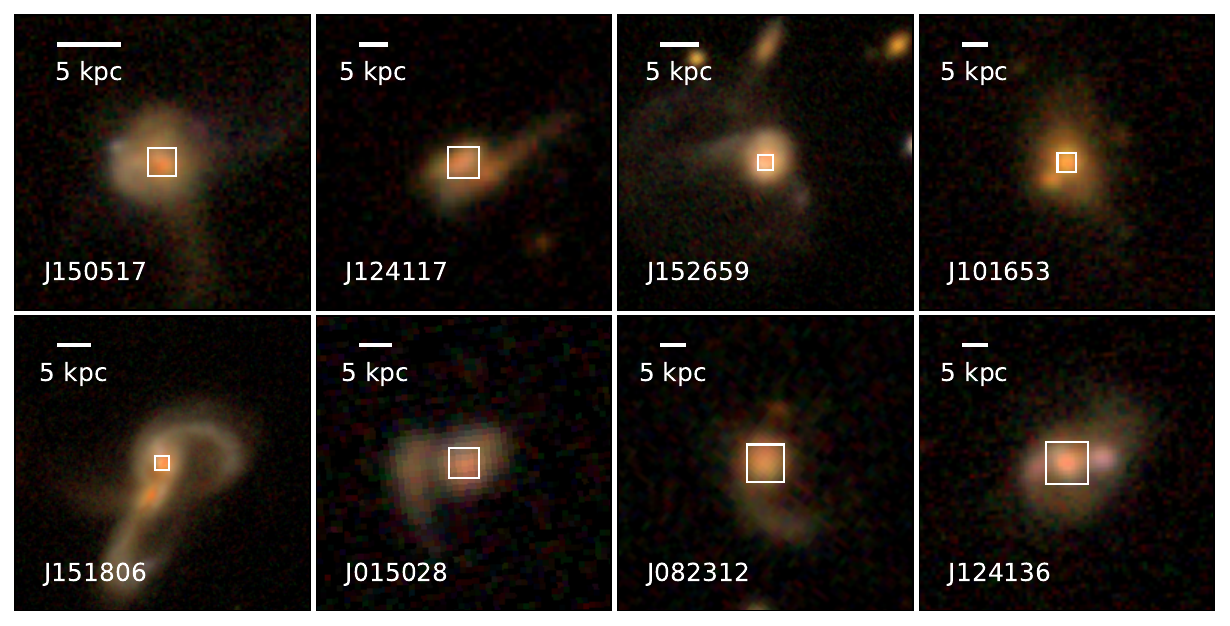}
\caption{
(Top) Keck/NIRC2 adaptive optics $K_p$-band cutout images of systems in the Primary Keck AO Merger Sample that contain compact secondary sources whose morphologies are consistent with star-forming regions rather than distinct nuclear components. (Bottom) Corresponding SDSS \emph{gri} color cutout images of the same targets. White squares indicate the field of view of the Keck/NIRC2 images. Images are oriented north up and east to the left, and the systems are arranged in order of increasing projected nuclear separation. Although J124136 exhibits apparent substructure, it does not satisfy our adopted flux-ratio threshold and is therefore not classified as a substructure detection. More detailed descriptions of individual targets are provided in Appendix~\ref{B}.
}
\label{keckstamps_sfregs}
\end{figure*}
\renewcommand{\thefigure}{\arabic{figure}} % restore default numbering
% ---------------- Figure 10c ----------------
\renewcommand{\thefigure}{9c}
\addtocounter{figure}{-1}  % Don't increment the counter
\begin{figure*}
% \captionsetup{labelformat=empty}
\centering
\includegraphics[width=1.0\linewidth]{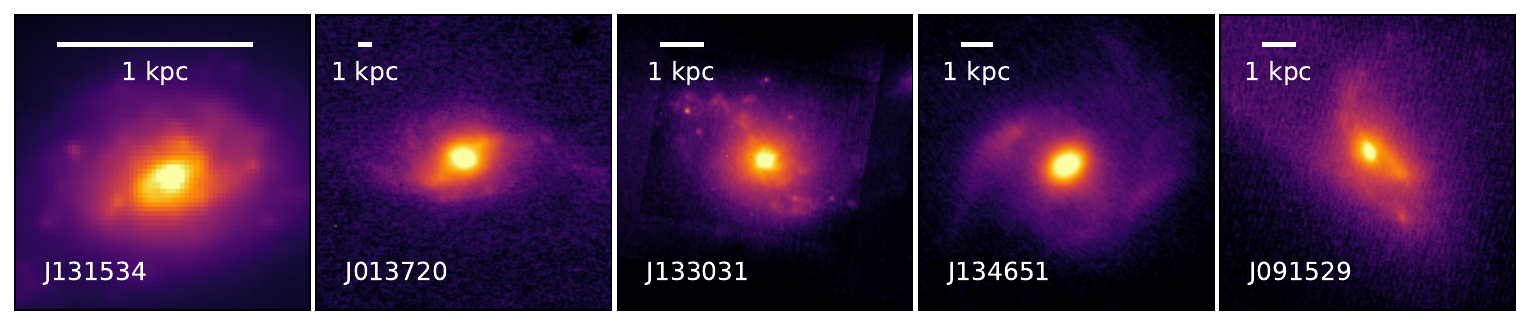}
\includegraphics[width=1.0\linewidth]{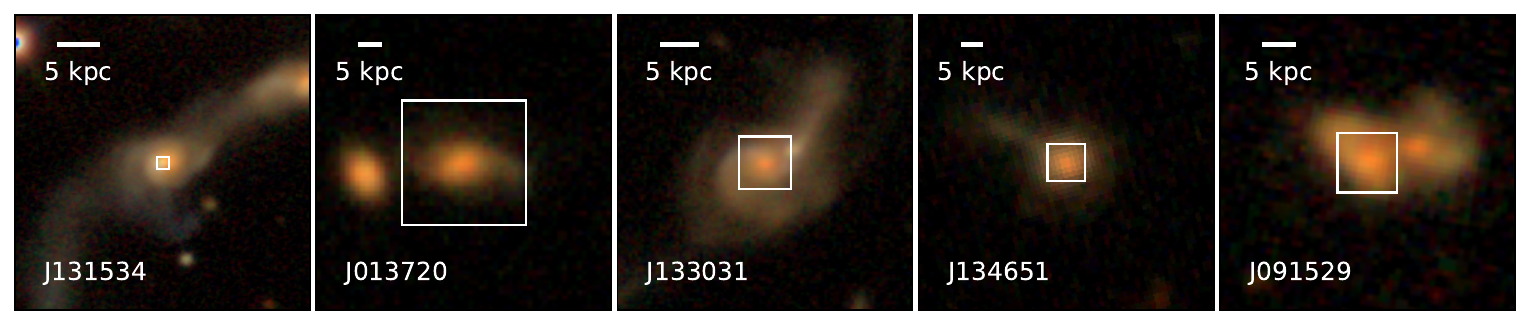}
\caption{
(Top) Keck/NIRC2 adaptive optics $K_p$-band cutout images of systems in the Primary Keck AO Merger Sample exhibiting morphological features consistent with spiral arms or tidal tails. (Bottom) Corresponding SDSS \emph{gri} color cutout images of the same targets. White squares indicate the field of view of the Keck/NIRC2 images. Images are oriented north up and east to the left, and the systems are arranged in order of increasing projected nuclear separation. More detailed descriptions of individual targets are provided in Appendix~\ref{B}.
}
\label{keckstamps_armstails}
\end{figure*}
\renewcommand{\thefigure}{\arabic{figure}} % restore default numbering
% ---------------- Figure 10d ----------------
\renewcommand{\thefigure}{9d}
\addtocounter{figure}{-1}  % Don't increment the counter
\begin{figure*}
% \captionsetup{labelformat=empty}
\centering
\includegraphics[width=1.0\linewidth]{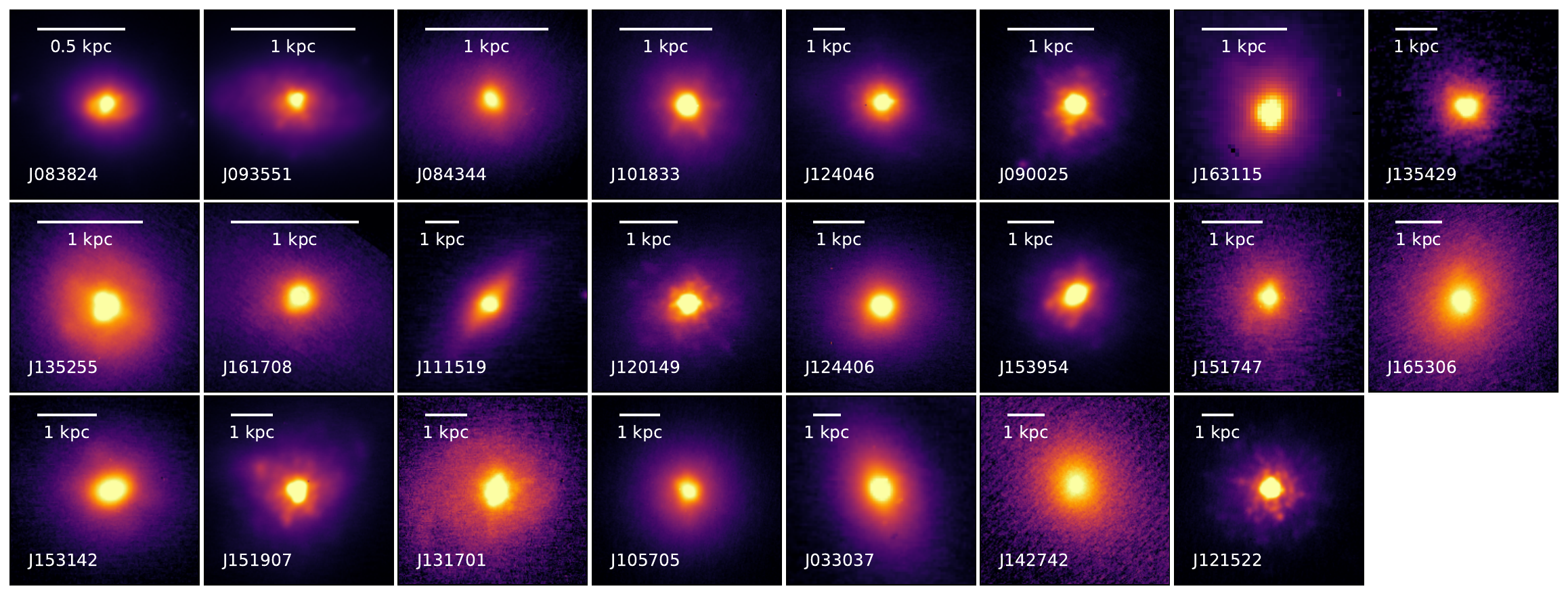}
\includegraphics[width=1.0\linewidth]{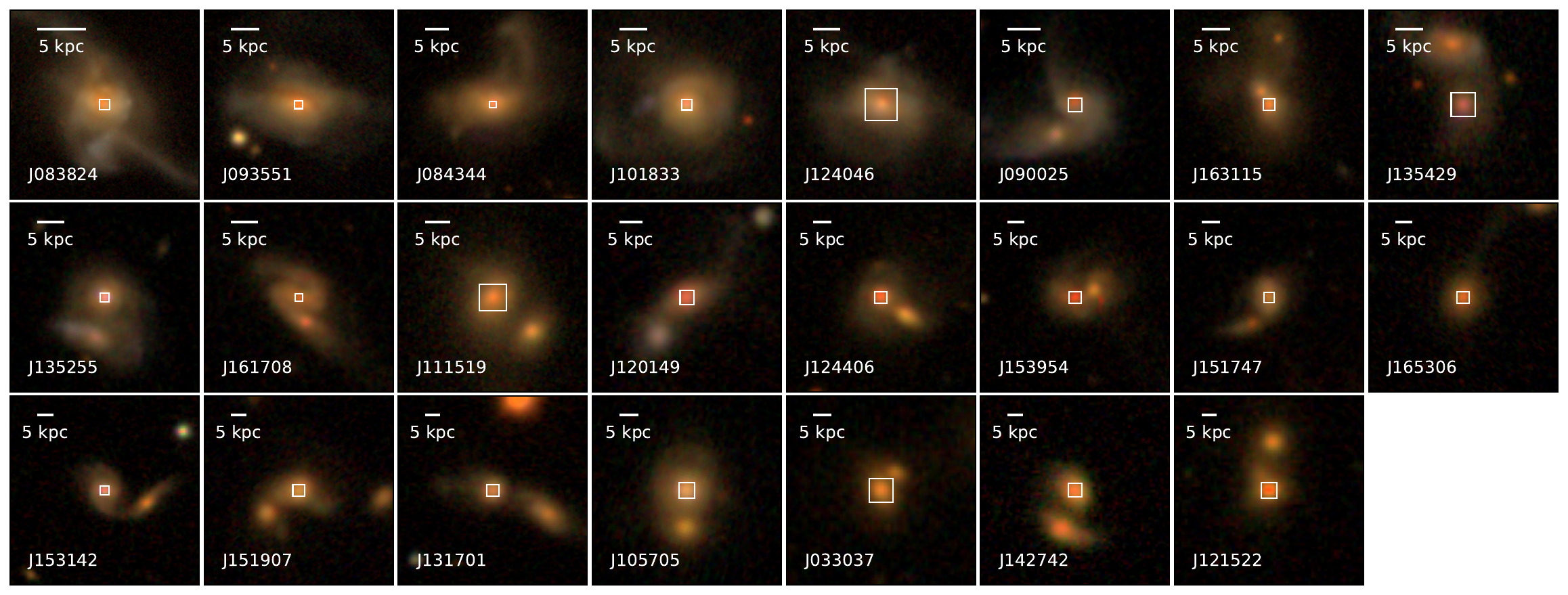}
\caption{
(Top) Keck/NIRC2 adaptive optics $K_p$-band cutout images of systems in the Primary Keck AO Merger Sample exhibiting morphological signatures consistent with single-nucleus systems. (Bottom) Corresponding SDSS \emph{gri} color cutout images of the same targets. White squares indicate the field of view of the Keck/NIRC2 images. Images are oriented north up and east to the left, and the systems are arranged in order of increasing redshift. More detailed descriptions of individual targets are provided in Appendix~\ref{B}.
}
\label{keckstamps_singles}
\end{figure*}
% ---- Restore normal figure numbering ----
% \setcounter{figure}{10}
\renewcommand{\thefigure}{\arabic{figure}}

\begin{table*}[t]
\centering{Photometric Properties of Substructure Pairs in the Primary Keck AO Merger Sample}
\centering
\begin{tabular}{lccccc}
\hline\hline

{Target} &
{Offset} &
{Separation} &
{Separation} &
{$Kp$ Flux} &
{Flux Ratio$^{a}$} \\

 &
('') &
('') &
(kpc) &
(mJy) &
($f_{\rm primary}/f_{\rm secondary}$) \\

\hline

J001707.95+011506.3 & \ldots & \ldots & \ldots & $5.630 \pm 1.591$ & \ldots \\
 & 0.511 & 0.986 & 1.910 & $1.826 \pm 0.516$ & 3.083 \\

J010942.21$-$004417.4 & \ldots & \ldots & \ldots & $0.076 \pm 0.001$ & \ldots \\
 & 0.443 & 1.129 & 1.871 & $0.052 \pm 0.0004$ & 1.474 \\

J015028.40+130858.4 & \ldots & \ldots & \ldots & $0.118 \pm 0.052$ & \ldots \\
 & 0.275 & 0.329 & 0.843 & $0.052 \pm 0.025$ & \ldots \\

J082312.61+275139.8 & \ldots & \ldots & \ldots & $0.115 \pm 0.005$ & \ldots \\
 & 0.294 & 0.564 & 1.617 & $0.014 \pm 0.001$ & \ldots \\

J091529.62+090601.5 & \ldots & \ldots & \ldots & $0.050 \pm 0.001$ & \ldots \\
 & 0.431 & 0.917 & 2.300 & $0.011 \pm 0.0003$ & \ldots \\

J093314.80+630945.0 & \ldots & \ldots & \ldots & $0.060 \pm 0.001$ & \ldots \\
 & 0.167 & 0.190 & 0.405 & $0.045 \pm 0.001$ & 1.323 \\

J101653.82+002857.0 & \ldots & \ldots & \ldots & $4.041 \pm 0.085$ & \ldots \\
 & 0.275 & 0.350 & 0.737 & $0.073 \pm 0.015$ & \ldots \\

J110213.01+645924.8 & \ldots & \ldots & \ldots & $0.982 \pm 0.020$ & \ldots \\
 & 0.276 & 0.357 & 0.524 & $0.573 \pm 0.012$ & 1.713 \\

J120408.82+132117.0 & \ldots & \ldots & \ldots & $0.063 \pm 0.001$ & \ldots \\
 & 0.486 & 0.740 & 1.070 & $0.059 \pm 0.001$ & 1.074 \\

J123915.40+531414.6$^{b}$ & 0.306 & 1.161 & 3.855 & \ldots & 1.163 \\

J124117.64+273252.3 & \ldots & \ldots & \ldots & $0.018 \pm 0.0003$ & \ldots \\
 & 0.411 & 0.266 & 0.655 & $0.019 \pm 0.0003$ & \ldots \\

J131517.26+442425.5$^{b}$ & 0.689 & 0.711 & 0.503 & \ldots & 3.225 \\

J131534.00+620728.6$^{b}$ & 0.585 & 0.287 & 0.177 & \ldots & \ldots \\

J132035.40+340821.5$^{b}$ & 1.209 & 0.784 & 0.365 & \ldots & 1.324 \\

J134442.16+555313.5$^{b}$ & 0.799 & 1.035 & 0.767 & \ldots & 2.430 \\

J134651.82+371231.0 & \ldots & \ldots & \ldots & $0.318 \pm 0.002$ & \ldots \\
 & 0.230 & 0.633 & 2.209 & $0.041 \pm 0.0003$ & \ldots \\

J135646.10+102609.0$^{b}$ & 0.258 & 1.253 & 2.771 & \ldots & 2.501 \\

J150517.88+080912.7 & \ldots & \ldots & \ldots & $0.309 \pm 0.003$ & \ldots \\
 & 0.194 & 0.347 & 0.270 & $0.079 \pm 0.001$ & \ldots \\

J151806.13+424445.0 & \ldots & \ldots & \ldots & $0.950 \pm 0.016$ & \ldots \\
 & 0.500 & 1.083 & 0.863 & $0.128 \pm 0.002$ & \ldots \\

J152659.44+355837.0 & \ldots & \ldots & \ldots & $0.531 \pm 0.048$ & \ldots \\
 & 0.376 & 0.631 & 0.677 & $0.145 \pm 0.013$ & \ldots \\

\hline\hline
\end{tabular}

\caption{
Photometric measurements of nuclear substructure pairs in the Primary Keck AO Merger Sample. Columns list (1) target identifier, (2) SDSS/WISE positional offset, (3–4) angular and physical separation between nuclear components, (5) $Kp$ flux density, and (6) flux ratio. $^{a}$ Flux ratios are reported only for systems with compact, well-resolved nuclear components and flux ratios consistent with major mergers (mass ratio $\lesssim4{:}1$), where aperture-corrected photometry is reliable (Section~\ref{sec:obsredux:phot}). Flux ratios are omitted for systems where photometry may be affected by extended emission, blending, or uncertain aperture corrections. $^{b}$ Archival Keck observations lack absolute flux calibration; only separations and flux ratios are reported. This table demonstrates that a subset of SDSS/WISE offset-selected mergers contain compact nuclear pairs with flux ratios consistent with major mergers.
}

\label{table:substructure}
\end{table*}

\begin{figure}[htbp]
\centering
\addtocounter{figure}{-1}  % Decrement first since counter is at 11
\phantomsection  % Create proper anchor point
\includegraphics[width = \linewidth]{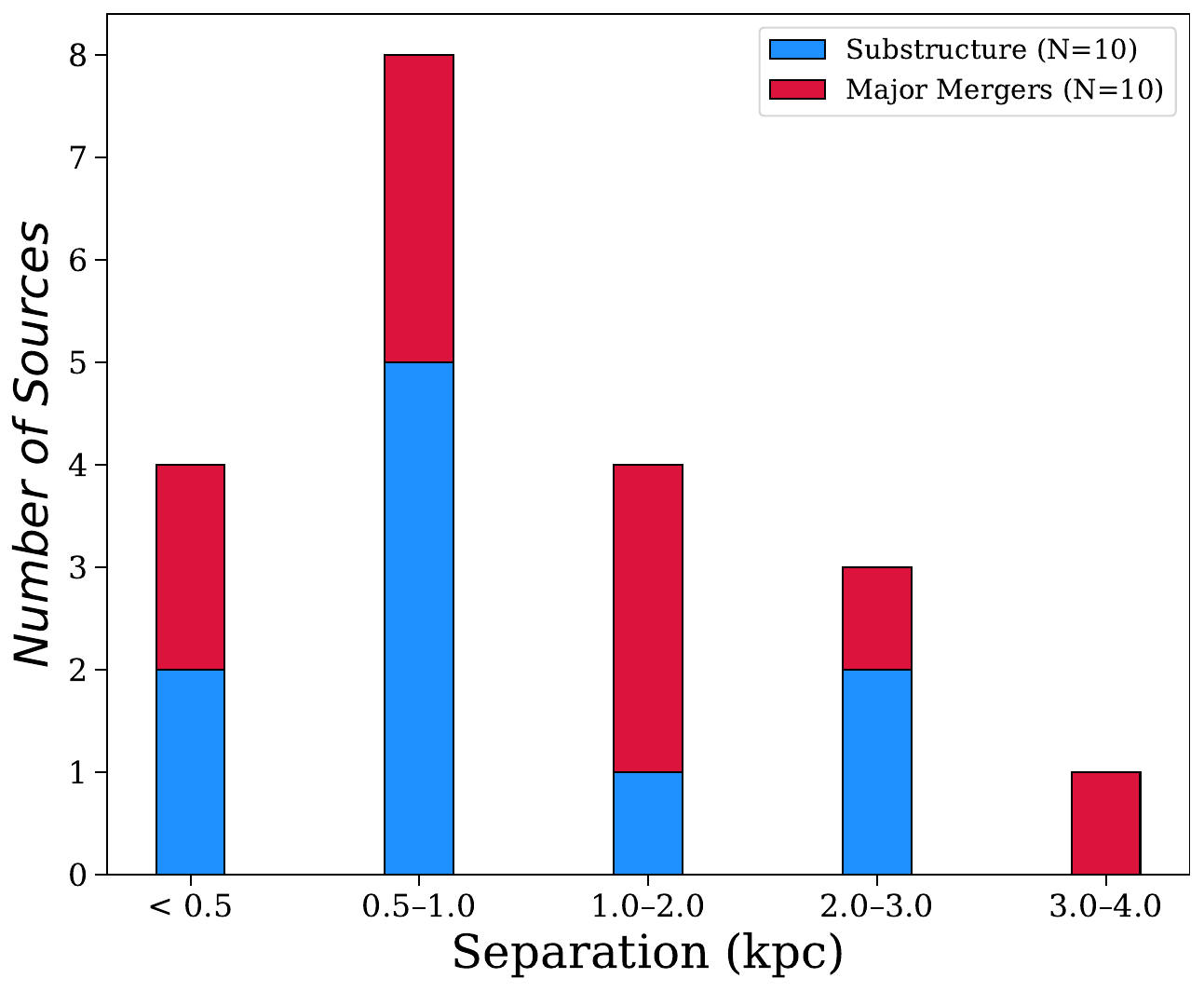}
\caption{
Distribution of nuclear separations for systems in the Primary Keck AO Merger Sample with detected near-infrared substructure. The blue bars show the total number of systems with substructure in each separation bin, while the red stacked component indicates the subset of systems with flux ratios consistent with major mergers ($f_{\rm secondary}/f_{\rm primary} > 0.25$; see Section~\ref{sec:results:subsID}). This figure quantifies the spatial scale of the nuclear substructure revealed by Keck AO imaging and demonstrates that the SDSS/WISE offset selection method identifies compact nuclear pairs across a range of separations from $\sim$0.2'' to $\sim$1.25'' (176~pc–3.85~kpc), including systems consistent with major mergers.}

\label{separations}
\end{figure}

\subsection{Substructure Detection Rates}
\label{sec:results:detectionrate}

We now quantify the incidence of substructure and examine how the detection rate depends on galaxy classification, obscuration, and SDSS/WISE offset size. Our analysis of the reduced and calibrated NIRC2 observations shows that 20 of the 46 systems (43.5\%) in the Primary Keck AO Merger Sample (SDSS/WISE offset $<1.5''$) exhibit substructure. We define substructure as the presence of a secondary compact source of emission unresolved by SDSS and with a flux ratio relative to the primary nucleus of $f_{\rm secondary}/f_{\rm primary} > 0.1$ (Section~\ref{sec:results:subsID}). 

Among the systems with detected substructure, 10 are optically classified as AGN, 9 as composite systems, and 1 as star-forming, based on their SDSS emission-line classifications (Table~\ref{table:sampleproperties}). Figures~\ref{bptmergers} and \ref{colorcolormergers} show the locations of these systems in optical emission-line diagnostic space and mid-infrared color space, respectively. Within the Primary Keck AO Merger Sample, substructure is detected in 34\% (10/29) of optically classified AGN, 64\% (9/14) of composite systems, and 50\% (1/2) of star-forming galaxies. Using mid-infrared selection criteria, substructure is detected in 42\% (8/19) of the Jarrett AGN. Although the absolute number of substructure detections in AGN and composite systems is similar, there is a higher detection fraction in composite targets.

Substructure detection is also strongly associated with red mid-infrared colors. In particular, 89\% (8/9) of systems with $W_2 - W_3 > 4.2$ contain substructure. Figure~\ref{ebvmergers} shows the distribution of optical extinction, parameterized by $E(B-V)$ from the MPA–JHU catalog, for systems with and without detected substructure. Substructure is detected across the full range of extinction values in the sample; however, the detection fraction increases from 36\% at lower extinction to 52\% for systems with $E(B-V) > 0.6$, indicating that highly obscured systems are particularly likely to host unresolved nuclear substructure.

To investigate how the detection rate depends on offset size, we also analyzed the Expanded Keck AO Merger Sample, which includes mergers with SDSS/WISE offsets extending to $6''$. Combining this sample with the UKIDSS Merger Sample, we identify substructure in an additional 51 systems. Across the full set of mergers with Keck AO and UKIDSS imaging, substructure is detected in 71 of 114 systems, corresponding to an overall detection rate of 62\%. This higher detection rate reflects the inclusion of systems with larger offsets, for which the separation between nuclear components is more readily resolved.

\subsection{Robustness of the SDSS/WISE Offset Method Across Offset Magnitudes}\label{sec:results:offsetrobustness}

A key question is whether the SDSS/WISE offset selection method remains reliable at the smallest offsets considered in this work, where astrometric uncertainties could potentially dominate over true physical separations. If the observed offsets were primarily driven by astrometric noise rather than genuine substructure, the success rate of detecting substructure in high-resolution imaging would be expected to decrease systematically toward smaller SDSS/WISE offsets.

To assess whether the SDSS/WISE offset method remains reliable across the full offset range, we measured the substructure detection rate as a function of SDSS/WISE offset using the combined sample of mergers with high-resolution near-infrared imaging, including the Primary Keck AO Merger Sample ($0.15''< $ offset $ <1.5''$) and the additional mergers in the Expanded Keck AO Merger Sample, and the UKIDSS Merger Sample (Section~\ref{sec:sampleselection:comparisonsamples}). These samples together span the full offset range of $0.15''<$ offset $<6''$ and provide a uniform test of the method across both sub-kpc and larger projected separations.

\begin{figure}
    \centering
    \includegraphics[width = \linewidth]{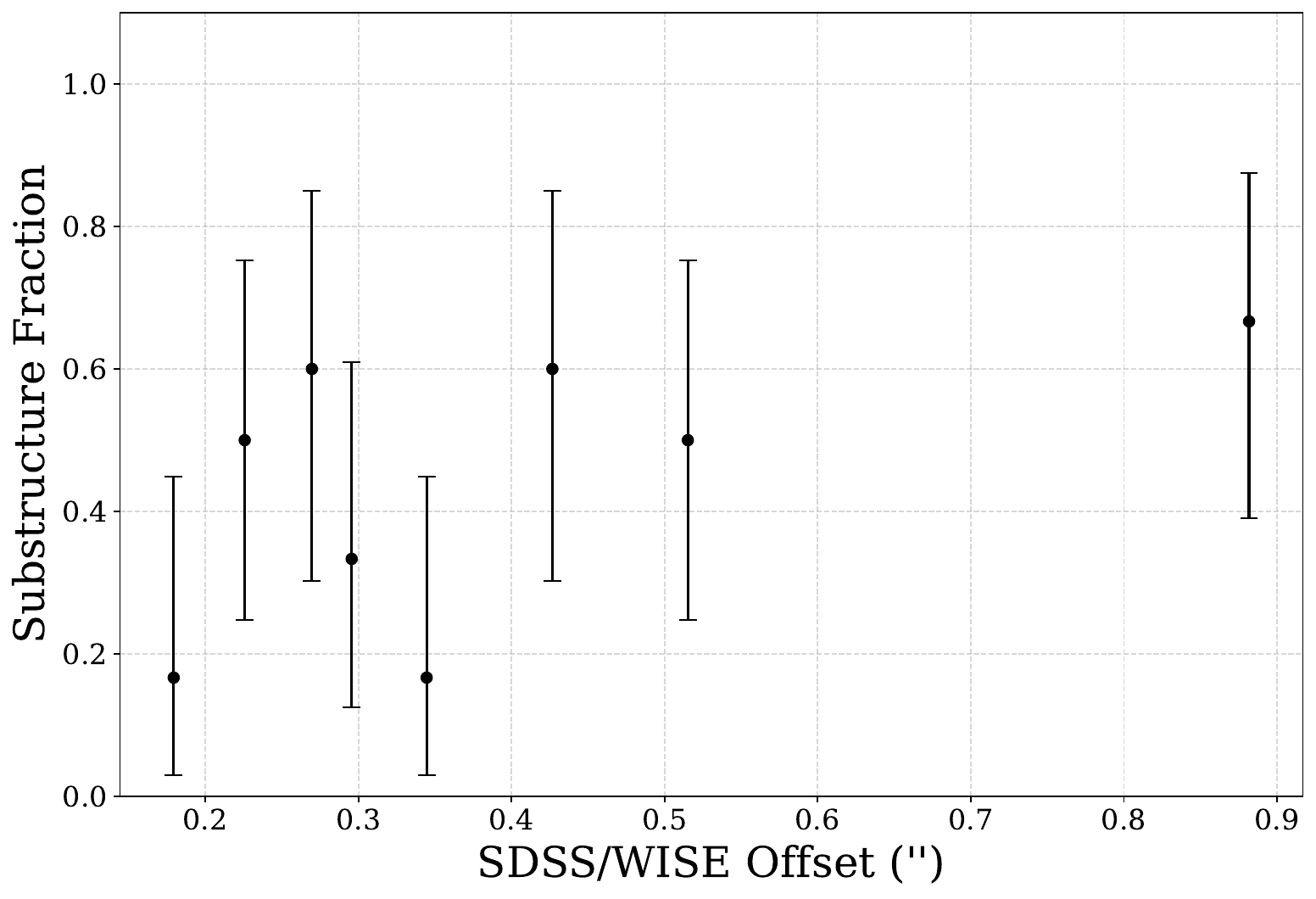}
    \caption{
Substructure detection rate as a function of SDSS/WISE positional offset for mergers with high-resolution near-infrared imaging, including the Primary Keck AO Merger Sample. The sample is divided into eight bins in offset, where the detection rate in each bin is defined as the fraction of systems in which Keck AO imaging reveals unresolved near-infrared substructure. Error bars represent 90\% binomial confidence intervals. This figure evaluates whether the SDSS/WISE offset selection method remains reliable at small offsets, where astrometric noise could potentially dominate. The absence of a systematic decline in detection rate toward smaller offsets indicates that the observed SDSS/WISE offsets are predominantly associated with genuine nuclear substructure rather than being driven by astrometric uncertainties.
}

    \label{subrate}
\end{figure}

We find no dependence on the substructure detection rate as a  function of offset over the full range of offsets. In Figure~\ref{subrate}, we show the substructure detection rate computed in eight bins spanning the sub-arcsecond range of offsets. In each bin, the detection rate is defined as the fraction of targets in which high-resolution imaging reveals nuclear substructure relative to the total number of targets in that offset interval. Error bars correspond to binomial statistics and represent the 90\% confidence interval.

We find no statistically significant decline in the detection rate toward smaller offsets. Instead, the detection rate remains approximately constant across the full range of SDSS/WISE offsets probed, including the smallest offsets where astrometric uncertainties would be expected to have the greatest impact. 

This result demonstrates that the effectiveness of the SDSS/WISE offset selection method does not degrade at small offsets and provides strong evidence that the observed offsets are predominantly driven by genuine unresolved nuclear substructure rather than random astrometric errors. It also confirms that the SDSS/WISE offset method remains reliable for identifying sub-kpc nuclear pairs across the full offset range considered in this work.

\subsection{Geometric Relationship Between SDSS/WISE Offsets and Nuclear Substructure}\label{sec:results:offsetvsepvpa}

A key test of the SDSS/WISE offset selection method is whether the measured positional offsets trace the geometry of the underlying nuclear substructure identified in high-resolution near-infrared imaging. If the SDSS and WISE centroids are displaced due to unresolved multiple sources of emission, the magnitude and orientation of the SDSS/WISE offset should correlate with the separation and position angle (PA) of the substructure pair.

Directly overlaying the SDSS and WISE coordinates on the Keck NIRC2 images is not possible because the NIRC2 images lack absolute astrometric registration. Instead, we test whether the relative geometry of the SDSS/WISE offsets is statistically correlated with the geometry of the substructure identified in the imaging.

This analysis was conducted on the 71 targets with detected substructure drawn from the combined sample of mergers with high-resolution near-infrared imaging, including the Primary Keck AO Merger Sample, the additional systems in the Expanded Keck AO Merger Sample, and the UKIDSS Merger Sample, spanning SDSS/WISE offsets of $0.15'' < $ offset $ < 6''$. For each target, we identified a substructure pair as defined in Section~\ref{sec:results:subsID}, and measured its angular separation and position angle.

For the Keck AO imaging, the lack of absolute astrometric registration prevents identifying which component corresponds directly to the SDSS or WISE centroid. To account for this ambiguity, we computed the SDSS/WISE offset PA in both possible directions and adopted the value that minimized the difference relative to the substructure pair PA. For the UKIDSS imaging, which has accurate absolute astrometry, we defined the substructure pair as the sources nearest to the SDSS and WISE coordinates, respectively.

\begin{figure}[]
\begin{flushleft} % Align to the left margin
\includegraphics[width = \linewidth]{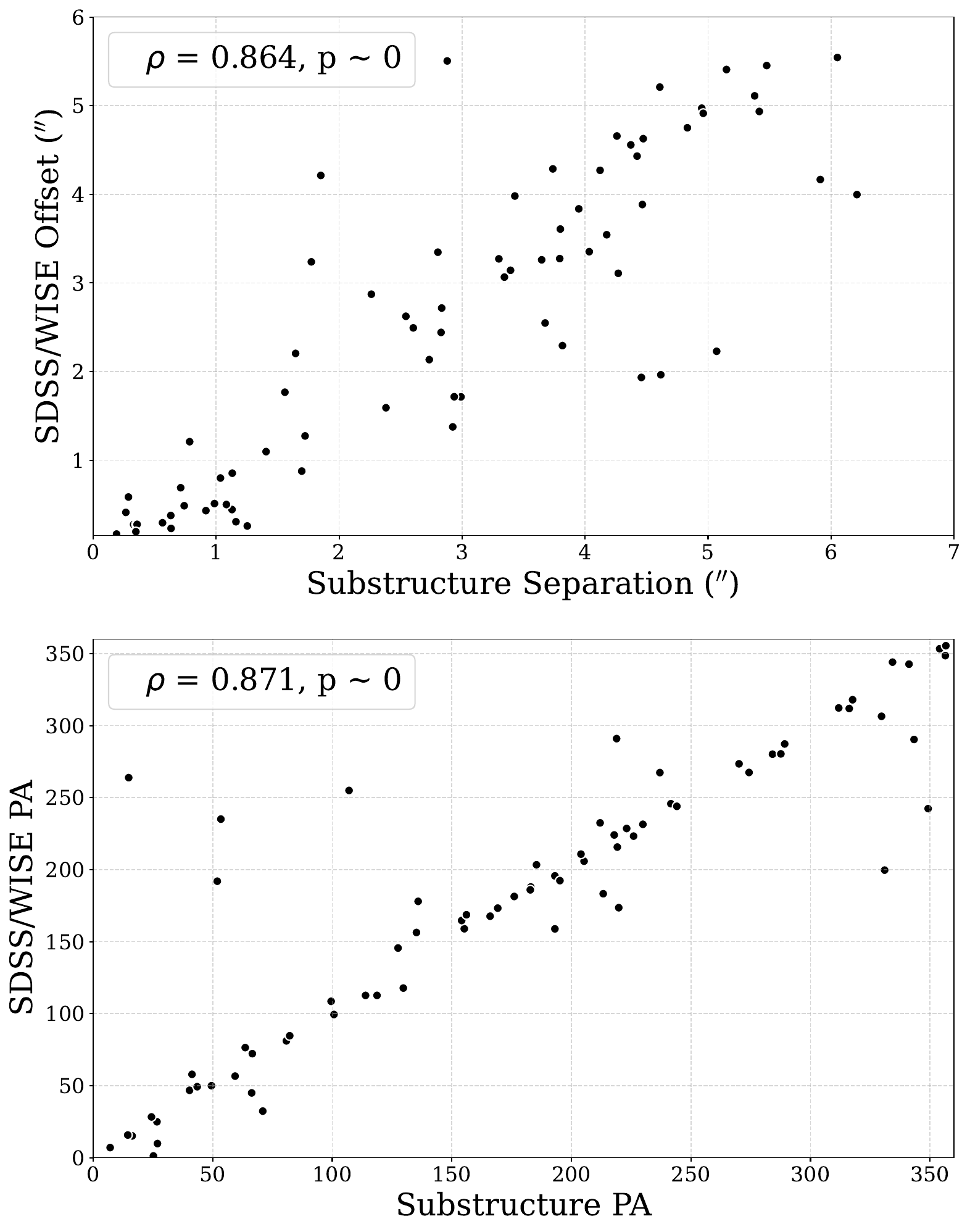}
\end{flushleft}
\caption{
Comparison between the SDSS/WISE offsets and the nuclear substructure geometry for the combined sample of mergers with detected substructure, including the Primary Keck AO Merger Sample, Expanded Keck AO Merger Sample, and UKIDSS Merger Sample. (Top) Angular separation of the SDSS/WISE positional offset versus the measured separation of the substructure pair. (Bottom) Position angle of the SDSS/WISE positional offset versus the position angle of the substructure pair. Spearman rank correlation coefficients ($\rho$) and corresponding $p$-values are shown in each panel. The strong correlations demonstrate that the SDSS/WISE positional offsets reliably trace both the orientation and characteristic spatial scale of the underlying nuclear substructure, providing direct geometric validation of the SDSS/WISE offset selection method.
}

\label{offsetpacomp}
\end{figure}

Figure~\ref{offsetpacomp} compares the SDSS/WISE positional offsets to the measured substructure separations and position angles. We find strong correlations between the offset magnitude and substructure separation (Spearman rank correlation: $\rho = 0.864$, $p \sim 0$) and between the offset PA and substructure PA (Spearman rank correlation: $\rho = 0.871$, $p \sim 0$).

The strong correlation in position angle demonstrates that the SDSS/WISE offsets reliably trace the orientation of the underlying nuclear substructure. The offset magnitudes are systematically smaller than the substructure separations, as expected, because the measured centroid shift depends on the relative brightness of the emission components. When both sources contribute significantly to the mid-infrared emission, the WISE centroids are displaced toward the flux-weighted midpoint between them. A similar effect occurs for the SDSS centroid in systems with separations smaller than $\sim1.5''$, where the components remain unresolved.

These results provide direct geometric validation that the SDSS/WISE offsets are physically associated with unresolved nuclear substructure and reliably trace both its orientation and characteristic spatial scale.

\subsection{Comparison Sample Results and Validation of Selection Criteria
}\label{sec:results:comparisonsamples}

To isolate the relative roles of merger classification and SDSS/WISE positional offsets in driving substructure detection, we analyzed two comparison samples described in Section~\ref{sec:sampleselection:comparisonsamples}: (1) the Keck Non-merger Comparison Sample, consisting of galaxies with significant SDSS/WISE offsets but no merger classification, and (2) the Keck Non-offset Comparison Sample, consisting of mergers that satisfy all selection criteria except the SDSS/WISE offset requirement.

\subsubsection{Keck Non-merger Comparison Sample}\label{sec:results:nonmergersample}

The Keck Non-merger Comparison Sample consists of 40 galaxies with low merger vote fractions ($f_m < 0.4$), red WISE colors ($W_1 - W_2 > 0.5$), $W_1 < 15$, SDSS/WISE positional offsets in the range $0.15'' < $ offset $ < 1.5''$, and archival NIRC2 imaging. Over the redshift range of the sample, these offsets correspond to projected physical separations $\lesssim 1$~kpc.

Within this sample, 10 of the 40 systems exhibit unresolved near-infrared substructure detected in Keck AO imaging, corresponding to a substructure detection rate of 25\%. Of these, 7 systems have nuclear flux ratios $>0.25$, consistent with major mergers under the assumption that K-band flux traces stellar mass.

Extending the offset range to include systems with $0.15'' < $ offset $ < 6''$, the detection rate becomes 14 out of 51 systems (27\%), indicating that SDSS/WISE positional offsets alone, independent of merger classification, are associated with an elevated probability of detecting unresolved near-infrared substructure.

However, the detection rate in this non-merger offset sample remains significantly lower than that observed in the Primary Keck AO Merger Sample (43.5\%; Section~\ref{sec:results:detectionrate}), demonstrating that merger classification substantially enhances the likelihood of detecting unresolved nuclear structure.

The detection of substructure, including systems consistent with major mergers, in galaxies not morphologically classified as mergers suggests that the SDSS/WISE offset method is capable of identifying nuclear pairs in systems where merger signatures are not readily apparent in SDSS imaging. This result is consistent with previous studies showing that late-stage mergers can lack clear morphological signatures at optical wavelengths \citep[e.g.,][]{koss+2018}.

\subsubsection{Keck Non-offset Comparison Sample}\label{sec:results:nonoffsetsample}

To assess whether SDSS/WISE positional offsets are necessary for identifying unresolved substructure, we also examined the Keck Non-offset Comparison Sample, consisting of 11 merger systems ($f_m > 0.4$) with red WISE colors ($W_1 - W_2 > 0.5$), $W_1 < 15$, archival NIRC2 imaging, and no statistically significant SDSS/WISE positional offset ($< 0.15''$).

Within this sample, only 1 out of 11 systems exhibits detectable substructure, corresponding to a detection rate of 9\%. The detected system has a flux ratio of 0.185, below the major merger threshold.

This substantially lower detection rate compared to both the Primary Keck AO Merger Sample (43.5\%) and the Keck Non-merger Comparison Sample (25\%) indicates that SDSS/WISE positional offsets are a critical component of the selection method. The combination of merger classification and positional offset selection yields the highest efficiency for identifying unresolved near-infrared substructure.

Although the sample size is limited, these results provide strong evidence that SDSS/WISE positional offsets are physically linked to unresolved nuclear structure rather than arising purely from astrometric uncertainties. Larger samples will be required to further quantify the relative contributions of merger classification and positional offsets to substructure detection.

\subsection{Role of Optical Obscuration in Producing SDSS/WISE Positional Offsets}\label{sec:results:obscuration}

To investigate the physical origin of SDSS/WISE positional offsets and assess the role of dust obscuration, we examined archival HST optical imaging for targets in the Primary Keck AO Merger Sample. Of the 46 mergers with SDSS/WISE offsets $<1.5''$, 16 have publicly available HST observations obtained with the WFC3 (FOV: ${162''\times162''}$, pixel scale: ${0.04''\,\mathrm{pixel}^{-1}}$), WFPC2 (FOV: ${35''\times35''}$, pixel scale: ${0.046''\,\mathrm{pixel}^{-1}}$), or ACS (FOV: ${202''\times202''}$, pixel scale: ${0.05''\,\mathrm{pixel}^{-1}}$) cameras.

These high-resolution optical images allow a direct comparison between the optical and near-infrared nuclear morphology. We find that the optical nuclear regions are frequently more complex and fragmented than their infrared counterparts, often exhibiting prominent dust lanes and patchy obscuration. Nuclear dust structures are commonly observed in nearby AGNs and merging systems \citep[e.g.,][]{malkan+1998,hunt_malkan+2004}.

\begin{figure*}

    \centering
    \overlay{(a)}{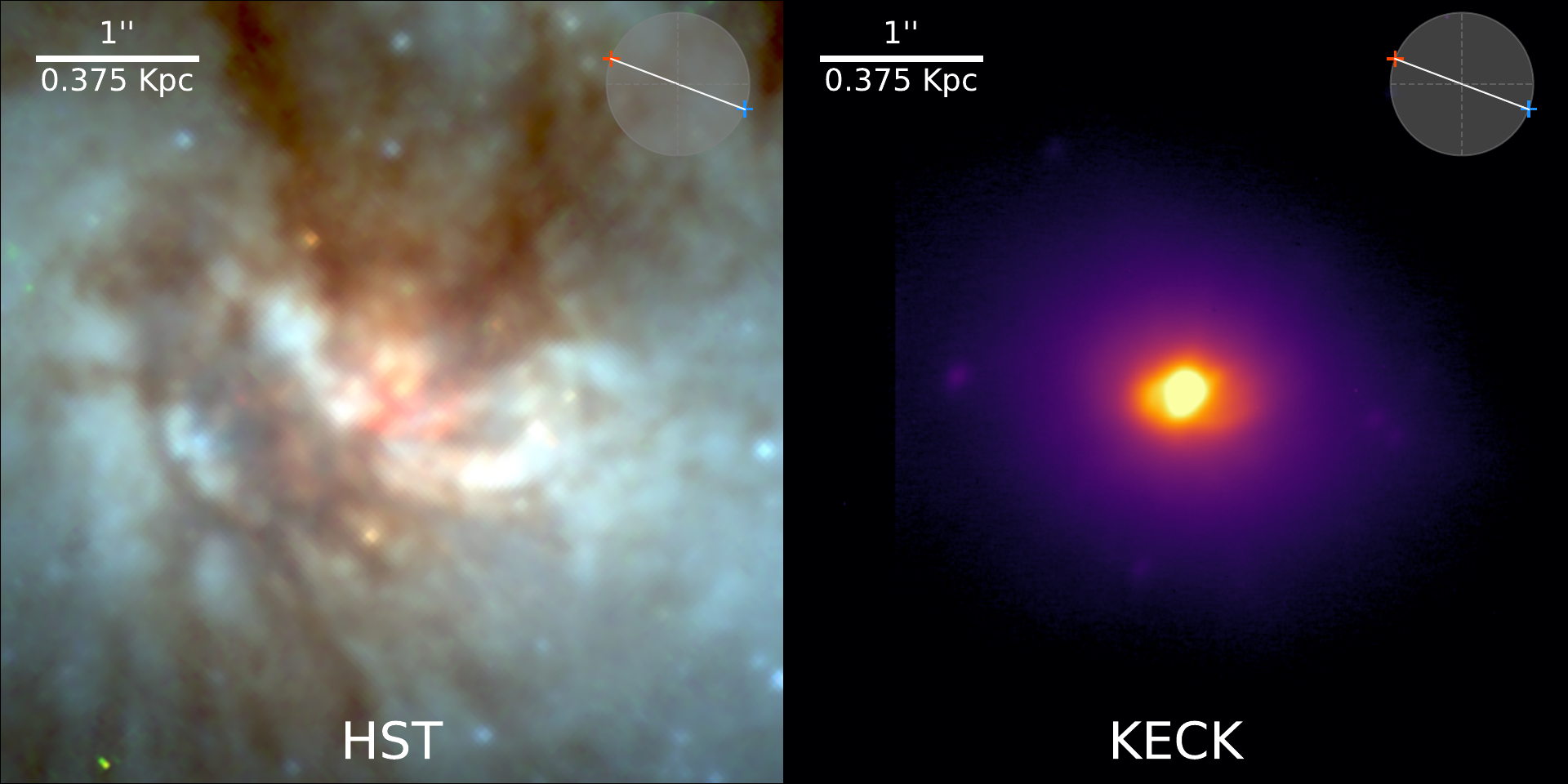}\hspace{0.0085\linewidth}\overlay{(b)}{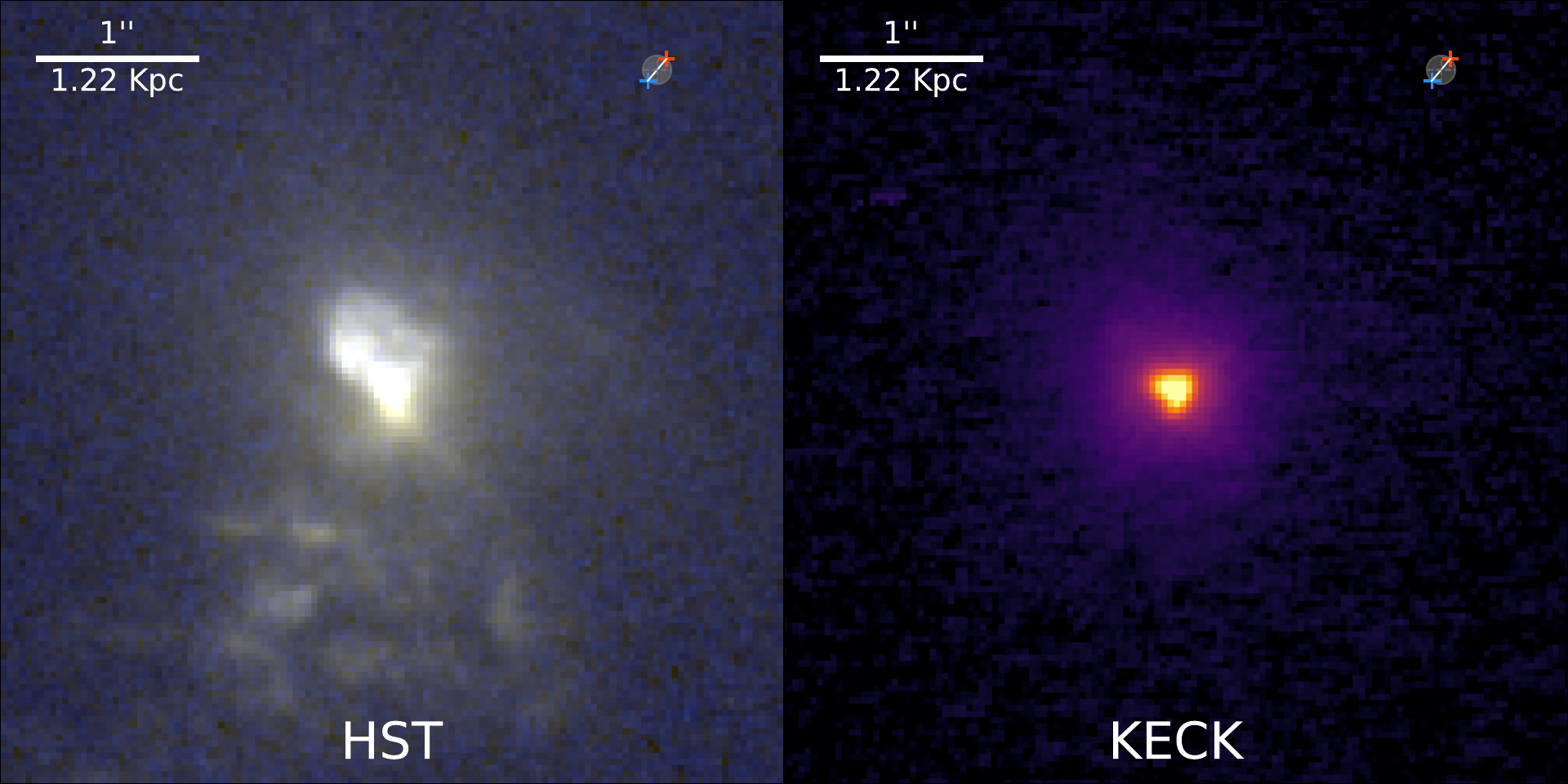}
    
    \vspace{0.005\linewidth}\overlay{(c)}{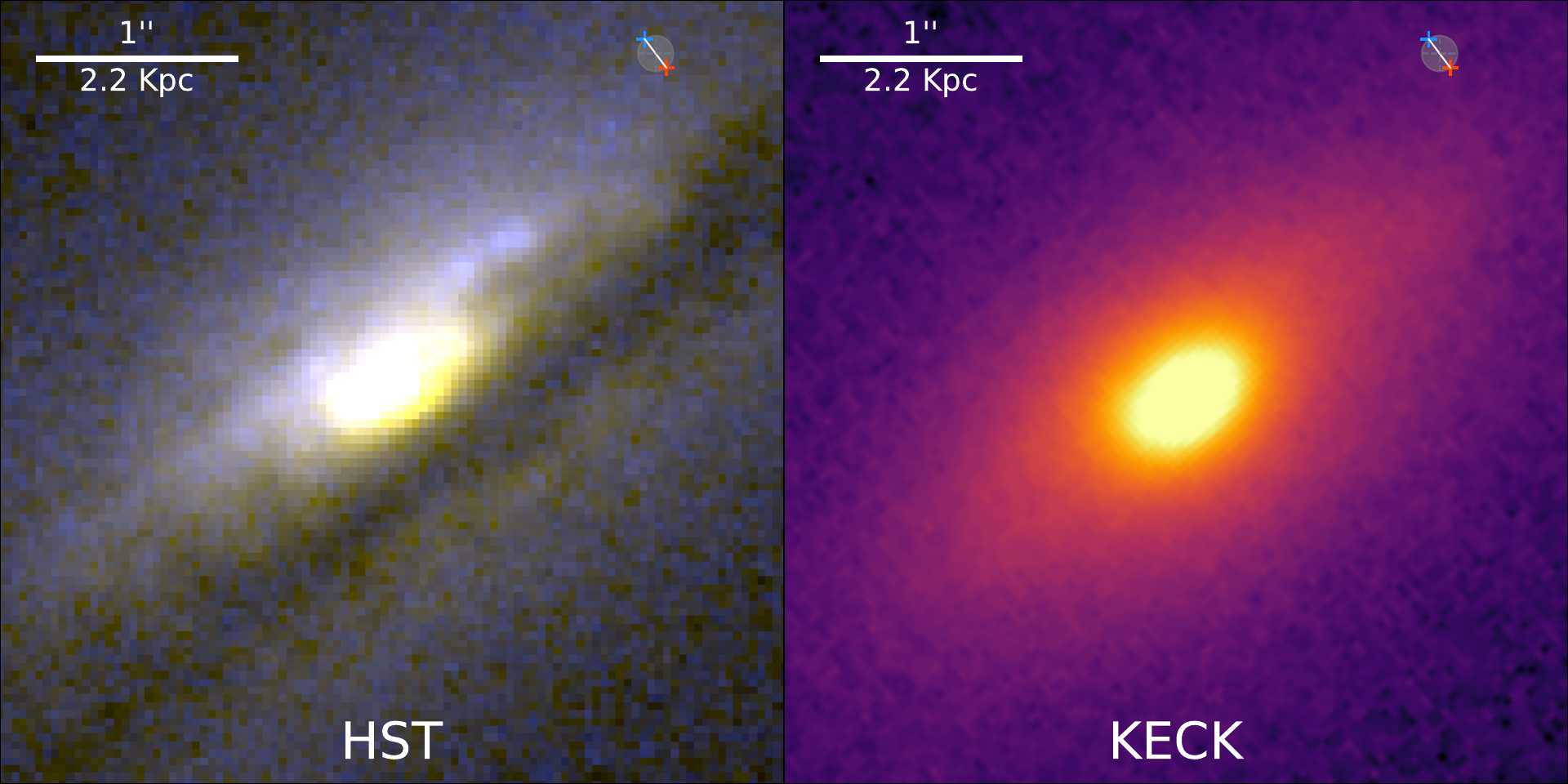}\hspace{0.0085\linewidth}\overlay{(d)}{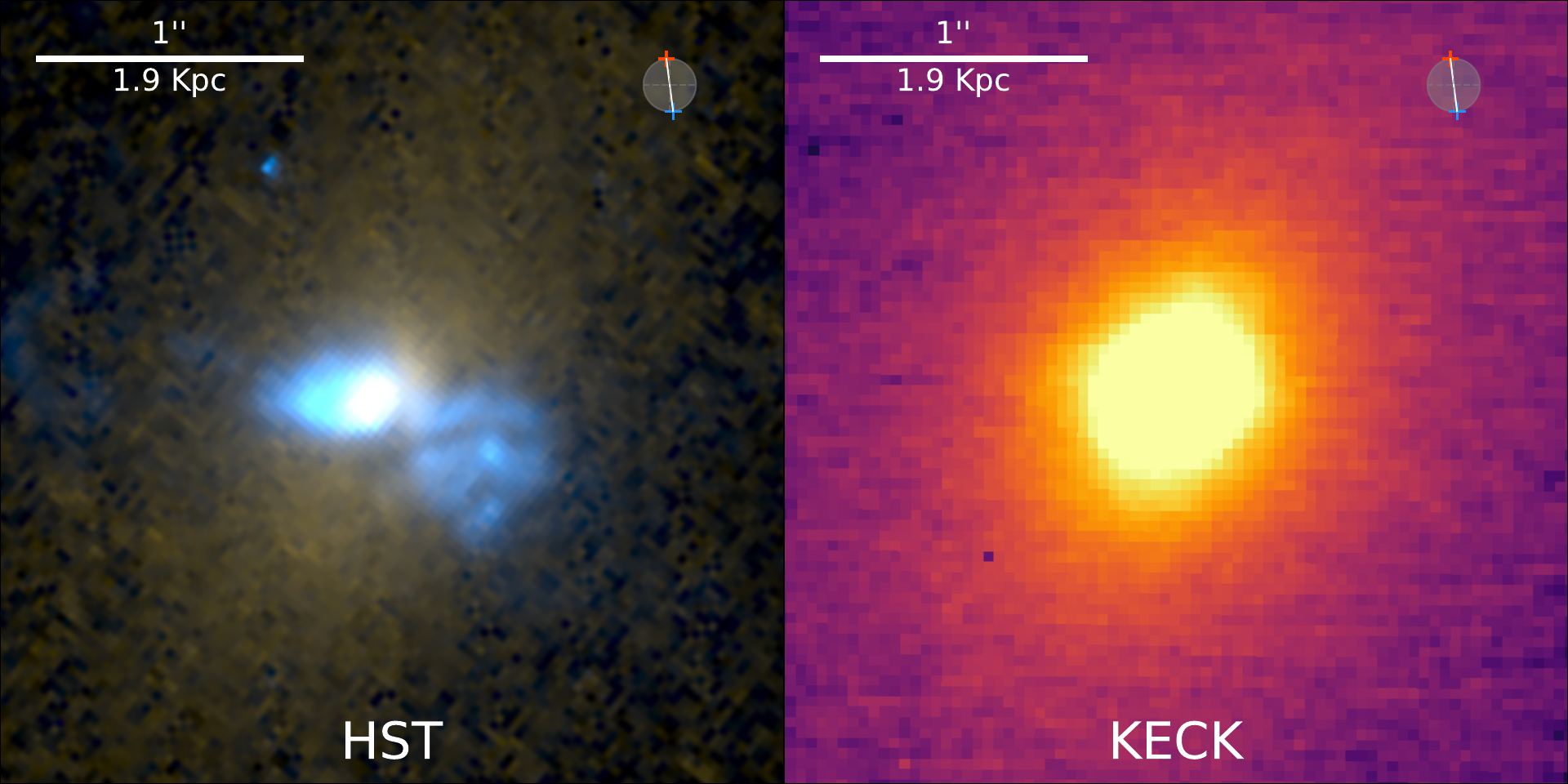}
    \caption{Examples of targets with no substructure whose SDSS/WISE offsets are likely caused by partial obscuration of their optical emission. For each target we include the corresponding HST and Keck images. In the top right of each HST and Keck image we include a representation of the SDSS/WISE offset which shows the orientation and length of the offset for that particular system. The top row of targets, J083824 (a: HST Program No. 9735) and J135429 (b: HST Program No. 13513), are from the Primary Keck AO Merger Sample and the bottom row of targets, J100654 (c: HST Program No. 12754, Keck Program ID C221OL) and J165315 (d: HST Program No. 13728, Keck Program ID H230N2L), are from the Keck Non-merger Comparison Sample.}
    \label{nonsubdust}
    
\end{figure*}

In several targets without detected near-infrared substructure in the Keck imaging, the HST images reveal significant nuclear obscuration that can plausibly explain the observed SDSS/WISE positional offsets. In these systems, the SDSS centroid is likely biased toward regions of unobscured optical emission, rather than tracing the true location of the nuclear infrared emission. As a result, dust obscuration can produce positional offsets even in the absence of multiple resolved infrared nuclei. Examples of such systems are shown in Figure~\ref{nonsubdust}, where the SDSS/WISE offsets are aligned with nuclear dust lanes and obscured regions.

\begin{figure*}
    \centering
    \stackinset{l}{2pt}{v}{4pt}{\color{white}\scalebox{1}{(a)}}{\includegraphics[width=0.16\linewidth]{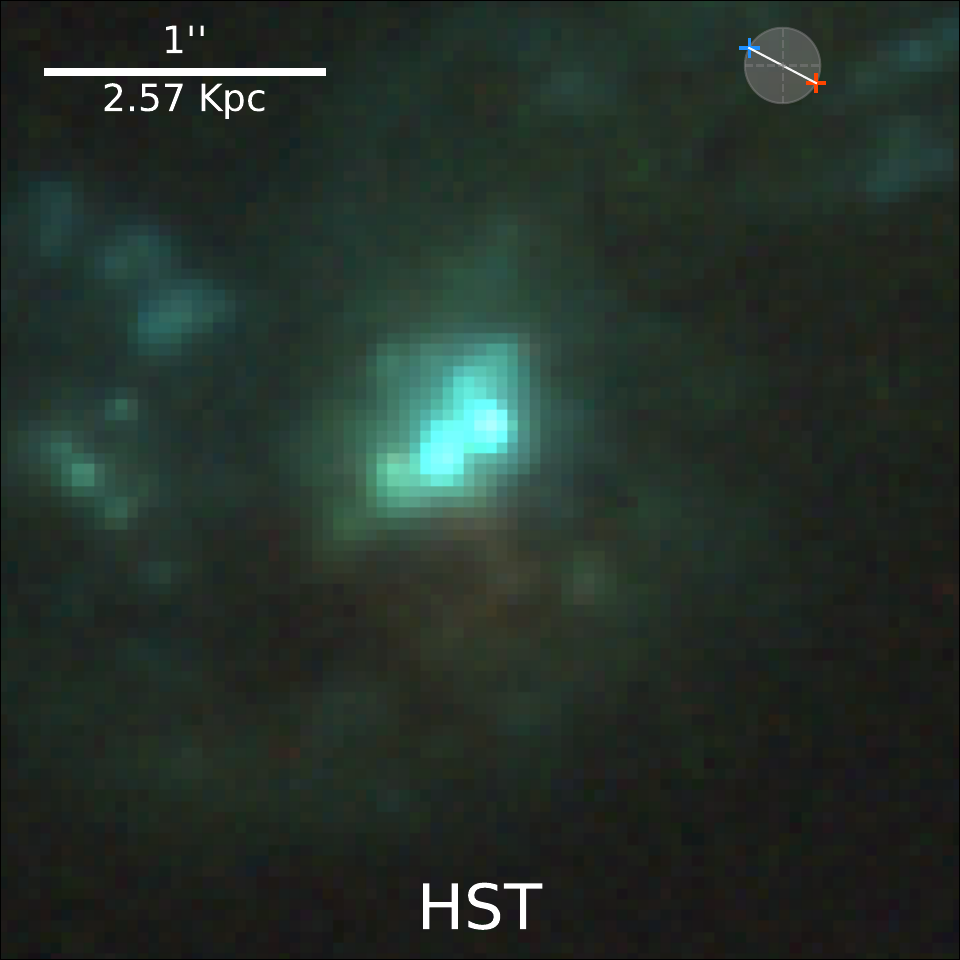}}\stackinset{l}{2pt}{v}{4pt}{\color{white}\scalebox{1}{}}{\includegraphics[width=0.16\linewidth]{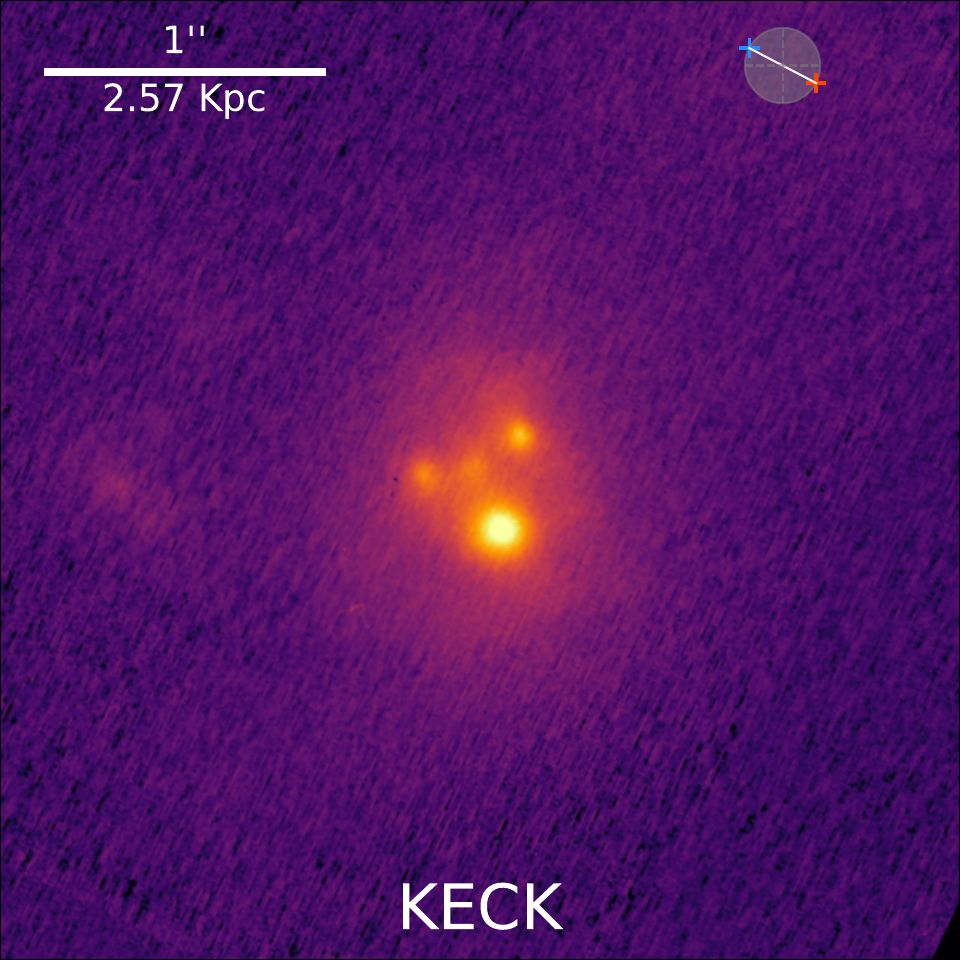}}\hspace{0.005\linewidth}\stackinset{l}{2pt}{v}{4pt}{\color{white}\scalebox{1}{(b)}}{\includegraphics[width=0.16\linewidth]{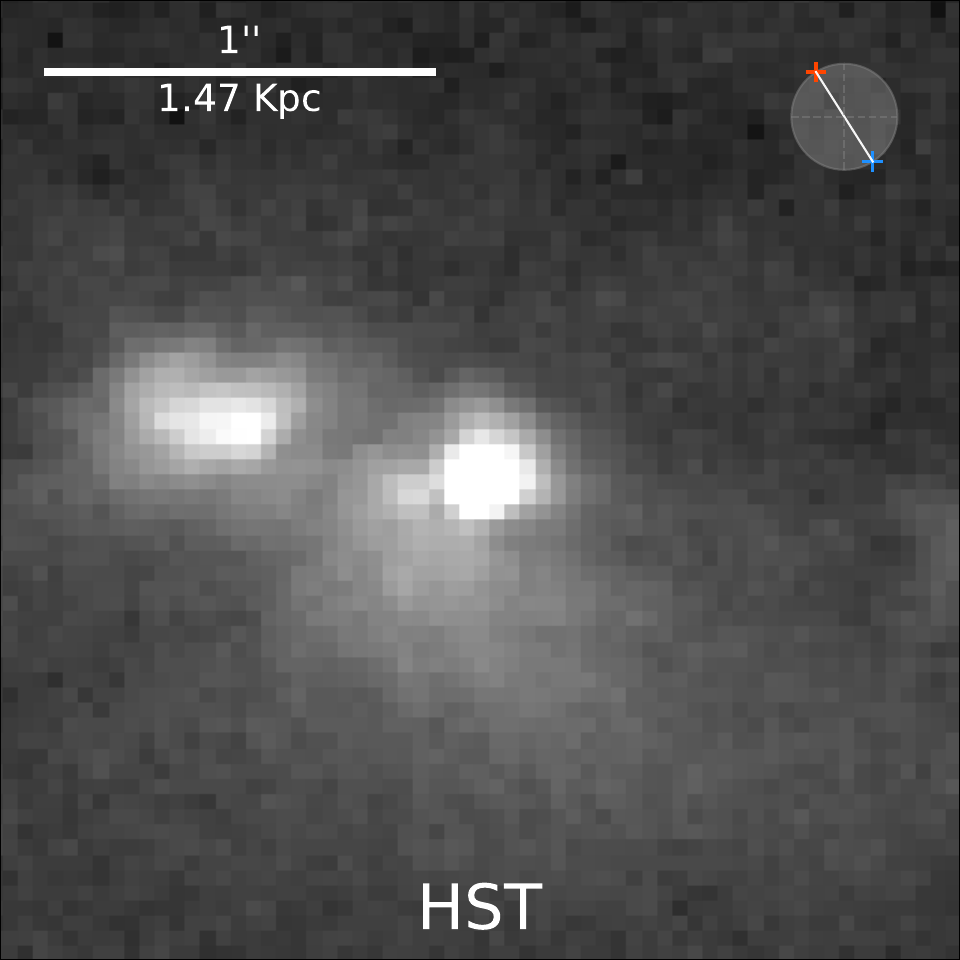}}\stackinset{l}{2pt}{v}{4pt}{\color{white}\scalebox{1}{}}{\includegraphics[width=0.16\linewidth]{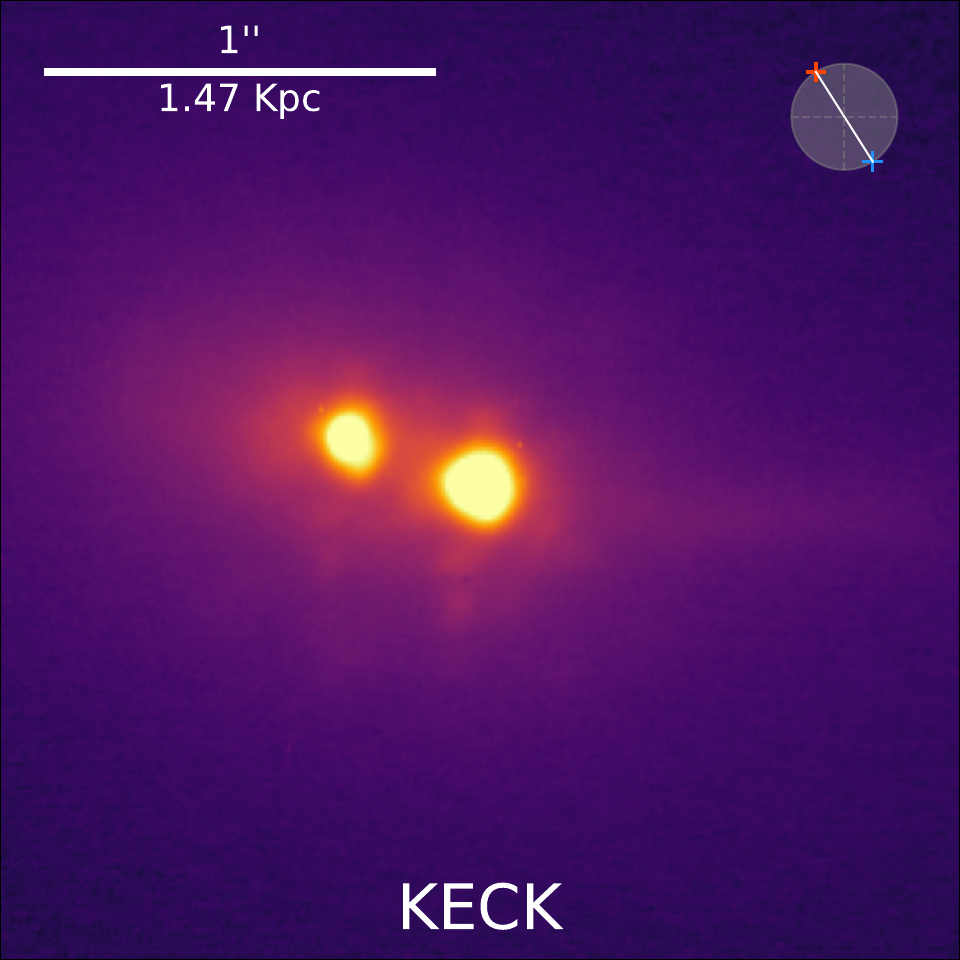}}\hspace{0.005\linewidth}\stackinset{l}{2pt}{v}{4pt}{\color{white}\scalebox{1}{(c)}}{\includegraphics[width=0.16\linewidth]{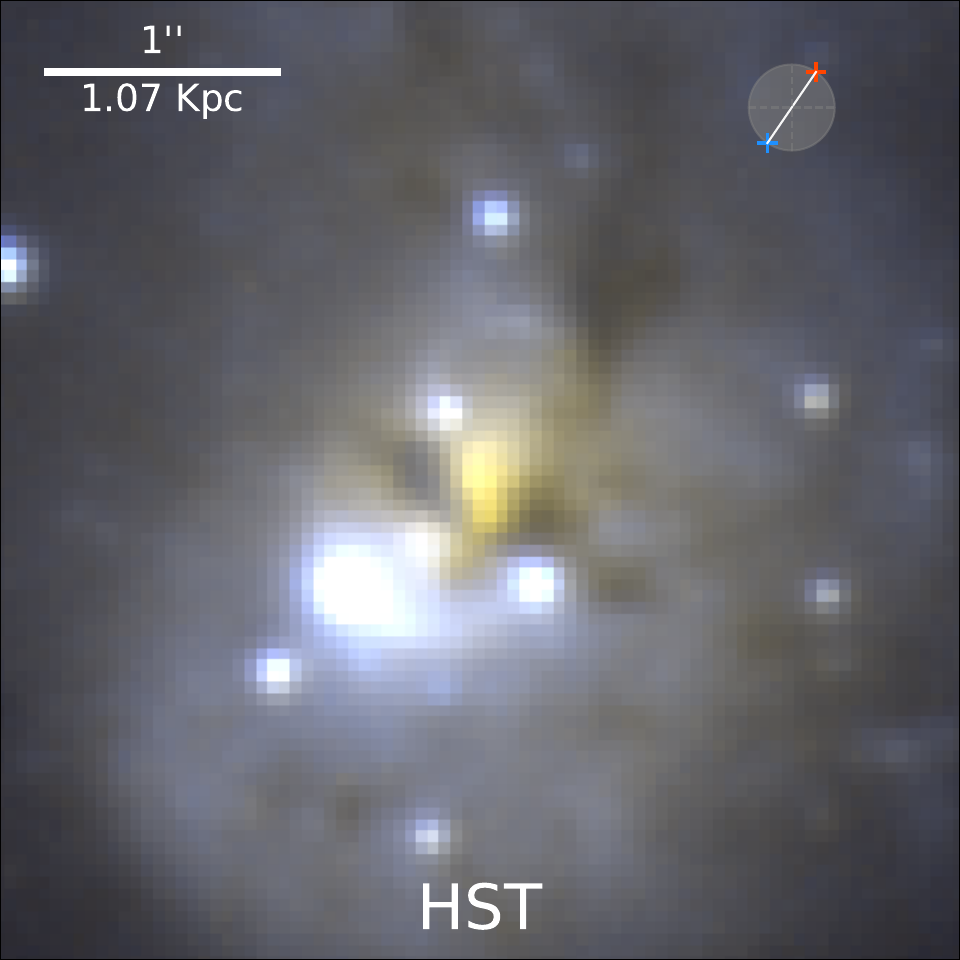}}\stackinset{l}{2pt}{v}{4pt}{\color{white}\scalebox{1}{}}{\includegraphics[width=0.16\linewidth]{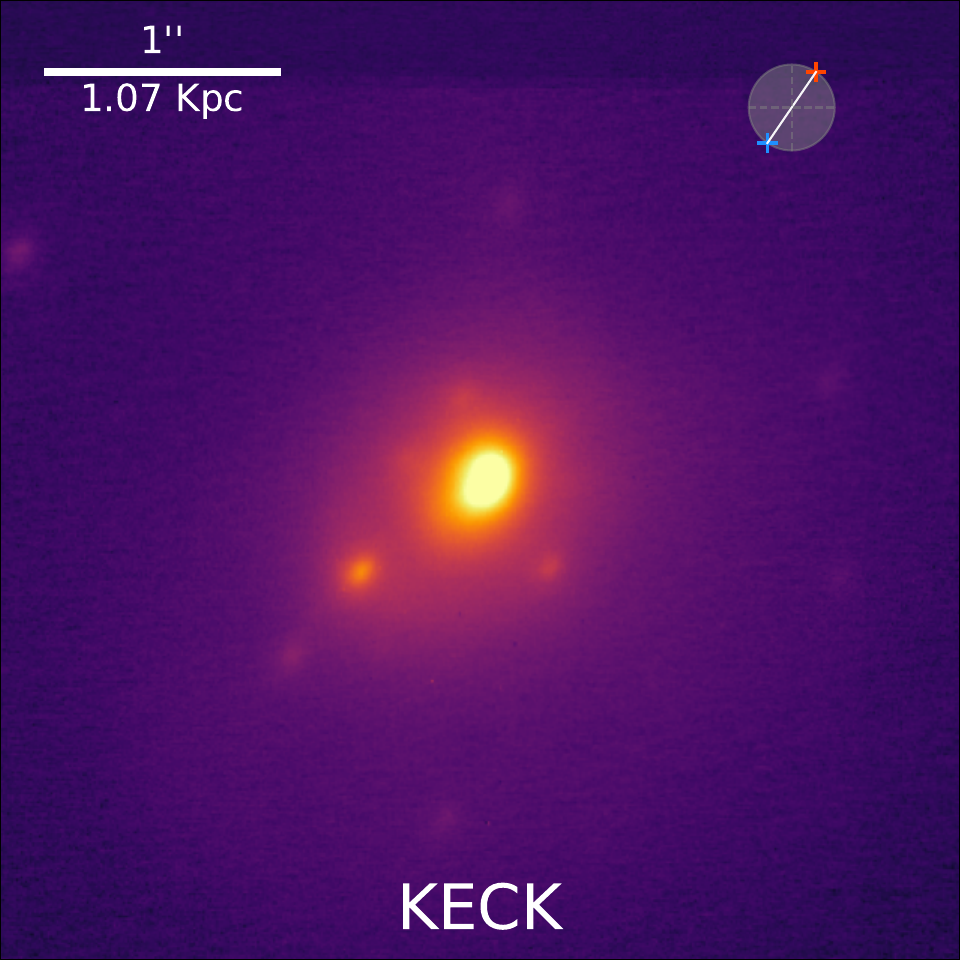}}
    \stackinset{l}{2pt}{v}{4pt}{\color{white}\scalebox{1}{(d)}}{\includegraphics[width=0.16\linewidth]{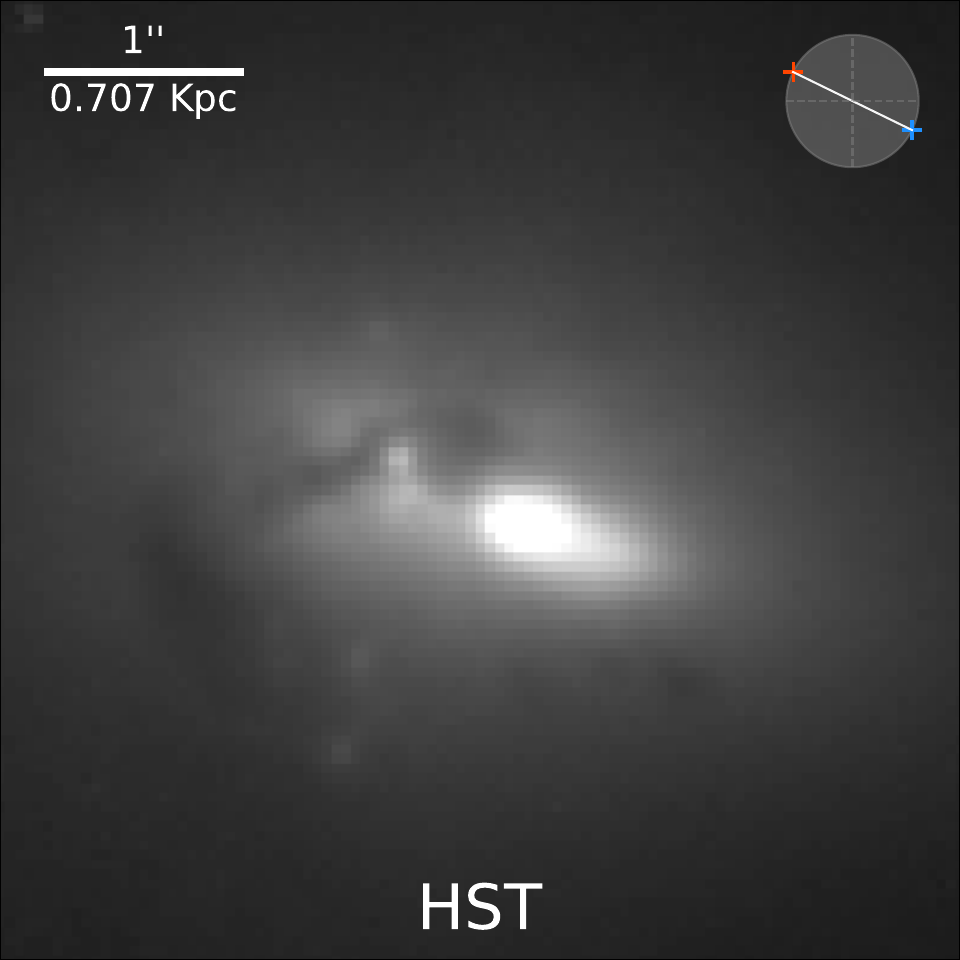}}\stackinset{l}{2pt}{v}{4pt}{\color{white}\scalebox{1}{}}{\includegraphics[width=0.16\linewidth]{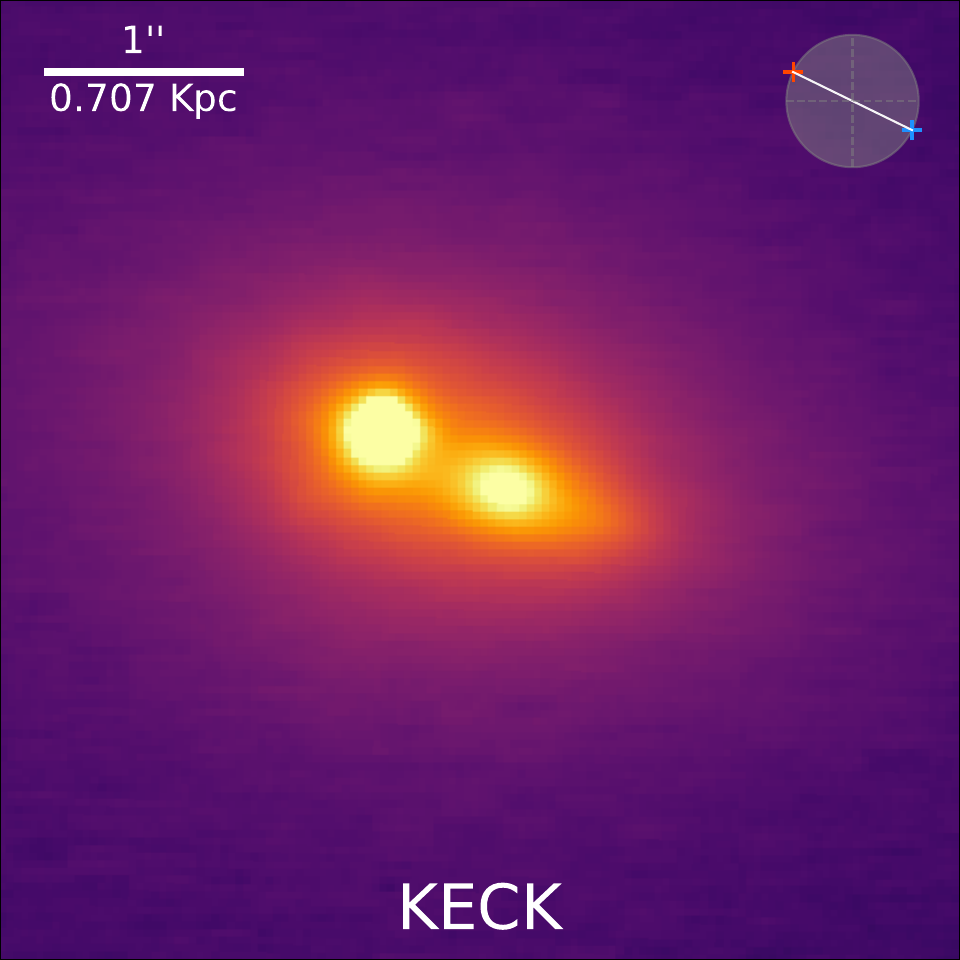}}\hspace{0.005\linewidth}\stackinset{l}{2pt}{v}{4pt}{\color{white}\scalebox{1}{(e)}}{\includegraphics[width=0.16\linewidth]{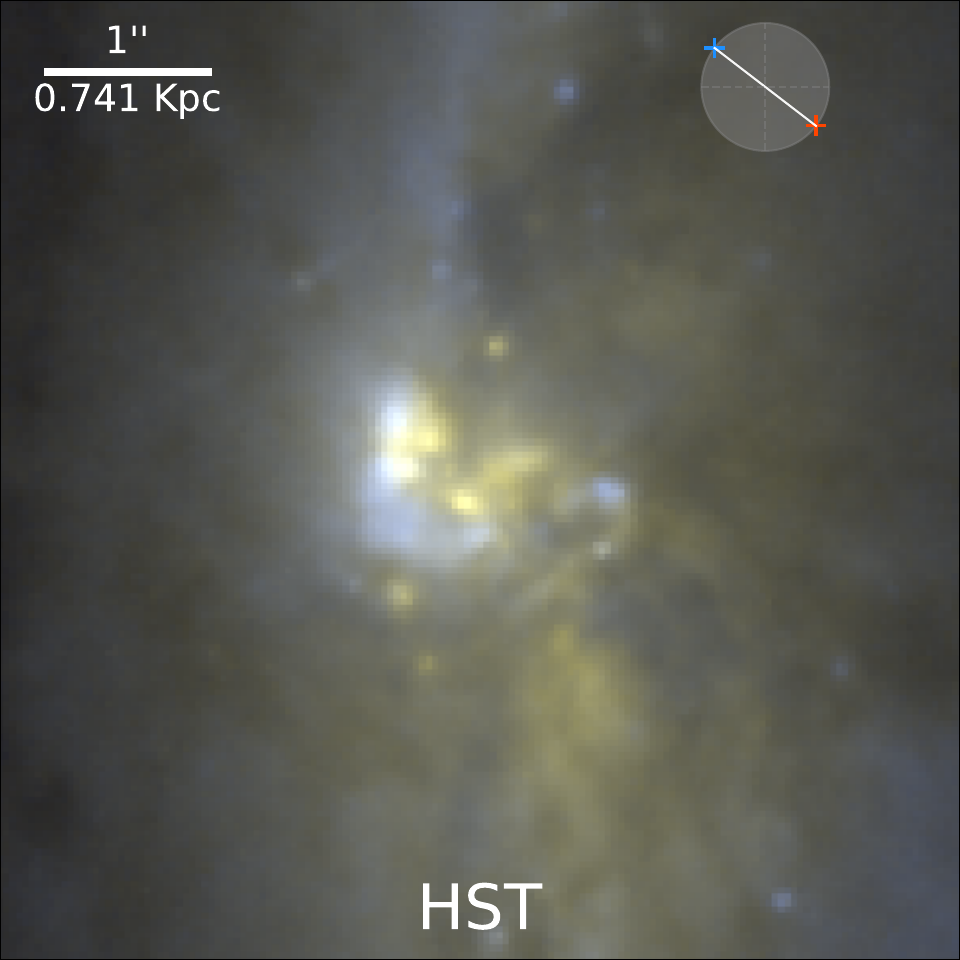}}\stackinset{l}{2pt}{v}{4pt}{\color{white}\scalebox{1}{}}{\includegraphics[width=0.16\linewidth]{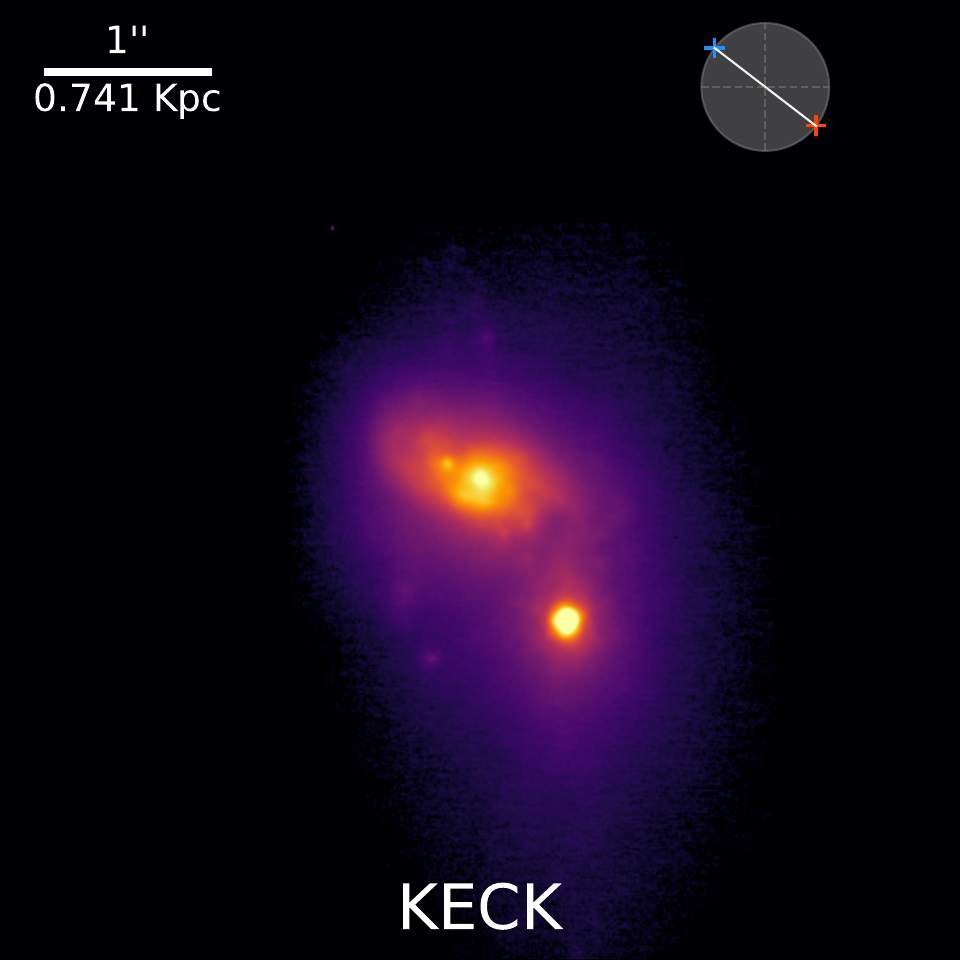}}\hspace{0.005\linewidth}\stackinset{l}{2pt}{v}{4pt}{\color{white}\scalebox{1}{(f)}}{\includegraphics[width=0.16\linewidth]{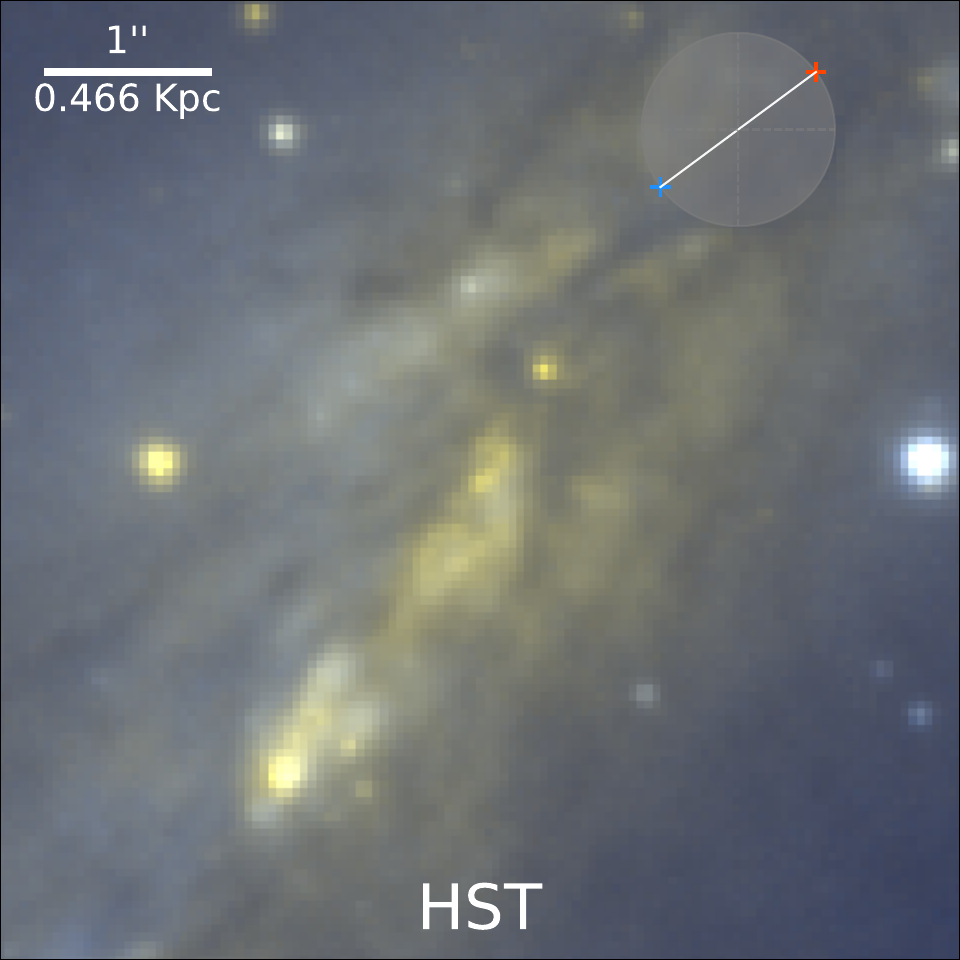}}\stackinset{l}{2pt}{v}{4pt}{\color{white}\scalebox{1}{}}{\includegraphics[width=0.16\linewidth]{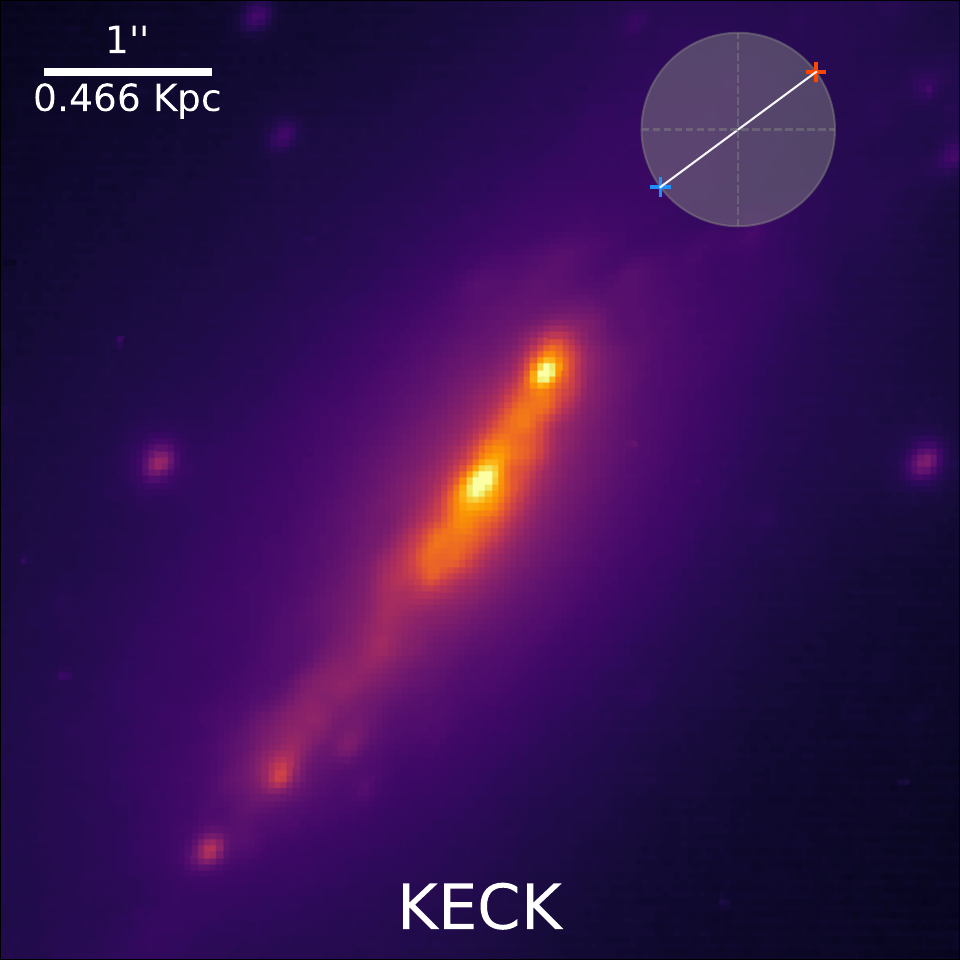}}
\caption{Examples of mergers with confirmed near-infrared substructure in Keck AO imaging but with one nucleus obscured or undetected in HST optical images. The SDSS/WISE positional offsets are consistent with one infrared nucleus being partially obscured at optical wavelengths, demonstrating that dust obscuration contributes to the observed offsets.}
\label{subdust}
\end{figure*}

Conversely, several targets with confirmed near-infrared substructure also exhibit strong nuclear obscuration in the optical HST images. In these systems, one of the near-infrared nuclei is partially or completely obscured at optical wavelengths. Examples are shown in Figure~\ref{subdust}, where the SDSS/WISE positional offset is consistent with one infrared nucleus being heavily obscured in the optical. This suggests that optical obscuration can contribute to the observed positional offset even when multiple genuine nuclei are present.

To further quantify the effect of obscuration, we identified 18 mergers from the Expanded Keck AO Merger Sample ($0.15'' < $ offset $ < 6''$) that have both Keck and HST imaging. For each system, we counted the number of nuclear sources detected in the near-infrared and optical images and computed the ratio of infrared to optical nuclear components.

\begin{figure}
    \includegraphics[width = \linewidth]{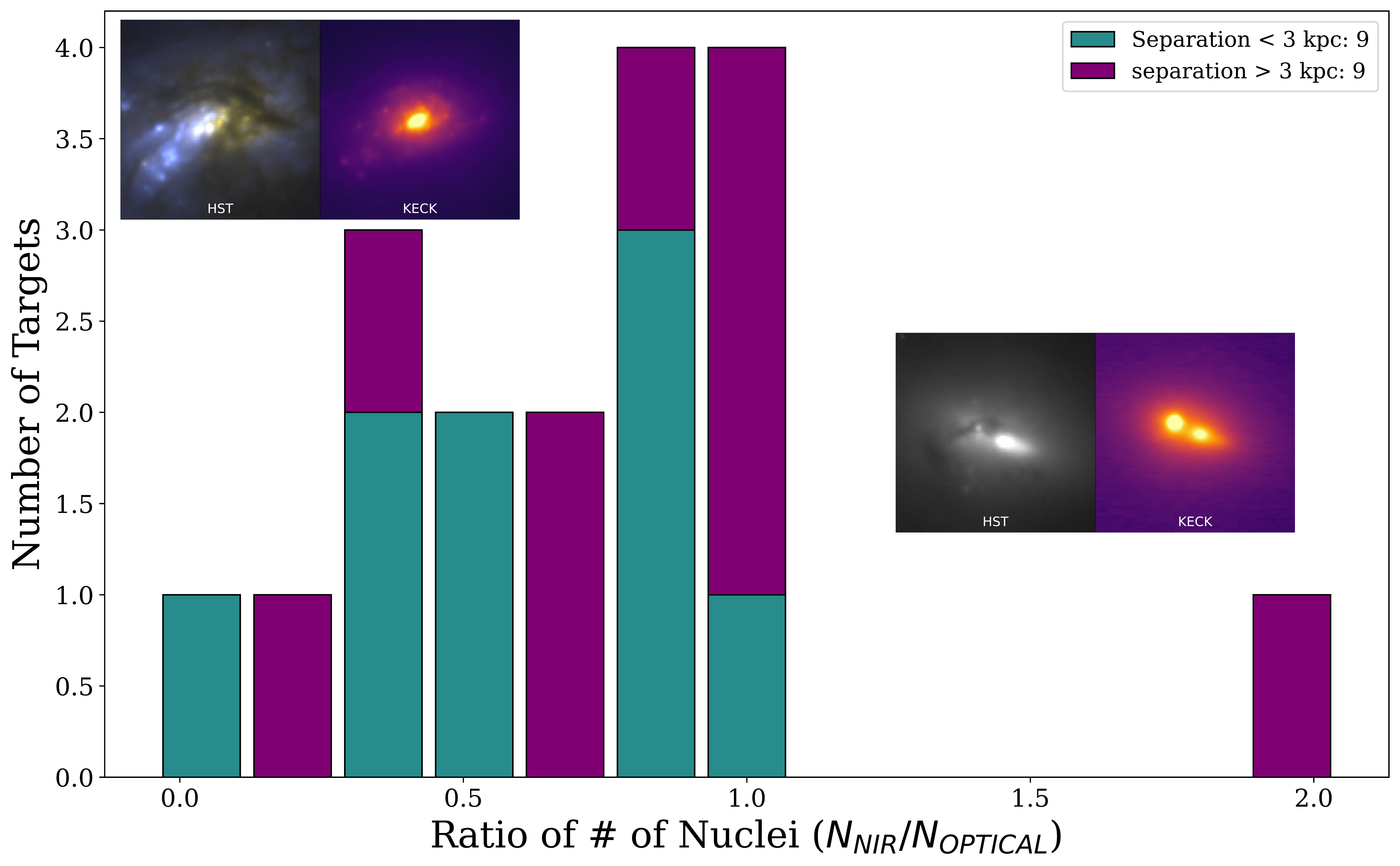}
\caption{
The distribution of the ratio of the number of apparent nuclear sources detected in the near-infrared (Keck NIRC2) to those detected in the optical (HST), separated based on merger separation (above and below 3~kpc) for targets in the Expanded Keck AO Merger Sample (53 systems) with both NIRC2 and HST imaging. This ratio quantifies the relative completeness of optical and near-infrared observations in identifying nuclear components. Ratios above unity indicate systems in which one or more nuclei detected in the near-infrared are obscured or unresolved in the optical, while ratios near or below unity indicate similar or greater numbers of optical sources. This figure tests whether optical imaging alone reliably identifies nuclear components across different merger stages. Example systems J131534 and J131517 are shown to illustrate cases with ratios above and below unity, respectively.}
    \label{iropticalratio}
\end{figure}

Figure~\ref{iropticalratio} shows the distribution of these ratios, separated by merger stage based on nuclear separation (above and below 3 kpc). We find that early-stage mergers (nuclear separations $>3$ kpc) more frequently exhibit ratios near unity, indicating similar nuclear structures in the optical and infrared. In contrast, late-stage mergers more commonly exhibit fewer detected infrared nuclei relative to apparent optical sources, consistent with stronger nuclear obscuration.

In many cases, dust lanes fragment the optical emission, producing apparent multiple optical components that do not correspond to distinct infrared nuclei. Conversely, genuine infrared nuclei may be entirely obscured at optical wavelengths. These results demonstrate that optical imaging alone may either miss genuine nuclear components or misidentify obscured structures.

\begin{figure}
    \includegraphics[width = \linewidth]{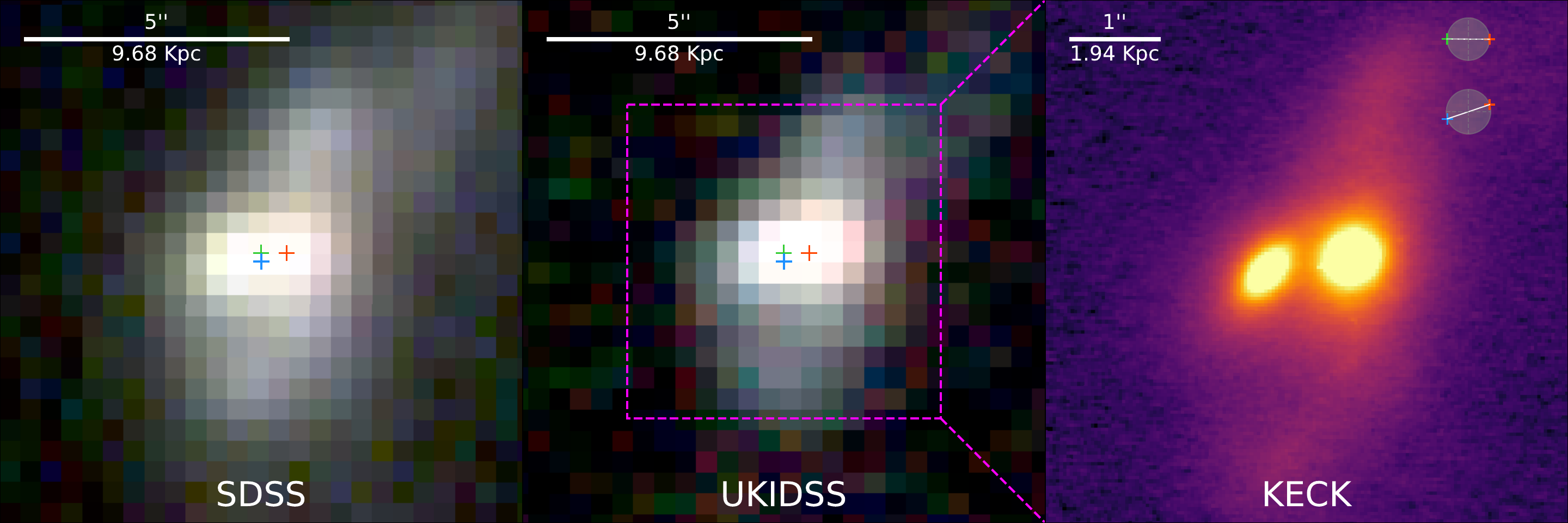}
    \caption{SDSS and Keck NIRC2 images of J001707, a system from the Primary Keck AO Merger Sample with sub-arcsecond offsets between its SDSS, UKIDSS, and WISE nuclear coordinates. The SDSS image shows the SDSS (blue), WISE (red), and UKIDSS (green) coordinates overlaid. In the top right of the NIRC2 image we include representations of the SDSS/WISE and UKIDSS/WISE offsets corresponding to their sizes and orientations. The Keck NIRC2 image reveals dual near-infrared nuclei unresolved in SDSS. Both offsets trace the orientation of the dual nuclei, demonstrating that astrometric offsets between surveys can identify unresolved nuclear structure. Because UKIDSS operates at near-infrared wavelengths and provides accurate absolute astrometry, its centroid is less affected by dust obscuration than SDSS, strengthening the interpretation that the observed offsets reflect genuine nuclear substructure rather than obscuration-induced centroid shifts, highlighting the importance of infrared observations for revealing compact nuclear components in late-stage galaxy mergers.
}

    \label{substructure_ukcoord}
\end{figure}

These findings provide important physical context for interpreting SDSS/WISE positional offsets. Optical obscuration can shift the SDSS centroid away from the true infrared nucleus, producing measurable positional offsets. However, infrared surveys such as UKIDSS and WISE are less affected by dust obscuration. As shown in Figure~\ref{substructure_ukcoord}, offsets measured between infrared surveys are more reliably associated with genuine dual nuclear components. This highlights the importance of near-infrared observations for identifying obscured nuclear substructure in studies of late stage galaxy mergers.

\subsection{Comparison to Other Works}\label{sec:results:papercomparisons}

To assess the effectiveness of our SDSS/WISE offset selection technique in identifying late-stage mergers with compact nuclear substructure, we compare our substructure-detection fraction with previous high-resolution near-infrared imaging studies of advanced mergers conducted with Keck AO and HST. To enable a consistent comparison across samples, we adopt a primary–secondary flux-ratio threshold of $>0.1$ and restrict the analysis to projected nuclear separations $<3$ kpc. We apply the flux-ratio threshold only to compact nuclear components identified in the near-infrared images, excluding diffuse tidal features or ambiguous substructure. This threshold is directly comparable to the criteria used by \citet{koss+2018}, who classified counterparts within 2.5 magnitudes of the primary nucleus, corresponding to a flux ratio of $\approx0.1$. We restrict the comparison to separations $<3$ kpc because the NIRC2 field of view limits our access to larger separations, ensuring that all samples are compared over a common and well-sampled physical separation range.

Under these criteria, our Primary Keck AO Merger Sample yields a substructure fraction of 34.7\%, substantially higher than reported in samples selected using alternative approaches. In comparison, the Swift/BAT hard X-ray–selected obscured AGN sample of \citet{koss+2018}, observed with Keck AO and HST NICMOS imaging, shows a substructure fraction of 17.6\%. Similarly, the adaptive-optics study of double-peaked [O~III] AGNs by \citet{fu+2011a}, which used GALFIT decomposition of near-infrared images to identify multiple nuclei, yields a substructure fraction of 8\%. We also include the (U)LIRG sample of \citet{haan+2011}, based on HST NICMOS imaging of infrared-luminous mergers, which shows a substructure fraction of 4.2\% over the same separation range. These comparison samples span a range of selection methods including hard X-ray emission, optical emission-line kinematics, and infrared luminosity, providing a broad context for evaluating the effectiveness of the SDSS/WISE offset selection technique.

We note, however, that this comparison carries important caveats. While \citet{koss+2018} employed a flux-ratio threshold equivalent to the one adopted here, and \citet{fu+2011a} measured nuclear flux ratios using GALFIT decomposition of adaptive-optics images, the photometric methodologies differ among studies. In our work, flux ratios are derived using aperture photometry centered on clearly resolved compact nuclei in the near-infrared images. In contrast, other studies employed two-dimensional surface-brightness modeling or morphological classifications to identify nuclear counterparts. For the \citet{haan+2011} ULIRG sample, individual nuclear flux ratios are not published, preventing us from applying an identical flux-ratio threshold; therefore, all morphologically identified nuclear pairs in their sample are included in our comparison. These methodological differences may introduce systematic uncertainties in the absolute substructure fractions. However, by restricting our analysis to compact nuclear components with flux ratios $>0.1$ and separations $<3$ kpc, we minimize differences in classification and ensure that the comparison focuses on physically comparable nuclear pairs.

\begin{figure}[]
\includegraphics[width = \linewidth]{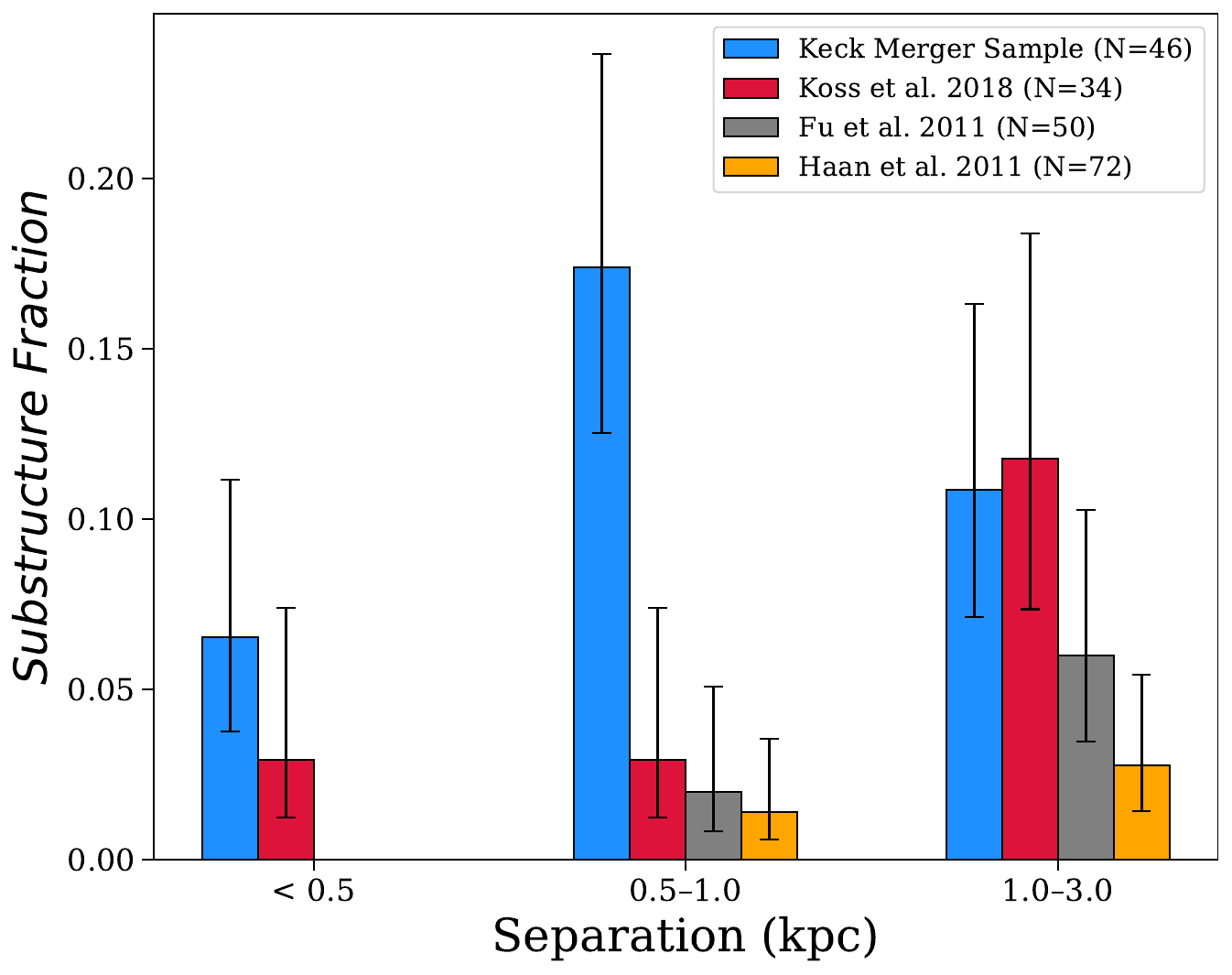}
\caption{Fraction of targets with compact nuclear substructure detected in high-resolution near-infrared imaging from this work and from \citet{koss+2018}, \citet{fu+2011a}, and \citet{haan+2011}, shown as a function of projected nuclear separation. To ensure a consistent comparison, only compact nuclear components with flux ratios $>0.1$ and separations $<3$ kpc are included for all samples where such measurements are available. For the \citet{haan+2011} sample, individual nuclear flux ratios are not published, and all morphologically identified nuclear pairs within this separation range are included. Error bars denote $1\sigma$ binomial confidence intervals. Despite differences in selection techniques and photometric methodologies, the SDSS/WISE offset–selected merger sample exhibits a higher fraction of compact nuclear substructure at separations below $1$ kpc, suggesting that the offset method is effective at identifying the most compact and obscured merger phases.}
\label{separationscomparison}
\end{figure}

Despite these limitations, the trend across all samples is suggestive. As shown in Figure~\ref{separationscomparison}, the SDSS/WISE-selected Primary Keck AO Merger Sample exhibits the highest substructure fraction at the smallest projected separations ($<0.5$ kpc and $<1$ kpc), where the effects of dust obscuration and nuclear coalescence are expected to be strongest. Unlike hard X-ray selected or emission-line–based selection methods, which depend on detecting radiative signatures from potentially obscured AGN activity, the SDSS/WISE offset technique is sensitive to centroid shifts caused by unresolved, spatially distinct sources of emission, allowing it to identify compact nuclear pairs even when they are heavily obscured or unresolved in individual surveys. Because late-stage mergers are expected to contain closely separated, heavily obscured nuclei whose emission contributes differently at optical and infrared wavelengths, centroid offsets between SDSS and WISE can potentially provide a direct observational signature of this phase, allowing the identification of compact nuclear pairs that trace the final stages of galaxy and SMBH coalescence.

While differences in photometric methodology, sample selection, and observational sensitivity prevent a strictly uniform comparison, the consistently higher substructure fraction observed in the SDSS/WISE-selected sample suggests that astrometric offset selection provides a powerful and complementary approach for identifying late-stage mergers containing closely separated nuclear components, particularly at sub-kpc separations where other selection methods may be incomplete.

\section{Summary and Conclusions}\label{conclusions}

The goal of this work was to test a novel selection technique that leverages the astrometric precision of SDSS and WISE to identify unresolved nuclear substructure in late-stage, MIR color-selected mergers using positional offsets between these two surveys. These systems probe a critical and poorly explored phase of galaxy evolution, where nuclear separations fall below $\sim$1 kpc and the central supermassive black holes are expected to form gravitationally bound pairs and eventually merge. Such systems are particularly important for understanding black hole growth, merger-driven fueling, and the formation of binary supermassive black holes, which are key targets for future gravitational-wave observatories such as LISA.

To evaluate the effectiveness of this method, we conducted high-resolution adaptive optics imaging of the Primary Keck AO Merger Sample, consisting of 46 Galaxy Zoo mergers with red WISE colors ($W_1 - W_2 > 0.5$; \citealt{blecha+2018}) and significant SDSS/WISE positional offsets ($0.15'' < $ offset $ < 1.5''$). Our primary objective was to determine whether this offset-based selection reliably identifies mergers containing compact nuclear substructure at sub-kpc scales.

The results of our analysis are summarized as follows:

\begin{enumerate}

\item \textbf{High substructure detection rate in the Primary Keck AO Merger Sample.}  
We detect unresolved near-infrared substructure in 20 of the 46 targets (43\%) in the Primary Keck AO Merger Sample. This demonstrates that SDSS/WISE positional offsets provide an efficient method for identifying mergers containing compact nuclear components unresolved in optical imaging. When restricting the analysis to compact nuclear components with flux ratios greater than 0.1 and separations below 3 kpc, consistent with criteria used in previous studies, our SDSS/WISE-selected sample exhibits a substructure fraction of 41.3\%, substantially higher than reported for hard X-ray--selected AGNs \citep{koss+2018}, double-peaked [O~III] emitters \citep{fu+2011a}, and ULIRGs \citep{haan+2011}. This comparison suggests that astrometric offset selection is particularly effective at identifying compact nuclear pairs at the smallest separations.

\item \textbf{SDSS/WISE offsets reliably trace genuine nuclear substructure.}  
Although the limited field of view of the NIRC2 images prevents direct absolute astrometric registration, we find a strong correlation between both the position angles and magnitudes of the SDSS/WISE offsets and the separations and orientations of the substructure pairs detected in the Keck imaging. Furthermore, the detection rate of substructure does not decline toward smaller offsets, demonstrating that the observed offsets are not dominated by astrometric noise but instead reflect genuine nuclear substructure over the full range of offsets explored.

\item \textbf{Optical obscuration contributes to SDSS/WISE positional offsets.}  
For targets in which no near-infrared substructure is detected, archival HST optical imaging reveals that many SDSS/WISE offsets arise from partial obscuration of the nuclear region. Dust lanes and obscured emission bias the SDSS optical centroid away from the true nucleus, while the infrared WISE centroid more accurately traces embedded emission. In several systems, nuclear components visible in the infrared are partially or completely absent in optical imaging, demonstrating that optical surveys alone can fail to identify closely separated nuclei in heavily obscured mergers.

\item \textbf{Substructure is preferentially associated with highly obscured systems.}  
Targets containing near-infrared substructure exhibit systematically higher extinction values, as measured by $E(B-V)$, compared to systems without detected substructure. This trend supports a physical picture in which the latest merger stages are characterized by high obscuration, which both facilitates nuclear inflows and obscures nuclear components at optical wavelengths.

\item \textbf{Comparison samples confirm the importance of both merger classification and astrometric offset selection.}  
We constructed a Keck Non-merger Comparison Sample of 40 galaxies selected using identical WISE color, magnitude, and SDSS/WISE offset criteria but not classified as mergers by Galaxy Zoo. We detect substructure in 25\% of these systems, demonstrating that SDSS/WISE offsets alone can reveal unresolved nuclear components even in systems lacking clear morphological merger signatures in optical imaging.

To isolate the role of the offset requirement itself, we also constructed a Non-offset Merger Comparison Sample consisting of mergers with identical WISE color and magnitude criteria but negligible SDSS/WISE offsets ($< 0.15''$). Only one system in this sample exhibits substructure, demonstrating that the offset requirement significantly enhances the efficiency of identifying unresolved nuclear components.

\item \textbf{Substructure is preferentially found in systems with signatures of nuclear activity.}  
Targets containing substructure are preferentially associated with AGN or composite classifications based on optical emission-line diagnostics and exhibit redder mid-infrared colors indicative of obscured nuclear activity. In particular, 89\% of targets with $W_2 - W_3 > 4.2$ exhibit near-infrared substructure, consistent with expectations that late-stage mergers are characterized by both enhanced accretion and obscured star formation activity.

\end{enumerate}

The combined optical and infrared observations presented in this work demonstrate that large-area optical surveys alone are not sufficient to identify compact nuclear pairs in heavily obscured mergers. Infrared imaging is essential for probing this critical stage of galaxy evolution. Future infrared missions, including the Nancy Grace Roman Space Telescope, will enable this astrometric offset technique to be applied over much larger volumes and to lower-mass systems, providing powerful new constraints on the late stages of galaxy mergers and the formation of binary supermassive black holes.

Follow-up spatially resolved spectroscopy is required to determine the nature of the individual nuclear components and confirm the presence of dual AGNs. Regardless of whether all systems host dual AGNs, identifying closely separated nuclei in late-stage mergers provides crucial insight into black hole growth, galaxy evolution, and the processes leading to gravitational-wave emission.

% \begin{acknowledgments}
% \acknowledgments
\section*{Acknowledgments}
We thank the anonymous referee for a careful and constructive report that significantly improved the clarity, organization, and presentation of this work. The referee’s detailed suggestions led us to strengthen the discussion of methodological caveats, streamline and reorganize the figures, clarify the interpretation of the photometric measurements and comparison samples, and improve the overall presentation of the results. We particularly appreciate the referee’s thoughtful comments regarding the separation of morphological classes, the treatment of flux ratios, and the framing of comparisons to previous studies, all of which helped us present the analysis in a clearer and more rigorous way. Camilo Vazquez acknowledges support from the Presidential Scholarship from George Mason University. R. W. P. gratefully acknowledges support through an appointment to the NASA Postdoctoral Program at Goddard Space Flight Center, administered by ORAU through a contract with NASA. Basic research in radio astronomy at the U.S. Naval Research Laboratory is supported by 6.1 Base Funding. J.M.C.’s work was supported by NASA under award number 80GSFC24M0006. F.M-S. acknowledges support from NASA through award 80NSSC23K1529. J.D.M's research was supported by an appointment to the NASA Postdoctoral Program at the NASA Goddard Space Flight Center, administered by Oak Ridge Associated Universities under contract with NASA.

This research made use of Astropy,\footnote{\url{http://www.astropy.org}} a community-developed core Python package for Astronomy \citep{2013A&A...558A..33A}, as well as \textsc{topcat} \citep{2005ASPC..347...29T}. The geometrical distortion correction code was run on ARGO and HOPPER, research computing clusters provided by the Office of Research Computing at George Mason University, VA. (\url{ http://orc.gmu.edu}).

Some of the data presented herein were obtained at Keck Observatory, which is a private 501(c)3 non-profit organization operated as a scientific partnership among the California Institute of Technology, the University of California, and the National Aeronautics and Space Administration. The Observatory was made possible by the generous financial support of the W. M. Keck Foundation. The authors wish to recognize and acknowledge the very significant cultural role and reverence that the summit of Maunakea has always had within the Native Hawaiian community. We are most fortunate to have the opportunity to conduct observations from this mountain. This research has made use of the Keck Observatory Archive (KOA), which is operated by the W. M. Keck Observatory and the NASA Exoplanet Science Institute (NExScI), under contract with the National Aeronautics and Space Administration.

Funding for the Sloan Digital Sky Survey (SDSS) and SDSS-II has been provided by the Alfred P. Sloan Foundation, the Participating Institutions, the National Science Foundation, the U.S. Department of Energy, the National Aeronautics and Space Administration, the Japanese Monbukagakusho, and the Max Planck Society, and the Higher Education Funding Council for England. The SDSS Web site is http://www.sdss.org/. The SDSS is managed by the Astrophysical Research Consortium (ARC) for the Participating Institutions. The Participating Institutions are the American Museum of Natural History, Astrophysical Institute Potsdam, University of Basel, University of Cambridge, Case Western Reserve University, The University of Chicago, Drexel University, Fermilab, the Institute for Advanced Study, the Japan Participation Group, The Johns Hopkins University, the Joint Institute for Nuclear Astrophysics, the Kavli Institute for Particle Astrophysics and Cosmology, the Korean Scientist Group, the Chinese Academy of Sciences (LAMOST), Los Alamos National Laboratory, the Max-Planck-Institute for Astronomy (MPIA), the Max-Planck-Institute for Astrophysics (MPA), New Mexico State University, Ohio State University, University of Pittsburgh, University of Portsmouth, Princeton University, the United States Naval Observatory, and the University of Washington. 

The HST data presented in this paper were obtained from the Mikulski Archive for Space Telescopes (MAST) at the Space Telescope Science Institute. The specific observations analyzed can be accessed via \dataset[doi:10.17909/dazg-en96]{doi:10.17909/dazg-en96}. STScI is operated by the Association of Universities for Research in Astronomy, Inc., under NASA contract NAS5–26555. Support to MAST for these data is provided by the NASA Office of Space Science via grant NAG5–7584 and by other grants and contracts.

This publication makes use of data products from the Wide-field Infrared Survey Explorer, which is a joint project of the University of California, Los Angeles, and the Jet Propulsion Laboratory/California Institute of Technology, funded by the National Aeronautics and Space Administration.

The UKIDSS project is defined in \cite{lawrence+2007}. UKIDSS uses the UKIRT Wide Field Camera \citep[WFCAM;][]{casali+2007} and a photometric system described in \cite{hewett+2006}. The pipeline processing and science archive are described in \citet{irwin+2008} and \citet{hambly+2008}. We have used data from the 9th data release, which is described in detail in \citet{lawrence+2007}.

% \end{acknowledgments}

% \facilities{Sloan, WISE, Keck}

\software{
% Astropy \citep{2013A&A...558A..33A}, 
Astropy \citep{astropy:2013, astropy:2018, astropy:2022}
NumPy,\citep{2020Natur.585..357H}
SciPy,\citep{2020NatMe..17..261V}
% \textit{emcee} \citep{2013PASP..125..306F},
% pPXF \citep{2017MNRAS.466..798C},
\textsc{topcat} \citep{2005ASPC..347...29T},
% \textsc{badass} \citep{sexton_2020},
% BIFR\"{O}ST (\href{https://github.com/Michael-Reefe/bifrost}{https://github.com/Michael-Reefe/bifrost})
}

% \nocite{1988AJ.....95...45A}
% \nocite{2018ApJ...861..142C}

%\end{acknowledgements}

\clearpage
% \begin{comment}
\twocolumngrid

\appendix
\renewcommand{\thefigure}{\thesection\arabic{figure}}
\makeatletter
\@addtoreset{figure}{section}
\makeatother

\section{Comparison of Sample Properties Between Primary Keck AO Merger Sample and Keck Non-merger Comparison Sample}\label{A}
\begin{figure*}[htbp]
    \centering
    \includegraphics[width=0.48\linewidth]{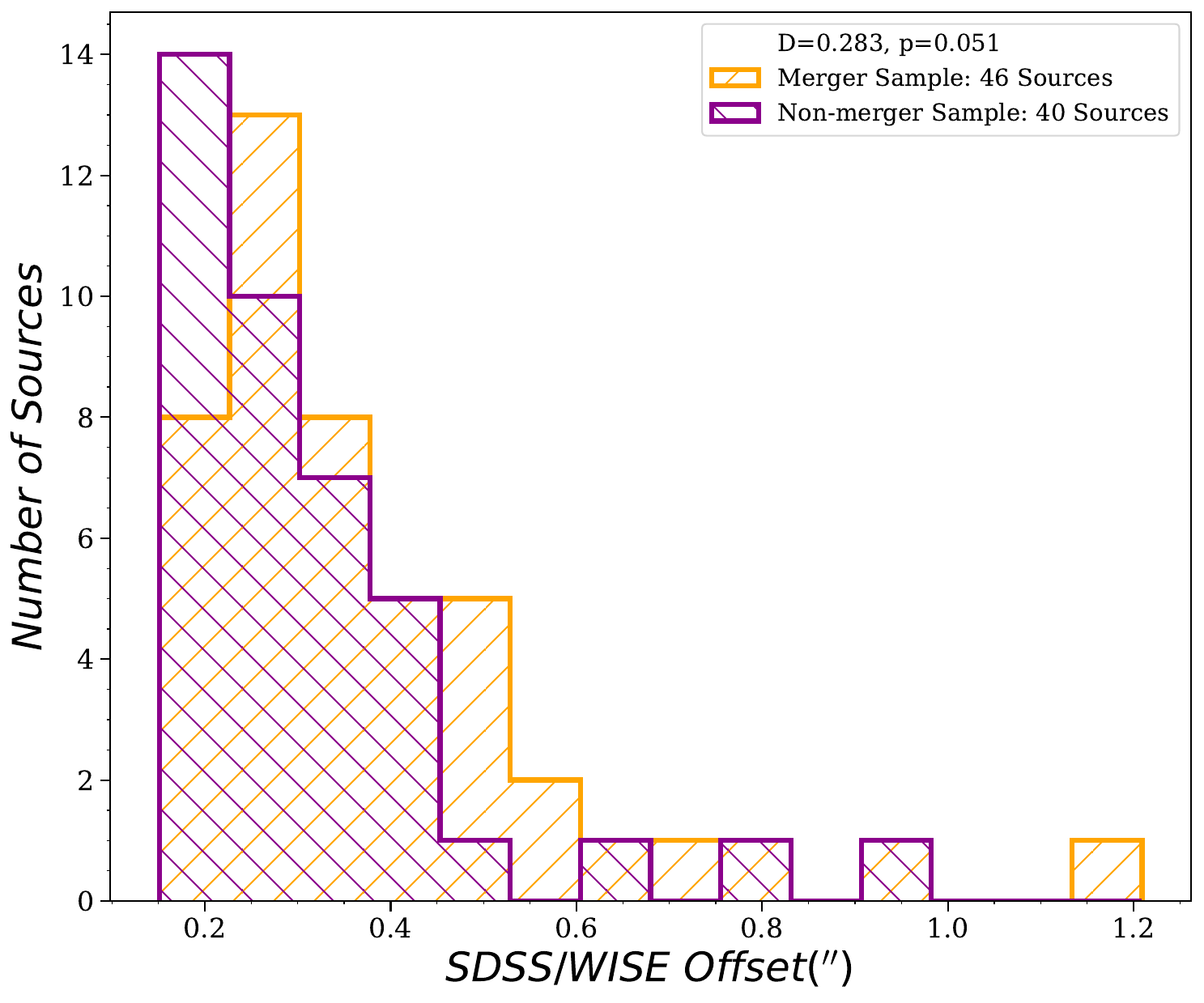}
    \hfill
    \includegraphics[width=0.48\linewidth]{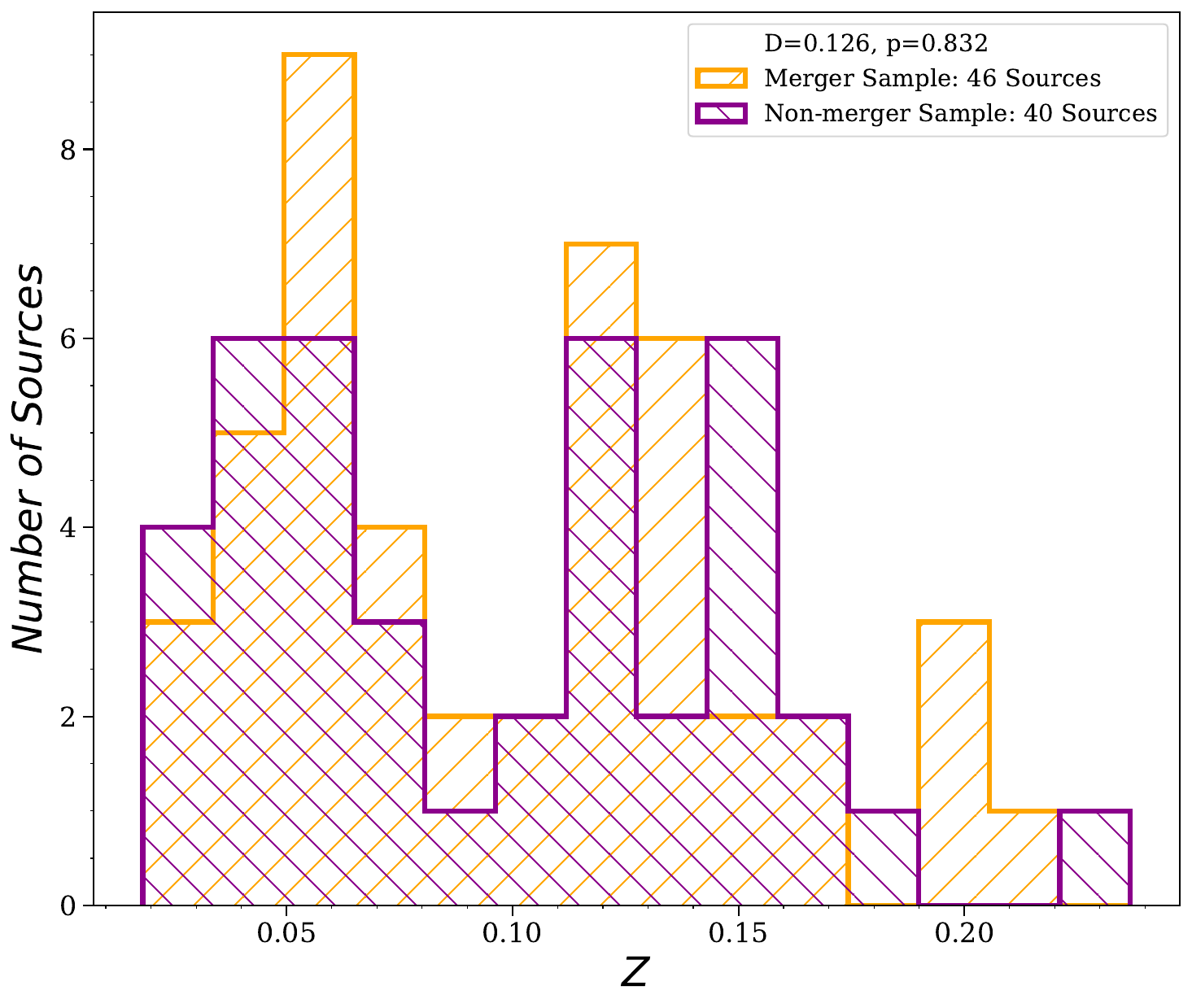}
    \caption{Comparison of the SDSS/WISE offset (left) and redshift (right) distributions for the Primary Keck AO Merger Sample (46 systems) and the Keck Non-merger Comparison Sample (40 systems with archival Keck AO imaging). This figure demonstrates that the comparison sample is well matched in both redshift and SDSS/WISE offset, ensuring that differences in substructure detection cannot be attributed to systematic differences in these parameters. The Kolmogorov–Smirnov statistic ($D$) and corresponding $p$-value for each comparison are listed in the legend.}
    \label{appendix:samplecomparison}

\end{figure*}

\section{Substructure Targets}\label{B}
In the following section, we present the results and analysis for all 20 targets with substructure detected with NIRC2 imaging. For each target, we show its SDSS \emph{gri} image, NIRC2 image, and when available, its UKIDSS \emph{YHK} image and/or HST image; for targets with HST observations in multiple filters we show a three color image. Each image is aligned with North pointing up and East to the left. For each SDSS and UKIDSS image we include a white scale bar -- corresponding to 5$''$-- in the top left and overlay the system's SDSS coordinate in blue and WISE coordinate in red. We also include a magenta square in either/both the SDSS and UKIDSS images that corresponds to the size and location of the Keck image cutout and HST image cutout when available. In each NIRC2 and HST image we include a white scale bar -- corresponding to 1$''$ -- in the top left and a visual representation of the SDSS/WISE offset in the top right, corresponding to the distance between the SDSS (blue) and WISE (red) coordinates and their orientation relative to each other. We do not present the location of the SDSS and WISE coordinates in these images because there are no stars within the FOV of the images and thus we are unable to spatially map between detector pixels and sky coordinates. 
\clearpage

\subsection{J001707+011506}
\begin{figure}[htbp]
\centering
\includegraphics[width = \linewidth]{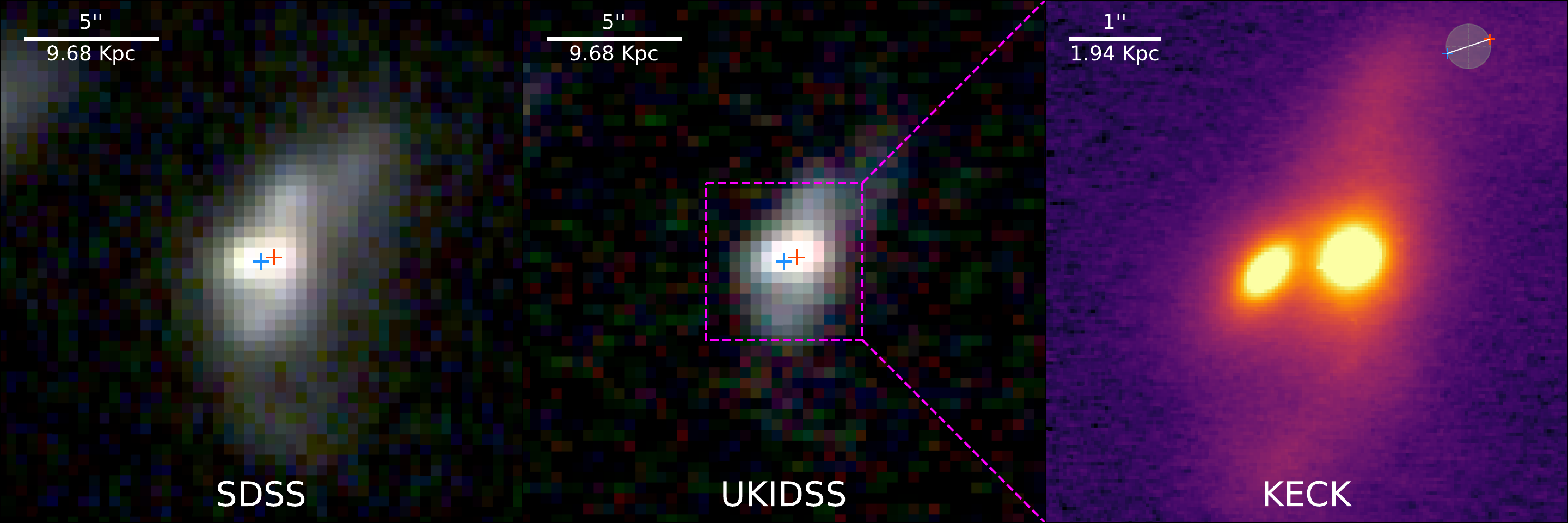}

\caption{Images provided above are described at the beginning of Appendix \ref{B}.}
\label{B:J001707}
\end{figure}

NIRC2 imaging of J001707 reveals that it contains clear dual nuclei separated by 0.986$''$ or 1.910 kpc with a PA of 99.51\degree. In both SDSS and UKIDSS these nuclei are blended into a single apparent nucleus elongated along the direction of the two nuclei. The host galaxy exhibits bipolar tidal tails extending out to $\sim$12.4 kpc in the North and $\sim$5.6 kpc in the South. The SDSS/WISE offset detected in this target has a separation of 0.511$''$ and PA of 108.6\degree. The SDSS/WISE offset detected in this merger system clearly corresponds to the dual nuclei as the PAs of the SDSS/WISE offset and nuclear pair are very similar, although the separation of the nuclei is under predicted by the SDSS/WISE offset (see Section \ref{sec:results:offsetvsepvpa}; $\Delta Sep. = 0.475''$ and $\Delta PA = 9.09\degree$). The host galaxy of this object is optically classified as a $K_{01}$ type AGN in the MPA-JHU catalog. Additionally, it lies within the AGN classification of our BPT diagram and within the Jarrett AGN region of our WISE color-color diagram suggesting it is a candidate dual AGN.

\subsection{J010942-004417}
\begin{figure}[htbp]
% \centering
\includegraphics[width = \linewidth]{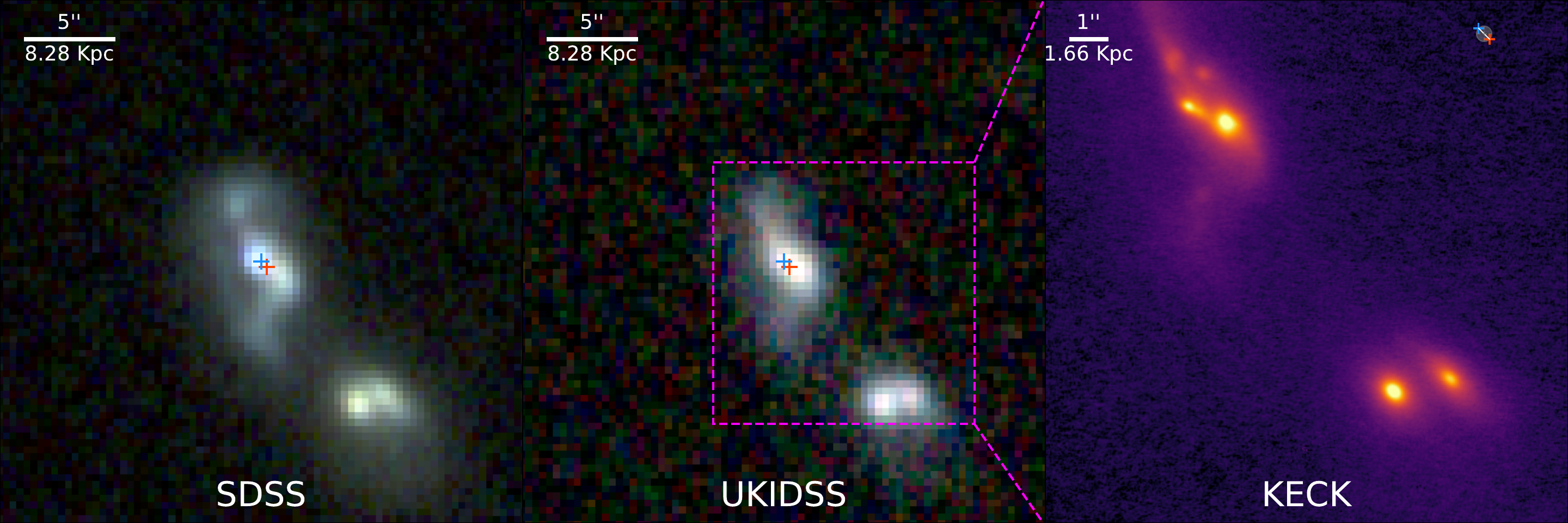}

\caption{Images provided above are described at the beginning of Appendix \ref{B}.}

\end{figure}

J010942 appears to be a highly disturbed galaxy undergoing a merger with a companion 10$''$ or 16.57 kpc to the Southwest. J010942 appears to contain a single elongated nucleus in both SDSS and UKIDSS whereas the companion contains two semi-resolved nuclei. High-resolution NIRC2 imaging of this target reveals two nuclei and confirms the presence of two nuclei in the companion. Inspection of SDSS DR16 spectra of J010942 and its companion reveals a velocity difference of $\sim$70 $km/s$ suggesting that they are associated with each other, although higher resolution spectroscopic follow-up is required for confirmation (McDonald et al. in prep.). The two nuclei in J010942 are separated by 1.129$''$ or 1.871 kpc with a PA of 246.3\degree. The SDSS/WISE offset detected in the target has a separation of 0.443$''$ and PA of 225\degree\  which under predicts the separation of the dual nuclei (see Section \ref{sec:results:offsetvsepvpa}) but more closely follows the orientation of the nuclei (i.e.  $\Delta Sep. = 0.686''$ and $\Delta PA = 21.30\degree$). Comparison between the SDSS and UKIDSS cutout images of this target shows that the apparent nucleus of the optical and near-infrared emission of this target are clearly offset from each other and thus driving the SDSS/WISE offset. However, high resolution optical imaging is required to determine whether this discrepancy is caused by partial obscuration or an inherent difference in the emission properties of the two nuclei. The host galaxy of J010942 is optically classified as a $K_{03}$ type AGN in the MPA-JHU catalog. Additionally, it lies within the composite classification of our BPT diagram and outside of the Jarrett AGN region of our WISE color-color diagram which may suggest that these two nuclei are not dual AGN but rather a single AGN and a non-AGN (i.e. inactive-SMBH or star forming region).

\subsection{J015028+130858}
\begin{figure}[htbp]
\centering
\includegraphics[width = 0.6666\linewidth]{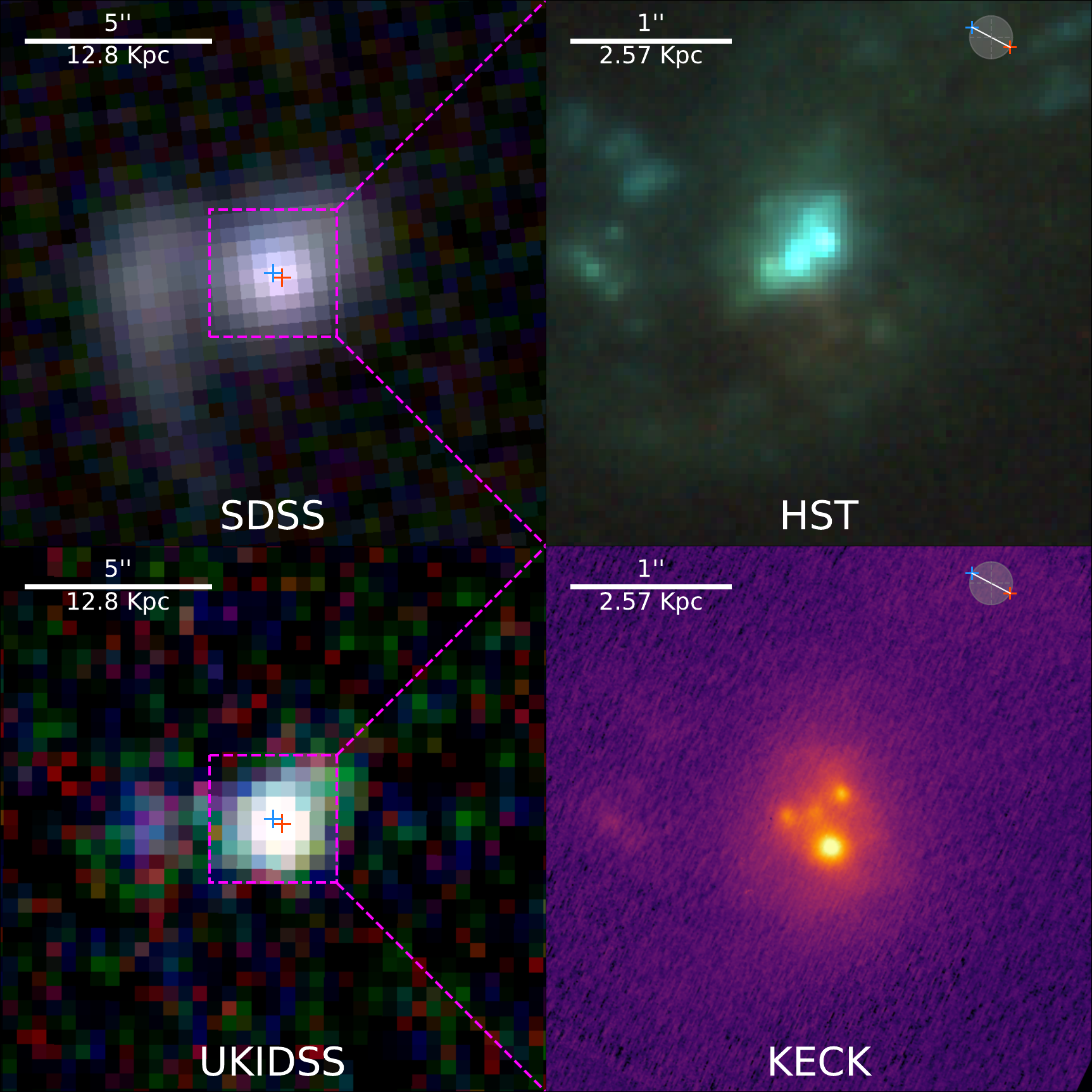}

\caption{Images provided above are described at the beginning of Appendix \ref{B}.}

\end{figure}

J015028 is a merger system with a galaxy pair in the East and West of the SDSS cutout image, only the Western galaxy is visible in UKIDSS, separated by 3.742$''$ or 9.60 kpc. We detect an SDSS/WISE offset in this Western galaxy with a separation of 0.275$''$ and PA of 62.35\degree\ and thus consider it to be the target galaxy for this analysis. High resolution NIRC2 imaging of the target reveals three sources of significant near-infrared emission in the central nuclear region forming a triangle with a bright southern source and two dimmer sources to the Northeast and Northwest. Inspection of the archival HST WFC3 imaging of this target (Program No. 12310; F390W, F475W and F850LP) reveals that the brightest source in the near-infrared is not visible at optical wavelengths, most likely due to obscuration from dust. Based on the orientation of the substructure in the near-infrared and optical it can also be seen that there is a fourth source of emission seen in the HST image which corresponds to the region between the two Northern sources in the NIRC2 image. The orientation of the SDSS/WISE offset visually suggests that the WISE coordinate is tracing the position of the bright Southern source of near-infrared emission and the SDSS coordinate is tracing the unobscured optical emission in the North, creating the SDSS/WISE offset we detect. Our comparison of the separation and PA of the SDSS/WISE offset with the substructure in this target yielded very similar separations but a significant difference in PA (i.e.  $\Delta Sep. = 0.054''$ and $\Delta PA = 106.8\degree$). This is due to the fact that the substructure pair is defined based on the brightest secondary source of near-infrared emission, which in this case is not necessarily the brightest secondary source of optical emission. The host galaxy of J015028 is optically classified as a $S_{06}$ type AGN in the MPA-JHU catalog. Additionally, it lies within the star-forming classification of our BPT diagram and outside of the Jarrett AGN region of our WISE color-color diagram which may suggest that the compact sources of emission in this target are star forming regions rather than galaxy nuclei or AGN.

\subsection{J082312+275140}
\begin{figure}[htbp]
\centering

\includegraphics[width = 0.6666\linewidth]{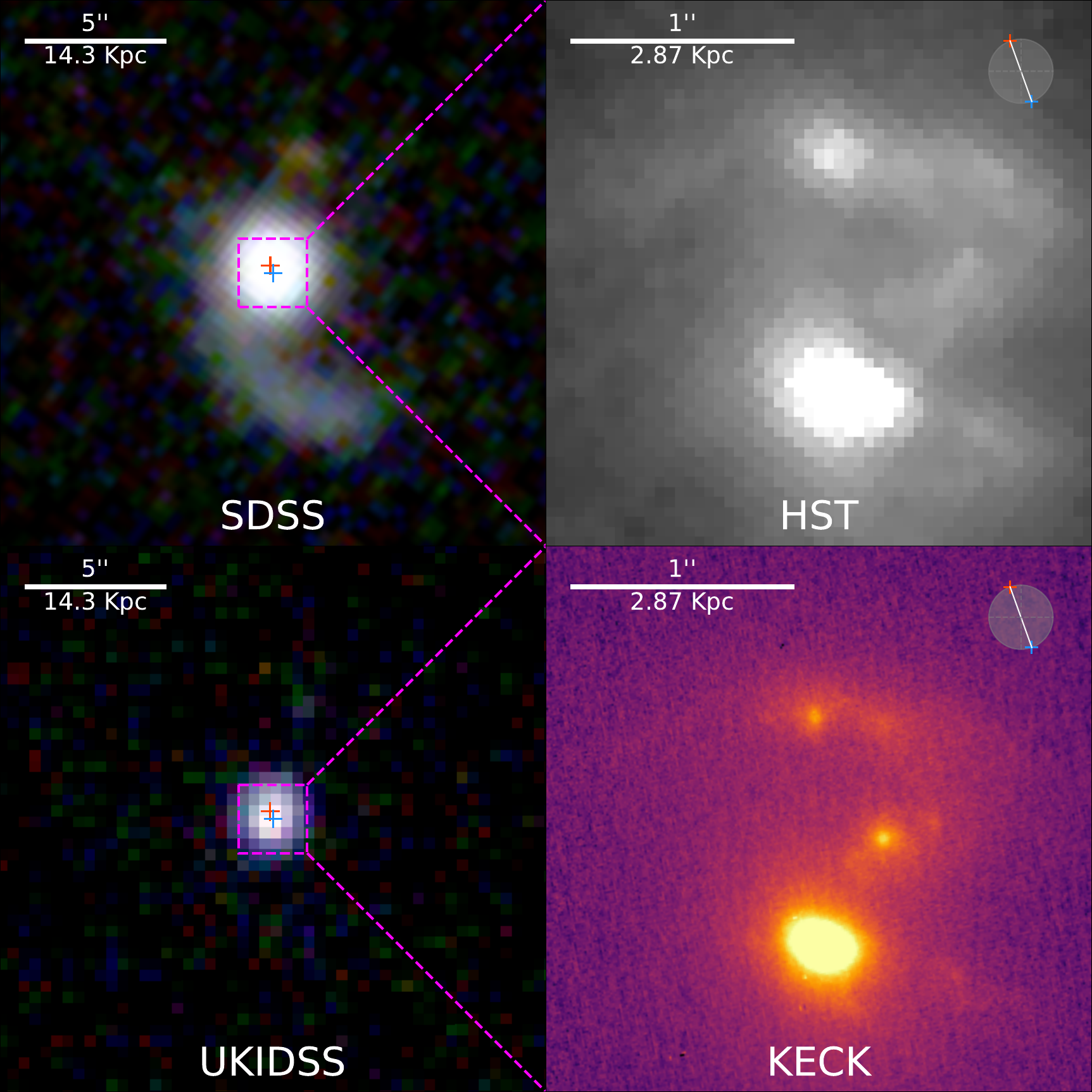}

\caption{Images provided above are described at the beginning of Appendix \ref{B}.}

\end{figure}
NIRC2 imaging of J082312 reveals three potential nuclei connected by an arc of emission. There is a clear primary nucleus and two secondary nuclei, northwest from the primary at a distance of 0.56$''$ and north from the primary at a distance of 1.06$''$ with PA's of 331.030\degree\ and 2.440\degree\ respectively. The host galaxy, as seen in SDSS, exhibits a tidal tail that extends out to $\sim 25$ kpc and curves towards the West, however in UKIDSS it is a point source. The HST WFPC2 imaging survey conducted by \cite{Borne+1999} included F814w observations of J082312. The same arc of emission and broad tidal tail can be seen in the HST image, however only two of the three nuclei are visible with a pair separation of $\sim1''$, indicating that the Northwestern nucleus is obscured. The Northwestern nucleus is the brightest of the secondary nuclei, thus we identify the primary and this secondary nucleus as the substructure pair in this target. The SDSS/WISE coordinate offset has a separation of 0.294$''$ and PA of 199.741\degree. Although the orientation of the SDSS/WISE offset does not seem to trace that of the substructure pair, it appears to have a similar orientation to that of the more separated secondary. The host galaxy of this object is optically classified as a $K_{03}$ type AGN in the MPA-JHU catalog. Additionally, it lies within the composite classification of our BPT diagram and outside of the Jarrett AGN region of our WISE color-color diagram.% suggesting that this is a single AGN system with two bright nuclear star formation regions.
% \vspace{-0.001\linewidth}
\clearpage

\subsection{J091529+090601}
\begin{figure}[htbp]
\centering
\includegraphics[width=\linewidth]{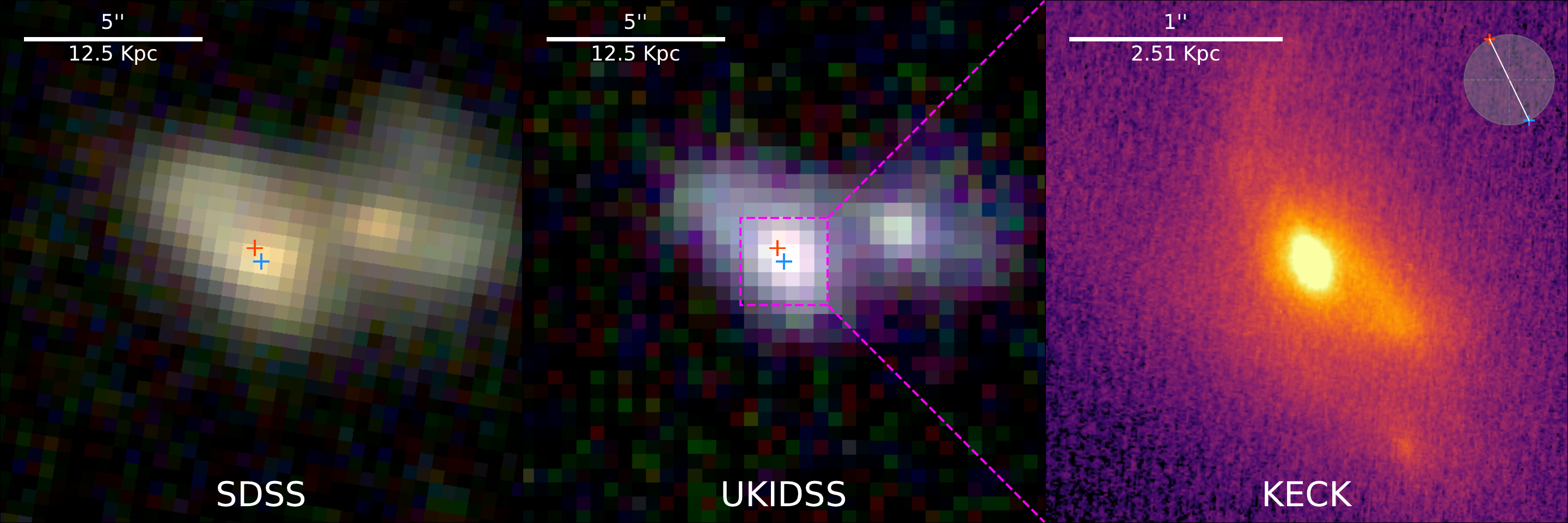}
\caption{Images provided above are described at the beginning of Appendix \ref{B}.}
\end{figure}
J091529 is a close galaxy pair separated by 3.344$''$ and 8.392 kpc; the nuclei of both galaxies are resolved in the SDSS, UKIDSS and NIRC2 images. The high resolution NIRC2 imaging of this target reveals previously unresolved substructure near the primary central nucleus of the Eastern galaxy. Surrounding this nucleus we can see faint emission consistent with bipolar tidal tails curving into the North and Southwest, most likely generated by the ongoing galaxy merger. There is one faint secondary source of emission within a 1.5$''$ radius of the primary central nucleus of the Eastern galaxy. This substructure lies at a distance of 0.917$''$ or 2.300 kpc Southwest from the primary with a PA of 205.302\degree. Although this substructure does not appear to be a secondary nucleus it does seem to trace the SDSS/WISE offset. The SDSS/WISE offset under predicts the separation of the substructure as expected (see Section \ref{sec:results:offsetvsepvpa}) with a separation of 0.431$''$ but the PA of 205.880\degree is very similar to that of the substructure pair. We note that there is another SDSS/WISE offset detected in this merger system with a separation of 3.066$''$ between the SDSS coordinate of the Eastern galaxy and the WISE coordinate which coincides with the position of the Western galaxy. Since this offset is resolvable by SDSS and UKIDSS and corresponds to a secondary galaxy resolvable by SDSS, we do not consider it a detection of substructure in the target galaxy. The host galaxy of this object is optically classified as a $K_{01}$ type AGN in the MPA-JHU catalog. Additionally, it lies within the AGN classification of our BPT diagram and within the Jarrett AGN region of our WISE color-color diagram.
\clearpage
\subsection{J093314+630945}
\begin{figure}[htbp]
\centering

\includegraphics[width=0.6666\linewidth]{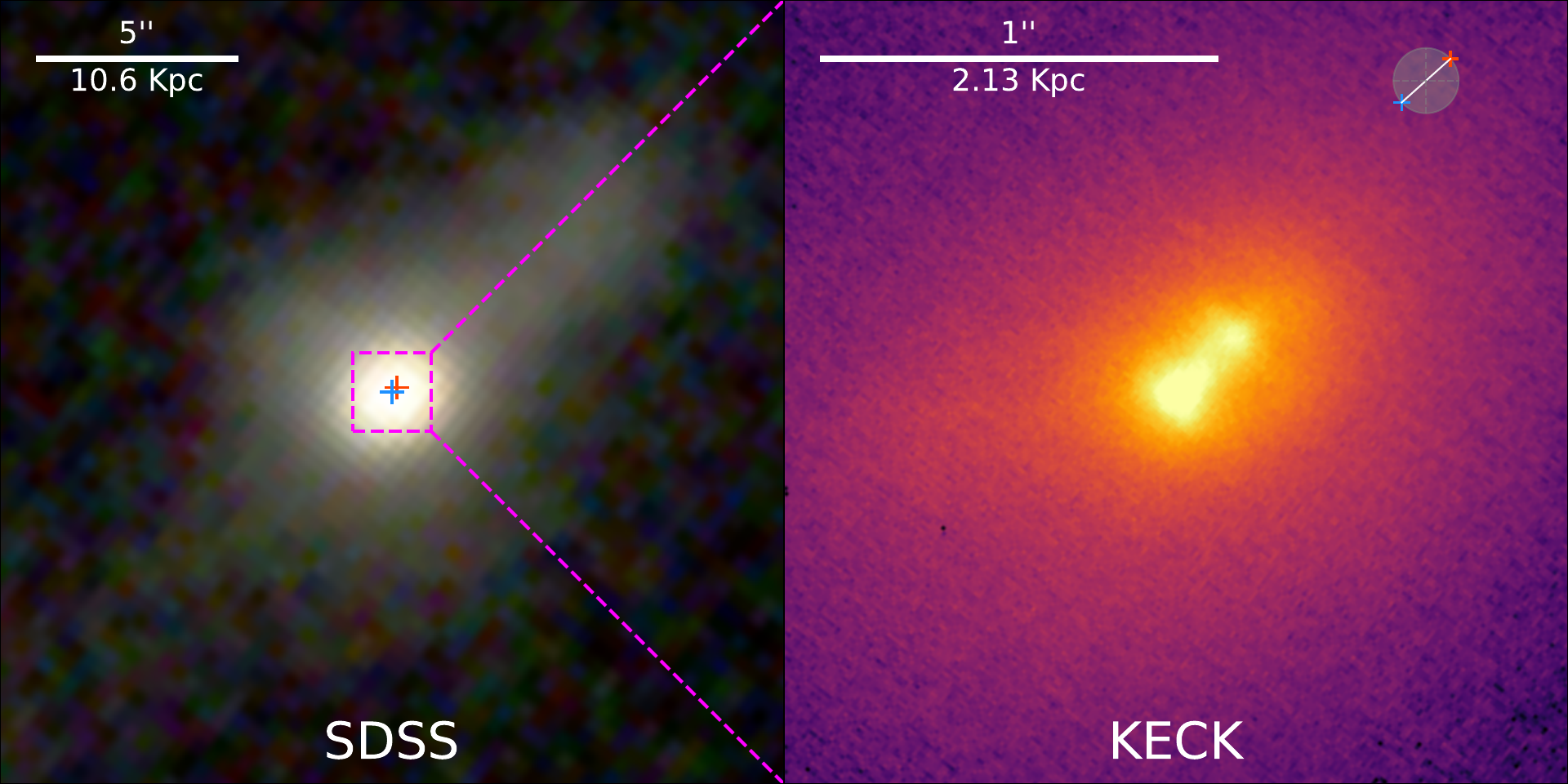}
\caption{Images provided above are described at the beginning of Appendix \ref{B}.}
\end{figure}
J093314 is an advanced merger system which appears to contain a single nucleus with multiple tidal tails in its SDSS image. The tidal tails extend in opposite directions without much curvature out to distances of $\sim 30$ kpc and $\sim 15$ kpc. The NIRC2 imaging reveals two apparent nuclei at a separation of 0.190$''$ or 0.405 kpc with a PA of 316.231\degree. The separation and PA of the SDSS/WISE offset detected in this target are 0.167$''$ and 311.876\degree. It is clear that our offset selection detected unresolved dual nuclei in this system but is also closely traced by the actual separation and PA between these nuclei (i.e. $\Delta Sep. = 0.023''$ and $\Delta PA = 4.355\degree$). The host galaxy of this object is optically classified as a $K_{01}$ type AGN in the MPA-JHU catalog. Additionally, it lies within the AGN classification of our BPT diagram and within the Jarrett AGN region of our WISE color-color diagram, suggesting this may be a candidate dual AGN system.

\subsection{J101653+002856}
\begin{figure}[htbp]
\centering

\includegraphics[width=\linewidth]{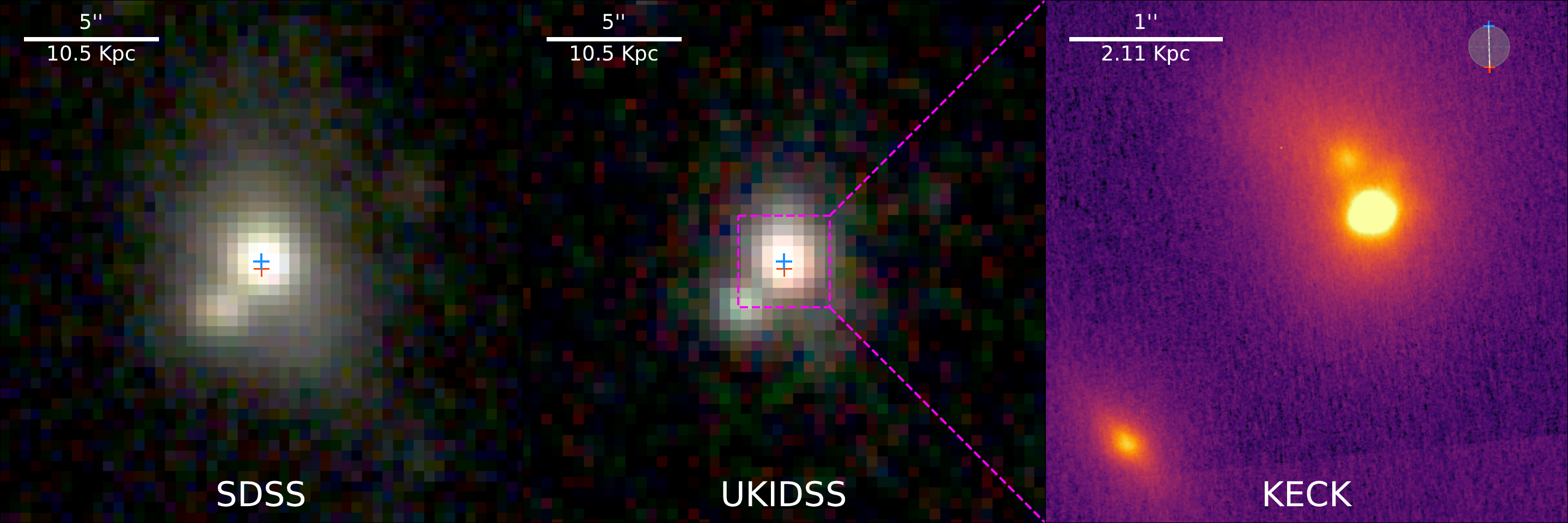}
\caption{Images provided above are described at the beginning of Appendix \ref{B}.}
\end{figure}
J101653 is a close galaxy pair separated by 2.26$''$ or 4.76 kpc, with the nuclei of both galaxies resolved in SDSS and UKIDSS. This target has been previously identified as an elliptical galaxy within a disturbed pre-coalescence system \citep[][]{Pierce+2023}. The detected SDSS/WISE offset is in the Northwestern galaxy of this pair. The NIRC2 imaging of the Northwestern galaxy reveals a third source of emission very close to its primary central nucleus. The separation between this close pair is 0.350$''$ or 0.737 kpc with a PA of 25.195\degree. The separation and PA of the SDSS/WISE offset are 0.275$''$ and 1.251\degree. The size and PA of the SDSS/WISE offset is similar to that of the substructure pair although with less accuracy than other targets in this sample. Additionally, we note that based on the orientation of the offset it appears that the WISE coordinate corresponds to the primary central nucleus and the SDSS coordinate corresponds to the secondary nucleus. This may suggest that the primary nucleus is obscured, however higher resolution optical imaging is required to confirm this. The host galaxy of this object is optically classified as a $K_{01}$ type AGN in the MPA-JHU catalog. Additionally, it lies within the AGN classification of our BPT diagram and within the Jarrett AGN region of our WISE color-color diagram.
% \clearpage
\subsection{J110213+645925}
\begin{figure}[htbp]
\centering
\includegraphics[width=\linewidth]{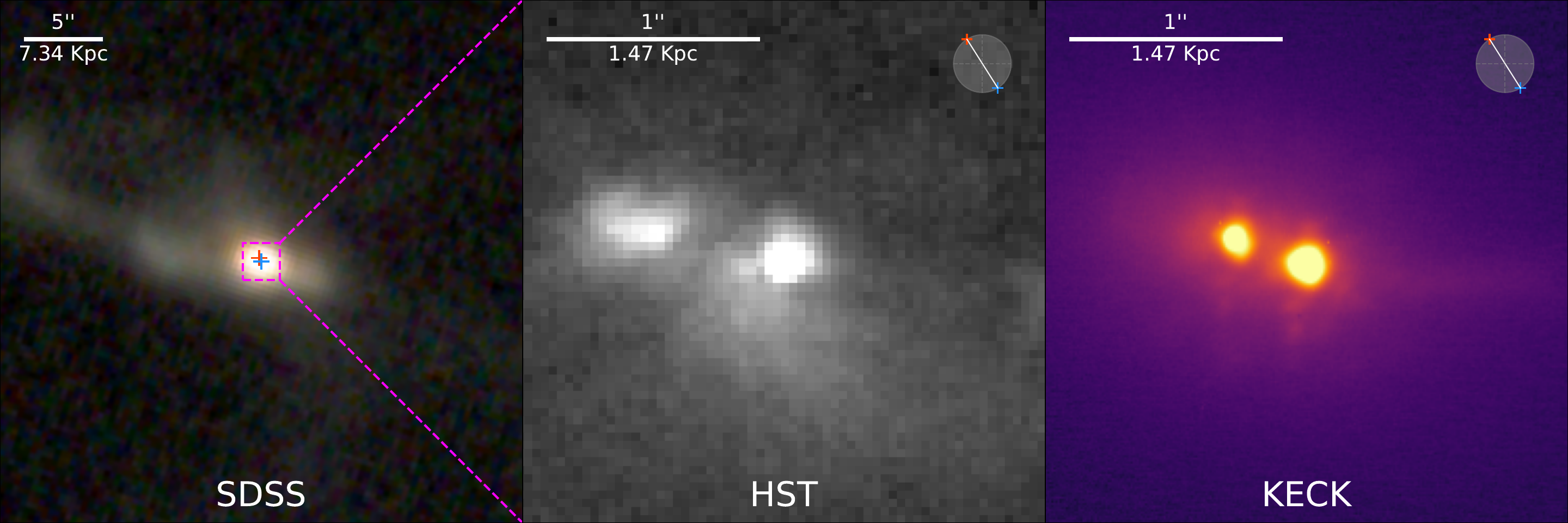}
\caption{Images provided above are described at the beginning of Appendix \ref{B}.}
\end{figure}
J110213 is a candidate dual AGN system containing clear sub-kpc dual nuclei with a separation of 0.357$''$ or 0.524 kpc and PA of 70.923\degree. This is a late-stage merging system with one apparent nucleus detected in SDSS. One prominent tidal tail extends 17.5$''$ out to the East and curves towards the North and back into the center of the galaxy forming a closed loop. \cite{zhao+19} imaged this target with the Hubble Space Telescope WFC3 camera (F105W and F438W), classifying it as a type 2 quasar within a merging system. Upon inspection of the HST images, two nuclei are visible with an apparent dust lane between them. However, these nuclei have much more irregular shapes than what is seen in our Keck image and the separation of the apparent nuclei in the optical F438W image is $\sim 0.67''$. It is possible that the dust lane is partially obscuring one of the nuclei in the optical image causing this discrepancy. If we assume the Western nucleus is the same source in both the NIRC2 and HST image then the location of the Eastern nucleus (in the NIRC2 image) coincides with the dust lane. Furthermore, we note that based on the orientation of the SDSS/WISE offset the WISE coordinate coincides with the obscured Eastern nucleus and the Galaxy Zoo position coincides with the Western nucleus. An obscuration of the Eastern nucleus would best account for the orientation and positioning of the offset since the Eastern nucleus would only be detected in WISE. Additionally, the separation and PA of the SDSS/WISE offset in J110213 is 0.276$''$ and 32.363\degree which is similar the separation and PA of the dual nuclei. The host galaxy of this object is optically classified as a $K_{01}$ type AGN in the MPA-JHU catalog. Additionally, it lies within the AGN classification of our BPT diagram and within the Jarrett AGN region of our WISE color-color diagram making this a strong candidate dual AGN system.

\clearpage
\subsection{J120408+132116}
\begin{figure}[htbp]
\centering

\includegraphics[width=0.6666\linewidth]{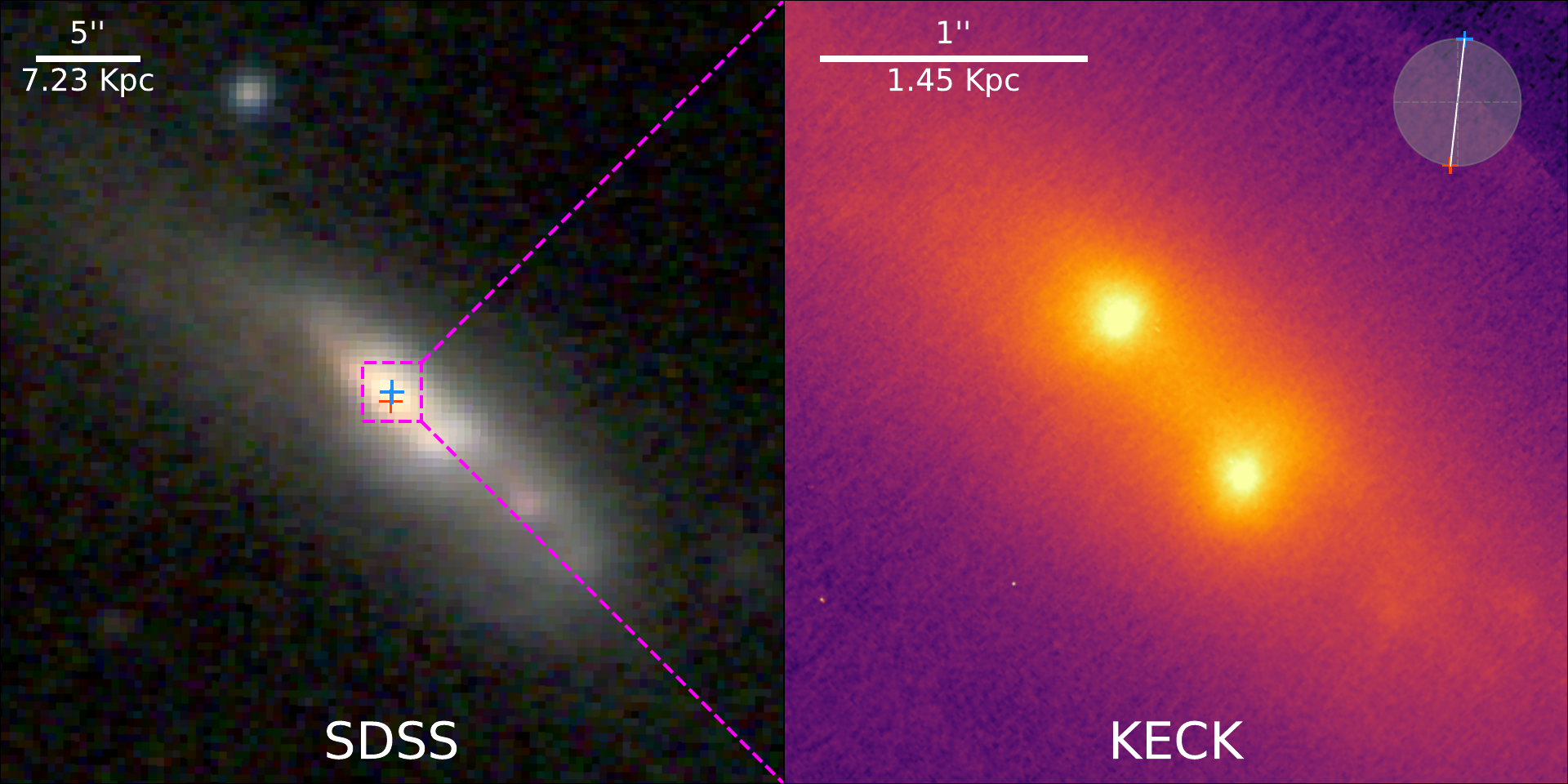}
\caption{Images provided above are described at the beginning of Appendix \ref{B}.}
\end{figure}
J120408 is a candidate dual AGN system containing two clear nuclei at a separation of 0.740$''$ or 1.070 kpc and PA of 219.846\degree. SDSS imaging of this target shows an elongated galaxy with one apparent optical nucleus and a highly extended ($> 10''$) tidal tail extending to the Northeast. The SDSS/WISE offset for this target has a separation of 0.486$''$ and PA of 173.630\degree. This does not trace the observed properties of the dual nuclei well, but nonetheless was able to detect them within an apparent single nuclei system. The host galaxy of this object is optically classified as a $K_{01}$ type AGN in the MPA-JHU catalog. Additionally, it lies within the AGN classification of our BPT diagram and outside of the Jarrett AGN region of our WISE color-color diagram.

\subsection{J123915+531414}
\begin{figure}[htbp]
\centering
\includegraphics[width = \linewidth]{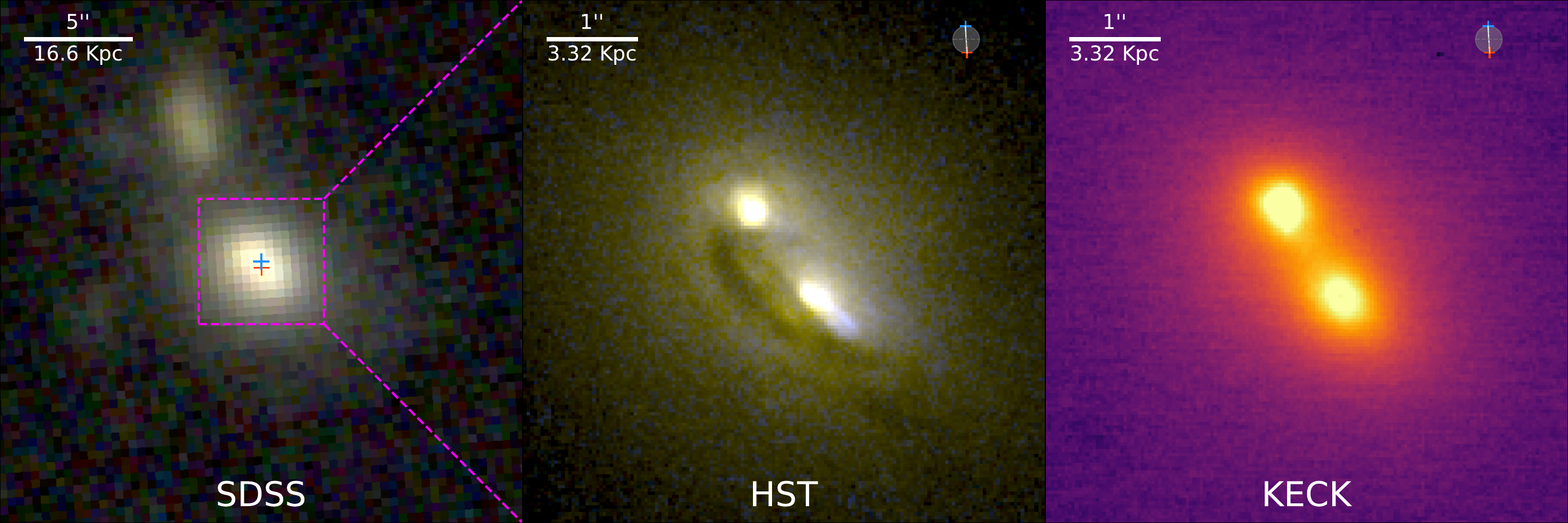}
\caption{Images provided above are described at the beginning of Appendix \ref{B}.}
\end{figure}

J123915 is one of the targets in the Primary Keck AO Merger Sample which has publicly available NIRC2 imaging available in the Keck archive (see Section \ref{sec:obsredux:archivalimaging}). \cite{fu+2011a} conducted Keck LGSAO observations of this target with the NIRC2 Wide-Field Camera (FOV: $40''\times 40''$, pixel scale: $0.039686''\rm{pixel}^{-1}$) in the $K_p$ filter as part of an observing program searching for kpc-scale dual AGNs in double-peaked [OIII] AGNs. Additionally, this target was included in HST F438W and F814W observations of AGN with double-peaked narrow emission lines presented in \cite{comerford+2015}, where they identify J123915 as a candidate dual AGN. The SDSS imaging of this target shows a galaxy with an unresolved and elongated bright central nucleus with a nearby galaxy in the Northeast likely to be unassociated ($\Delta v \sim 14800$ km/s). In both the NIRC2 and HST imaging we see clear dual nuclei at a separation of 1.161$''$ or 3.855 kpc and PA of 213.349\degree, as measured in the NIRC2 image. In the HST image there is a clear dust lane appearing to encircle and partially obscuring the Southwestern nucleus. The SDSS/WISE offset detected in this system has a separation of 0.306$''$ and PA of 183.300\degree, poorly tracing the orientation and separation of the two nuclei (i.e. $\Delta Sep. = 0.855''$ and $\Delta PA = 30.049\degree$). The SDSS/WISE offset may be tracing the obscuring dust lane with the SDSS coordinate coinciding with the region between the two nuclei but above the obscuring dust and the WISE coordinate coinciding with the region between the two nuclei being obscured by the dust lane. Interestingly, \cite{comerford+2015} show that the double-peaked emission lines seen in J123915 correspond to two [OIII] $\lambda5007$ emission components separated by 0.39$''$, which is similar to the separation of the SDSS/WISE offset we detect. The host galaxy of this object is optically classified as a $K_{01}$ type AGN in the MPA-JHU catalog. Additionally, it lies within the AGN classification of our BPT diagram and outside of the Jarrett AGN region of our WISE color-color diagram.
% \clearpage

\subsection{J124117+273251}
\begin{figure}[htbp]
\centering

\includegraphics[width=\linewidth]{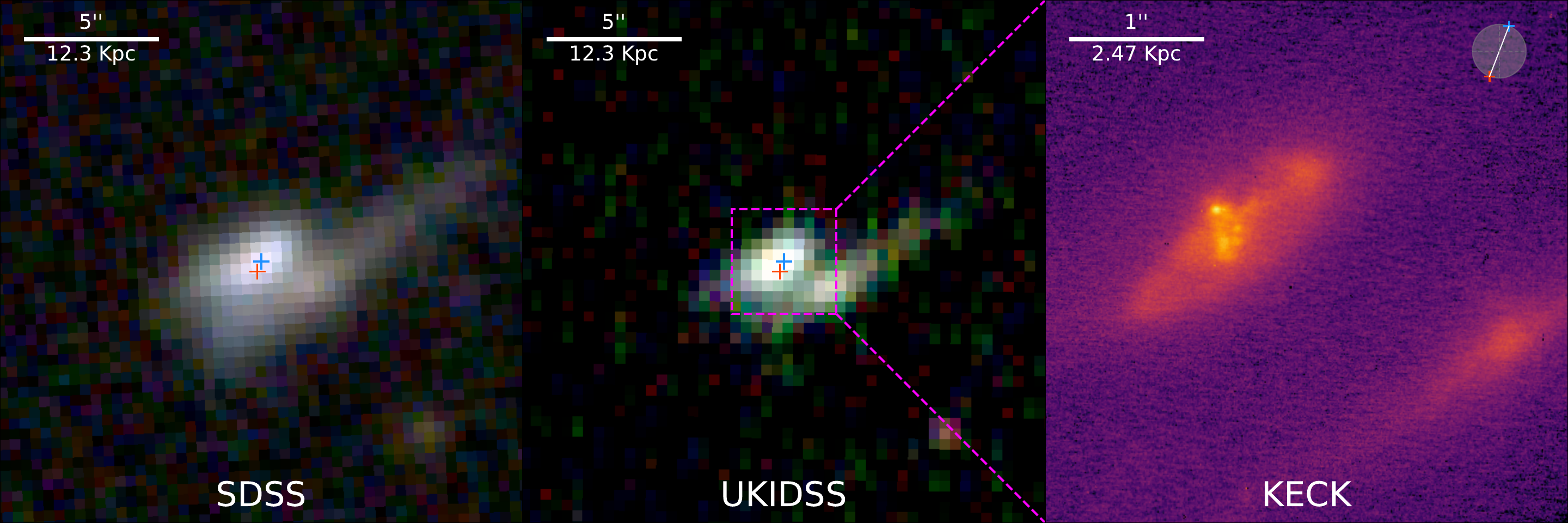}
\caption{Images provided above are described at the beginning of Appendix \ref{B}.}
\end{figure}
J124117 is the Northeastern galaxy in a close galaxy pair separated by 2.44$''$ or 6 kpc, in which we detect a coordinate offset. Both galaxies are visible in the NIRC2 imaging, with the galaxy of interest (Northeast) containing a complex nuclear region with multiple compact sources of emission within the central arcsecond of the galaxy. The galaxy pair can be seen in SDSS and UKIDSS but the substructure in J124117 is unresolved in both surveys. The resolved substructure does not appear to be clear nuclei such as those in J110213 or J120408, the emission seems more consistent with extended nuclear emission that is partially obscured by a dust-lane. Although this target does not contain clear dual nuclei it does contain sub-arcsecond structure that seems to be traced by the offset measurement. The SDSS/WISE offset detected in this system has a separation of 0.411$''$ and have a PA of 158.881\degree. Within the nuclear region of the primary galaxy we see two prominent ``blobs'' forming a substructure pair with a separation of 0.266$''$ or 0.655 kpc and PA of 193.076\degree. The orientation of the SDSS/WISE offset does not appear to trace that of the candidate pair very well but it may be that the SDSS coordinate is biased towards the more extended emission in the galaxy (in the Northwest) and the WISE coordinate is more accurately detecting the bright nuclear emission in the South. The host galaxy of this object is optically classified as a $K_{03}$ type AGN in the MPA-JHU catalog. Additionally, it lies within the composite classification of our BPT diagram and outside of the Jarrett AGN region of our WISE color-color diagram.
\clearpage
\subsection{J131517+442425}
\begin{figure}[htbp]
\centering
\includegraphics[width = \linewidth]{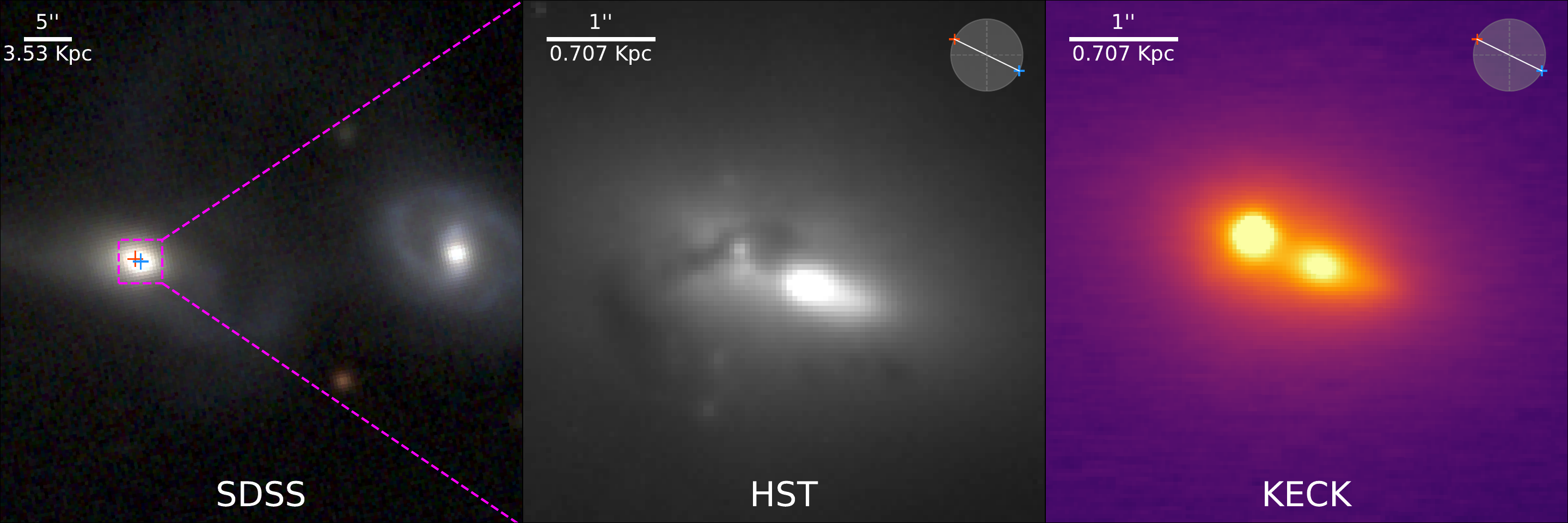}
\caption{Images provided above are described at the beginning of Appendix \ref{B}.}
\end{figure}
J131517 is one of the targets in the Primary Keck AO Merger Sample which has publicly available NIRC2 imaging in the Keck archive (see Section \ref{sec:obsredux:archivalimaging}). Keck LGSAO observations of this target were conducted with the NIRC2 Wide-Field Camera (FOV: $40''\times 40''$, pixel scale: $0.039686''\rm{pixel}^{-1}$) in the $K_p$ filter as part of Keck observing program Y076 in 2022.  Additionally, this target was included in HST F814W observations of Swift-BAT AGN presented in \cite{kim+2021}. In the SDSS cutout image we see that this target is part of an early stage merger system at a separation of 36.438$''$ or 25.744 kpc, with the Eastern galaxy being the galaxy in which we detect an SDSS/WISE offset. The high resolution NIRC2 imaging reveals clear dual nuclei separated by 0.711$''$ or 0.503 kpc and PA of 244.142\degree. The SDSS/WISE offset, with a separation of 0.689$''$ and PA of 243.980\degree, very closely traces the orientation of the nuclear pair (i.e. $\Delta Sep. = 0.022''$ and $\Delta PA = 0.162\degree$). Inspection of the HST F814W image of J131517 reveals that the Eastern nucleus of this pair is almost entirely obscured by a dust lane, which appears to be causing the significant offset between the WISE and SDSS nuclear coordinates. Furthermore, the orientation of the SDSS/WISE offset suggests that the WISE coordinate coincides with the optically obscured Eastern nucleus and the SDSS coordinate coincides with the unobscured Western nucleus. The host galaxy of this object is optically classified as a $K_{03}$ type AGN in the MPA-JHU catalog. Additionally, it lies within the composite classification of our BPT diagram and within the Jarrett AGN region of our WISE color-color diagram.
\clearpage
\subsection{J131534+620728}
\begin{figure}[htbp]
\centering
\includegraphics[width = \linewidth]{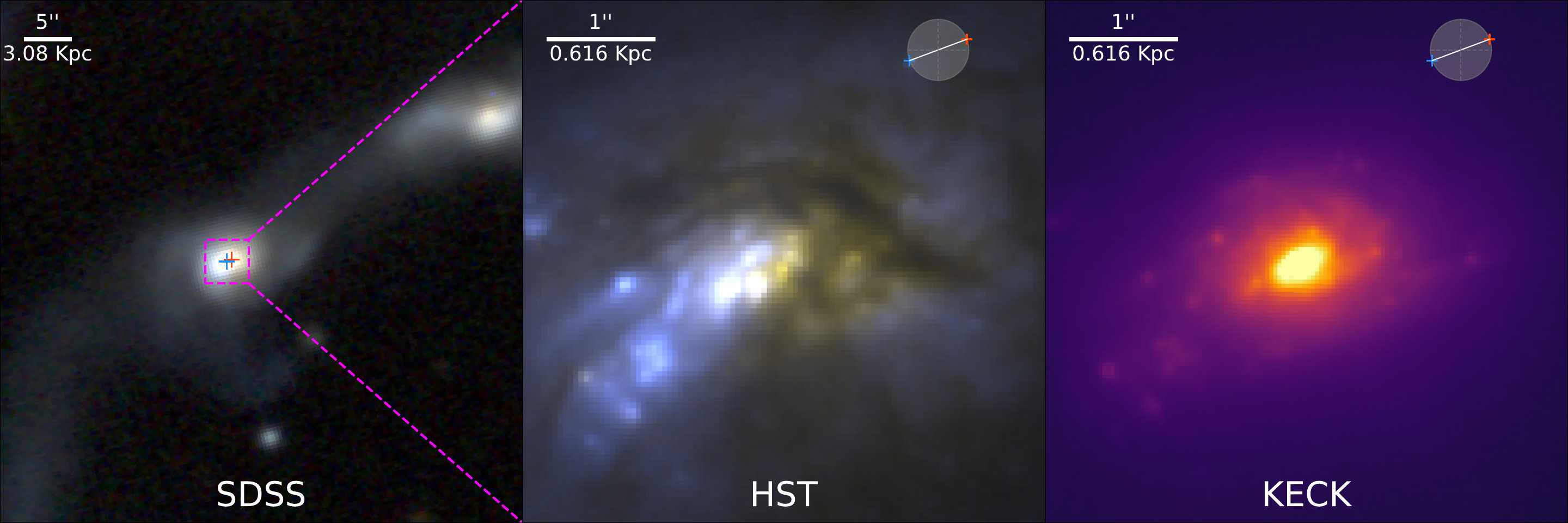}
\caption{Images provided above are described at the beginning of Appendix \ref{B}.}
\end{figure}
J131534 is one of the targets in the Primary Keck AO Merger Sample which has publicly available NIRC2 imaging in the Keck archive (see Section \ref{sec:obsredux:archivalimaging}). Keck LGSAO observations of this target were conducted with the NIRC2 Wide-Field Camera (FOV: $40''\times 40''$, pixel scale: $0.039686''\rm{pixel}^{-1}$) in the $K_s$ ($\lambda_c = 2.146 \mu\rm{m}$) filter as part of Keck observing program Z229 in 2017.  Additionally, this target was included in the The Great Observatories All-Sky LIRG Survey HST sample \citep[GOALS;][]{armus+2009} where it was imaged with the ACS/WFC using the F435W and F814W filters (GO program 10592, PI: A. Evans). In the SDSS image we see that this target is part of an early stage merger system at a separation of 35$''$ or 21.553 kpc, with the Eastern galaxy being the galaxy in which we detect an SDSS/WISE offset. The Eastern galaxy has two prominent tidal tails, one extending 24.6 kpc to the Southeast and the other connecting the two galaxies. In the NIRC2 imaging of this target we see that there is a single bright central nucleus or AGN surrounded by many sources of faint secondary emission, or substructure. Based on the criteria laid out in Section \ref{sec:results:subsID} the substructure pair we identify in this target for comparison with the SDSS/WISE offset consists of the central nucleus and a small source of emission North of the nucleus at a separation of 0.287$''$ or 0.177 kpc and PA of 343.328\degree. The SDSS/WISE offset, with a separation of 0.585$''$ and PA of 290.428\degree, does not appear to trace the substructure pair we define but comparison between the HST and Keck imaging reveals that the SDSS/WISE offset reflects a difference in the near-infrared and optical emission of J131517 rather than real substructure revealed by near-infrared imaging. The HST image shows that there is a significant amount of dust obscuring the Western half of the optical nuclear emission. By identifying matching sources of secondary emission visible in both the Keck and HST images we find that the central near-infrared nucleus is likely spatially coincident with the large dust cloud in the Western region of the optical nuclear emission and not the bright optical emission we see. This suggests that the SDSS/WISE offset we detect is caused by the central nucleus or AGN being obscured by dust and thus biasing the SDSS coordinate away from the true location of the AGN. The host galaxy of this object is optically classified as a $K_{03}$ type AGN in the MPA-JHU catalog. Additionally, it lies within the composite classification of our BPT diagram and outside of the Jarrett AGN region of our WISE color-color diagram.

\clearpage
\subsection{J132035+340822}
\begin{figure}[htbp]
\centering
\includegraphics[width = 0.6666\linewidth]{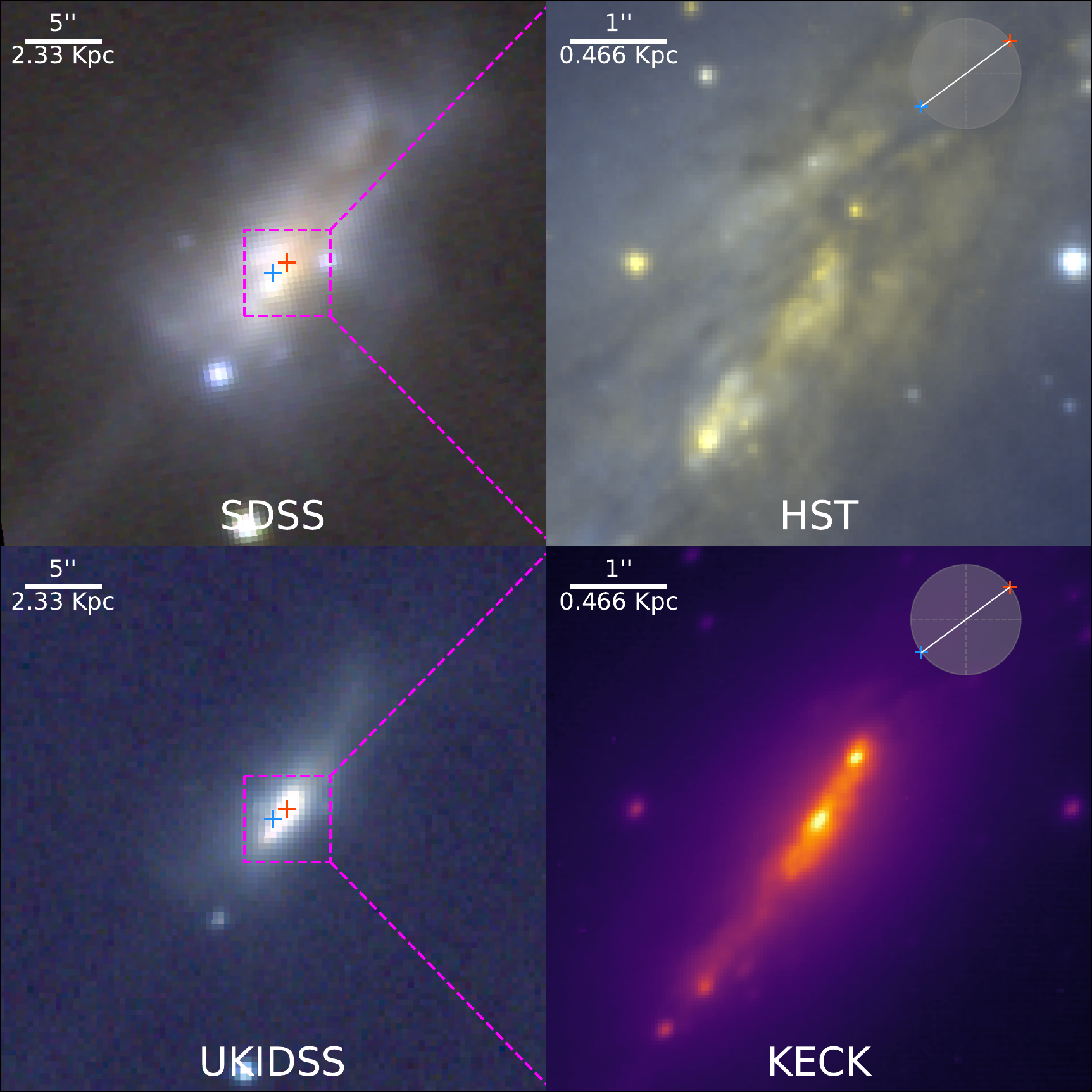}
\caption{Images provided above are described at the beginning of Appendix \ref{B}. Note that this target does not have a UKIDSS  Y band cutout image available, thus the UKIDSS three color image is comprised of K band, H band and J band images.}
\end{figure}
J132035 is one of the targets in the Primary Keck AO Merger Sample which has publicly available NIRC2 imaging in the Keck archive (see Section \ref{sec:obsredux:archivalimaging}). Keck LGSAO observations of this target were conducted with the NIRC2 Wide-Field Camera (FOV: $40''\times 40''$, pixel scale: $0.039686''\rm{pixel}^{-1}$) in the $K_s$($\lambda_c = 2.146 \mu\rm{m}$) filter as part of Keck observing program Z229 in 2017.  Additionally, this target was included in the GOALS HST sample \citep[][]{armus+2009} where it was imaged with the ACS/WFC using the F435W and F814W filters (GO program 10592, PI: A. Evans). The SDSS and UKIDSS imaging of this target shows it is a highly disturbed late-stage merger with a bright and elongated central nucleus. An extended tidal tail stretching towards the Northwest is visible in both the optical and IR. Additionally, outside of the field of view of the SDSS image there are two perpendicular tidal tails, one extending 23 kpc to the Southeast and the other 29 kpc to the Southwest. NIRC2 imaging of J132035 reveals dual nuclei separated by 0.784$''$ or 0.365 kpc and PA of 329.747\degree. \cite{romero+2017} confirmed the AGN nature of the central nucleus in J132035, reporting the presence of a pc-scale jet based on high-resolution VLBI observations. The secondary nucleus located to the Northwest of the central nucleus has not been confirmed as an AGN. The SDSS/WISE offset detected in the nucleus of J132035 has a separation of 1.209$''$ and PA 306.487\degree\ which overpredicts the separation of the dual nuclei but has a similar orientation. The high-resolution optical HST imaging of J132035 shows that both nuclei are heavily obscured by a large dust lane visible in the SDSS image and reveals an additional bright source of optical emission in the Southeast of the HST image. It is likely that the partial obscuration of the nuclear region in J132035 is driving the SDSS/WISE offset we detect. The host galaxy of this object is optically classified as a $K_{03}$ type AGN in the MPA-JHU catalog. Additionally, it lies within the composite classification of our BPT diagram and outside of the Jarrett AGN region of our WISE color-color diagram.

\subsection{J134442+555313}
\begin{figure}[H]
\centering
\includegraphics[width = \linewidth]{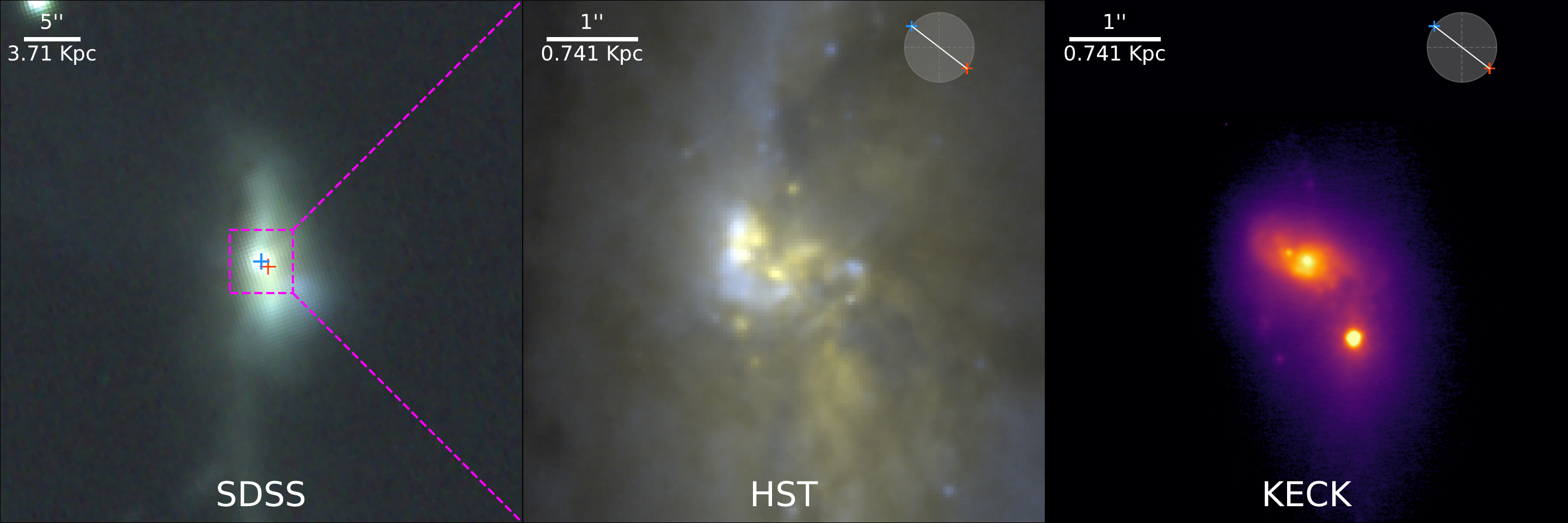}
\caption{Images provided above are described at the beginning of Appendix \ref{B}.}
\end{figure}
J134442 is one of the targets in the Primary Keck AO Merger Sample which has publicly available NIRC2 imaging in the Keck archive (see Section \ref{sec:obsredux:archivalimaging}). This target is an ultra-luminous infrared galaxy (ULIRG) and late-stage merger system with a very extended optical tidal tail stretching 40 kpc to the South in the SDSS image. \cite{U+2013} conducted Keck LGSAO observations of this target with the NIRC2 Narrow-Field Camera (FOV: $10''\times 10''$, pixel scale: $0.009942''\rm{pixel}^{-1}$) in the H ($\lambda_c = 1.633 \mu\rm{m}$) and $K_p$ filters. Here we show the $K_p$ filter image which reveals two nuclei: a nucleus surrounded by a rotating gaseous disk in the Northeast and a prominent nucleus in the Southwest at a separation of 1.035$''$ or 0.767 kpc and PA of 211.969\degree. J134442 was confirmed to be a dual AGN system by \cite{iwasawa+2018} with follow-up NuSTAR observations which are best explained by a dual AGN system. Additionally, this target was included in the GOALS HST sample \citep[][]{armus+2009} where it was imaged with the ACS/WFC using the F435W and F814W filters (GO program 10592, PI: A. Evans). The optical HST imaging of J134442 shows a much more disturbed morphology and significant obscuration, with the Northeast AGN appearing partially obscured and the Southwest AGN being completely obscured. The SDSS/WISE offset detected in this system roughly traces the separation and orientation of the dual AGN system with a separation of 0.799$''$ and PA of 232.468\degree (i.e. $\Delta Sep. = 0.236''$ and $\Delta PA = 20.499\degree$). Due to the relative orientation of the SDSS/WISE offset we can see that the WISE coordinate likely coincides with the obscured Southwestern AGN and the SDSS with the Northeastern AGN, showing that the SDSS/WISE offset is likely caused by the obscuration of the optical emission of the Southwestern AGN. Additionally, due to the partial obscuration of the optical emission from the Northeastern AGN it is likely that the SDSS coordinate is not accurate to the position of the AGN and thus causing the discrepancy between the PA of the SDSS/WISE offset and dual AGN system. The host galaxy of this object is optically classified as a $K_{01}$ type AGN in the MPA-JHU catalog. Additionally, it lies within the AGN classification of our BPT diagram and within the Jarrett AGN region of our WISE color-color diagram.

\clearpage
\subsection{J134651+371231}
\begin{figure}[htbp]
\centering

\includegraphics[width=0.6666\linewidth]{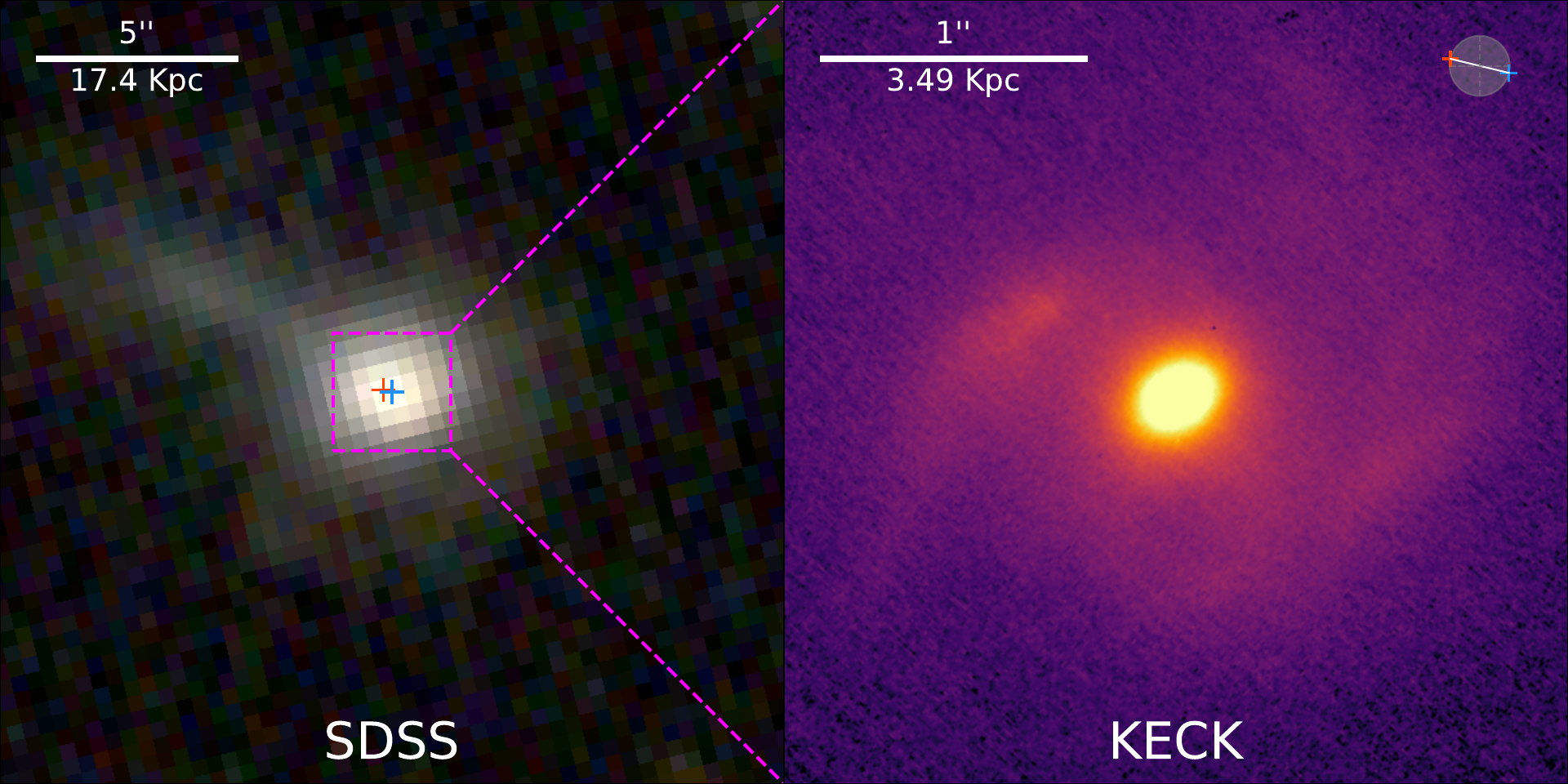}
\caption{Images provided above are described at the beginning of Appendix \ref{B}.}
\end{figure}
J134651 appears to be a post-merger system with an extended tidal tail extending 5.5$''$ or 19.19 kpc to the Northeast. The NIRC2 imaging reveals substructure in the form of apparent spiral arms in the Northeast and Southwest. In the Northeast arm we identify concentrated emission as the brightest secondary source of emission, forming a substructure pair with the central primary central nucleus with a separation of 0.633$''$ or 2.209 kpc and PA of 63.589\degree. The SDSS/WISE offset detected in this system has a separation of 0.230$''$ and PA of 76.457\degree\ which under predicts the separation of the substructure as expected (see Section \ref{sec:results:offsetvsepvpa} but appears accurate to the orientation of the spiral arm relative to the central nucleus. The host galaxy of this object is optically classified as a $K_{01}$ type AGN in the MPA-JHU catalog. Additionally, it lies within the AGN classification of our BPT diagram and outside of the Jarrett AGN region of our WISE color-color diagram.
\clearpage
\subsection{J135646+102608}
\begin{figure}[htbp]
\centering
\includegraphics[width = 0.6666\linewidth]{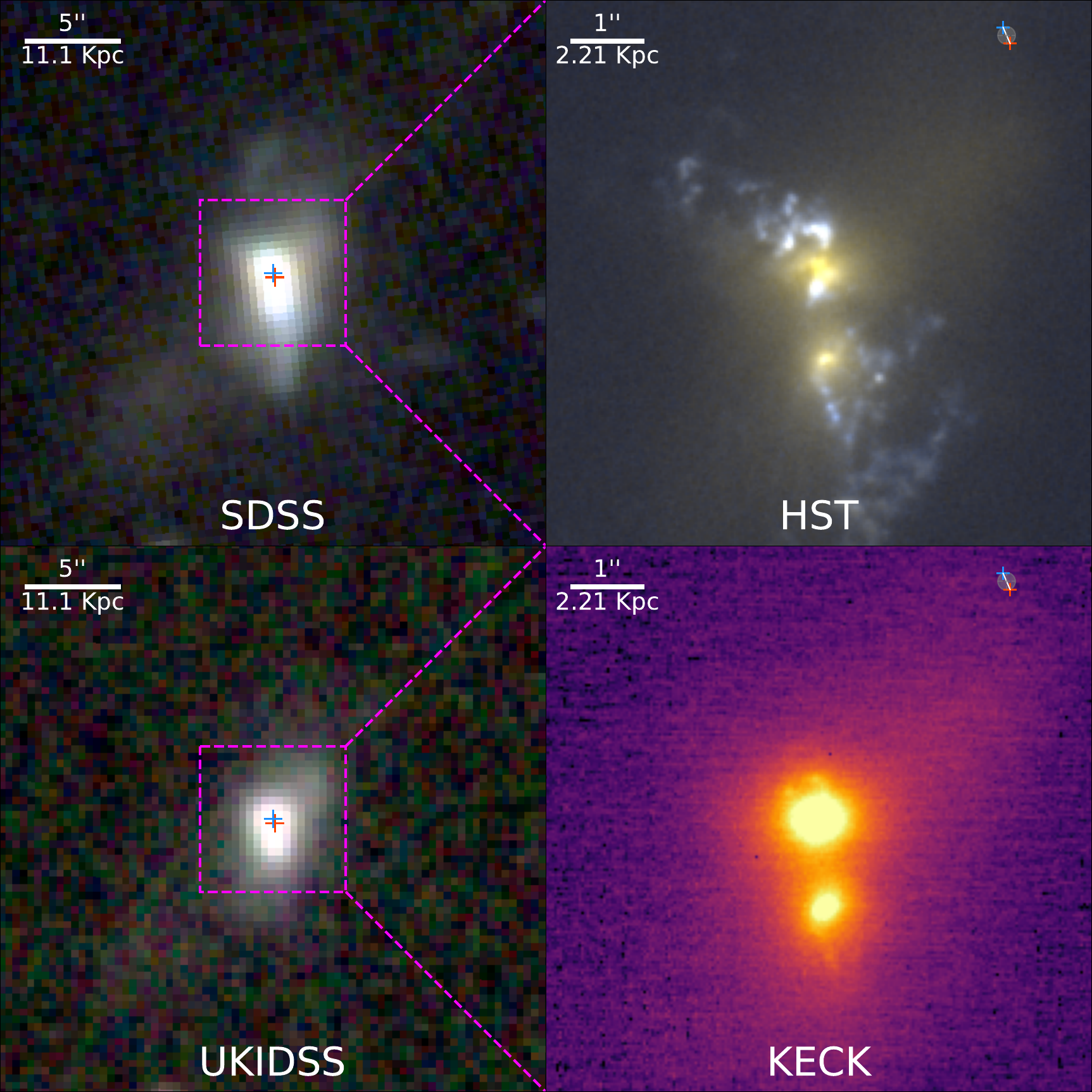}
\caption{Images provided above are described at the beginning of Appendix \ref{B}.}
\end{figure}

J135646 is one of the targets in the Primary Keck AO Merger Sample which has publicly available NIRC2 imaging available in the Keck archive (see Section \ref{sec:obsredux:archivalimaging}). \cite{fu+2011a} conducted Keck LGSAO observations of this target with the NIRC2 Wide-Field Camera (FOV: $40''\times 40''$, pixel scale: $0.039686''\rm{pixel}^{-1}$) in the $K_p$ filter as part of an observing program searching for kpc-scale dual AGNs in double-peaked [OIII] AGNs. Additionally, this target was included in HST F438W and F814W observations of AGN with double-peaked narrow emission lines presented in \cite{comerford+2015}, where they identify J135646 as a candidate dual AGN. J135646 exhibits bipolar extended emission stretching $\sim$15 kpc to the North and $\sim$12 kpc to the South in SDSS but not in UKIDSS. Closer observations of this extended emission has found that it is likely to be a $\sim$20 kpc scale outflow being driven by radiative feedback from an AGN in J135646 \citep[e.g. ][]{greene+2012}. The NIRC2 imaging of this target reveals a very bright Northern nucleus and a dimmer Southern nucleus forming a clear nuclear pair at a separation of 1.253$''$ or 2.771 kpc and PA of 185.403\degree. The HST imaging of J135646 similarly shows two nuclei, however the optical emission of both nuclei is significantly broken up and distorted by the outflow identified by \cite{greene+2012}. The SDSS/WISE offset detected in this system has a separation of 0.258$''$ and PA of 203.430\degree, which is not accurate to the orientation of the dual nuclei we see in the NIRC2 and HST imaging. Based on the location of the SDSS and WISE coordinates in the SDSS and UKIDSS cutout images it is likely that the offset corresponds to the immediate vicinity of the Norther nucleus. Assuming this, it is likely that the SDSS/WISE offset is tracing the complex optical emission of the outflow near the Northern nucleus, with the WISE coordinate corresponding to the nucleus and the SDSS coordinate corresponding to the outflow. The host galaxy of this object is optically classified as a $K_{01}$ type AGN in the MPA-JHU catalog. Additionally, it lies within the AGN classification of our BPT diagram and within the Jarrett AGN region of our WISE color-color diagram.
\clearpage
\subsection{J150517+080912}
\begin{figure}[htbp]
\centering
\includegraphics[width=\linewidth]{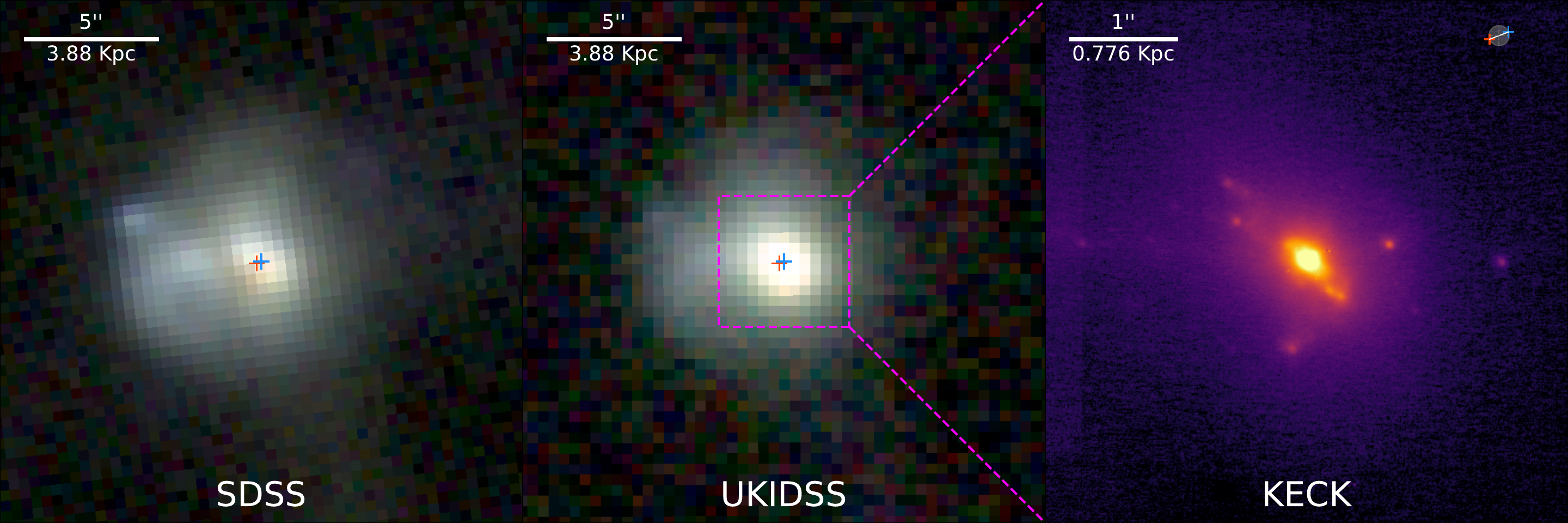}
\caption{Images provided above are described at the beginning of Appendix \ref{B}.}
\label{B:J1505}
\end{figure}
J1505 is a late-stage merger system with significant tidal features and an elongated nucleus. In its SDSS image we see two tidal tails, one extending Southward out to 11 kpc and a Western tail curving Northward with an approximate length of 13 kpc. Additionally, there is a highly curved tail of emission coming out of the Western side of the nucleus looping back into the galaxy towards the South and another curved tail coming out of the Northern side of the nucleus. The NIRC2 image of this target shows multiple sources of secondary emission at sub-arcsecond separations from the central primary nucleus. The brightest of the secondaries lies at a separation of 0.347$''$ or 0.269 kpc and PA of 218.893\degree, Southeast from the primary nucleus. The SDSS/WISE offset separation and PA are 0.194$''$ and 290.978\degree, which does not appear to be significantly correlated to the orientation of the substructure pair. There is a dimmer secondary in the image which is separated from the primary nucleus by 0.797$''$ and has a position angle of 282.239\degree\ which appears to be more consistent with the offset orientation. The host galaxy of this object is optically classified as a $K_{03}$ type AGN in the MPA-JHU catalog. Additionally, it lies within the composite classification of our BPT diagram and outside of the Jarrett AGN region of our WISE color-color diagram.

\subsection{J151806+424444}
\begin{figure}[htbp]
\centering

\includegraphics[width=\linewidth]{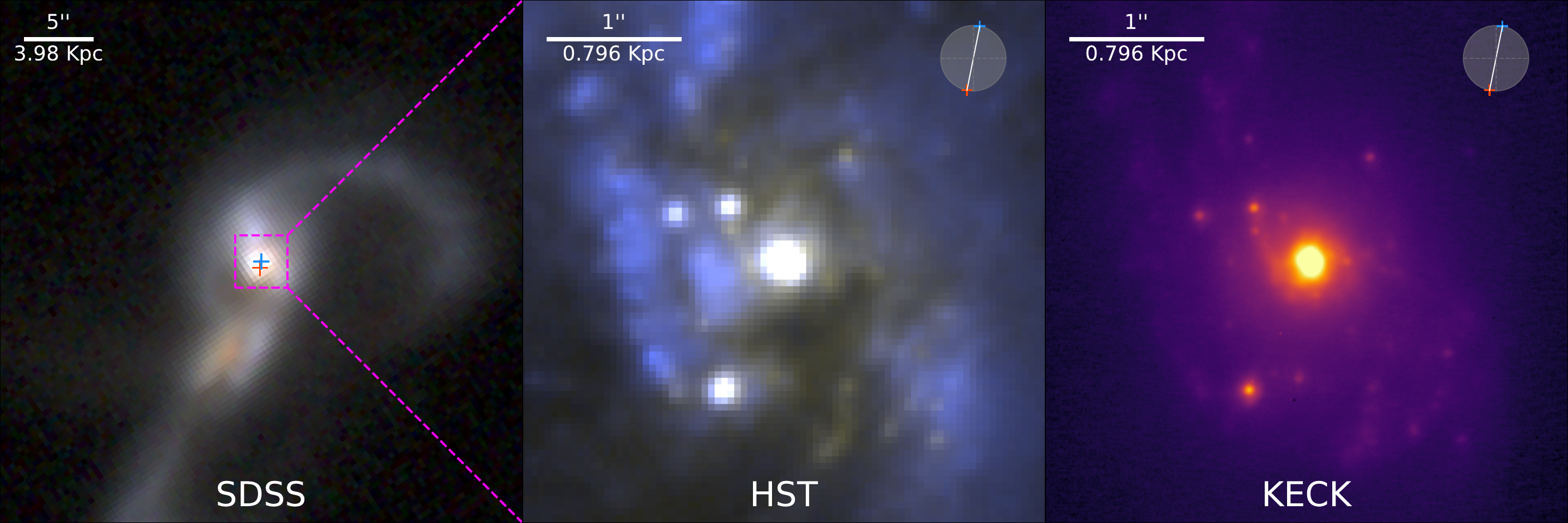}
\caption{Images provided above are described at the beginning of Appendix \ref{B}.}
\end{figure}
J151806 is an extensively studied target also known as VV 705. The SDSS image shows that this is clearly a merging system which appears to have gone through its first pass, indicated by its very prominent tidal tails and bridge of emission between the two galaxies. This is a previously identified candidate dual AGN at a separation of 5.8 kpc based on the detection of X-ray emission in both nuclei \citep[see][]{Comerford+2022,He+2023}. This target is also included in the GOALS HST sample \citep{armus+2009} where it was imaged with the ACS/WFC using the F435W and F814W filters (GO program 10592, PI: A. Evans). With our NIRC2 imaging we target the Northwest galaxy in which we detect an SDSS/WISE offset with a separation of 0.500$''$ and PA of 168.679\degree. In the high-resolution NIRC2 imaging of this target, we see a bright central primary nucleus and 4 additional secondary sources of emission in the immediate nuclear region of the galaxy. Of the four secondaries the brightest lies 1.083$''$ or 0.863 kpc in the Southeast from the primary with a PA of 156.010\degree. Together with the central primary nucleus this forms the substructure pair in J151806. In the HST images of this target we can see significant obscuration throughout the host galaxy but the same substructure can still be seen in the optical. The PA of the substructure pair is very similar to that of the SDSS/WISE offset, suggesting that the detected offset is accurately tracing the substructure in this system. The host galaxy of this object is optically classified as a $K_{03}$ type AGN in the MPA-JHU catalog. Additionally, it lies within the composite classification of our BPT diagram and outside of the Jarrett AGN region of our WISE color-color diagram.

\subsection{J152659+355837}
\begin{figure}[htbp]
\centering

\includegraphics[width=\linewidth]{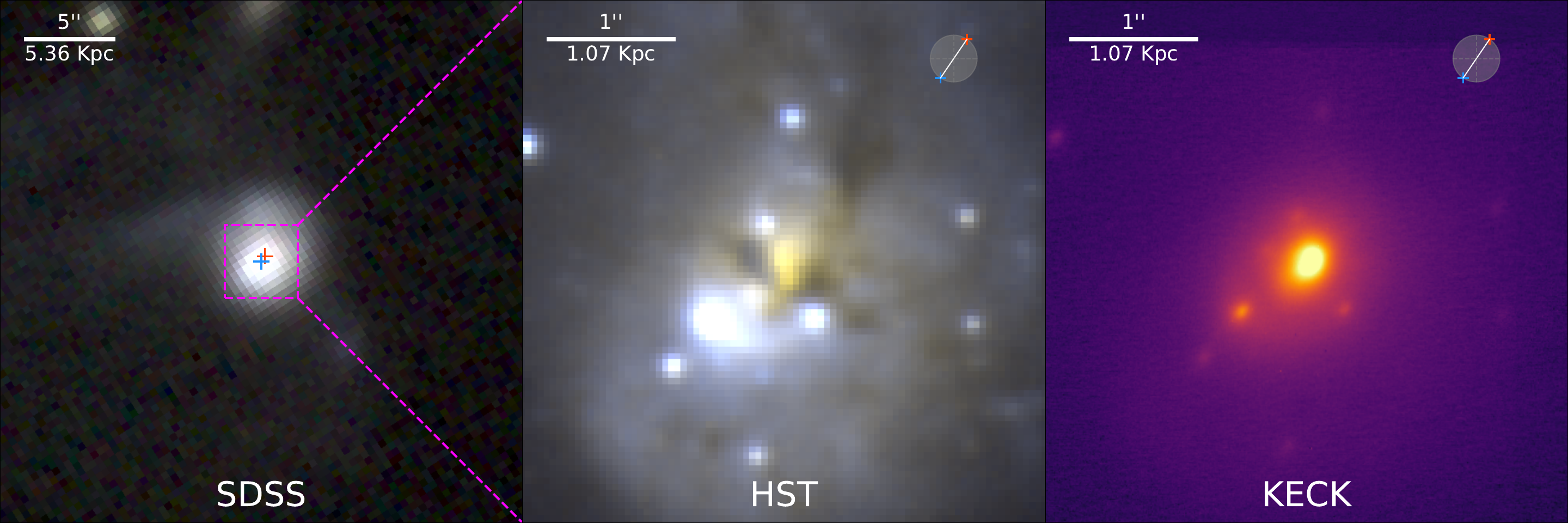}
\caption{Images provided above are described at the beginning of Appendix \ref{B}.}
\end{figure}
J152659 is an advanced merger system with tidal tails in the Southwest and East connected by an arc of emission with an approximate radius of 12$''$ or 13 kpc. In the NIRC2 image we can see multiple sources of substructure within the nuclear region of this target which were unresolved in SDSS. There is a prominent bright primary nucleus in the center of the galaxy and we identify 9 different secondary sources of near-infrared emission. The brightest of these secondaries lies 0.631$''$ or 0.677 kpc away from the primary nucleus with a PA of 127.543\degree. The SDSS/WISE offset that we detect in this system has a separation of 0.376$''$ and PA of 145.649\degree\ which is relatively similar to the orientation of the substructure pair. J152659 has been studied extensively in the past at multiple wavelengths and lower resolutions with some identifying the same substructure seen in the NIRC2 image. \cite{scoville+2000} presented HST NICMOS imaging of this galaxy at $1.1 \mu m$, $1.6 \mu m$, and $2.2 \mu m$ and identify the same pair of candidate nuclei. \cite{scoville+2000} do not classify the secondary source as a nucleus or SF region, though it is noted that the near-infrared colors of the secondary source are very red. Additionally, this target was included in the GOALS HST sample \citep[][]{armus+2009} where it was imaged with the ACS/WFC using the F435W and F814W filters (GO program 10592, PI: A. Evans). The same sources of secondary near-infrared emission can be seen in the optical HST imaging except for the primary central nucleus, where it is significantly obscured. When analyzed in conjunction with the orientation of the SDSS/WISE offset it appears that the WISE coordinate traces the location of the primary nucleus and the SDSS coordinate of the unobscured secondary. The host galaxy of this object is optically classified as a $K_{03}$ type AGN in the MPA-JHU catalog. Additionally, it lies within the composite classification of our BPT diagram but is not included in the WISE color-color diagram due to it not having a listed $W_3$ magnitude.
% \end{comment}

\bibliographystyle{yahapj}
\bibliography{main}

@ARTICLE{assef+2013,
       author = {{Assef}, R.~J. and {Stern}, D. and {Kochanek}, C.~S. and {Blain}, A.~W. and {Brodwin}, M. and {Brown}, M.~J.~I. and {Donoso}, E. and {Eisenhardt}, P.~R.~M. and {Jannuzi}, B.~T. and {Jarrett}, T.~H. and {Stanford}, S.~A. and {Tsai}, C.-W. and {Wu}, J. and {Yan}, L.},
        title = "{Mid-infrared Selection of Active Galactic Nuclei with the Wide-field Infrared Survey Explorer. II. Properties of WISE-selected Active Galactic Nuclei in the NDWFS Bo{\"o}tes Field}",
      journal = {\apj},
         year = 2013,
        month = jul,
       volume = {772},
       number = {1},
          eid = {26},
        pages = {26},
          doi = {10.1088/0004-637X/772/1/26},
archivePrefix = {arXiv},
       eprint = {1209.6055},
 primaryClass = {astro-ph.CO},
       adsurl = {https://ui.adsabs.harvard.edu/abs/2013ApJ...772...26A}
}

@dataset{assef+2018,
       author = {{Assef}, R.~J. and {Stern}, D. and {Noirot}, G. and {Jun}, H.~D. and {Cutri}, R.~M. and {Eisenhardt}, P.~R.~M.},
        title = "{VizieR Online Data Catalog: The WISE AGN candidates catalogs (Assef+, 2018)}",
 howpublished = {VizieR On-line Data Catalog: J/ApJS/234/23. Originally published in: 2018ApJS..234...23A},
         year = 2018,
        month = mar,
          eid = {J/ApJS/234/23},
          doi = {10.26093/cds/vizier.22340023},
       adsurl = {https://ui.adsabs.harvard.edu/abs/2018yCat..22340023A}
}

@ARTICLE{abazajian+2009,
       author = {{Abazajian}, Kevork N. and {Adelman-McCarthy}, Jennifer K. and {Ag{\"u}eros}, Marcel A. and {Allam}, Sahar S. and {Allende Prieto}, Carlos and {An}, Deokkeun and {Anderson}, Kurt S.~J. and {Anderson}, Scott F. and {Annis}, James and {Bahcall}, Neta A. and {Bailer-Jones}, C.~A.~L. and {Barentine}, J.~C. and {Bassett}, Bruce A. and {Becker}, Andrew C. and {Beers}, Timothy C. and {Bell}, Eric F. and {Belokurov}, Vasily and {Berlind}, Andreas A. and {Berman}, Eileen F. and {Bernardi}, Mariangela and {Bickerton}, Steven J. and {Bizyaev}, Dmitry and {Blakeslee}, John P. and {Blanton}, Michael R. and {Bochanski}, John J. and {Boroski}, William N. and {Brewington}, Howard J. and {Brinchmann}, Jarle and {Brinkmann}, J. and {Brunner}, Robert J. and {Budav{\'a}ri}, Tam{\'a}s and {Carey}, Larry N. and {Carliles}, Samuel and {Carr}, Michael A. and {Castander}, Francisco J. and {Cinabro}, David and {Connolly}, A.~J. and {Csabai}, Istv{\'a}n and {Cunha}, Carlos E. and {Czarapata}, Paul C. and {Davenport}, James R.~A. and {de Haas}, Ernst and {Dilday}, Ben and {Doi}, Mamoru and {Eisenstein}, Daniel J. and {Evans}, Michael L. and {Evans}, N.~W. and {Fan}, Xiaohui and {Friedman}, Scott D. and {Frieman}, Joshua A. and {Fukugita}, Masataka and {G{\"a}nsicke}, Boris T. and {Gates}, Evalyn and {Gillespie}, Bruce and {Gilmore}, G. and {Gonzalez}, Belinda and {Gonzalez}, Carlos F. and {Grebel}, Eva K. and {Gunn}, James E. and {Gy{\"o}ry}, Zsuzsanna and {Hall}, Patrick B. and {Harding}, Paul and {Harris}, Frederick H. and {Harvanek}, Michael and {Hawley}, Suzanne L. and {Hayes}, Jeffrey J.~E. and {Heckman}, Timothy M. and {Hendry}, John S. and {Hennessy}, Gregory S. and {Hindsley}, Robert B. and {Hoblitt}, J. and {Hogan}, Craig J. and {Hogg}, David W. and {Holtzman}, Jon A. and {Hyde}, Joseph B. and {Ichikawa}, Shin-ichi and {Ichikawa}, Takashi and {Im}, Myungshin and {Ivezi{\'c}}, {\v{Z}}eljko and {Jester}, Sebastian and {Jiang}, Linhua and {Johnson}, Jennifer A. and {Jorgensen}, Anders M. and {Juri{\'c}}, Mario and {Kent}, Stephen M. and {Kessler}, R. and {Kleinman}, S.~J. and {Knapp}, G.~R. and {Konishi}, Kohki and {Kron}, Richard G. and {Krzesinski}, Jurek and {Kuropatkin}, Nikolay and {Lampeitl}, Hubert and {Lebedeva}, Svetlana and {Lee}, Myung Gyoon and {Lee}, Young Sun and {French Leger}, R. and {L{\'e}pine}, S{\'e}bastien and {Li}, Nolan and {Lima}, Marcos and {Lin}, Huan and {Long}, Daniel C. and {Loomis}, Craig P. and {Loveday}, Jon and {Lupton}, Robert H. and {Magnier}, Eugene and {Malanushenko}, Olena and {Malanushenko}, Viktor and {Mandelbaum}, Rachel and {Margon}, Bruce and {Marriner}, John P. and {Mart{\'\i}nez-Delgado}, David and {Matsubara}, Takahiko and {McGehee}, Peregrine M. and {McKay}, Timothy A. and {Meiksin}, Avery and {Morrison}, Heather L. and {Mullally}, Fergal and {Munn}, Jeffrey A. and {Murphy}, Tara and {Nash}, Thomas and {Nebot}, Ada and {Neilsen}, Jr., Eric H. and {Newberg}, Heidi Jo and {Newman}, Peter R. and {Nichol}, Robert C. and {Nicinski}, Tom and {Nieto-Santisteban}, Maria and {Nitta}, Atsuko and {Okamura}, Sadanori and {Oravetz}, Daniel J. and {Ostriker}, Jeremiah P. and {Owen}, Russell and {Padmanabhan}, Nikhil and {Pan}, Kaike and {Park}, Changbom and {Pauls}, George and {Peoples}, Jr., John and {Percival}, Will J. and {Pier}, Jeffrey R. and {Pope}, Adrian C. and {Pourbaix}, Dimitri and {Price}, Paul A. and {Purger}, Norbert and {Quinn}, Thomas and {Raddick}, M. Jordan and {Re Fiorentin}, Paola and {Richards}, Gordon T. and {Richmond}, Michael W. and {Riess}, Adam G. and {Rix}, Hans-Walter and {Rockosi}, Constance M. and {Sako}, Masao and {Schlegel}, David J. and {Schneider}, Donald P. and {Scholz}, Ralf-Dieter and {Schreiber}, Matthias R. and {Schwope}, Axel D. and {Seljak}, Uro{\v{s}} and {Sesar}, Branimir and {Sheldon}, Erin and {Shimasaku}, Kazu and {Sibley}, Valena C. and {Simmons}, A.~E. and {Sivarani}, Thirupathi and {Allyn Smith}, J. and {Smith}, Martin C. and {Smol{\v{c}}i{\'c}}, Vernesa and {Snedden}, Stephanie A. and {Stebbins}, Albert and {Steinmetz}, Matthias and {Stoughton}, Chris and {Strauss}, Michael A. and {SubbaRao}, Mark and {Suto}, Yasushi and {Szalay}, Alexander S. and {Szapudi}, Istv{\'a}n and {Szkody}, Paula and {Tanaka}, Masayuki and {Tegmark}, Max and {Teodoro}, Luis F.~A. and {Thakar}, Aniruddha R. and {Tremonti}, Christy A. and {Tucker}, Douglas L. and {Uomoto}, Alan and {Vanden Berk}, Daniel E. and {Vandenberg}, Jan and {Vidrih}, S. and {Vogeley}, Michael S. and {Voges}, Wolfgang and {Vogt}, Nicole P. and {Wadadekar}, Yogesh and {Watters}, Shannon and {Weinberg}, David H. and {West}, Andrew A. and {White}, Simon D.~M. and {Wilhite}, Brian C. and {Wonders}, Alainna C. and {Yanny}, Brian and {Yocum}, D.~R.},
       title = "{The Seventh Data Release of the Sloan Digital Sky Survey}",
       journal = {\apjs},
       year = 2009,
       month = jun,
       volume = {182},
       number = {2},
       pages = {543-558},
       doi = {10.1088/0067-0049/182/2/543},
archivePrefix = {arXiv},
       eprint = {0812.0649},
 primaryClass = {astro-ph},
       adsurl = {https://ui.adsabs.harvard.edu/abs/2009ApJS..182..543A}
}

@ARTICLE{astropy:2022,
       author = {{Astropy Collaboration} and {Price-Whelan}, Adrian M. and {Lim}, Pey Lian and {Earl}, Nicholas and {Starkman}, Nathaniel and {Bradley}, Larry and {Shupe}, David L. and {Patil}, Aarya A. and {Corrales}, Lia and {Brasseur}, C.~E. and {N{\"o}the}, Maximilian and {Donath}, Axel and {Tollerud}, Erik and {Morris}, Brett M. and {Ginsburg}, Adam and {Vaher}, Eero and {Weaver}, Benjamin A. and {Tocknell}, James and {Jamieson}, William and {van Kerkwijk}, Marten H. and {Robitaille}, Thomas P. and {Merry}, Bruce and {Bachetti}, Matteo and {G{\"u}nther}, H. Moritz and {Aldcroft}, Thomas L. and {Alvarado-Montes}, Jaime A. and {Archibald}, Anne M. and {B{\'o}di}, Attila and {Bapat}, Shreyas and {Barentsen}, Geert and {Baz{\'a}n}, Juanjo and {Biswas}, Manish and {Boquien}, M{\'e}d{\'e}ric and {Burke}, D.~J. and {Cara}, Daria and {Cara}, Mihai and {Conroy}, Kyle E. and {Conseil}, Simon and {Craig}, Matthew W. and {Cross}, Robert M. and {Cruz}, Kelle L. and {D'Eugenio}, Francesco and {Dencheva}, Nadia and {Devillepoix}, Hadrien A.~R. and {Dietrich}, J{\"o}rg P. and {Eigenbrot}, Arthur Davis and {Erben}, Thomas and {Ferreira}, Leonardo and {Foreman-Mackey}, Daniel and {Fox}, Ryan and {Freij}, Nabil and {Garg}, Suyog and {Geda}, Robel and {Glattly}, Lauren and {Gondhalekar}, Yash and {Gordon}, Karl D. and {Grant}, David and {Greenfield}, Perry and {Groener}, Austen M. and {Guest}, Steve and {Gurovich}, Sebastian and {Handberg}, Rasmus and {Hart}, Akeem and {Hatfield-Dodds}, Zac and {Homeier}, Derek and {Hosseinzadeh}, Griffin and {Jenness}, Tim and {Jones}, Craig K. and {Joseph}, Prajwel and {Kalmbach}, J. Bryce and {Karamehmetoglu}, Emir and {Ka{\l}uszy{\'n}ski}, Miko{\l}aj and {Kelley}, Michael S.~P. and {Kern}, Nicholas and {Kerzendorf}, Wolfgang E. and {Koch}, Eric W. and {Kulumani}, Shankar and {Lee}, Antony and {Ly}, Chun and {Ma}, Zhiyuan and {MacBride}, Conor and {Maljaars}, Jakob M. and {Muna}, Demitri and {Murphy}, N.~A. and {Norman}, Henrik and {O'Steen}, Richard and {Oman}, Kyle A. and {Pacifici}, Camilla and {Pascual}, Sergio and {Pascual-Granado}, J. and {Patil}, Rohit R. and {Perren}, Gabriel I. and {Pickering}, Timothy E. and {Rastogi}, Tanuj and {Roulston}, Benjamin R. and {Ryan}, Daniel F. and {Rykoff}, Eli S. and {Sabater}, Jose and {Sakurikar}, Parikshit and {Salgado}, Jes{\'u}s and {Sanghi}, Aniket and {Saunders}, Nicholas and {Savchenko}, Volodymyr and {Schwardt}, Ludwig and {Seifert-Eckert}, Michael and {Shih}, Albert Y. and {Jain}, Anany Shrey and {Shukla}, Gyanendra and {Sick}, Jonathan and {Simpson}, Chris and {Singanamalla}, Sudheesh and {Singer}, Leo P. and {Singhal}, Jaladh and {Sinha}, Manodeep and {Sip{\H{o}}cz}, Brigitta M. and {Spitler}, Lee R. and {Stansby}, David and {Streicher}, Ole and {{\v{S}}umak}, Jani and {Swinbank}, John D. and {Taranu}, Dan S. and {Tewary}, Nikita and {Tremblay}, Grant R. and {de Val-Borro}, Miguel and {Van Kooten}, Samuel J. and {Vasovi{\'c}}, Zlatan and {Verma}, Shresth and {de Miranda Cardoso}, Jos{\'e} Vin{\'\i}cius and {Williams}, Peter K.~G. and {Wilson}, Tom J. and {Winkel}, Benjamin and {Wood-Vasey}, W.~M. and {Xue}, Rui and {Yoachim}, Peter and {Zhang}, Chen and {Zonca}, Andrea and {Astropy Project Contributors}},
        title = "{The Astropy Project: Sustaining and Growing a Community-oriented Open-source Project and the Latest Major Release (v5.0) of the Core Package}",
      journal = {\apj},
         year = 2022,
        month = aug,
       volume = {935},
       number = {2},
          eid = {167},
        pages = {167},
          doi = {10.3847/1538-4357/ac7c74},
archivePrefix = {arXiv},
       eprint = {2206.14220},
 primaryClass = {astro-ph.IM},
       adsurl = {https://ui.adsabs.harvard.edu/abs/2022ApJ...935..167A}
}

@ARTICLE{astropy:2018,
       author = {{Astropy Collaboration} and {Price-Whelan}, A.~M. and {Sip{\H{o}}cz}, B.~M. and {G{\"u}nther}, H.~M. and {Lim}, P.~L. and {Crawford}, S.~M. and {Conseil}, S. and {Shupe}, D.~L. and {Craig}, M.~W. and {Dencheva}, N. and {Ginsburg}, A. and {VanderPlas}, J.~T. and {Bradley}, L.~D. and {P{\'e}rez-Su{\'a}rez}, D. and {de Val-Borro}, M. and {Aldcroft}, T.~L. and {Cruz}, K.~L. and {Robitaille}, T.~P. and {Tollerud}, E.~J. and {Ardelean}, C. and {Babej}, T. and {Bach}, Y.~P. and {Bachetti}, M. and {Bakanov}, A.~V. and {Bamford}, S.~P. and {Barentsen}, G. and {Barmby}, P. and {Baumbach}, A. and {Berry}, K.~L. and {Biscani}, F. and {Boquien}, M. and {Bostroem}, K.~A. and {Bouma}, L.~G. and {Brammer}, G.~B. and {Bray}, E.~M. and {Breytenbach}, H. and {Buddelmeijer}, H. and {Burke}, D.~J. and {Calderone}, G. and {Cano Rodr{\'\i}guez}, J.~L. and {Cara}, M. and {Cardoso}, J.~V.~M. and {Cheedella}, S. and {Copin}, Y. and {Corrales}, L. and {Crichton}, D. and {D'Avella}, D. and {Deil}, C. and {Depagne}, {\'E}. and {Dietrich}, J.~P. and {Donath}, A. and {Droettboom}, M. and {Earl}, N. and {Erben}, T. and {Fabbro}, S. and {Ferreira}, L.~A. and {Finethy}, T. and {Fox}, R.~T. and {Garrison}, L.~H. and {Gibbons}, S.~L.~J. and {Goldstein}, D.~A. and {Gommers}, R. and {Greco}, J.~P. and {Greenfield}, P. and {Groener}, A.~M. and {Grollier}, F. and {Hagen}, A. and {Hirst}, P. and {Homeier}, D. and {Horton}, A.~J. and {Hosseinzadeh}, G. and {Hu}, L. and {Hunkeler}, J.~S. and {Ivezi{\'c}}, {\v{Z}}. and {Jain}, A. and {Jenness}, T. and {Kanarek}, G. and {Kendrew}, S. and {Kern}, N.~S. and {Kerzendorf}, W.~E. and {Khvalko}, A. and {King}, J. and {Kirkby}, D. and {Kulkarni}, A.~M. and {Kumar}, A. and {Lee}, A. and {Lenz}, D. and {Littlefair}, S.~P. and {Ma}, Z. and {Macleod}, D.~M. and {Mastropietro}, M. and {McCully}, C. and {Montagnac}, S. and {Morris}, B.~M. and {Mueller}, M. and {Mumford}, S.~J. and {Muna}, D. and {Murphy}, N.~A. and {Nelson}, S. and {Nguyen}, G.~H. and {Ninan}, J.~P. and {N{\"o}the}, M. and {Ogaz}, S. and {Oh}, S. and {Parejko}, J.~K. and {Parley}, N. and {Pascual}, S. and {Patil}, R. and {Patil}, A.~A. and {Plunkett}, A.~L. and {Prochaska}, J.~X. and {Rastogi}, T. and {Reddy Janga}, V. and {Sabater}, J. and {Sakurikar}, P. and {Seifert}, M. and {Sherbert}, L.~E. and {Sherwood-Taylor}, H. and {Shih}, A.~Y. and {Sick}, J. and {Silbiger}, M.~T. and {Singanamalla}, S. and {Singer}, L.~P. and {Sladen}, P.~H. and {Sooley}, K.~A. and {Sornarajah}, S. and {Streicher}, O. and {Teuben}, P. and {Thomas}, S.~W. and {Tremblay}, G.~R. and {Turner}, J.~E.~H. and {Terr{\'o}n}, V. and {van Kerkwijk}, M.~H. and {de la Vega}, A. and {Watkins}, L.~L. and {Weaver}, B.~A. and {Whitmore}, J.~B. and {Woillez}, J. and {Zabalza}, V. and {Astropy Contributors}},
        title = "{The Astropy Project: Building an Open-science Project and Status of the v2.0 Core Package}",
      journal = {\aj},
         year = 2018,
        month = sep,
       volume = {156},
       number = {3},
          eid = {123},
        pages = {123},
          doi = {10.3847/1538-3881/aabc4f},
archivePrefix = {arXiv},
       eprint = {1801.02634},
 primaryClass = {astro-ph.IM},
       adsurl = {https://ui.adsabs.harvard.edu/abs/2018AJ....156..123A}
}

@ARTICLE{astropy:2013,
       author = {{Astropy Collaboration} and {Robitaille}, Thomas P. and {Tollerud}, Erik J. and {Greenfield}, Perry and {Droettboom}, Michael and {Bray}, Erik and {Aldcroft}, Tom and {Davis}, Matt and {Ginsburg}, Adam and {Price-Whelan}, Adrian M. and {Kerzendorf}, Wolfgang E. and {Conley}, Alexander and {Crighton}, Neil and {Barbary}, Kyle and {Muna}, Demitri and {Ferguson}, Henry and {Grollier}, Fr{\'e}d{\'e}ric and {Parikh}, Madhura M. and {Nair}, Prasanth H. and {Unther}, Hans M. and {Deil}, Christoph and {Woillez}, Julien and {Conseil}, Simon and {Kramer}, Roban and {Turner}, James E.~H. and {Singer}, Leo and {Fox}, Ryan and {Weaver}, Benjamin A. and {Zabalza}, Victor and {Edwards}, Zachary I. and {Azalee Bostroem}, K. and {Burke}, D.~J. and {Casey}, Andrew R. and {Crawford}, Steven M. and {Dencheva}, Nadia and {Ely}, Justin and {Jenness}, Tim and {Labrie}, Kathleen and {Lim}, Pey Lian and {Pierfederici}, Francesco and {Pontzen}, Andrew and {Ptak}, Andy and {Refsdal}, Brian and {Servillat}, Mathieu and {Streicher}, Ole},
        title = "{Astropy: A community Python package for astronomy}",
      journal = {\aap},
         year = 2013,
        month = oct,
       volume = {558},
          eid = {A33},
        pages = {A33},
          doi = {10.1051/0004-6361/201322068},
archivePrefix = {arXiv},
       eprint = {1307.6212},
 primaryClass = {astro-ph.IM},
       adsurl = {https://ui.adsabs.harvard.edu/abs/2013A&A...558A..33A}
}

@ARTICLE{agazie+2023,
       author = {{Agazie}, Gabriella and {Anumarlapudi}, Akash and {Archibald}, Anne M. and {Arzoumanian}, Zaven and {Baker}, Paul T. and {B{\'e}csy}, Bence and {Blecha}, Laura and {Brazier}, Adam and {Brook}, Paul R. and {Burke-Spolaor}, Sarah and {Burnette}, Rand and {Case}, Robin and {Charisi}, Maria and {Chatterjee}, Shami and {Chatziioannou}, Katerina and {Cheeseboro}, Belinda D. and {Chen}, Siyuan and {Cohen}, Tyler and {Cordes}, James M. and {Cornish}, Neil J. and {Crawford}, Fronefield and {Cromartie}, H. Thankful and {Crowter}, Kathryn and {Cutler}, Curt J. and {Decesar}, Megan E. and {Degan}, Dallas and {Demorest}, Paul B. and {Deng}, Heling and {Dolch}, Timothy and {Drachler}, Brendan and {Ellis}, Justin A. and {Ferrara}, Elizabeth C. and {Fiore}, William and {Fonseca}, Emmanuel and {Freedman}, Gabriel E. and {Garver-Daniels}, Nate and {Gentile}, Peter A. and {Gersbach}, Kyle A. and {Glaser}, Joseph and {Good}, Deborah C. and {G{\"u}ltekin}, Kayhan and {Hazboun}, Jeffrey S. and {Hourihane}, Sophie and {Islo}, Kristina and {Jennings}, Ross J. and {Johnson}, Aaron D. and {Jones}, Megan L. and {Kaiser}, Andrew R. and {Kaplan}, David L. and {Kelley}, Luke Zoltan and {Kerr}, Matthew and {Key}, Joey S. and {Klein}, Tonia C. and {Laal}, Nima and {Lam}, Michael T. and {Lamb}, William G. and {Lazio}, T. Joseph W. and {Lewandowska}, Natalia and {Littenberg}, Tyson B. and {Liu}, Tingting and {Lommen}, Andrea and {Lorimer}, Duncan R. and {Luo}, Jing and {Lynch}, Ryan S. and {Ma}, Chung-Pei and {Madison}, Dustin R. and {Mattson}, Margaret A. and {McEwen}, Alexander and {McKee}, James W. and {McLaughlin}, Maura A. and {McMann}, Natasha and {Meyers}, Bradley W. and {Meyers}, Patrick M. and {Mingarelli}, Chiara M.~F. and {Mitridate}, Andrea and {Natarajan}, Priyamvada and {Ng}, Cherry and {Nice}, David J. and {Ocker}, Stella Koch and {Olum}, Ken D. and {Pennucci}, Timothy T. and {Perera}, Benetge B.~P. and {Petrov}, Polina and {Pol}, Nihan S. and {Radovan}, Henri A. and {Ransom}, Scott M. and {Ray}, Paul S. and {Romano}, Joseph D. and {Sardesai}, Shashwat C. and {Schmiedekamp}, Ann and {Schmiedekamp}, Carl and {Schmitz}, Kai and {Schult}, Levi and {Shapiro-Albert}, Brent J. and {Siemens}, Xavier and {Simon}, Joseph and {Siwek}, Magdalena S. and {Stairs}, Ingrid H. and {Stinebring}, Daniel R. and {Stovall}, Kevin and {Sun}, Jerry P. and {Susobhanan}, Abhimanyu and {Swiggum}, Joseph K. and {Taylor}, Jacob and {Taylor}, Stephen R. and {Turner}, Jacob E. and {Unal}, Caner and {Vallisneri}, Michele and {van Haasteren}, Rutger and {Vigeland}, Sarah J. and {Wahl}, Haley M. and {Wang}, Qiaohong and {Witt}, Caitlin A. and {Young}, Olivia and {Nanograv Collaboration}},
        title = "{The NANOGrav 15 yr Data Set: Evidence for a Gravitational-wave Background}",
      journal = {\apjl},
         year = 2023,
        month = jul,
       volume = {951},
       number = {1},
          eid = {L8},
        pages = {L8},
          doi = {10.3847/2041-8213/acdac6},
archivePrefix = {arXiv},
       eprint = {2306.16213},
 primaryClass = {astro-ph.HE},
       adsurl = {https://ui.adsabs.harvard.edu/abs/2023ApJ...951L...8A}
}

@ARTICLE{armus+2009,
       author = {{Armus}, L. and {Mazzarella}, J.~M. and {Evans}, A.~S. and {Surace}, J.~A. and {Sanders}, D.~B. and {Iwasawa}, K. and {Frayer}, D.~T. and {Howell}, J.~H. and {Chan}, B. and {Petric}, A. and {Vavilkin}, T. and {Kim}, D.~C. and {Haan}, S. and {Inami}, H. and {Murphy}, E.~J. and {Appleton}, P.~N. and {Barnes}, J.~E. and {Bothun}, G. and {Bridge}, C.~R. and {Charmandaris}, V. and {Jensen}, J.~B. and {Kewley}, L.~J. and {Lord}, S. and {Madore}, B.~F. and {Marshall}, J.~A. and {Melbourne}, J.~E. and {Rich}, J. and {Satyapal}, S. and {Schulz}, B. and {Spoon}, H.~W.~W. and {Sturm}, E. and {U}, V. and {Veilleux}, S. and {Xu}, K.},
        title = "{GOALS: The Great Observatories All-Sky LIRG Survey}",
      journal = {\pasp},
         year = 2009,
        month = jun,
       volume = {121},
       number = {880},
        pages = {559},
          doi = {10.1086/600092},
archivePrefix = {arXiv},
       eprint = {0904.4498},
 primaryClass = {astro-ph.CO},
       adsurl = {https://ui.adsabs.harvard.edu/abs/2009PASP..121..559A}
}

@ARTICLE{2013A&A...558A..33A,
       author = {{Astropy Collaboration} and {Robitaille}, Thomas P. and {Tollerud}, Erik J. and {Greenfield}, Perry and {Droettboom}, Michael and {Bray}, Erik and {Aldcroft}, Tom and {Davis}, Matt and {Ginsburg}, Adam and {Price-Whelan}, Adrian M. and {Kerzendorf}, Wolfgang E. and {Conley}, Alexander and {Crighton}, Neil and {Barbary}, Kyle and {Muna}, Demitri and {Ferguson}, Henry and {Grollier}, Fr{\'e}d{\'e}ric and {Parikh}, Madhura M. and {Nair}, Prasanth H. and {Unther}, Hans M. and {Deil}, Christoph and {Woillez}, Julien and {Conseil}, Simon and {Kramer}, Roban and {Turner}, James E.~H. and {Singer}, Leo and {Fox}, Ryan and {Weaver}, Benjamin A. and {Zabalza}, Victor and {Edwards}, Zachary I. and {Azalee Bostroem}, K. and {Burke}, D.~J. and {Casey}, Andrew R. and {Crawford}, Steven M. and {Dencheva}, Nadia and {Ely}, Justin and {Jenness}, Tim and {Labrie}, Kathleen and {Lim}, Pey Lian and {Pierfederici}, Francesco and {Pontzen}, Andrew and {Ptak}, Andy and {Refsdal}, Brian and {Servillat}, Mathieu and {Streicher}, Ole},
        title = "{Astropy: A community Python package for astronomy}",
      journal = {\aap},
         year = 2013,
        month = oct,
       volume = {558},
          eid = {A33},
        pages = {A33},
          doi = {10.1051/0004-6361/201322068},
archivePrefix = {arXiv},
       eprint = {1307.6212},
 primaryClass = {astro-ph.IM},
       adsurl = {https://ui.adsabs.harvard.edu/abs/2013A&A...558A..33A}
}

@ARTICLE{brinchmann+2004,
       author = {{Brinchmann}, J. and {Charlot}, S. and {White}, S.~D.~M. and {Tremonti}, C. and {Kauffmann}, G. and {Heckman}, T. and {Brinkmann}, J.},
        title = "{The physical properties of star-forming galaxies in the low-redshift Universe}",
      journal = {\mnras},
         year = 2004,
        month = jul,
       volume = {351},
       number = {4},
        pages = {1151-1179},
          doi = {10.1111/j.1365-2966.2004.07881.x},
archivePrefix = {arXiv},
       eprint = {astro-ph/0311060},
 primaryClass = {astro-ph},
       adsurl = {https://ui.adsabs.harvard.edu/abs/2004MNRAS.351.1151B}
}

@ARTICLE{baldwin+1981,
       author = {{Baldwin}, J.~A. and {Phillips}, M.~M. and {Terlevich}, R.},
        title = "{Classification parameters for the emission-line spectra of extragalactic objects.}",
      journal = {\pasp},
         year = 1981,
        month = feb,
       volume = {93},
        pages = {5-19},
          doi = {10.1086/130766},
       adsurl = {https://ui.adsabs.harvard.edu/abs/1981PASP...93....5B}
}

@ARTICLE{blecha+2018,
       author = {{Blecha}, Laura and {Snyder}, Gregory F. and {Satyapal}, Shobita and {Ellison}, Sara L.},
        title = "{The power of infrared AGN selection in mergers: a theoretical study}",
      journal = {\mnras},
         year = 2018,
        month = aug,
       volume = {478},
       number = {3},
        pages = {3056-3071},
          doi = {10.1093/mnras/sty1274},
archivePrefix = {arXiv},
       eprint = {1711.02094},
 primaryClass = {astro-ph.GA},
       adsurl = {https://ui.adsabs.harvard.edu/abs/2018MNRAS.478.3056B}
}

@ARTICLE{bianchi+2008,
       author = {{Bianchi}, Stefano and {Chiaberge}, Marco and {Piconcelli}, Enrico and {Guainazzi}, Matteo and {Matt}, Giorgio},
        title = "{Chandra unveils a binary active galactic nucleus in Mrk 463}",
      journal = {\mnras},
         year = 2008,
        month = may,
       volume = {386},
       number = {1},
        pages = {105-110},
          doi = {10.1111/j.1365-2966.2008.13078.x},
archivePrefix = {arXiv},
       eprint = {0802.0825},
 primaryClass = {astro-ph},
       adsurl = {https://ui.adsabs.harvard.edu/abs/2008MNRAS.386..105B}
}

@INPROCEEDINGS{Borne+1999,
       author = {{Borne}, K.~D. and {Bushouse}, H. and {Colina}, L. and {Lucas}, R.~A. and {Baker}, A. and {Clements}, D. and {Lawrence}, A. and {Oliver}, S. and {Rowan-Robinson}, M.},
        title = "{NICMOS and WFPC2 Imaging of Ultraluminous Galaxies}",
    booktitle = {Astrophysics with Infrared Surveys: A Prelude to SIRTF},
         year = 1999,
       editor = {{Bicay}, Michael D. and {Cutri}, Roc M. and {Madore}, Barry F.},
       series = {Astronomical Society of the Pacific Conference Series},
       volume = {177},
        month = jan,
        pages = {167},
          doi = {10.48550/arXiv.astro-ph/9809040},
archivePrefix = {arXiv},
       eprint = {astro-ph/9809040},
 primaryClass = {astro-ph},
       adsurl = {https://ui.adsabs.harvard.edu/abs/1999ASPC..177..167B}
}

@ARTICLE{conselice+2006,
       author = {{Conselice}, Christopher J.},
        title = "{Early and Rapid Merging as a Formation Mechanism of Massive Galaxies: Empirical Constraints}",
      journal = {\apj},
         year = 2006,
        month = feb,
       volume = {638},
       number = {2},
        pages = {686-702},
          doi = {10.1086/499067},
archivePrefix = {arXiv},
       eprint = {astro-ph/0507146},
 primaryClass = {astro-ph},
       adsurl = {https://ui.adsabs.harvard.edu/abs/2006ApJ...638..686C}
}

@ARTICLE{casali+2007,
       author = {{Casali}, M. and {Adamson}, A. and {Alves de Oliveira}, C. and {Almaini}, O. and {Burch}, K. and {Chuter}, T. and {Elliot}, J. and {Folger}, M. and {Foucaud}, S. and {Hambly}, N. and {Hastie}, M. and {Henry}, D. and {Hirst}, P. and {Irwin}, M. and {Ives}, D. and {Lawrence}, A. and {Laidlaw}, K. and {Lee}, D. and {Lewis}, J. and {Lunney}, D. and {McLay}, S. and {Montgomery}, D. and {Pickup}, A. and {Read}, M. and {Rees}, N. and {Robson}, I. and {Sekiguchi}, K. and {Vick}, A. and {Warren}, S. and {Woodward}, B.},
        title = "{The UKIRT wide-field camera}",
      journal = {\aap},
         year = 2007,
        month = may,
       volume = {467},
       number = {2},
        pages = {777-784},
          doi = {10.1051/0004-6361:20066514},
       adsurl = {https://ui.adsabs.harvard.edu/abs/2007A&A...467..777C}
}

@ARTICLE{chen+2023b,
       author = {{Chen}, Yu-Ching and {Liu}, Xin and {Lazio}, Joseph and {Breiding}, Peter and {Burke-Spolaor}, Sarah and {Hwang}, Hsiang-Chih and {Shen}, Yue and {Zakamska}, Nadia L.},
        title = "{Varstrometry for Off-nucleus and Dual Sub-kiloparsec Active Galactic Nuclei (VODKA): Very Long Baseline Array Searches for Dual or Off-nucleus Quasars and Small-scale Jets}",
      journal = {\apj},
         year = 2023,
        month = nov,
       volume = {958},
       number = {1},
          eid = {29},
        pages = {29},
          doi = {10.3847/1538-4357/ad00b3},
archivePrefix = {arXiv},
       eprint = {2307.06363},
 primaryClass = {astro-ph.GA},
       adsurl = {https://ui.adsabs.harvard.edu/abs/2023ApJ...958...29C}
}

@ARTICLE{chen+2022,
       author = {{Chen}, Yu-Ching and {Hwang}, Hsiang-Chih and {Shen}, Yue and {Liu}, Xin and {Zakamska}, Nadia L. and {Yang}, Qian and {Li}, Jennifer I.},
        title = "{Varstrometry for Off-nucleus and Dual Subkiloparsec AGN (VODKA): Hubble Space Telescope Discovers Double Quasars}",
      journal = {\apj},
         year = 2022,
        month = feb,
       volume = {925},
       number = {2},
          eid = {162},
        pages = {162},
          doi = {10.3847/1538-4357/ac401b},
archivePrefix = {arXiv},
       eprint = {2108.01672},
 primaryClass = {astro-ph.HE},
       adsurl = {https://ui.adsabs.harvard.edu/abs/2022ApJ...925..162C}
}

@ARTICLE{comerford+2018,
       author = {{Comerford}, Julia M. and {Nevin}, Rebecca and {Stemo}, Aaron and {M{\"u}ller-S{\'a}nchez}, Francisco and {Barrows}, R. Scott and {Cooper}, Michael C. and {Newman}, Jeffrey A.},
        title = "{The Origin of Double-peaked Narrow Lines in Active Galactic Nuclei. IV. Association with Galaxy Mergers}",
      journal = {\apj},
         year = 2018,
        month = nov,
       volume = {867},
       number = {1},
          eid = {66},
        pages = {66},
          doi = {10.3847/1538-4357/aae2b4},
archivePrefix = {arXiv},
       eprint = {1810.11543},
 primaryClass = {astro-ph.GA},
       adsurl = {https://ui.adsabs.harvard.edu/abs/2018ApJ...867...66C}
}

@ARTICLE{comerford+2013,
       author = {{Comerford}, Julia M. and {Schluns}, Kyle and {Greene}, Jenny E. and {Cool}, Richard J.},
        title = "{Dual Supermassive Black Hole Candidates in the AGN and Galaxy Evolution Survey}",
      journal = {\apj},
         year = 2013,
        month = nov,
       volume = {777},
       number = {1},
          eid = {64},
        pages = {64},
          doi = {10.1088/0004-637X/777/1/64},
archivePrefix = {arXiv},
       eprint = {1309.2284},
 primaryClass = {astro-ph.CO},
       adsurl = {https://ui.adsabs.harvard.edu/abs/2013ApJ...777...64C}
}

@ARTICLE{comerford+2011,
       author = {{Comerford}, Julia M. and {Pooley}, David and {Gerke}, Brian F. and {Madejski}, Greg M.},
        title = "{Chandra Observations of a 1.9 kpc Separation Double X-Ray Source in a Candidate Dual Active Galactic Nucleus Galaxy at z = 0.16}",
      journal = {\apjl},
         year = 2011,
        month = aug,
       volume = {737},
       number = {1},
          eid = {L19},
        pages = {L19},
          doi = {10.1088/2041-8205/737/1/L19},
archivePrefix = {arXiv},
       eprint = {1106.0746},
 primaryClass = {astro-ph.CO},
       adsurl = {https://ui.adsabs.harvard.edu/abs/2011ApJ...737L..19C}
}

@ARTICLE{chen+2023a,
       author = {{Chen}, Nianyi and {Di Matteo}, Tiziana and {Ni}, Yueying and {Tremmel}, Michael and {DeGraf}, Colin and {Shen}, Yue and {Holgado}, A. Miguel and {Bird}, Simeon and {Croft}, Rupert and {Feng}, Yu},
        title = "{Properties and evolution of dual and offset AGN in the ASTRID simulation at z   2}",
      journal = {\mnras},
         year = 2023,
        month = jun,
       volume = {522},
       number = {2},
        pages = {1895-1913},
          doi = {10.1093/mnras/stad834},
archivePrefix = {arXiv},
       eprint = {2208.04970},
 primaryClass = {astro-ph.GA},
       adsurl = {https://ui.adsabs.harvard.edu/abs/2023MNRAS.522.1895C}
}

@ARTICLE{capelo+2017,
       author = {{Capelo}, Pedro R. and {Dotti}, Massimo and {Volonteri}, Marta and {Mayer}, Lucio and {Bellovary}, Jillian M. and {Shen}, Sijing},
        title = "{A survey of dual active galactic nuclei in simulations of galaxy mergers: frequency and properties}",
      journal = {\mnras},
         year = 2017,
        month = aug,
       volume = {469},
       number = {4},
        pages = {4437-4454},
          doi = {10.1093/mnras/stx1067},
archivePrefix = {arXiv},
       eprint = {1611.09244},
 primaryClass = {astro-ph.GA},
       adsurl = {https://ui.adsabs.harvard.edu/abs/2017MNRAS.469.4437C}
}

@ARTICLE{capelo+2015,
       author = {{Capelo}, Pedro R. and {Volonteri}, Marta and {Dotti}, Massimo and {Bellovary}, Jillian M. and {Mayer}, Lucio and {Governato}, Fabio},
        title = "{Growth and activity of black holes in galaxy mergers with varying mass ratios}",
      journal = {\mnras},
         year = 2015,
        month = mar,
       volume = {447},
       number = {3},
        pages = {2123-2143},
          doi = {10.1093/mnras/stu2500},
archivePrefix = {arXiv},
       eprint = {1409.0004},
 primaryClass = {astro-ph.GA},
       adsurl = {https://ui.adsabs.harvard.edu/abs/2015MNRAS.447.2123C}
}

@ARTICLE{comerford+2015,
       author = {{Comerford}, Julia M. and {Pooley}, David and {Barrows}, R. Scott and {Greene}, Jenny E. and {Zakamska}, Nadia L. and {Madejski}, Greg M. and {Cooper}, Michael C.},
        title = "{Merger-driven Fueling of Active Galactic Nuclei: Six Dual and Offset AGNs Discovered with Chandra and Hubble Space Telescope Observations}",
      journal = {\apj},
         year = 2015,
        month = jun,
       volume = {806},
       number = {2},
          eid = {219},
        pages = {219},
          doi = {10.1088/0004-637X/806/2/219},
archivePrefix = {arXiv},
       eprint = {1504.01391},
 primaryClass = {astro-ph.GA},
       adsurl = {https://ui.adsabs.harvard.edu/abs/2015ApJ...806..219C}
}

@ARTICLE{Comerford+2022,
       author = {{Comerford}, Julia M. and {Negus}, James and {Barrows}, R. Scott and {Wylezalek}, Dominika and {Greene}, Jenny E. and {M{\"u}ller-S{\'a}nchez}, Francisco and {Nevin}, Rebecca},
        title = "{Toward a More Complete Optical Census of Active Galactic Nuclei via Spatially Resolved Spectroscopy}",
      journal = {\apj},
         year = 2022,
        month = mar,
       volume = {927},
       number = {1},
          eid = {23},
        pages = {23},
          doi = {10.3847/1538-4357/ac496a},
archivePrefix = {arXiv},
       eprint = {2201.07250},
 primaryClass = {astro-ph.GA},
       adsurl = {https://ui.adsabs.harvard.edu/abs/2022ApJ...927...23C}
}

@ARTICLE{darg+2010,
       author = {{Darg}, D.~W. and {Kaviraj}, S. and {Lintott}, C.~J. and {Schawinski}, K. and {Sarzi}, M. and {Bamford}, S. and {Silk}, J. and {Proctor}, R. and {Andreescu}, D. and {Murray}, P. and {Nichol}, R.~C. and {Raddick}, M.~J. and {Slosar}, A. and {Szalay}, A.~S. and {Thomas}, D. and {Vandenberg}, J.},
        title = "{Galaxy Zoo: the fraction of merging galaxies in the SDSS and their morphologies}",
      journal = {\mnras},
         year = 2010,
        month = jan,
       volume = {401},
       number = {2},
        pages = {1043-1056},
          doi = {10.1111/j.1365-2966.2009.15686.x},
archivePrefix = {arXiv},
       eprint = {0903.4937},
 primaryClass = {astro-ph.GA},
       adsurl = {https://ui.adsabs.harvard.edu/abs/2010MNRAS.401.1043D}
}

@ARTICLE{fu+2011a,
       author = {{Fu}, Hai and {Myers}, Adam D. and {Djorgovski}, S.~G. and {Yan}, Lin},
        title = "{Mergers in Double-peaked [O III] Active Galactic Nuclei}",
      journal = {\apj},
         year = 2011,
        month = jun,
       volume = {733},
       number = {2},
          eid = {103},
        pages = {103},
          doi = {10.1088/0004-637X/733/2/103},
archivePrefix = {arXiv},
       eprint = {1009.0767},
 primaryClass = {astro-ph.CO},
       adsurl = {https://ui.adsabs.harvard.edu/abs/2011ApJ...733..103F}
}

@ARTICLE{fu+2011b,
       author = {{Fu}, Hai and {Zhang}, Zhi-Yu and {Assef}, Roberto J. and {Stockton}, Alan and {Myers}, Adam D. and {Yan}, Lin and {Djorgovski}, S.~G. and {Wrobel}, J.~M. and {Riechers}, Dominik A.},
        title = "{A Kiloparsec-scale Binary Active Galactic Nucleus Confirmed by the Expanded Very Large Array}",
      journal = {\apjl},
         year = 2011,
        month = oct,
       volume = {740},
       number = {2},
          eid = {L44},
        pages = {L44},
          doi = {10.1088/2041-8205/740/2/L44},
archivePrefix = {arXiv},
       eprint = {1109.0008},
 primaryClass = {astro-ph.CO},
       adsurl = {https://ui.adsabs.harvard.edu/abs/2011ApJ...740L..44F}
}

@ARTICLE{fu+2012,
       author = {{Fu}, Hai and {Yan}, Lin and {Myers}, Adam D. and {Stockton}, Alan and {Djorgovski}, S.~G. and {Aldering}, G. and {Rich}, Jeffrey A.},
        title = "{The Nature of Double-peaked [O III] Active Galactic Nuclei}",
      journal = {\apj},
         year = 2012,
        month = jan,
       volume = {745},
       number = {1},
          eid = {67},
        pages = {67},
          doi = {10.1088/0004-637X/745/1/67},
archivePrefix = {arXiv},
       eprint = {1107.3564},
 primaryClass = {astro-ph.CO},
       adsurl = {https://ui.adsabs.harvard.edu/abs/2012ApJ...745...67F}
}

@ARTICLE{greene+2012,
       author = {{Greene}, Jenny E. and {Zakamska}, Nadia L. and {Smith}, Paul S.},
        title = "{A Spectacular Outflow in an Obscured Quasar}",
      journal = {\apj},
         year = 2012,
        month = feb,
       volume = {746},
       number = {1},
          eid = {86},
        pages = {86},
          doi = {10.1088/0004-637X/746/1/86},
archivePrefix = {arXiv},
       eprint = {1112.3358},
 primaryClass = {astro-ph.CO},
       adsurl = {https://ui.adsabs.harvard.edu/abs/2012ApJ...746...86G}
}

@ARTICLE{hernandeztoledo+2005,
       author = {{Hern{\'a}ndez-Toledo}, H.~M. and {Avila-Reese}, V. and {Conselice}, C.~J. and {Puerari}, I.},
        title = "{The Structural Properties of Isolated Galaxies, Spiral-Spiral Pairs, and Mergers: The Robustness of Galaxy Morphology during Secular Evolution}",
      journal = {\aj},
         year = 2005,
        month = feb,
       volume = {129},
       number = {2},
        pages = {682-697},
          doi = {10.1086/427134},
archivePrefix = {arXiv},
       eprint = {astro-ph/0410722},
 primaryClass = {astro-ph},
       adsurl = {https://ui.adsabs.harvard.edu/abs/2005AJ....129..682H}
}

@ARTICLE{haan+2011,
       author = {{Haan}, S. and {Surace}, J.~A. and {Armus}, L. and {Evans}, A.~S. and {Howell}, J.~H. and {Mazzarella}, J.~M. and {Kim}, D.~C. and {Vavilkin}, T. and {Inami}, H. and {Sanders}, D.~B. and {Petric}, A. and {Bridge}, C.~R. and {Melbourne}, J.~L. and {Charmandaris}, V. and {Diaz-Santos}, T. and {Murphy}, E.~J. and {U}, V. and {Stierwalt}, S. and {Marshall}, J.~A.},
        title = "{The Nuclear Structure in Nearby Luminous Infrared Galaxies: Hubble Space Telescope NICMOS Imaging of the GOALS Sample}",
      journal = {\aj},
         year = 2011,
        month = mar,
       volume = {141},
       number = {3},
          eid = {100},
        pages = {100},
          doi = {10.1088/0004-6256/141/3/100},
archivePrefix = {arXiv},
       eprint = {1012.4012},
 primaryClass = {astro-ph.CO},
       adsurl = {https://ui.adsabs.harvard.edu/abs/2011AJ....141..100H}
}

@ARTICLE{hambly+2008,
       author = {{Hambly}, N.~C. and {Collins}, R.~S. and {Cross}, N.~J.~G. and {Mann}, R.~G. and {Read}, M.~A. and {Sutorius}, E.~T.~W. and {Bond}, I. and {Bryant}, J. and {Emerson}, J.~P. and {Lawrence}, A. and {Rimoldini}, L. and {Stewart}, J.~M. and {Williams}, P.~M. and {Adamson}, A. and {Hirst}, P. and {Dye}, S. and {Warren}, S.~J.},
        title = "{The WFCAM Science Archive}",
      journal = {\mnras},
         year = 2008,
        month = feb,
       volume = {384},
       number = {2},
        pages = {637-662},
          doi = {10.1111/j.1365-2966.2007.12700.x},
archivePrefix = {arXiv},
       eprint = {0711.3593},
 primaryClass = {astro-ph},
       adsurl = {https://ui.adsabs.harvard.edu/abs/2008MNRAS.384..637H}
}

@ARTICLE{hewett+2006,
       author = {{Hewett}, P.~C. and {Warren}, S.~J. and {Leggett}, S.~K. and {Hodgkin}, S.~T.},
        title = "{The UKIRT Infrared Deep Sky Survey ZY JHK photometric system: passbands and synthetic colours}",
      journal = {\mnras},
         year = 2006,
        month = apr,
       volume = {367},
       number = {2},
        pages = {454-468},
          doi = {10.1111/j.1365-2966.2005.09969.x},
archivePrefix = {arXiv},
       eprint = {astro-ph/0601592},
 primaryClass = {astro-ph},
       adsurl = {https://ui.adsabs.harvard.edu/abs/2006MNRAS.367..454H}
}

@ARTICLE{hopkins+2006,
       author = {{Hopkins}, Philip F. and {Hernquist}, Lars},
        title = "{Fueling Low-Level AGN Activity through Stochastic Accretion of Cold Gas}",
      journal = {\apjs},
         year = 2006,
        month = sep,
       volume = {166},
       number = {1},
        pages = {1-36},
          doi = {10.1086/505753},
archivePrefix = {arXiv},
       eprint = {astro-ph/0603180},
 primaryClass = {astro-ph},
       adsurl = {https://ui.adsabs.harvard.edu/abs/2006ApJS..166....1H}
}

@ARTICLE{hwang+2020,
       author = {{Hwang}, Hsiang-Chih and {Shen}, Yue and {Zakamska}, Nadia and {Liu}, Xin},
        title = "{Varstrometry for Off-nucleus and Dual Subkiloparsec AGN (VODKA): Methodology and Initial Results with Gaia DR2}",
      journal = {\apj},
         year = 2020,
        month = jan,
       volume = {888},
       number = {2},
          eid = {73},
        pages = {73},
          doi = {10.3847/1538-4357/ab5c1a},
archivePrefix = {arXiv},
       eprint = {1908.02292},
 primaryClass = {astro-ph.GA},
       adsurl = {https://ui.adsabs.harvard.edu/abs/2020ApJ...888...73H}
}

@ARTICLE{hou+2019,
       author = {{Hou}, Meicun and {Liu}, Xin and {Guo}, Hengxiao and {Li}, Zhiyuan and {Shen}, Yue and {Green}, Paul J.},
        title = "{Active Galactic Nucleus Pairs from the Sloan Digital Sky Survey. III. Chandra X-Ray Observations Unveil Obscured Double Nuclei}",
      journal = {\apj},
         year = 2019,
        month = sep,
       volume = {882},
       number = {1},
          eid = {41},
        pages = {41},
          doi = {10.3847/1538-4357/ab3225},
archivePrefix = {arXiv},
       eprint = {1904.12998},
 primaryClass = {astro-ph.GA},
       adsurl = {https://ui.adsabs.harvard.edu/abs/2019ApJ...882...41H}
}

@ARTICLE{He+2023,
       author = {{He}, Lin and {Hou}, Meicun and {Li}, Zhiyuan and {Feng}, Shuai and {Liu}, Xin},
        title = "{A Chandra X-Ray Survey of Optically Selected Close Galaxy Pairs: Unexpectedly Low Occupation of Active Galactic Nuclei}",
      journal = {\apj},
         year = 2023,
        month = jun,
       volume = {949},
       number = {2},
          eid = {49},
        pages = {49},
          doi = {10.3847/1538-4357/acc5e1},
archivePrefix = {arXiv},
       eprint = {2303.08388},
 primaryClass = {astro-ph.GA},
       adsurl = {https://ui.adsabs.harvard.edu/abs/2023ApJ...949...49H}
}

@ARTICLE{2020Natur.585..357H,
       author = {{Harris}, Charles R. and {Millman}, K. Jarrod and {van der Walt}, St{\'e}fan J. and {Gommers}, Ralf and {Virtanen}, Pauli and {Cournapeau}, David and {Wieser}, Eric and {Taylor}, Julian and {Berg}, Sebastian and {Smith}, Nathaniel J. and {Kern}, Robert and {Picus}, Matti and {Hoyer}, Stephan and {van Kerkwijk}, Marten H. and {Brett}, Matthew and {Haldane}, Allan and {del R{\'\i}o}, Jaime Fern{\'a}ndez and {Wiebe}, Mark and {Peterson}, Pearu and {G{\'e}rard-Marchant}, Pierre and {Sheppard}, Kevin and {Reddy}, Tyler and {Weckesser}, Warren and {Abbasi}, Hameer and {Gohlke}, Christoph and {Oliphant}, Travis E.},
        title = "{Array programming with NumPy}",
      journal = {\nat},
         year = 2020,
        month = sep,
       volume = {585},
       number = {7825},
        pages = {357-362},
          doi = {10.1038/s41586-020-2649-2},
archivePrefix = {arXiv},
       eprint = {2006.10256},
 primaryClass = {cs.MS},
       adsurl = {https://ui.adsabs.harvard.edu/abs/2020Natur.585..357H}
}

@INPROCEEDINGS{irwin+2008,
       author = {{Irwin}, M.~J.},
        title = "{Processing Wide Field Imaging Data}",
    booktitle = {2007 ESO Instrument Calibration Workshop},
         year = 2008,
       editor = {{Kaufer}, Andreas and {Kerber}, Florian},
        month = jan,
        pages = {541},
          doi = {10.1007/978-3-540-76963-7_74},
       adsurl = {https://ui.adsabs.harvard.edu/abs/2008eic..work..541I}
}

@ARTICLE{iwasawa+2018,
       author = {{Iwasawa}, K. and {U}, V. and {Mazzarella}, J.~M. and {Medling}, A.~M. and {Sanders}, D.~B. and {Evans}, A.~S.},
        title = "{Testing a double AGN hypothesis for Mrk 273}",
      journal = {\aap},
         year = 2018,
        month = apr,
       volume = {611},
          eid = {A71},
        pages = {A71},
          doi = {10.1051/0004-6361/201731662},
archivePrefix = {arXiv},
       eprint = {1711.01750},
 primaryClass = {astro-ph.GA},
       adsurl = {https://ui.adsabs.harvard.edu/abs/2018A&A...611A..71I}
}

@ARTICLE{jarrett+2011,
       author = {{Jarrett}, T.~H. and {Cohen}, M. and {Masci}, F. and {Wright}, E. and {Stern}, D. and {Benford}, D. and {Blain}, A. and {Carey}, S. and {Cutri}, R.~M. and {Eisenhardt}, P. and {Lonsdale}, C. and {Mainzer}, A. and {Marsh}, K. and {Padgett}, D. and {Petty}, S. and {Ressler}, M. and {Skrutskie}, M. and {Stanford}, S. and {Surace}, J. and {Tsai}, C.~W. and {Wheelock}, S. and {Yan}, D.~L.},
        title = "{The Spitzer-WISE Survey of the Ecliptic Poles}",
      journal = {\apj},
         year = 2011,
        month = jul,
       volume = {735},
       number = {2},
          eid = {112},
        pages = {112},
          doi = {10.1088/0004-637X/735/2/112},
       adsurl = {https://ui.adsabs.harvard.edu/abs/2011ApJ...735..112J}
}

@ARTICLE{koss+2018,
       author = {{Koss}, Michael J. and {Blecha}, Laura and {Bernhard}, Phillip and {Hung}, Chao-Ling and {Lu}, Jessica R. and {Trakhtenbrot}, Benny and {Treister}, Ezequiel and {Weigel}, Anna and {Sartori}, Lia F. and {Mushotzky}, Richard and {Schawinski}, Kevin and {Ricci}, Claudio and {Veilleux}, Sylvain and {Sanders}, David B.},
        title = "{A population of luminous accreting black holes with hidden mergers}",
      journal = {\nat},
         year = 2018,
        month = nov,
       volume = {563},
       number = {7730},
        pages = {214-216},
          doi = {10.1038/s41586-018-0652-7},
archivePrefix = {arXiv},
       eprint = {1811.03641},
 primaryClass = {astro-ph.GA},
       adsurl = {https://ui.adsabs.harvard.edu/abs/2018Natur.563..214K}
}

@ARTICLE{kim+2020,
       author = {{Kim}, D. -C. and {Yoon}, Ilsang and {Evans}, A.~S. and {Kim}, Minjin and {Momjian}, E. and {Kim}, Ji Hoon},
        title = "{Dual AGN Candidates with Double-peaked [O III] Lines Matching that of Confirmed Dual AGNs}",
      journal = {\apj},
         year = 2020,
        month = nov,
       volume = {904},
       number = {1},
          eid = {23},
        pages = {23},
          doi = {10.3847/1538-4357/abb9a0},
archivePrefix = {arXiv},
       eprint = {2011.09961},
 primaryClass = {astro-ph.GA},
       adsurl = {https://ui.adsabs.harvard.edu/abs/2020ApJ...904...23K}
}

@ARTICLE{kelley+2017,
       author = {{Kelley}, Luke Zoltan and {Blecha}, Laura and {Hernquist}, Lars},
        title = "{Massive black hole binary mergers in dynamical galactic environments}",
      journal = {\mnras},
         year = 2017,
        month = jan,
       volume = {464},
       number = {3},
        pages = {3131-3157},
          doi = {10.1093/mnras/stw2452},
archivePrefix = {arXiv},
       eprint = {1606.01900},
 primaryClass = {astro-ph.HE},
       adsurl = {https://ui.adsabs.harvard.edu/abs/2017MNRAS.464.3131K}
}

@ARTICLE{kim+2021,
       author = {{Kim}, Minjin and {Barth}, Aaron J. and {Ho}, Luis C. and {Son}, Suyeon},
        title = "{A Hubble Space Telescope Imaging Survey of Low-redshift Swift-BAT Active Galaxies}",
      journal = {\apjs},
         year = 2021,
        month = oct,
       volume = {256},
       number = {2},
          eid = {40},
        pages = {40},
          doi = {10.3847/1538-4365/ac133e},
archivePrefix = {arXiv},
       eprint = {2107.04213},
 primaryClass = {astro-ph.GA},
       adsurl = {https://ui.adsabs.harvard.edu/abs/2021ApJS..256...40K}
}

@ARTICLE{kauffmann+2003b,
       author = {{Kauffmann}, Guinevere and {Heckman}, Timothy M. and {White}, Simon D.~M. and {Charlot}, St{\'e}phane and {Tremonti}, Christy and {Brinchmann}, Jarle and {Bruzual}, Gustavo and {Peng}, Eric W. and {Seibert}, Mark and {Bernardi}, Mariangela and {Blanton}, Michael and {Brinkmann}, Jon and {Castander}, Francisco and {Cs{\'a}bai}, Istvan and {Fukugita}, Masataka and {Ivezic}, Zeljko and {Munn}, Jeffrey A. and {Nichol}, Robert C. and {Padmanabhan}, Nikhil and {Thakar}, Aniruddha R. and {Weinberg}, David H. and {York}, Donald},
        title = "{Stellar masses and star formation histories for {}10$^{5}$ galaxies from the Sloan Digital Sky Survey}",
      journal = {\mnras},
         year = 2003,
        month = may,
       volume = {341},
       number = {1},
        pages = {33-53},
          doi = {10.1046/j.1365-8711.2003.06291.x},
archivePrefix = {arXiv},
       eprint = {astro-ph/0204055},
 primaryClass = {astro-ph},
       adsurl = {https://ui.adsabs.harvard.edu/abs/2003MNRAS.341...33K}
}

@ARTICLE{komossa+2003,
       author = {{Komossa}, S. and {Burwitz}, V. and {Hasinger}, G. and {Predehl}, P. and {Kaastra}, J.~S. and {Ikebe}, Y.},
        title = "{Discovery of a Binary Active Galactic Nucleus in the Ultraluminous Infrared Galaxy NGC 6240 Using Chandra}",
      journal = {\apjl},
         year = 2003,
        month = jan,
       volume = {582},
       number = {1},
        pages = {L15-L19},
          doi = {10.1086/346145},
archivePrefix = {arXiv},
       eprint = {astro-ph/0212099},
 primaryClass = {astro-ph},
       adsurl = {https://ui.adsabs.harvard.edu/abs/2003ApJ...582L..15K}
}

@ARTICLE{koss+2023,
       author = {{Koss}, Michael J. and {Treister}, Ezequiel and {Kakkad}, Darshan and {Casey-Clyde}, J. Andrew and {Kawamuro}, Taiki and {Williams}, Jonathan and {Foord}, Adi and {Trakhtenbrot}, Benny and {Bauer}, Franz E. and {Privon}, George C. and {Ricci}, Claudio and {Mushotzky}, Richard and {Barcos-Munoz}, Loreto and {Blecha}, Laura and {Connor}, Thomas and {Harrison}, Fiona and {Liu}, Tingting and {Magno}, Macon and {Mingarelli}, Chiara M.~F. and {Muller-Sanchez}, Francisco and {Oh}, Kyuseok and {Shimizu}, T. Taro and {Smith}, Krista Lynne and {Stern}, Daniel and {Tello}, Miguel Parra and {Urry}, C. Megan},
        title = "{UGC 4211: A Confirmed Dual Active Galactic Nucleus in the Local Universe at 230 pc Nuclear Separation}",
      journal = {\apjl},
         year = 2023,
        month = jan,
       volume = {942},
       number = {1},
          eid = {L24},
        pages = {L24},
          doi = {10.3847/2041-8213/aca8f0},
archivePrefix = {arXiv},
       eprint = {2301.03609},
 primaryClass = {astro-ph.GA},
       adsurl = {https://ui.adsabs.harvard.edu/abs/2023ApJ...942L..24K}
}

@ARTICLE{koss+2012,
       author = {{Koss}, Michael and {Mushotzky}, Richard and {Treister}, Ezequiel and {Veilleux}, Sylvain and {Vasudevan}, Ranjan and {Trippe}, Margaret},
        title = "{Understanding Dual Active Galactic Nucleus Activation in the nearby Universe}",
      journal = {\apjl},
         year = 2012,
        month = feb,
       volume = {746},
       number = {2},
          eid = {L22},
        pages = {L22},
          doi = {10.1088/2041-8205/746/2/L22},
archivePrefix = {arXiv},
       eprint = {1201.2944},
 primaryClass = {astro-ph.HE},
       adsurl = {https://ui.adsabs.harvard.edu/abs/2012ApJ...746L..22K}
}

@ARTICLE{kauffmann+2003a,
       author = {{Kauffmann}, Guinevere and {Heckman}, Timothy M. and {Tremonti}, Christy and {Brinchmann}, Jarle and {Charlot}, St{\'e}phane and {White}, Simon D.~M. and {Ridgway}, Susan E. and {Brinkmann}, Jon and {Fukugita}, Masataka and {Hall}, Patrick B. and {Ivezi{\'c}}, {\v{Z}}eljko and {Richards}, Gordon T. and {Schneider}, Donald P.},
        title = "{The host galaxies of active galactic nuclei}",
      journal = {\mnras},
         year = 2003,
        month = dec,
       volume = {346},
       number = {4},
        pages = {1055-1077},
          doi = {10.1111/j.1365-2966.2003.07154.x},
archivePrefix = {arXiv},
       eprint = {astro-ph/0304239},
 primaryClass = {astro-ph},
       adsurl = {https://ui.adsabs.harvard.edu/abs/2003MNRAS.346.1055K}
}

@ARTICLE{kewley+2001,
       author = {{Kewley}, L.~J. and {Dopita}, M.~A. and {Sutherland}, R.~S. and {Heisler}, C.~A. and {Trevena}, J.},
        title = "{Theoretical Modeling of Starburst Galaxies}",
      journal = {\apj},
         year = 2001,
        month = jul,
       volume = {556},
       number = {1},
        pages = {121-140},
          doi = {10.1086/321545},
archivePrefix = {arXiv},
       eprint = {astro-ph/0106324},
 primaryClass = {astro-ph},
       adsurl = {https://ui.adsabs.harvard.edu/abs/2001ApJ...556..121K}
}

@ARTICLE{lawrence+2007,
       author = {{Lawrence}, A. and {Warren}, S.~J. and {Almaini}, O. and {Edge}, A.~C. and {Hambly}, N.~C. and {Jameson}, R.~F. and {Lucas}, P. and {Casali}, M. and {Adamson}, A. and {Dye}, S. and {Emerson}, J.~P. and {Foucaud}, S. and {Hewett}, P. and {Hirst}, P. and {Hodgkin}, S.~T. and {Irwin}, M.~J. and {Lodieu}, N. and {McMahon}, R.~G. and {Simpson}, C. and {Smail}, I. and {Mortlock}, D. and {Folger}, M.},
        title = "{The UKIRT Infrared Deep Sky Survey (UKIDSS)}",
      journal = {\mnras},
         year = 2007,
        month = aug,
       volume = {379},
       number = {4},
        pages = {1599-1617},
          doi = {10.1111/j.1365-2966.2007.12040.x},
archivePrefix = {arXiv},
       eprint = {astro-ph/0604426},
 primaryClass = {astro-ph},
       adsurl = {https://ui.adsabs.harvard.edu/abs/2007MNRAS.379.1599L}
}

@ARTICLE{liu+2010b,
       author = {{Liu}, Xin and {Greene}, Jenny E. and {Shen}, Yue and {Strauss}, Michael A.},
        title = "{Discovery of Four kpc-scale Binary Active Galactic Nuclei}",
      journal = {\apjl},
         year = 2010,
        month = may,
       volume = {715},
       number = {1},
        pages = {L30-L34},
          doi = {10.1088/2041-8205/715/1/L30},
archivePrefix = {arXiv},
       eprint = {1003.3467},
 primaryClass = {astro-ph.CO},
       adsurl = {https://ui.adsabs.harvard.edu/abs/2010ApJ...715L..30L}
}

@ARTICLE{lyu_liu+2016,
       author = {{Lyu}, Yang and {Liu}, Xin},
        title = "{A high fraction of double-peaked narrow emission lines in powerful active galactic nuclei}",
      journal = {\mnras},
         year = 2016,
        month = nov,
       volume = {463},
       number = {1},
        pages = {24-36},
          doi = {10.1093/mnras/stw1945},
archivePrefix = {arXiv},
       eprint = {1606.06742},
 primaryClass = {astro-ph.GA},
       adsurl = {https://ui.adsabs.harvard.edu/abs/2016MNRAS.463...24L}
}

@ARTICLE{li+2022,
       author = {{Li}, Kunyang and {Bogdanovi{\'c}}, Tamara and {Ballantyne}, David R. and {Bonetti}, Matteo},
        title = "{Massive Black Hole Binaries from the TNG50-3 Simulation. I. Coalescence and LISA Detection Rates}",
      journal = {\apj},
         year = 2022,
        month = jul,
       volume = {933},
       number = {1},
          eid = {104},
        pages = {104},
          doi = {10.3847/1538-4357/ac74b5},
archivePrefix = {arXiv},
       eprint = {2201.11088},
 primaryClass = {astro-ph.GA},
       adsurl = {https://ui.adsabs.harvard.edu/abs/2022ApJ...933..104L}
}

@ARTICLE{li+2021,
       author = {{Li}, Kunyang and {Ballantyne}, David R. and {Bogdanovi{\'c}}, Tamara},
        title = "{The Detectability of Kiloparsec-scale Dual Active Galactic Nuclei: The Impact of Galactic Structure and Black Hole Orbital Properties}",
      journal = {\apj},
         year = 2021,
        month = aug,
       volume = {916},
       number = {2},
          eid = {110},
        pages = {110},
          doi = {10.3847/1538-4357/ac06a0},
archivePrefix = {arXiv},
       eprint = {2103.02862},
 primaryClass = {astro-ph.GA},
       adsurl = {https://ui.adsabs.harvard.edu/abs/2021ApJ...916..110L}
}

@ARTICLE{liu+2010a,
       author = {{Liu}, Xin and {Shen}, Yue and {Strauss}, Michael A. and {Greene}, Jenny E.},
        title = "{Type 2 Active Galactic Nuclei with Double-Peaked [O III] Lines: Narrow-Line Region Kinematics or Merging Supermassive Black Hole Pairs?}",
      journal = {\apj},
         year = 2010,
        month = jan,
       volume = {708},
       number = {1},
        pages = {427-434},
          doi = {10.1088/0004-637X/708/1/427},
archivePrefix = {arXiv},
       eprint = {0908.2426},
 primaryClass = {astro-ph.CO},
       adsurl = {https://ui.adsabs.harvard.edu/abs/2010ApJ...708..427L}
}

@ARTICLE{lintott+2008,
       author = {{Lintott}, Chris J. and {Schawinski}, Kevin and {Slosar}, An{\v{z}}e and {Land}, Kate and {Bamford}, Steven and {Thomas}, Daniel and {Raddick}, M. Jordan and {Nichol}, Robert C. and {Szalay}, Alex and {Andreescu}, Dan and {Murray}, Phil and {Vandenberg}, Jan},
        title = "{Galaxy Zoo: morphologies derived from visual inspection of galaxies from the Sloan Digital Sky Survey}",
      journal = {\mnras},
         year = 2008,
        month = sep,
       volume = {389},
       number = {3},
        pages = {1179-1189},
          doi = {10.1111/j.1365-2966.2008.13689.x},
archivePrefix = {arXiv},
       eprint = {0804.4483},
 primaryClass = {astro-ph},
       adsurl = {https://ui.adsabs.harvard.edu/abs/2008MNRAS.389.1179L}
}

@ARTICLE{liu+2013,
       author = {{Liu}, Xin and {Civano}, Francesca and {Shen}, Yue and {Green}, Paul and {Greene}, Jenny E. and {Strauss}, Michael A.},
        title = "{Chandra X-Ray and Hubble Space Telescope Imaging of Optically Selected Kiloparsec-scale Binary Active Galactic Nuclei. I. Nature of the Nuclear Ionizing Sources}",
      journal = {\apj},
         year = 2013,
        month = jan,
       volume = {762},
       number = {2},
          eid = {110},
        pages = {110},
          doi = {10.1088/0004-637X/762/2/110},
archivePrefix = {arXiv},
       eprint = {1209.5418},
 primaryClass = {astro-ph.CO},
       adsurl = {https://ui.adsabs.harvard.edu/abs/2013ApJ...762..110L}
}

@ARTICLE{hunt_malkan+2004,
       author = {{Hunt}, L.~K. and {Malkan}, M.~A.},
        title = "{Circumnuclear Structure and Black Hole Fueling: Hubble Space Telescope NICMOS Imaging of 250 Active and Normal Galaxies}",
      journal = {\apj},
         year = 2004,
        month = dec,
       volume = {616},
       number = {2},
        pages = {707-729},
          doi = {10.1086/424958},
archivePrefix = {arXiv},
       eprint = {astro-ph/0408469},
 primaryClass = {astro-ph},
       adsurl = {https://ui.adsabs.harvard.edu/abs/2004ApJ...616..707H}
}

@ARTICLE{malkan+1998,
       author = {{Malkan}, Matthew A. and {Gorjian}, Varoujan and {Tam}, Raymond},
        title = "{A Hubble Space Telescope Imaging Survey of Nearby Active Galactic Nuclei}",
      journal = {\apjs},
         year = 1998,
        month = jul,
       volume = {117},
       number = {1},
        pages = {25-88},
          doi = {10.1086/313110},
archivePrefix = {arXiv},
       eprint = {astro-ph/9803123},
 primaryClass = {astro-ph},
       adsurl = {https://ui.adsabs.harvard.edu/abs/1998ApJS..117...25M}
}

@ARTICLE{marocco+2021,
       author = {{Marocco}, Federico and {Eisenhardt}, Peter R.~M. and {Fowler}, John W. and {Kirkpatrick}, J. Davy and {Meisner}, Aaron M. and {Schlafly}, Edward F. and {Stanford}, S.~A. and {Garcia}, Nelson and {Caselden}, Dan and {Cushing}, Michael C. and {Cutri}, Roc M. and {Faherty}, Jacqueline K. and {Gelino}, Christopher R. and {Gonzalez}, Anthony H. and {Jarrett}, Thomas H. and {Koontz}, Renata and {Mainzer}, Amanda and {Marchese}, Elijah J. and {Mobasher}, Bahram and {Schlegel}, David J. and {Stern}, Daniel and {Teplitz}, Harry I. and {Wright}, Edward L.},
        title = "{The CatWISE2020 Catalog}",
      journal = {\apjs},
         year = 2021,
        month = mar,
       volume = {253},
       number = {1},
          eid = {8},
        pages = {8},
          doi = {10.3847/1538-4365/abd805},
archivePrefix = {arXiv},
       eprint = {2012.13084},
 primaryClass = {astro-ph.IM},
       adsurl = {https://ui.adsabs.harvard.edu/abs/2021ApJS..253....8M}
}

@ARTICLE{maschmann+2020,
       author = {{Maschmann}, Daniel and {Melchior}, Anne-Laure and {Mamon}, Gary A. and {Chilingarian}, Igor V. and {Katkov}, Ivan Yu.},
        title = "{Double-peak emission line galaxies in the SDSS catalogue. A minor merger sequence}",
      journal = {\aap},
         year = 2020,
        month = sep,
       volume = {641},
          eid = {A171},
        pages = {A171},
          doi = {10.1051/0004-6361/202037868},
archivePrefix = {arXiv},
       eprint = {2007.14410},
 primaryClass = {astro-ph.GA},
       adsurl = {https://ui.adsabs.harvard.edu/abs/2020A&A...641A.171M}
}

@ARTICLE{mcgurk2015,
       author = {{McGurk}, R.~C. and {Max}, C.~E. and {Medling}, A.~M. and {Shields}, G.~A. and {Comerford}, J.~M.},
        title = "{Spatially Resolved Imaging and Spectroscopy of Candidate Dual Active Galactic Nuclei}",
      journal = {\apj},
         year = 2015,
        month = sep,
       volume = {811},
       number = {1},
          eid = {14},
        pages = {14},
          doi = {10.1088/0004-637X/811/1/14},
       adsurl = {https://ui.adsabs.harvard.edu/abs/2015ApJ...811...14M}
}

@ARTICLE{mullersanchez+2015,
       author = {{M{\"u}ller-S{\'a}nchez}, F. and {Comerford}, J.~M. and {Nevin}, R. and {Barrows}, R.~S. and {Cooper}, M.~C. and {Greene}, J.~E.},
        title = "{The Origin of Double-peaked Narrow Lines in Active Galactic Nuclei. I. Very Large Array Detections of Dual AGNs and AGN Outflows}",
      journal = {\apj},
         year = 2015,
        month = nov,
       volume = {813},
       number = {2},
          eid = {103},
        pages = {103},
          doi = {10.1088/0004-637X/813/2/103},
archivePrefix = {arXiv},
       eprint = {1509.04291},
 primaryClass = {astro-ph.GA},
       adsurl = {https://ui.adsabs.harvard.edu/abs/2015ApJ...813..103M}
}

@ARTICLE{merritandmilosavljevic+2005,
       author = {{Merritt}, David and {Milosavljevi{\'c}}, Milos},
        title = "{Massive Black Hole Binary Evolution}",
      journal = {Living Reviews in Relativity},
         year = 2005,
        month = nov,
       volume = {8},
        pages = {8},
          doi = {10.12942/lrr-2005-8},
archivePrefix = {arXiv},
       eprint = {astro-ph/0410364},
 primaryClass = {astro-ph},
       adsurl = {https://ui.adsabs.harvard.edu/abs/2005LRR.....8....8M}
}

@ARTICLE{pfeifle+2024,
       author = {{Pfeifle}, Ryan W. and {Weaver}, Kimberly A. and {Secrest}, Nathan J. and {Rothberg}, Barry and {Patton}, David R.},
        title = "{Super-Size Me: The Big Multi-AGN Catalog (The Big MAC), Data Release 1: The Source Catalog}",
      journal = {arXiv e-prints},
         year = 2024,
        month = nov,
          eid = {arXiv:2411.12799},
        pages = {arXiv:2411.12799},
          doi = {10.48550/arXiv.2411.12799},
archivePrefix = {arXiv},
       eprint = {2411.12799},
 primaryClass = {astro-ph.GA},
       adsurl = {https://ui.adsabs.harvard.edu/abs/2024arXiv241112799P}
}

@ARTICLE{pfeifle+2019b,
       author = {{Pfeifle}, Ryan W. and {Satyapal}, Shobita and {Secrest}, Nathan J. and {Gliozzi}, Mario and {Ricci}, Claudio and {Ellison}, Sara L. and {Rothberg}, Barry and {Cann}, Jenna and {Blecha}, Laura and {Williams}, James K. and {Constantin}, Anca},
        title = "{Buried Black Hole Growth in IR-selected Mergers: New Results from Chandra}",
      journal = {\apj},
         year = 2019,
        month = apr,
       volume = {875},
       number = {2},
          eid = {117},
        pages = {117},
          doi = {10.3847/1538-4357/ab07bc},
archivePrefix = {arXiv},
       eprint = {1904.10955},
 primaryClass = {astro-ph.GA},
       adsurl = {https://ui.adsabs.harvard.edu/abs/2019ApJ...875..117P}
}

@ARTICLE{pfeifle+2019a,
       author = {{Pfeifle}, Ryan W. and {Satyapal}, Shobita and {Manzano-King}, Christina and {Cann}, Jenna and {Sexton}, Remington O. and {Rothberg}, Barry and {Canalizo}, Gabriela and {Ricci}, Claudio and {Blecha}, Laura and {Ellison}, Sara L. and {Gliozzi}, Mario and {Secrest}, Nathan J. and {Constantin}, Anca and {Harvey}, Jenna B.},
        title = "{A Triple AGN in a Mid-infrared Selected Late-stage Galaxy Merger}",
      journal = {\apj},
         year = 2019,
        month = oct,
       volume = {883},
       number = {2},
          eid = {167},
        pages = {167},
          doi = {10.3847/1538-4357/ab3a9b},
archivePrefix = {arXiv},
       eprint = {1908.01732},
 primaryClass = {astro-ph.GA},
       adsurl = {https://ui.adsabs.harvard.edu/abs/2019ApJ...883..167P}
}

@ARTICLE{Pierce+2023,
       author = {{Pierce}, J.~C.~S. and {Tadhunter}, C. and {Ramos Almeida}, C. and {Bessiere}, P. and {Heaton}, J.~V. and {Ellison}, S.~L. and {Speranza}, G. and {Gordon}, Y. and {O'Dea}, C. and {Grimmett}, L. and {Makrygianni}, L.},
        title = "{Galaxy interactions are the dominant trigger for local type 2 quasars}",
      journal = {\mnras},
         year = 2023,
        month = jun,
       volume = {522},
       number = {2},
        pages = {1736-1751},
          doi = {10.1093/mnras/stad455},
archivePrefix = {arXiv},
       eprint = {2303.15506},
 primaryClass = {astro-ph.GA},
       adsurl = {https://ui.adsabs.harvard.edu/abs/2023MNRAS.522.1736P}
}

@ARTICLE{ricci+2017,
       author = {{Ricci}, C. and {Bauer}, F.~E. and {Treister}, E. and {Schawinski}, K. and {Privon}, G.~C. and {Blecha}, L. and {Arevalo}, P. and {Armus}, L. and {Harrison}, F. and {Ho}, L.~C. and {Iwasawa}, K. and {Sanders}, D.~B. and {Stern}, D.},
        title = "{Growing supermassive black holes in the late stages of galaxy mergers are heavily obscured}",
      journal = {\mnras},
         year = 2017,
        month = jun,
       volume = {468},
       number = {2},
        pages = {1273-1299},
          doi = {10.1093/mnras/stx173},
archivePrefix = {arXiv},
       eprint = {1701.04825},
 primaryClass = {astro-ph.HE},
       adsurl = {https://ui.adsabs.harvard.edu/abs/2017MNRAS.468.1273R}
}

@ARTICLE{romero+2017,
       author = {{Romero-Ca{\~n}izales}, C. and {Alberdi}, A. and {Ricci}, C. and {Ar{\'e}valo}, P. and {P{\'e}rez-Torres}, M. {\'A}. and {Conway}, J.~E. and {Beswick}, R.~J. and {Bondi}, M. and {Muxlow}, T.~W.~B. and {Argo}, M.~K. and {Bauer}, F.~E. and {Efstathiou}, A. and {Herrero-Illana}, R. and {Mattila}, S. and {Ryder}, S.~D.},
        title = "{Unveiling the AGN in IC 883: discovery of a parsec-scale radio jet}",
      journal = {\mnras},
         year = 2017,
        month = may,
       volume = {467},
       number = {2},
        pages = {2504-2516},
          doi = {10.1093/mnras/stx224},
archivePrefix = {arXiv},
       eprint = {1701.07025},
 primaryClass = {astro-ph.GA},
       adsurl = {https://ui.adsabs.harvard.edu/abs/2017MNRAS.467.2504R}
}

@ARTICLE{rosario+2010,
       author = {{Rosario}, D.~J. and {Shields}, G.~A. and {Taylor}, G.~B. and {Salviander}, S. and {Smith}, K.~L.},
        title = "{The Jet-driven Outflow in the Radio Galaxy SDSS J1517+3353: Implications for Double-peaked Narrow-line Active Galactic Nucleus}",
      journal = {\apj},
         year = 2010,
        month = jun,
       volume = {716},
       number = {1},
        pages = {131-143},
          doi = {10.1088/0004-637X/716/1/131},
archivePrefix = {arXiv},
       eprint = {1005.0021},
 primaryClass = {astro-ph.CO},
       adsurl = {https://ui.adsabs.harvard.edu/abs/2010ApJ...716..131R}
}

@ARTICLE{stern+2012,
       author = {{Stern}, Daniel and {Assef}, Roberto J. and {Benford}, Dominic J. and {Blain}, Andrew and {Cutri}, Roc and {Dey}, Arjun and {Eisenhardt}, Peter and {Griffith}, Roger L. and {Jarrett}, T.~H. and {Lake}, Sean and {Masci}, Frank and {Petty}, Sara and {Stanford}, S.~A. and {Tsai}, Chao-Wei and {Wright}, E.~L. and {Yan}, Lin and {Harrison}, Fiona and {Madsen}, Kristin},
        title = "{Mid-infrared Selection of Active Galactic Nuclei with the Wide-Field Infrared Survey Explorer. I. Characterizing WISE-selected Active Galactic Nuclei in COSMOS}",
      journal = {\apj},
         year = 2012,
        month = jul,
       volume = {753},
       number = {1},
          eid = {30},
        pages = {30},
          doi = {10.1088/0004-637X/753/1/30},
archivePrefix = {arXiv},
       eprint = {1205.0811},
 primaryClass = {astro-ph.CO},
       adsurl = {https://ui.adsabs.harvard.edu/abs/2012ApJ...753...30S}
}

@ARTICLE{snyder+2013,
       author = {{Snyder}, Gregory F. and {Hayward}, Christopher C. and {Sajina}, Anna and {Jonsson}, Patrik and {Cox}, Thomas J. and {Hernquist}, Lars and {Hopkins}, Philip F. and {Yan}, Lin},
        title = "{Modeling Mid-infrared Diagnostics of Obscured Quasars and Starbursts}",
      journal = {\apj},
         year = 2013,
        month = may,
       volume = {768},
       number = {2},
          eid = {168},
        pages = {168},
          doi = {10.1088/0004-637X/768/2/168},
archivePrefix = {arXiv},
       eprint = {1210.6347},
 primaryClass = {astro-ph.CO},
       adsurl = {https://ui.adsabs.harvard.edu/abs/2013ApJ...768..168S}
}

@ARTICLE{satyapal+2014,
       author = {{Satyapal}, Shobita and {Ellison}, Sara L. and {McAlpine}, William and {Hickox}, Ryan C. and {Patton}, David R. and {Mendel}, J. Trevor},
        title = "{Galaxy pairs in the Sloan Digital Sky Survey - IX. Merger-induced AGN activity as traced by the Wide-field Infrared Survey Explorer}",
      journal = {\mnras},
         year = 2014,
        month = jun,
       volume = {441},
       number = {2},
        pages = {1297-1304},
          doi = {10.1093/mnras/stu650},
archivePrefix = {arXiv},
       eprint = {1403.7531},
 primaryClass = {astro-ph.GA},
       adsurl = {https://ui.adsabs.harvard.edu/abs/2014MNRAS.441.1297S}
}

@ARTICLE{schwartzman+2024,
       author = {{Schwartzman}, Emma and {Clarke}, Tracy E. and {Nyland}, Kristina and {Secrest}, Nathan J. and {Pfeifle}, Ryan W. and {Schmitt}, Henrique and {Satyapal}, Shobita and {Rothberg}, Barry},
        title = "{VaDAR: Varstrometry for Dual AGN Using Radio Interferometry}",
      journal = {\apj},
         year = 2024,
        month = feb,
       volume = {961},
       number = {2},
          eid = {233},
        pages = {233},
          doi = {10.3847/1538-4357/ad0ed0},
archivePrefix = {arXiv},
       eprint = {2306.13219},
 primaryClass = {astro-ph.GA},
       adsurl = {https://ui.adsabs.harvard.edu/abs/2024ApJ...961..233S}
}

@ARTICLE{shen+2019,
       author = {{Shen}, Yue and {Hwang}, Hsiang-Chih and {Zakamska}, Nadia and {Liu}, Xin},
        title = "{Varstrometry for Off-nucleus and Dual Sub-Kpc AGN (VODKA): How Well Centered Are Low-z AGN?}",
      journal = {\apjl},
         year = 2019,
        month = nov,
       volume = {885},
       number = {1},
          eid = {L4},
        pages = {L4},
          doi = {10.3847/2041-8213/ab4b54},
archivePrefix = {arXiv},
       eprint = {1910.02969},
 primaryClass = {astro-ph.GA},
       adsurl = {https://ui.adsabs.harvard.edu/abs/2019ApJ...885L...4S}
}

@ARTICLE{shen+2013,
       author = {{Shen}, Yue and {Liu}, Xin and {Loeb}, Abraham and {Tremaine}, Scott},
        title = "{Constraining Sub-parsec Binary Supermassive Black Holes in Quasars with Multi-epoch Spectroscopy. I. The General Quasar Population}",
      journal = {\apj},
         year = 2013,
        month = sep,
       volume = {775},
       number = {1},
          eid = {49},
        pages = {49},
          doi = {10.1088/0004-637X/775/1/49},
archivePrefix = {arXiv},
       eprint = {1306.4330},
 primaryClass = {astro-ph.CO},
       adsurl = {https://ui.adsabs.harvard.edu/abs/2013ApJ...775...49S}
}

@ARTICLE{smith+2010,
       author = {{Smith}, K.~L. and {Shields}, G.~A. and {Bonning}, E.~W. and {McMullen}, C.~C. and {Rosario}, D.~J. and {Salviander}, S.},
        title = "{A Search for Binary Active Galactic Nuclei: Double-peaked [O III] AGNs in the Sloan Digital Sky Survey}",
      journal = {\apj},
         year = 2010,
        month = jun,
       volume = {716},
       number = {1},
        pages = {866-877},
          doi = {10.1088/0004-637X/716/1/866},
archivePrefix = {arXiv},
       eprint = {0908.1998},
 primaryClass = {astro-ph.CO},
       adsurl = {https://ui.adsabs.harvard.edu/abs/2010ApJ...716..866S}
}

@ARTICLE{satyapal+2017,
       author = {{Satyapal}, Shobita and {Secrest}, Nathan J. and {Ricci}, Claudio and {Ellison}, Sara L. and {Rothberg}, Barry and {Blecha}, Laura and {Constantin}, Anca and {Gliozzi}, Mario and {McNulty}, Paul and {Ferguson}, Jason},
        title = "{Buried AGNs in Advanced Mergers: Mid-infrared Color Selection as a Dual AGN Candidate Finder}",
      journal = {\apj},
         year = 2017,
        month = oct,
       volume = {848},
       number = {2},
          eid = {126},
        pages = {126},
          doi = {10.3847/1538-4357/aa88ca},
archivePrefix = {arXiv},
       eprint = {1707.03921},
 primaryClass = {astro-ph.GA},
       adsurl = {https://ui.adsabs.harvard.edu/abs/2017ApJ...848..126S}
}

@ARTICLE{scoville+2000,
       author = {{Scoville}, N.~Z. and {Evans}, A.~S. and {Thompson}, R. and {Rieke}, M. and {Hines}, D.~C. and {Low}, F.~J. and {Dinshaw}, N. and {Surace}, J.~A. and {Armus}, L.},
        title = "{NICMOS Imaging of Infrared-Luminous Galaxies}",
      journal = {\aj},
         year = 2000,
        month = mar,
       volume = {119},
       number = {3},
        pages = {991-1061},
          doi = {10.1086/301248},
archivePrefix = {arXiv},
       eprint = {astro-ph/9912246},
 primaryClass = {astro-ph},
       adsurl = {https://ui.adsabs.harvard.edu/abs/2000AJ....119..991S}
}

@ARTICLE{Service+2016,
       author = {{Service}, M. and {Lu}, J.~R. and {Campbell}, R. and {Sitarski}, B.~N. and {Ghez}, A.~M. and {Anderson}, J.},
        title = "{A New Distortion Solution for NIRC2 on the Keck II Telescope}",
      journal = {\pasp},
         year = 2016,
        month = sep,
       volume = {128},
       number = {967},
        pages = {095004},
          doi = {10.1088/1538-3873/128/967/095004},
       adsurl = {https://ui.adsabs.harvard.edu/abs/2016PASP..128i5004S}
}

@ARTICLE{treister+2018,
       author = {{Treister}, Ezequiel and {Privon}, George C. and {Sartori}, Lia F. and {Nagar}, Neil and {Bauer}, Franz E. and {Schawinski}, Kevin and {Messias}, Hugo and {Ricci}, Claudio and {U}, Vivian and {Casey}, Caitlin and {Comerford}, Julia M. and {Muller-Sanchez}, Francisco and {Evans}, Aaron S. and {Finlez}, Carolina and {Koss}, Michael and {Sanders}, David B. and {Urry}, C. Megan},
        title = "{Optical, Near-IR, and Sub-mm IFU Observations of the Nearby Dual Active Galactic Nuclei MRK 463}",
      journal = {\apj},
         year = 2018,
        month = feb,
       volume = {854},
       number = {2},
          eid = {83},
        pages = {83},
          doi = {10.3847/1538-4357/aaa963},
archivePrefix = {arXiv},
       eprint = {1801.06190},
 primaryClass = {astro-ph.GA},
       adsurl = {https://ui.adsabs.harvard.edu/abs/2018ApJ...854...83T}
}

@ARTICLE{toomre+1972,
       author = {{Toomre}, Alar and {Toomre}, Juri},
        title = "{Galactic Bridges and Tails}",
      journal = {\apj},
         year = 1972,
        month = dec,
       volume = {178},
        pages = {623-666},
          doi = {10.1086/151823},
       adsurl = {https://ui.adsabs.harvard.edu/abs/1972ApJ...178..623T}
}

@ARTICLE{troncoso+2025,
       author = {{Troncoso}, Marco and {Treister}, Ezequiel and {Rojas}, Alejandra and {Boquien}, M{\'e}d{\'e}ric and {Bauer}, Franz and {Koss}, Michael J. and {Assef}, Roberto J. and {Parra Tello}, Miguel and {del Moral-Castro}, Ignacio and {Ricci}, Claudio and {Dai}, Sophia and {Oh}, Kyuseok and {Ricci}, Federica and {Peca}, Alessandro and {Urry}, C. Megan and {Gupta}, Kriti Kamal and {Venturi}, Giacomo and {Signorini}, Matilde and {Mushotzky}, Richard and {Sanders}, David},
        title = "{BASS. LIV. Physical Properties of AGN-hosting Galaxy Mergers from Multiwavelength SED Fitting}",
      journal = {\apj},
         year = 2025,
        month = dec,
       volume = {994},
       number = {2},
          eid = {181},
        pages = {181},
          doi = {10.3847/1538-4357/ae0a4b},
archivePrefix = {arXiv},
       eprint = {2509.17720},
 primaryClass = {astro-ph.GA},
       adsurl = {https://ui.adsabs.harvard.edu/abs/2025ApJ...994..181T}
}

@ARTICLE{tremonti+2004,
       author = {{Tremonti}, Christy A. and {Heckman}, Timothy M. and {Kauffmann}, Guinevere and {Brinchmann}, Jarle and {Charlot}, St{\'e}phane and {White}, Simon D.~M. and {Seibert}, Mark and {Peng}, Eric W. and {Schlegel}, David J. and {Uomoto}, Alan and {Fukugita}, Masataka and {Brinkmann}, Jon},
        title = "{The Origin of the Mass-Metallicity Relation: Insights from 53,000 Star-forming Galaxies in the Sloan Digital Sky Survey}",
      journal = {\apj},
         year = 2004,
        month = oct,
       volume = {613},
       number = {2},
        pages = {898-913},
          doi = {10.1086/423264},
archivePrefix = {arXiv},
       eprint = {astro-ph/0405537},
 primaryClass = {astro-ph},
       adsurl = {https://ui.adsabs.harvard.edu/abs/2004ApJ...613..898T}
}

@INPROCEEDINGS{2005ASPC..347...29T,
       author = {{Taylor}, M.~B.},
        title = "{TOPCAT \& STIL: Starlink Table/VOTable Processing Software}",
    booktitle = {Astronomical Data Analysis Software and Systems XIV},
         year = 2005,
       editor = {{Shopbell}, P. and {Britton}, M. and {Ebert}, R.},
       series = {Astronomical Society of the Pacific Conference Series},
       volume = {347},
        month = dec,
        pages = {29},
       adsurl = {https://ui.adsabs.harvard.edu/abs/2005ASPC..347...29T}
}

@ARTICLE{uppal+2024,
       author = {{Uppal}, Anavi and {Ward}, Charlotte and {Gezari}, Suvi and {Natarajan}, Priyamvada and {Chen}, Nianyi and {LaChance}, Patrick and {Di Matteo}, Tiziana},
        title = "{Astrometric Jitter as a Detection Diagnostic for Recoiling and Slingshot Supermassive Black Hole Candidates}",
      journal = {\apj},
         year = 2024,
        month = nov,
       volume = {975},
       number = {2},
          eid = {286},
        pages = {286},
          doi = {10.3847/1538-4357/ad7ff0},
archivePrefix = {arXiv},
       eprint = {2405.11026},
 primaryClass = {astro-ph.GA},
       adsurl = {https://ui.adsabs.harvard.edu/abs/2024ApJ...975..286U}
}

@ARTICLE{U+2013,
       author = {{U}, Vivian and {Medling}, Anne and {Sanders}, David and {Max}, Claire and {Armus}, Lee and {Iwasawa}, Kazushi and {Evans}, Aaron and {Kewley}, Lisa and {Fazio}, Giovanni},
        title = "{The Inner Kiloparsec of Mrk 273 with Keck Adaptive Optics}",
      journal = {\apj},
         year = 2013,
        month = oct,
       volume = {775},
       number = {2},
          eid = {115},
        pages = {115},
          doi = {10.1088/0004-637X/775/2/115},
archivePrefix = {arXiv},
       eprint = {1307.8440},
 primaryClass = {astro-ph.CO},
       adsurl = {https://ui.adsabs.harvard.edu/abs/2013ApJ...775..115U}
}

@ARTICLE{Van_Wassenhove+2012,
       author = {{Van Wassenhove}, Sandor and {Volonteri}, Marta and {Mayer}, Lucio and {Dotti}, Massimo and {Bellovary}, Jillian and {Callegari}, Simone},
        title = "{Observability of Dual Active Galactic Nuclei in Merging Galaxies}",
      journal = {\apjl},
         year = 2012,
        month = mar,
       volume = {748},
       number = {1},
          eid = {L7},
        pages = {L7},
          doi = {10.1088/2041-8205/748/1/L7},
archivePrefix = {arXiv},
       eprint = {1111.0223},
 primaryClass = {astro-ph.CO},
       adsurl = {https://ui.adsabs.harvard.edu/abs/2012ApJ...748L...7V}
}

@ARTICLE{voggel+2022,
       author = {{Voggel}, Karina T. and {Seth}, Anil C. and {Baumgardt}, Holger and {Husemann}, Bernd and {Neumayer}, Nadine and {Hilker}, Michael and {Pechetti}, Renuka and {Mieske}, Steffen and {Dumont}, Antoine and {Georgiev}, Iskren},
        title = "{First direct dynamical detection of a dual supermassive black hole system at sub-kiloparsec separation}",
      journal = {\aap},
         year = 2022,
        month = feb,
       volume = {658},
          eid = {A152},
        pages = {A152},
          doi = {10.1051/0004-6361/202140827},
archivePrefix = {arXiv},
       eprint = {2111.14854},
 primaryClass = {astro-ph.GA},
       adsurl = {https://ui.adsabs.harvard.edu/abs/2022A&A...658A.152V}
}

@ARTICLE{2020NatMe..17..261V,
       author = {{Virtanen}, Pauli and {Gommers}, Ralf and {Oliphant}, Travis E. and {Haberland}, Matt and {Reddy}, Tyler and {Cournapeau}, David and {Burovski}, Evgeni and {Peterson}, Pearu and {Weckesser}, Warren and {Bright}, Jonathan and {van der Walt}, St{\'e}fan J. and {Brett}, Matthew and {Wilson}, Joshua and {Millman}, K. Jarrod and {Mayorov}, Nikolay and {Nelson}, Andrew R.~J. and {Jones}, Eric and {Kern}, Robert and {Larson}, Eric and {Carey}, C.~J. and {Polat}, {\.I}lhan and {Feng}, Yu and {Moore}, Eric W. and {VanderPlas}, Jake and {Laxalde}, Denis and {Perktold}, Josef and {Cimrman}, Robert and {Henriksen}, Ian and {Quintero}, E.~A. and {Harris}, Charles R. and {Archibald}, Anne M. and {Ribeiro}, Ant{\^o}nio H. and {Pedregosa}, Fabian and {van Mulbregt}, Paul and {SciPy 1. 0 Contributors}},
        title = "{SciPy 1.0: fundamental algorithms for scientific computing in Python}",
      journal = {Nature Methods},
         year = 2020,
        month = feb,
       volume = {17},
        pages = {261-272},
          doi = {10.1038/s41592-019-0686-2},
archivePrefix = {arXiv},
       eprint = {1907.10121},
 primaryClass = {cs.MS},
       adsurl = {https://ui.adsabs.harvard.edu/abs/2020NatMe..17..261V}
}

@ARTICLE{wang+2023,
       author = {{Wang}, Hao-Chen and {Wang}, Jun-Xian and {Gu}, Min-Feng and {Liao}, Mai},
        title = "{Varstrometry selected radio-loud candidates of dual and off-nucleus quasars at sub-kpc scales}",
      journal = {\mnras},
         year = 2023,
        month = sep,
       volume = {524},
       number = {1},
        pages = {L38-L44},
          doi = {10.1093/mnrasl/slad069},
archivePrefix = {arXiv},
       eprint = {2306.03357},
 primaryClass = {astro-ph.GA},
       adsurl = {https://ui.adsabs.harvard.edu/abs/2023MNRAS.524L..38W}
}

@ARTICLE{wang+2009,
       author = {{Wang}, Jian-Min and {Chen}, Yan-Mei and {Hu}, Chen and {Mao}, Wei-Ming and {Zhang}, Shu and {Bian}, Wei-Hao},
        title = "{Active Galactic Nuclei with Double-Peaked Narrow Lines: Are they Dual Active Galactic Nuclei?}",
      journal = {\apjl},
         year = 2009,
        month = nov,
       volume = {705},
       number = {1},
        pages = {L76-L80},
          doi = {10.1088/0004-637X/705/1/L76},
archivePrefix = {arXiv},
       eprint = {0910.0580},
 primaryClass = {astro-ph.CO},
       adsurl = {https://ui.adsabs.harvard.edu/abs/2009ApJ...705L..76W}
}

@ARTICLE{WISE+2010,
       author = {{Wright}, Edward L. and {Eisenhardt}, Peter R.~M. and {Mainzer}, Amy K. and {Ressler}, Michael E. and {Cutri}, Roc M. and {Jarrett}, Thomas and {Kirkpatrick}, J. Davy and {Padgett}, Deborah and {McMillan}, Robert S. and {Skrutskie}, Michael and {Stanford}, S.~A. and {Cohen}, Martin and {Walker}, Russell G. and {Mather}, John C. and {Leisawitz}, David and {Gautier}, Thomas N., III and {McLean}, Ian and {Benford}, Dominic and {Lonsdale}, Carol J. and {Blain}, Andrew and {Mendez}, Bryan and {Irace}, William R. and {Duval}, Valerie and {Liu}, Fengchuan and {Royer}, Don and {Heinrichsen}, Ingolf and {Howard}, Joan and {Shannon}, Mark and {Kendall}, Martha and {Walsh}, Amy L. and {Larsen}, Mark and {Cardon}, Joel G. and {Schick}, Scott and {Schwalm}, Mark and {Abid}, Mohamed and {Fabinsky}, Beth and {Naes}, Larry and {Tsai}, Chao-Wei},
        title = "{The Wide-field Infrared Survey Explorer (WISE): Mission Description and Initial On-orbit Performance}",
      journal = {\aj},
         year = 2010,
        month = dec,
       volume = {140},
       number = {6},
        pages = {1868-1881},
          doi = {10.1088/0004-6256/140/6/1868},
archivePrefix = {arXiv},
       eprint = {1008.0031},
 primaryClass = {astro-ph.IM},
       adsurl = {https://ui.adsabs.harvard.edu/abs/2010AJ....140.1868W}
}

@ARTICLE{zhao+19,
       author = {{Zhao}, Dongyao and {Ho}, Luis C. and {Zhao}, Yulin and {Shangguan}, Jinyi and {Kim}, Minjin},
        title = "{The Role of Major Mergers and Nuclear Star Formation in Nearby Obscured Quasars}",
      journal = {\apj},
         year = 2019,
        month = may,
       volume = {877},
       number = {1},
          eid = {52},
        pages = {52},
          doi = {10.3847/1538-4357/ab1921},
archivePrefix = {arXiv},
       eprint = {1904.06734},
 primaryClass = {astro-ph.GA},
       adsurl = {https://ui.adsabs.harvard.edu/abs/2019ApJ...877...52Z}
}

@ARTICLE{Wizinowich+2000,
       author = {{Wizinowich}, P. and {Acton}, D.~S. and {Shelton}, C. and {Stomski}, P. and {Gathright}, J. and {Ho}, K. and {Lupton}, W. and {Tsubota}, K. and {Lai}, O. and {Max}, C. and {Brase}, J. and {An}, J. and {Avicola}, K. and {Olivier}, S. and {Gavel}, D. and {Macintosh}, B. and {Ghez}, A. and {Larkin}, J.},
        title = "{First Light Adaptive Optics Images from the Keck II Telescope: A New Era of High Angular Resolution Imagery}",
      journal = {\pasp},
         year = 2000,
        month = mar,
       volume = {112},
       number = {769},
        pages = {315-319},
          doi = {10.1086/316543},
       adsurl = {https://ui.adsabs.harvard.edu/abs/2000PASP..112..315W}
}

@ARTICLE{Wizinowich+2006,
       author = {{Wizinowich}, Peter L. and {Le Mignant}, David and {Bouchez}, Antonin H. and {Campbell}, Randy D. and {Chin}, Jason C.~Y. and {Contos}, Adam R. and {van Dam}, Marcos A. and {Hartman}, Scott K. and {Johansson}, Erik M. and {Lafon}, Robert E. and {Lewis}, Hilton and {Stomski}, Paul J. and {Summers}, Douglas M. and {Brown}, Curtis G. and {Danforth}, Pamela M. and {Max}, Claire E. and {Pennington}, Deanna M.},
        title = "{The W. M. Keck Observatory Laser Guide Star Adaptive Optics System: Overview}",
      journal = {\pasp},
         year = 2006,
        month = feb,
       volume = {118},
       number = {840},
        pages = {297-309},
          doi = {10.1086/499290},
       adsurl = {https://ui.adsabs.harvard.edu/abs/2006PASP..118..297W}
}

@ARTICLE{vanDam+2006,
       author = {{van Dam}, Marcos A. and {Bouchez}, Antonin H. and {Le Mignant}, David and {Johansson}, Erik M. and {Wizinowich}, Peter L. and {Campbell}, Randy D. and {Chin}, Jason C.~Y. and {Hartman}, Scott K. and {Lafon}, Robert E. and {Stomski}, Jr., Paul J. and {Summers}, Douglas M.},
        title = "{The W. M. Keck Observatory Laser Guide Star Adaptive Optics System: Performance Characterization}",
      journal = {\pasp},
         year = 2006,
        month = feb,
       volume = {118},
       number = {840},
        pages = {310-318},
          doi = {10.1086/499498},
       adsurl = {https://ui.adsabs.harvard.edu/abs/2006PASP..118..310V}
}

% \begin{thebibliography}{}
%\newpage
% \providecommand\natexlab[1]{#1}
% \providecommand\JournalTitle[1]{#1}

% \bibitem[Appenzeller \& Oestreicher(1988)]{appenzeller1988} Appenzeller, I. \& Oestreicher, R.\ 1988, \aj, 95, 45. doi:10.1086/114611

% \bibitem[Cann et al.(2018)]{cann2018} Cann, J.~M., Satyapal, S., Abel, N.~P., et al.\ 2018, \apj, 861, 142. doi:10.3847/1538-4357/aac64a

%\input{./bibliography}

% \end{thebibliography}

\end{document}